\documentclass[12pt,a4paper]{JHEP3}
\usepackage{amsmath,epsfig}
\usepackage{amssymb,amsfonts}
\usepackage{latexsym}
\usepackage[latin1]{inputenc}
\usepackage{slashed}
\usepackage{empheq}
\numberwithin{equation}{section}
\usepackage{dsfont}
\usepackage{cancel}
\usepackage{overpic}
\usepackage{subcaption}
\usepackage{subeqnarray}
\usepackage{xcolor}
\let\normalcolor\relax
\usepackage{longtable}
\usepackage{color}
\usepackage{bbm}
\usepackage{multirow}
\usepackage{epstopdf}
\def\nn{\nonumber}

\relax
\def\hri#1#2{\href{http://arxiv.org/abs/#1}{[ArXiv:#1]#2}}
\def\hre#1#2{\href{http://arxiv.org/abs/#1/#2}{[ArXiv:#1/#2]}}

\def\hree#1#2{\href{https://doi.org/#1}{#2}}
\def\hrj#1#2{\href{www.doi.org/#1}{#2}}
\renewcommand{\theequation}{\arabic{section}.\arabic{equation}}

\def\be{\begin{equation}}
\def\ee{\end{equation}}

\newcommand{\bear}{\begin{eqnarray}}
\newcommand{\bea}{\begin{eqnarray}}
\newcommand{\eear}{\end{eqnarray}}
\newcommand{\eea}{\end{eqnarray}}
\def\hre#1#2{\href{http://arxiv.org/abs/#1/#2}{[ArXiv:#1/#2]}}

\newbox\pippobox

\renewcommand{\b}[1]{\textbf{#1}}

\def\II{\relax{\rm I\kern-.18em I}}

\def\m{\mu}
\def\n{\nu}

\def\de{\partial}

\def\f{\varphi}
\def\z{\zeta}
\def\a{\alpha}
\def\b{\beta}

\def\l{\lambda}
\def\g{\gamma}

\def\II{{\cal I}}

\def\ar{\Rightarrow}

\def\bsq{\begin{subequations}}
\def\esq{\end{subequations}}

\def\D{\Delta}

\title{Holographic QFTs on AdS$_d$, wormholes and holographic interfaces}
\author{ A. Ghodsi$^a$, J.K. Ghosh$^{b,c}$, E. Kiritsis$^{d,e}$, F. Nitti$^{d}$, V. Nourry$^{d}$ \\

$^a$
 Department of Physics, Faculty of Science,	Ferdowsi University of Mashhad,  Mashhad, Iran.
~\\

$^b$  \href{http://www.iub.edu.bd/}{Independent University Bangladesh (IUB)}, Bashundhara RA, Dhaka 1229, Bangladesh ~\\

$c$ \href{https://ccds.ai/compas/} {Center for Computational and Data Sciences}, Independent University, Bangladesh, Bashundhara RA, Dhaka 1229, Bangladesh ~\\

$^d$ \href{http://www.apc.univ-paris7.fr}{Universit\'e Paris Cit\' e, CNRS, Astroparticule et Cosmologie}, F-75013 Paris, France.
~\\

$^e$ \href{http://hep.physics.uoc.gr}{Crete Center for Theoretical Physics}, Institute for Theoretical and Computational Physics,
Department of Physics\\
University of Crete, Heraklion, Greece
}

\preprint{CCTP-2022-5\\
ITCP-2022/4\\
}

\abstract{We consider three related topics:  (a) Holographic quantum field theories on AdS spaces.
(b) Holographic interfaces of flat space QFTs.
(c) Wormholes connecting generically different QFTs.
We investigate in a concrete example  how the related classical solutions explore the space of QFTs and we construct the general solutions
that interpolate between the same or different CFTs with arbitrary couplings.
The solution space contains many exotic flow solutions that realize unusual asymptotics, as boundaries of different regions in the space of solutions.
We find  phenomena like  ``walking" flows and the generation of extra boundaries via {\em flow fragmentation}.
}


\begin{document}

\section{Introduction and summary}

Gauge/gravity duality relates a quantum field theory (QFT) in $d$-dimensions to a higher-dimensional bulk theory containing dynamical gravity \cite{Malda,GKP,Witten98,review}. In suitable situations (namely among others, when the QFT is strongly coupled and has a large-$N$ expansion) the latter can be approximated by semiclassical general relativity coupled to other fields, which are dual to QFT  operators.

Classical solutions of the bulk gravitational theory with an asymptotically AdS$_{d+1}$ boundary correspond to (the ground state of)  QFT renormalization-group flows (RG flows), and evolution along the radial bulk coordinate has the dual interpretation as evolution with energy scale in the QFT. Bulk fields which have a non-trivial dependence on the radial coordinate correspond to running couplings. Solutions of this type are called {\em holographic RG flows}, \cite{Freedman:1999gp,deBoer:1999tgo,Bianchi:2001de}

 Fully AdS-invariant solutions correspond to special points in solution space and are dual to conformal fixed points on the field theory side. More generally, the space of conformal fixed points (CFTs) and the RG flows connecting them, is mapped on the gravity side to the space of regular solutions of the bulk theory.

Holographic RG flows have been extensively explored, in both top-down models and from the bottom-up point of view. The simplest bulk theory which gives this type of solution is Einstein gravity minimally coupled to a single real scalar field, which we shall call {\em dilaton} by an abuse of language, and whose potential admits one or more extrema with a negative cosmological constant.  Although this setup is minimal, it provides a very rich structure: depending on the scalar potential, it offers a very rich space of solution which reproduce all features one expects in usual field theories (e.g. RG flows between a UV and IR fixed points, the generation of an IR scale with a mass gap, the  possibility of confinement and phase transitions, etc) as well as certain types of {\em exotic} RG flows which cannot occur in perturbative field theories \cite{exotic,Gur}.
Including more fields does not change the story qualitatively but allows for multiscale RG flows, \cite{exotic2}.

The correspondence between QFT RG flows and gravity solutions is not limited to QFTs on flat space-time but extends to the situation where the QFT is defined on a curved manifold. This is because, asymptotically,  one can write the near-boundary AdS metric as a foliation whose $d$-dimensional radial sections admit an arbitrary metric. This near-boundary expansion is what defines the metric of the spacetime on which the QFT lives. This opens the way to studying holographic RG flows of QFTs on curved space-times.

A systematic analysis of curved space-time holographic RG flows in Einstein-dilaton theories has been initiated in \cite{C} in the case when the boundary field theory is defined on an Einstein space with positive or negative curvature. For positive curvature, the picture in terms of RG flows is not very different from that of flat space field theories, except for the fact that the curvature dominates in the  IR and gaps the theory before the deep IR regime is reached. In the case of negative curvature, however, the field theory interpretation of the resulting solutions is {\em very} different from that of an RG flow: the reason is that, when the bulk is foliated by constant negative curvature $d$-dimensional radial slices, the solution has generically {\em two} asymptotically AdS$_{d+1}$ boundaries. Rather than an RG flow, this corresponds to two UV CFTs which are interacting in a non-local way through the bulk.

Such solutions in string theory, with asymptotic boundary metrics being AdS,  have been found and studied for some time, \cite{Bak}-\cite{C2}.
They have two apparently distinct conformal boundaries at the two end-points of the holographic coordinate. However, as the slices involve a non-compact manifold which also has a conformal boundary, the two asymptotic boundaries are connected, resulting in a single conformal boundary. This is represented in the left part of figure \ref{fig:CFT}: the total conformal boundary is composed of the two gray caps plus the red surface joining them.

If the bulk is $d+1$ dimensional, and the slices are AdS$_d$, the total boundary
is conformal to two pieces of $S^d$ separated by a defect on the equator\footnote{In the context of holography, this description is most appropriate when the bulk AdS$_{d+1}$ is written in global coordinates (or in spherical slicing), as we will discuss in more detail in section \ref{sec:results}. }  $S^{d-1}$. As the two endpoints of the flow may have different sources, the two theories can have different couplings and they are separated by an interface, prompting the name Janus solutions.
A related class of solutions contains a single boundary and is delimited in the bulk by a brane that ends on "the boundary of the boundary". They are also AdS-sliced and the first example was discussed in \cite{KR}. They have been advocated as holographic duals of boundary CFTs, \cite{BCFT1,BCFT2}. Holographic RG flows in this context  have  been  considered in  \cite{Gutperle:2012hy}.

There is however another incarnation of such solutions. In Euclidean cases, where the slice manifold is a constant negative curvature manifold of finite volume and no boundary, then such a solution is an example of  a Euclidean wormhole,
an object that still holds mysteries for the holographic correspondence, \cite{MM,BKP}.
The holographic interpretation of such solutions is still debated and therefore the study of a large class of such solutions may be interesting in assessing their generic properties.

There are three interesting physics problems, the solution of which involves partly such AdS-sliced solutions. We discuss them in the next three subsections.

\subsection{QFT$_d$ on AdS$_d$}

It was argued already in \cite{CW} that, when placing a  QFT  on AdS$_d$,  the  IR dynamics of the QFT is drastically affected. The reason is that even
massless particles have propagators that fall off exponentially with distance in AdS. This is a consequence of the fact that Laplacians on AdS have a gap.
It was argued in \cite{CW} that this could be used to regulate the strong IR divergences of QCD perturbation theory.
Moreover, it would be also useful for the same reason in critical theories, as one would expect AdS to suppress the strong IR fluctuations.  In general, the expectation
is that AdS is expected to quench strong IR physics.

A similar approach to regulating the IR has also been applied in string theory, \cite{KK}. In that case, the spatial geometry was that of $S^3$ but the running dilaton produced AdS-like effects on the spectrum including a universal mass gap for massless particles.

Compared to Minkowski space, AdS has a different set of isometries which are in the same number as in flat space. However, in AdS, the boundary conditions are much more important for the physics than in Minkowski.
In particular, for gauge theories, there are roughly two types of boundary conditions\footnote{One can have more complicated boundary conditions on subgroups of the gauge group.}:
 Dirichlet (or electric) boundary conditions and Neumann (or magnetic) boundary conditions (bcs).
With electric bcs, gluons are allowed in the spectrum (as in flat space), they are gapped, and there is a global $SU(N)$ symmetry, \cite{A2}. However, it has only boundary currents.

With magnetic bcs, electric charges are not allowed in the bulk of AdS, there are ${\cal O}(1)$ degrees of freedom and there is (bc-induced) confinement.
It was argued in \cite{A2} that for asymptotically-free gauge theories with electric bcs, a confinement/deconfinement (quantum) phase transition was expected.
If we denote by $\Lambda$ the scale of the gauge theory, and  by $L_{ADS}$ the radius of AdS, we expect the following two phases
\begin{enumerate}

\item When $\Lambda L_{AdS}\gtrsim 1$ we expect a confining phase, with strong interaction in the IR before we reach the AdS mass gap.

\item When $\Lambda L_{AdS}\lesssim 1$ we expect a deconfined weakly-coupled phase, where above the AdS mass gap the theory is weakly-coupled.

\end{enumerate}
The two phases are expected to be separated by a phase transition whose details are not known.

With magnetic boundary conditions, one expects confinement at all scales,
and a free energy of ${\cal O}(1)$. This is a kind of trivial confinement as no electric
charges are allowed in the bulk.

So far, the only clear criterion for confinement is the order of magnitude
of the free energy: either  ${\cal O}(1)$ or ${\cal O}(N^2)$ when $N\to \infty$.
 Wilson loops do not provide an easy criterion for confinement, as for
large Wilson loops, the area and the perimeter scale the same way, in
global coordinates. But in Poincar\'e coordinates, there are two classes of loops with different
behavior for length and area.  However, QFT on AdS in different coordinates gives rise to a different
quantum theory.

Another important expected difference, discussed in \cite{CW} is the nature of the instanton gas.
As already argued in \cite{W1}-\cite{W2}, instantons in flat space YM form a liquid. Only above the deconfinement phase transition, instantons form a gas, \cite{rev}.
However, it is expected that for YM on AdS instantons form a gas in most cases, \cite{CW}.

The fate of CFTs on AdS is also an interesting problem. The prime example in four dimensions, $N=4$ SYM  was analyzed in some detail,
\cite{GW1,A1,A2}.  Boundary conditions on $R^4_+$ that { preserve supersymmetry} have been classified, and there are many, \cite{GW1,GW2}.
Upon a conformal transformation, the theory can be put on { AdS$_4$ in Poincar\'e coordinates}.
 Dirichlet bcs generically involve {non-trivial vevs for three of the six scalars}.
 At {weak coupling} the theory is generically { non-confining}.
 But at { strong coupling} some boundary conditions { induce confinement}.

 For example, using { S-duality}, the $g\gg 1$ theory with a Higgs condensate, is mapped to a $g\ll 1$ theory with a magnetic condensate that should be { confining}.
 In particular, S-duality interchanges (among others) Dirichlet and Neumann bcs.
 With Neumann bcs { no order parameter exists that distinguishes a confining from a non-confining phase}.
Therefore, no sharp transition is expected in accordance with the presence of a large amount of supersymmetry.
On the other hand, the { finite-temperature behavior} is not understood.

Boundary RG flows connecting CFTs in AdS subject to different boundary conditions have been studied in  \cite{Giombi:2020rmc}.

\subsection{Proximity and QFT interfaces}
There is a general question that involves the notion of ``proximity" in quantum field theory. It can be formulated in several different ways which may not be equivalent but which are expected to be qualitatively similar

One form of the question asks: ``When two CFTs can be defined in the same Hilbert space?'', or ``when the states of one CFT can be written in terms of the states of the other?''
There is a (partial) answer to this question: this is possible if the two CFTs are connected by a Renormalization Group (RG)  flow. This seems intuitively correct, but it is not known if this is mathematically correct. Moreover, it is not known if this remains correct in the whole family tree of theories that belong to an RG cascade.

Another form of the question, formulated and analysed in \cite{holoween} in the holographic context is: ``when can a (semiclassical) state of CFT$_1$ described by the geometry $M$  be approximated by a semiclassical state of another CFT$_2$?'' The ``approximation''  is interpreted as the two geometries being the same on an arbitrarily large causal patch of $M$.
Reference \cite{holoween} argued using ideas from \cite{BCFT1,BCFT2} that such a notion of proximity is possible if the two CFTs can be connected with an interface.

It was also argued in \cite{holoween} that this is consistent with the idea that the precise degrees of freedom and Hamiltonian of a holographic CFT, is only important in fixing the asymptotic behavior of a dual space-time. On the other hand,  the interior space-time of a region, space-like-separated from a boundary time slice, is determined by more universal properties (such as the entanglement structure) of the quantum state at this time slice.
This picture requires that low-energy gravitational theories associated with CFTs that it is possible to non-trivially couple at an interface, are part of the same non-perturbative theory of quantum gravity.
\subsection{Wormholes and quantum gravity}
\begin{figure}[ht!]
\begin{center}
\includegraphics[width = 9cm]{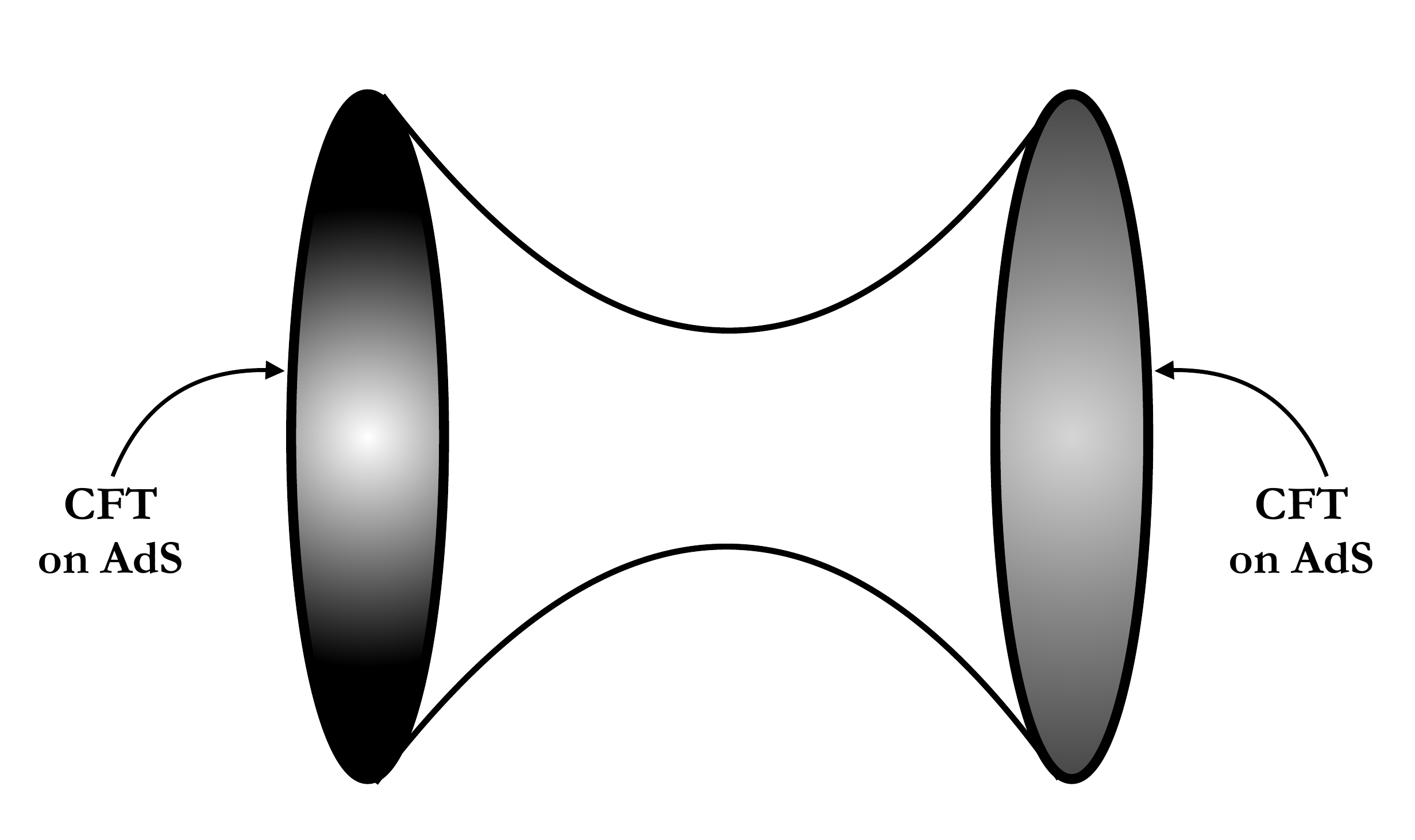}
\end{center}
\end{figure}

As mentioned, any semiclassical gravitational solution that represents an interface,  can be immediately transformed into a solution describing a Euclidean wormhole with two boundaries, as in the picture below.
The step is to replace the AdS$_d$ slice geometry with any $d$-dimensional Euclidean, finite-volume, constant-negative-curvature manifold.
This is automatically a solution to the second-order gravitational equations without further ado.

In  $d=2$, finite volume negative constant curvature manifolds are the $g > 2$ Riemann surfaces,
or  Schottky manifolds in $d > 2$. Examples have been discussed in \cite{3c} for $d=3$ and \cite{4c} for $d=4$.

The two end-points are asymptotically AdS boundaries that can in principle belong to different CFTs. Moreover, the connection is mediated by running scalars
and is therefore similar to RG flow geometries.
The puzzles of negative curvature AdS boundary metrics were discussed in \cite{WY,Ander}, and the associated wormhole puzzles were first discussed in \cite{MM}.
It was pointed out that Euclidean wormholes can easily contain unstable modes. Moreover, their holographic interpretation is not clear.
A more recent analysis of correlators and Wilson loops indicated some generic properties, like UV-soft cross-correlators, as well as generic confining behavior (that in many cases rimes with the results of \cite{SW}), \cite{BKP}. These results prompted some dual models for wormholes, \cite{VR,BKP2}.

The issues of wormholes and their interpretation/use in quantum gravity have obtained a new twist with the advent of the SYK model and its black-hole-like
 interpretation in the large-N limit, \cite{Kita,MSYK}. The suggestion is that they are a part of the gravitational path-integral when some form of averaging is involved.
  This is an issue of current debate.

\subsection{Results} \label{sec:results}

In this paper, we start a detailed study, in the context of holography, of the three questions mentioned above. Here we will study the structure of the two-boundary solutions in a concrete bottom-up model.
Although the feature of two boundaries is generic, only the simplest solutions of this type (in which the bulk geometry has a $Z_2$ symmetry along the radial direction) were studied in \cite{C}. Here we shall perform a systematic search for solutions in Einstein-Dilaton gravity that have constant negative curvature slices and map the space and properties of such solutions.
In particular, we would like to explore how the two boundary theories explore the space of extrema of the scalar potential.

We should emphasize here that there  are three interpretation for the solutions we  find:
\begin{enumerate}
\item As interfaces between two QFTs with a shared common boundary (Janus type solutions) when the slice geometry has infinite volume and a boundary, \cite{Bak}-\cite{C2}.
In this case, the full solution is split in two parts: One corresponds to a flow from one boundary to a minimum of the scale factor, and is an RG flow of one of the two QFTs.
The other half of the solution, is a flow from the other boundary to the minimum of the scale factor. This is again an  RG flow for  the second QFT.
\item As wormholes, if the slice has finite volume and no boundary. In such a case the holographic interpretation is debated in the literature.
One interpretation   (see    \cite{BKP,VR,BKP2}) is as two Euclidean QFTs coupled in the UV via soft interactions, but this needs further investigation and checks.
\item As single boundary theories after orbifolding symmetric solutions. In that case the flow is monotonic and consistent with a standard QFT on AdS  interpretation.
  \end{enumerate}
We shall call our solutions {\em flows}, although in some cases such flows are not necessarily RG flows in the strict sense.

 In the rest of this introduction, we briefly summarise our results and discuss the questions which are left open.

\subsubsection{Pure gravity}
The simplest theory one can consider is pure $(d+1)$-dimensional gravity with a negative cosmological constant. In this case, the solutions corresponding to (conformal) field theories on AdS$_d$ are well known: the bulk space-time is just AdS$_{d+1}$ foliated by AdS$_{d}$:
\be \label{intro1}
ds^2 = du^2 + \cosh^2[(u-u_0)/\ell] ds_d^2\,,
\ee
where $\ell$ is the AdS$_{d+1}$  length, $u_0$ is an arbitrary constant and $ds_d^2$ is the metric on AdS$_{d}$. The metric above is obtained via a diffeomorphism from the  global AdS$_{d+1}$ metric,
\be \label{intro2}
ds^2(global) = \ell^2 \left(\cosh^2\rho \,d\tau^2 +  d\rho^2 + \sinh^2 \rho\,  d\Omega_{d-1}^2 \right),
\ee
but such a diffeomorphism acts non-trivially on the boundary: the latter is conformal to $R\times S^{d-1}$ for global AdS, and to  $AdS_d$ if one uses the coordinates (\ref{intro1}).

\begin{figure}[ht!]
\begin{center}
\includegraphics[width = 4.6cm]{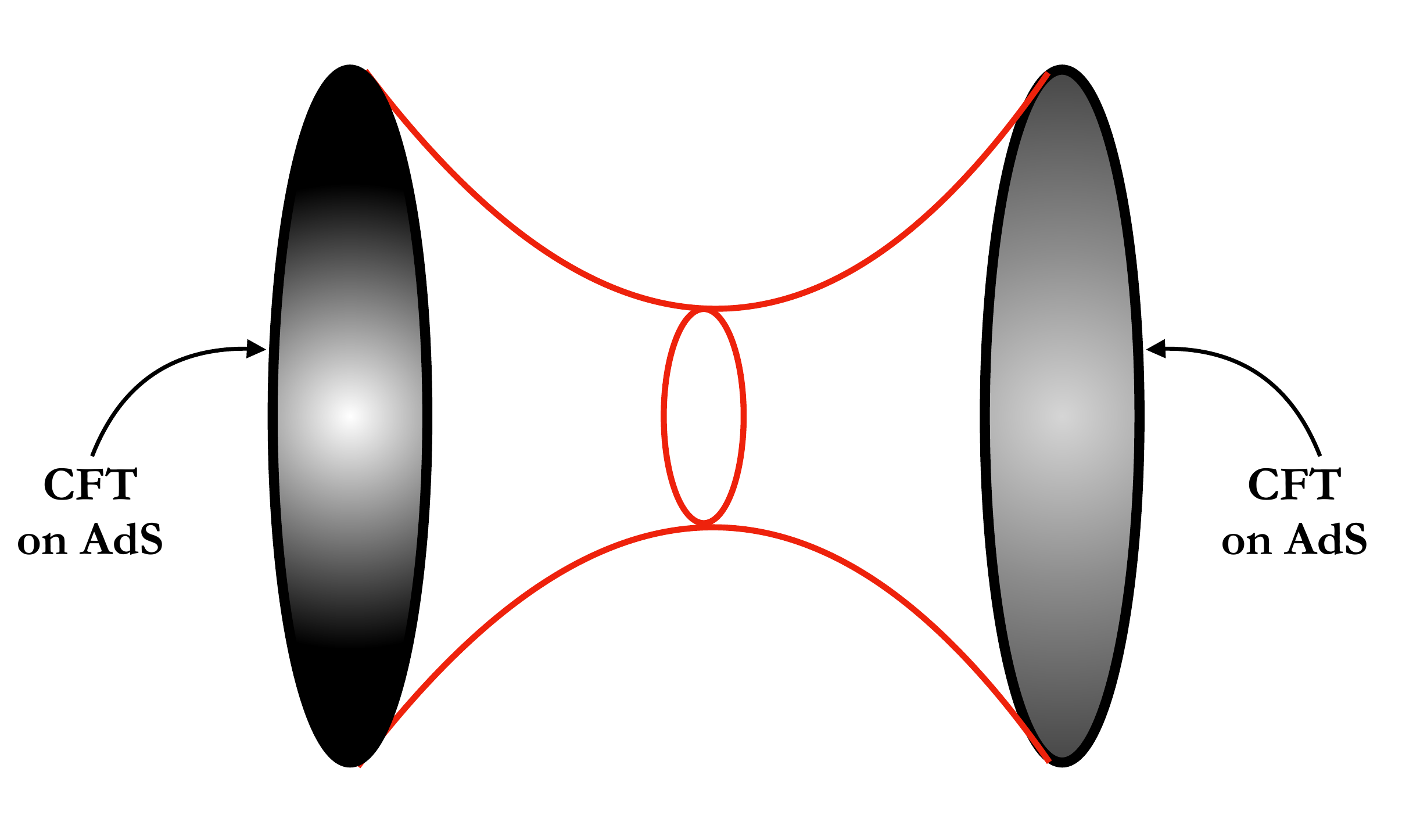}\hspace{0.5cm}
\includegraphics[width = 4.6cm]{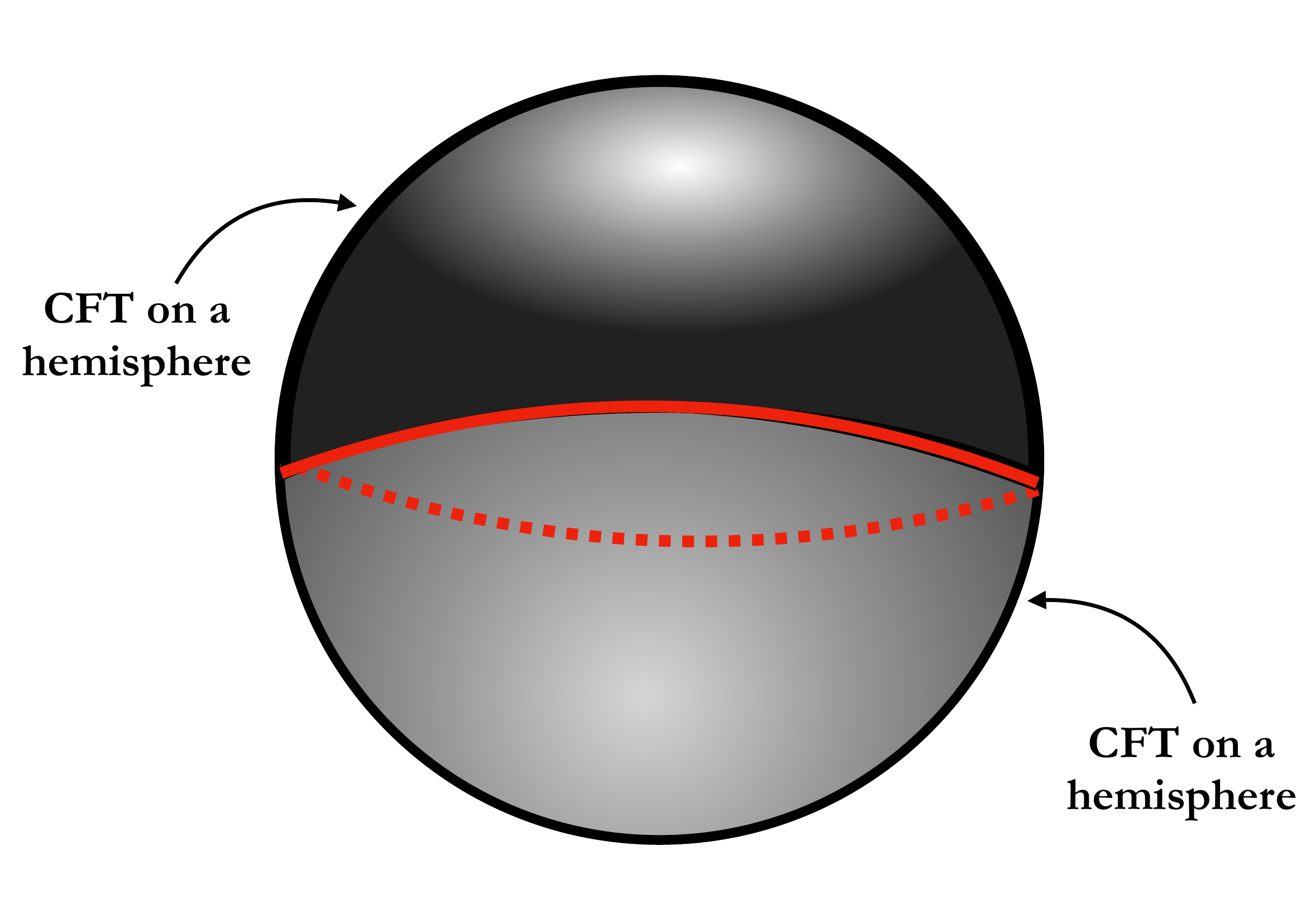}\hspace{0.5cm}
\includegraphics[width = 4.6cm]{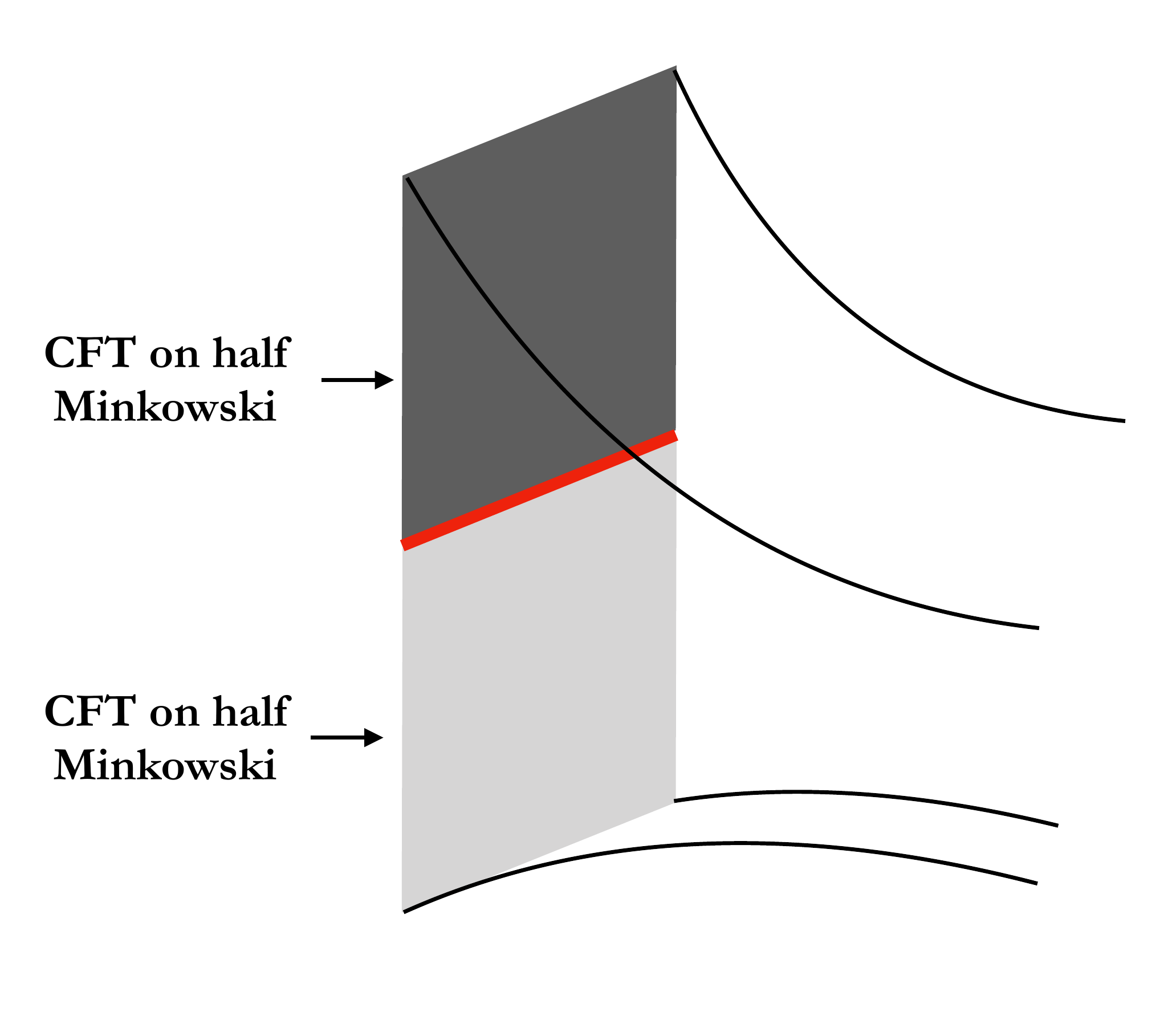}
\caption{\footnotesize{Fixed-time $d$-dimensional sections of  AdS$_{d+1}$ written in different coordinates. Left:  AdS-slicing wormhole. The boundary is composed of two  AdS spaces and a  ``side'' boundary (red). Center:   Global AdS interface. The boundary is made of two hemispheres joined at the equator,  which is identified with the whole ``side'' boundary of the left figure. Right:  Poincar\'e AdS interface. The boundary is made of two halves of flat space joined on a straight line. These space-times  are all locally diffeomorphic, but the coordinate transformation from one to the other changes the geometry of the conformal boundary.}}\label{fig:CFT}
\end{center}
\end{figure}

Therefore, in the standard holographic dictionary, the dual field theory to the metric (\ref{intro1}) lives in a different spacetime than the one dual to global AdS with coordinates (\ref{intro2}).  The metric   (\ref{intro1}) has two  asymptotic boundaries at $u\to \pm \infty$, joined by a third asymptotic boundary reached as $r\to \infty$ with $u$ arbitrary.   In the global AdS picture, the two boundaries at $u\to \pm \infty$  are mapped to two halves of the boundary $S^{d-1}$. They are  glued together at the equator $S^{d-2}$: this  is the image of the  asymptotic boundary of the {\em slice} $AdS_d$ at $r\to + \infty$ (see figure \ref{fig:CFT}) which, under this map, becomes a co-dimension two surface in global coordinates. This is why this solution is usually called an {\em interface} CFT: in global coordinates, it is dual to two copies of the  CFT, each defined on a half-space, then glued along a codimension-one hypersurface \cite{CF}. We should insist that the situation described in the coordinates (\ref{intro1}) is different: the conformal boundary consists of two  AdS$_d$, at $u\to \pm\infty$ plus the third boundary at $r\to +\infty$. However, there is no known holographic interpretation of this third boundary in terms of QFT degrees of freedom living there: more precisely,   there is no known  way of turning on sources and giving a consistent prescription to compute holographic correlators of operators inserted at   $r\to \infty$ and finite $u$. For this reason, we will think of the CFTs as living on the opposite $AdS_d$  at $u\to \pm infty$ and being coupled in a non-local (in the sense that there are non-local cross-correlators) via the bulk dynamics\footnote{The situation is similar to the  case asymptotic infinity in holography with flat slices: in Poincar\'e coordinates, strictly speaking, the surface consisting of spacelike infinity in Minkowski space  ($|\vec{x}| \to \infty$)  and arbitrary  holographic coordinate $u$) is part of the AdS boundary, but one usually does not consider CFT degrees of freedom living there: rather, one imposes boundary conditions so that the fields vanish at spacelike infinity.}.

  From the bulk point of view, the geometry (\ref{intro1}) describes a wormhole with two asymptotic boundaries. A similar story applies if one uses Poincar\'e (rather than global) coordinates: in this case, the two CFTs live in two halves of flat $d$-dimensional Minkowski spaces, joined by a flat $(d-1)$-dimensional hyperplane (rightmost picture in figure \ref{fig:CFT}).

Notice that the two conformal boundaries in (\ref{intro1})  do not have the same curvature:
 the ratio of the resulting Ricci scalars $R_{L}$ and$ R_{R}$ on the two boundaries\footnote{With the boundary metrics defined in terms of the Fefferman-Graham expansion, \cite{PG}.} at $u\to \pm \infty$  is:
\be \label{intro4}
\frac{R_{L}}{R_{R}}  =   e^{2 u_0/\ell}.
\ee
Equation (\ref{intro4}) gives the holographic interpretation of the constant $u_0$ appearing in (\ref{intro1}): it fixes the single dimensionless parameter one can construct out of the two curvature scales on the boundaries.

\subsubsection{Einstein-dilaton gravity}
Adding a bulk scalar field gives, as we shall see,  a much richer space of solutions, whose analysis is the main purpose of this paper. For definiteness, we consider a bulk theory described by the action:
\be\label{intro5}
S_{Bulk}=M_P^{d-1}\int du\, d^d x \sqrt{-g}\Big(R -\frac12 g^{ab}\partial_a\f\partial_b\f-V(\f)\Big)\,,
\ee
where $u$ is the radial bulk coordinate. We take the potential $V(\f)$ to be negative and to admit several extrema, each corresponding to an RG fixed point in the dual picture. We then study solutions which generalize (\ref{intro1}):
\be \label{intro6}
ds^2 = du^2 + e^{2A(u)} ds_d^2\,, \qquad \f = \f(u)\,,
\ee
where again $ds_d^2$ is a fiducial metric on AdS$_d$ (whose curvature scale is arbitrary, but fixed). The solution is characterized by the scalar field profile $\f(u)$ and by the scale factor $A(u)$, which are related via the bulk Einstein equations.

For regular solutions, the bulk geometry looks qualitatively similar to the one on the left in figure \ref{fig:CFT}. The non-trivial running of the scalar field, however, leads to a much richer space of solutions (and the corresponding dual QFT pictures) than in the case of pure gravity, as we briefly describe below.

\begin{figure}[h!]
\begin{center}
\includegraphics[width = 10cm]{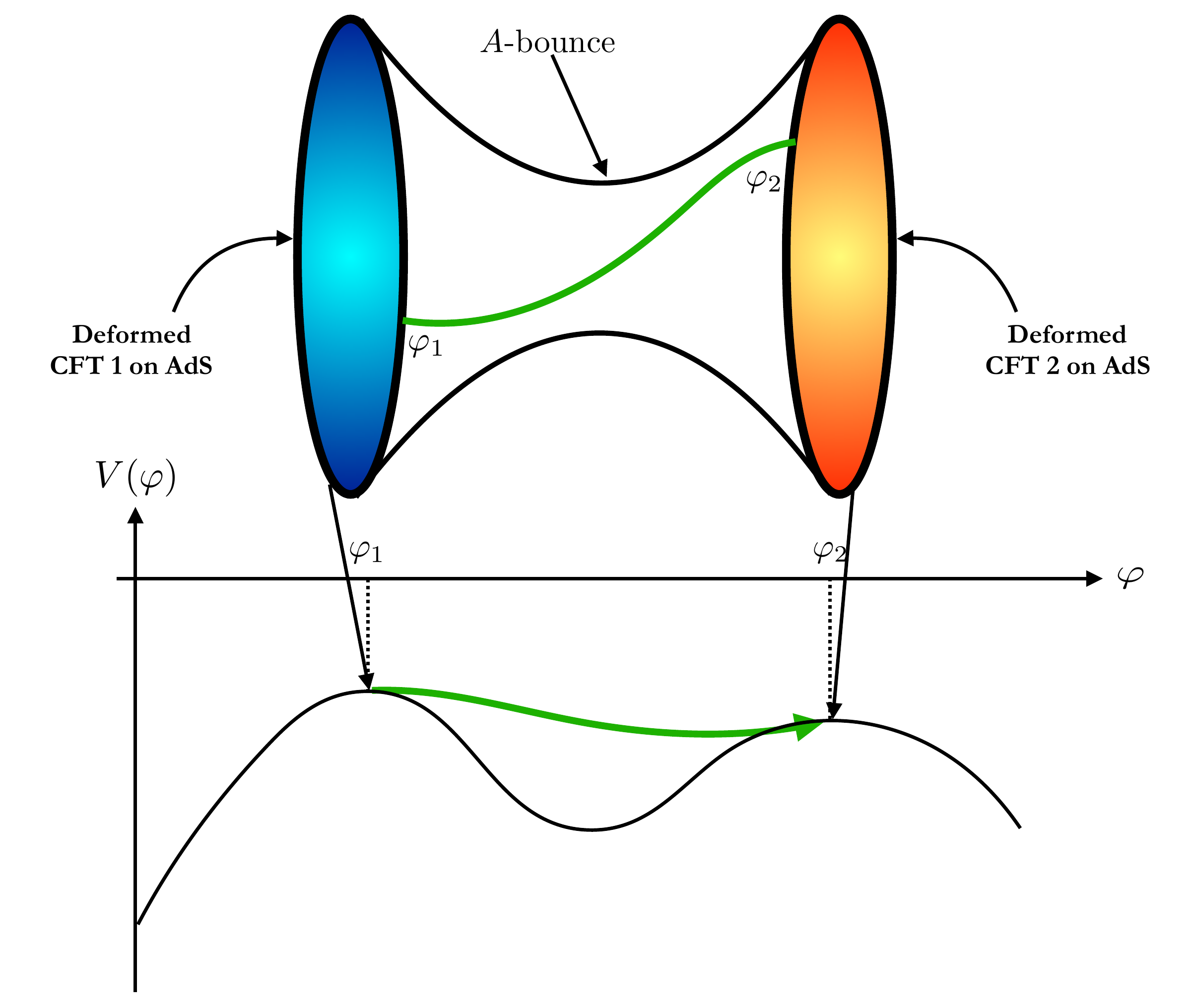}
\caption{\footnotesize{Wormhole connecting two UV CFTs on AdS$_d$ living at different maxima of $V(\f)$. The solution has one $A$-bounce and no $\f$-bounce.}}\label{fig:RGflow}
\end{center}
\end{figure}

\begin{figure}[ht!]
\begin{center}
\includegraphics[width = 10cm]{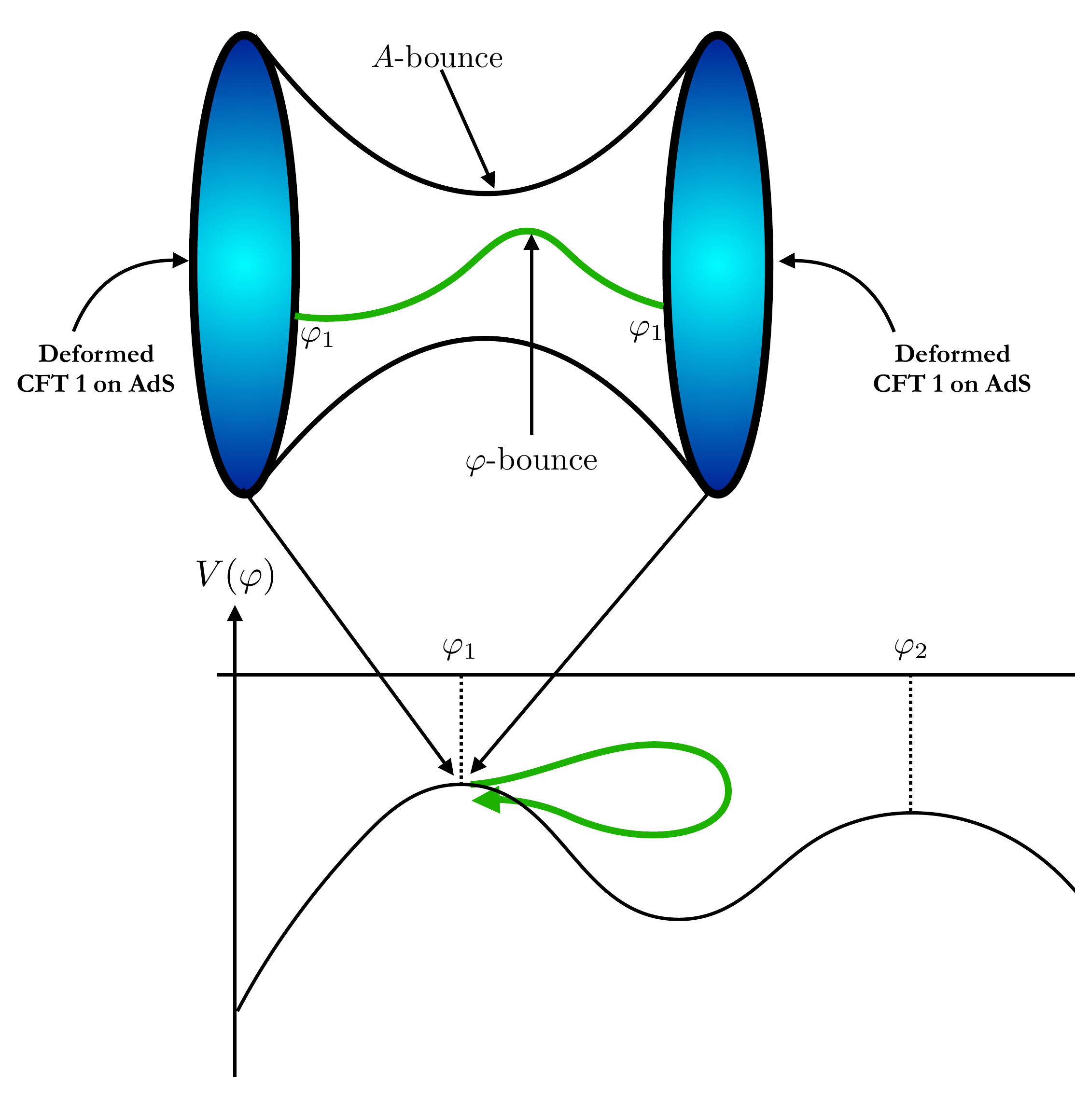}
\caption{\footnotesize{Wormhole connecting two versions of the same CFT living at one maximum of $V(\f)$. The solution has one  $A$-bounce and one $\f$-bounce. }  }\label{fig:RGflow-bounce}
\end{center}
\end{figure}

The generic solution is a two-sided  wormhole connecting two separate asymptotic boundaries.  As we approach each boundary, the scale factor diverges and the scalar field approaches one of the  extrema of the potential.   As each extremum corresponds to a distinct dual CFT in the same flow landscape, we have several possibilities:
\begin{enumerate}
\item The wormhole connects two different extrema of $V(\f)$. In this case, the dual description is that of two distinct CFTs, each  living  on an AdS$_d$  space-time, and each deformed in the UV by a relevant operator. The two CFTs are coupled by a non-local interaction mediated by the bulk.
\item The wormhole connects an  extremum of $V(\f)$ with itself. In this case, we have two copies of the same CFT, each living on two distinct AdS$_d$ space-time (possibly with different curvature scales).
\end{enumerate}
In figure \ref{fig:RGflow} we display schematically a solution of the first type, which connects two different extrema  of the bulk potential; a solution of the second kind, connecting one extremum with itself, is shown in figure \ref{fig:RGflow-bounce}. In these figures the reader can identify,
as two salient   features,  points in the bulk where either the scale factor turns around, or the scalar field turns around, as a function of the radial bulk coordinate. We call the former feature an {\em  $A$-bounce}, and the latter a {\em $\f$-bounce}.  The solution in figure    \ref{fig:RGflow} has one $A$-bounce and no $\f$-bounce, whereas the one in figure  \ref{fig:RGflow-bounce} has both an $A$-bounce and a $\f$-bounce.
In general, since the generic  solution connects two asymptotic UV boundaries, it  must have at least one $A$-bounce (and in any case, an odd number of them). In addition,  the solution can have any number (including zero) of $\f$-bounces.

{ We note that solutions of the first type connecting two different CFTs have  recently been constructed in \cite{Arav:2020asu} in the particular example of deformed $N=4$ SYM, in the special case when the CFTs on the two sides are the UV and IR endpoints of a single flow. This results in a so-called {\em RG interface} \cite{Arav:2020asu}.}

\subsubsection{Space of solutions}
The bulk solutions  are determined (up to diffeomorphisms)  by the values of the UV sources at each boundary. For the solutions under consideration, there are generically\footnote{There are also special solutions which one finds on measure-zero set of parameter space, which we  shall discuss later}  four independent dimensionful boundary parameters. These are the two scalar curvatures $R_i^{UV}$ and $R_f^{UV}$  of the ``initial'' and ``final'' boundary\footnote{We assign the attributes ``initial'' and ``final'' by making an arbitrary choice in the direction of the bulk radial coordinates},  defined by the leading term in the Fefferman-Graham expansion of the metric close to each boundary. There are also  the two scalar sources $\f_-^{(L)}$ and  $\f_-^{(R)}$ defined by the near-boundary expansions of the scalar field,
\bsq
\begin{align} \label{intro7}
& \f(u) \simeq \f_-^{(i)} \ell_i^{\Delta_-^i}\exp \Big[\frac{\Delta_-^i u}{\ell_i}\Big]\,, \, \quad\qquad  u\to -\infty\,, \\
& \f(u) \simeq \f_-^{(f)}\ell_f^{\Delta_-^f} \exp \Big[-\frac{\Delta_-^f u}{\ell_{f}}\Big]\,,  \qquad  u\to +\infty,
\end{align}
\esq
where $\Delta_-^{i,f}$ are the dimension of the couplings of the operator dual to $\f$  in the UV CFT on each side (which are positive, as the UV is at an extremum of $V$) and $\ell_{i,f}$ are the AdS length of each boundary.

Out of these four quantities, it is convenient to construct three dimensionless combinations, which we take to be:
\be \label{intro8}
{\cal R}_i = \frac{R^{UV}_i }{ \left(\f_-^{(i)}\right)^{2/\Delta_-^{i}}}\,, \qquad {\cal R}_f = \frac{R^{UV}_f }{ \left(\f_-^{(f)}\right)^{2/\Delta_-^{f}}}\,, \qquad \xi =   \frac{\left(\f_-^{(i)}\right)^{1/\Delta_-^{i}} }{ \left(\f_-^{(f)}\right)^{1/\Delta_-^{f}}}\,.
\ee
Constructing the full four-dimensional space of (regular) solutions with arbitrary values of the boundary parameters is not straightforward.
In fact, the solutions of the form (\ref{intro6}) come in a three-parameter family: fixing  ${\cal R}_i$, ${\cal R}_f$  and $\xi$  is enough to fix all integration constants of the bulk equations.  This can be understood from the fact that the bulk equations form a third-order system. The two dimensionless boundary quantities ${\cal R}_i$ and ${\cal R}_f$ can be mapped to the bulk integration constants.  A convenient parametrization of these constants  is  to take them as the value of $\f$ and its radial derivative at the position of the $A$-bounce, plus the position of the $A$-bounce in the radial direction\footnote{The position of the $A$-bounce can be shifted by a bulk diffeomorphism which, however, acts non-trivially on the boundary parameter, therefore leading to a physically inequivalent solution.}.

Therefore, if one considers the ansatz (\ref{intro6}),   there seems to be no place for solutions with a fourth boundary parameter.

In fact, we shall show that the space of solutions can be extended  by considering piece-wise solutions, obtained by gluing two  solutions of the type (\ref{intro6}) across an interface at a fixed $u$, in such a way that  the induced metric, its extrinsic curvature, the scalar field and its radial derivative  are all continuous: this way, Israel's junction conditions are satisfied without the need to introduce localized sources on the interface. The corresponding geometries are qualitatively similar to the three-parameter family of solutions (\ref{intro6}), but now one can change independently all four boundary sources. Since the piecewise solutions are obtained by gluing the solutions  of the type (\ref{intro6}), studying these ``global''  solutions  is enough to have an exhaustive classification from the qualitative standpoint.

As we mentioned above, the solutions (\ref{intro6}) may have any number of bounces. We show that for a fixed number of $A$-bounces, the  two-dimensional space of solutions is divided into continuous regions,  containing solutions connecting the same two maxima of $V$ and  with a fixed number of $\f$-bounces. When  crossing the boundaries between these regions,the following phenomena can take place:
\begin{enumerate}
\item The number of $\f$-bounces changes.
\item One endpoint of the solution changes from one maximum of $V$ to another.
\end{enumerate}
Solutions which sit exactly on the boundaries of this parameter space are particularly interesting. Across boundaries of the first of the two types above, one of the sources $\f_-$ changes sign. Therefore, when we are exactly on the boundary the source is zero, and the solution corresponds to a {\em vev} deformation of the UV CFT (aka a Coulomb branch solution). These holographic flows are known to have one parameter less than relevant coupling deformations \cite{exotic,Gur}.

Solutions which sit on a boundary of the second kind are even more peculiar: when exactly at the boundary, the $A$-bounce  approaches asymptotically  an intermediate minimum of the potential. The solution  splits into two full space-times:

1) A negative curvature flow joining a UV maximum and a UV minimum of $V$, for which the flow is driven by the vev of an {\em irrelevant} operator.

2) A  {\em flat} flow describing a standard holographic RG flow from the  maximum of $V$ in the UV to the minimum of $V$ in the IR (i.e. on this ``half'' of the solution the scale factor asymptotes to zero as we approach the minimum).
We call this phenomenon {\em flow fragmentation}.

\subsubsection{Single-sided  solutions}
All the solutions  described above, have  two asymptotic boundaries, and their holographic interpretation is in terms of two non-locally coupled CFTs on different AdS$_d$ space-times and deformed by relevant couplings (or, in special cases, by vevs of operators), \cite{BKP,VR,BKP2}. This begs the question of whether one can construct, in this simple setup, the holographic dual of a {\em single} QFT (conformal or running)  living on AdS$_d$. Here,  we explore a tentative answer to this question suggested already in \cite{A1}.

In  the space of solutions, there is a special one-parameter subspace: solutions which  connect the one fixed-point with itself (as in figure (\ref{fig:RGflow-bounce}) {\em and} are such that the $A$-bounce and the $\f$-bounce occur at the same radial position. These solutions are $Z_2$-symmetric about the bounce and are those which were analyzed in \cite{C}.

To obtain a single-sided solution, one can terminate such a symmetric solution at the bounce. Since both the scale factor and the dilaton have an extremum at that point, this can be done without adding any sources at the end-point. One can think of this as cutting off the space by an empty and tensionless end-of-the-world brane positioned at the bounce. The resulting geometry has a single boundary. The theory is further specified by imposing boundary conditions for the fluctuations  at the brane, which may be either Neumann or Dirichlet if one wants no energy loss.
 An alternative way to think of these solutions is as a Z$_2$ orbifold that changes the direction of the flow and identifies the two boundaries.

Such branes at the end of the world have been considered before in the context of {\em Boundary CFTs} (BCFTs), \cite{BCFT1,BCFT2}: in the case of pure gravity, when changing coordinates  from AdS$_d$ radial slicing   to flat radial slicing (i.e. Poincar\'e AdS$_{d+1}$ metric), the bulk endpoint is mapped to a codimension-one surface which intersects the AdS$_{d+1}$ boundary on a codimension-two hypersurface: the resulting boundary CFT lives on a $d$-dimensional flat space-time whose boundary is this ($d-1$)-dimensional flat hyperplane. As in the case of the interface CFT interpretation, one must be careful with this interpretation: the change of coordinates used to go from the ``half-wormhole'' solution to the BCFT description is such that it changes the boundary geometry, and therefore from a holographic standpoint the two situations should not be thought of as equivalent: for example, in the case of the single-boundary geometry with negative curvature radial slices, the conformal boundary is geodesically complete (it is AdS$_d$), unlike in the BCFT case.

To test the features of these single-boundary geometries, we consider the simple case of a probe-free scalar on a pure-gravity background and analyze the boundary two-point function of the corresponding operator in the dual CFT on AdS$_d$. These correlators were discussed extensively in \cite{hinter}. From their results,  it follows that neither Neumann nor Dirichlet boundary conditions for the probe scalar at the IR endpoint leads to a {\em conformal} correlator on AdS$_d$, that one would obtain from the conformal-invariant  two point function in Minkowski space by performing a Weyl rescaling.  Instead, one obtains an additional contribution which breaks conformal invariance and can be understood as originating from the fact that AdS$_d$ is conformal to half  of Minkowski space.

One  can nevertheless ask the question of whether a definition of the theory exists such that the correlator preserves the full conformal symmetry. One possibility to obtain the conformal two-point function  is to change the boundary conditions at the IR, by adding a quadratic action for the scalar on the end-of-the-world brane. This  results in Robin boundary conditions,
\be \label{intro9}
\de_u\f(u_0,\nu) = \mu(\nu) \f(u_0,\nu)\,,
\ee
where $u_0$ is the position of the brane, and $\nu$ is a quantum number which is the AdS version of the Minkowski $\sqrt{k^2}$ (it is defined  by expanding the probe scalar in eigenfunctions of the AdS$_d$ Laplacian). We  determine the unique function  $\mu(\nu)$ which leads to the conformal form  of the boundary two-point function on AdS$_d$. { This two-point function is the same as one would obtain in the full two-boundary geometry if one set all source terms to zero on one of the two sides. }
However, the corresponding quadratic action  one should add on the  brane at $u_0$ to obtain (\ref{intro9}) is non-local in the brane coordinates . Thus, it seems impossible to obtain the full conformally invariant correlator for a single CFT on AdS$_d$ from a bulk theory with a local classical action.

\begin{figure}[ht!]
\begin{center}
\includegraphics[width = 6cm]{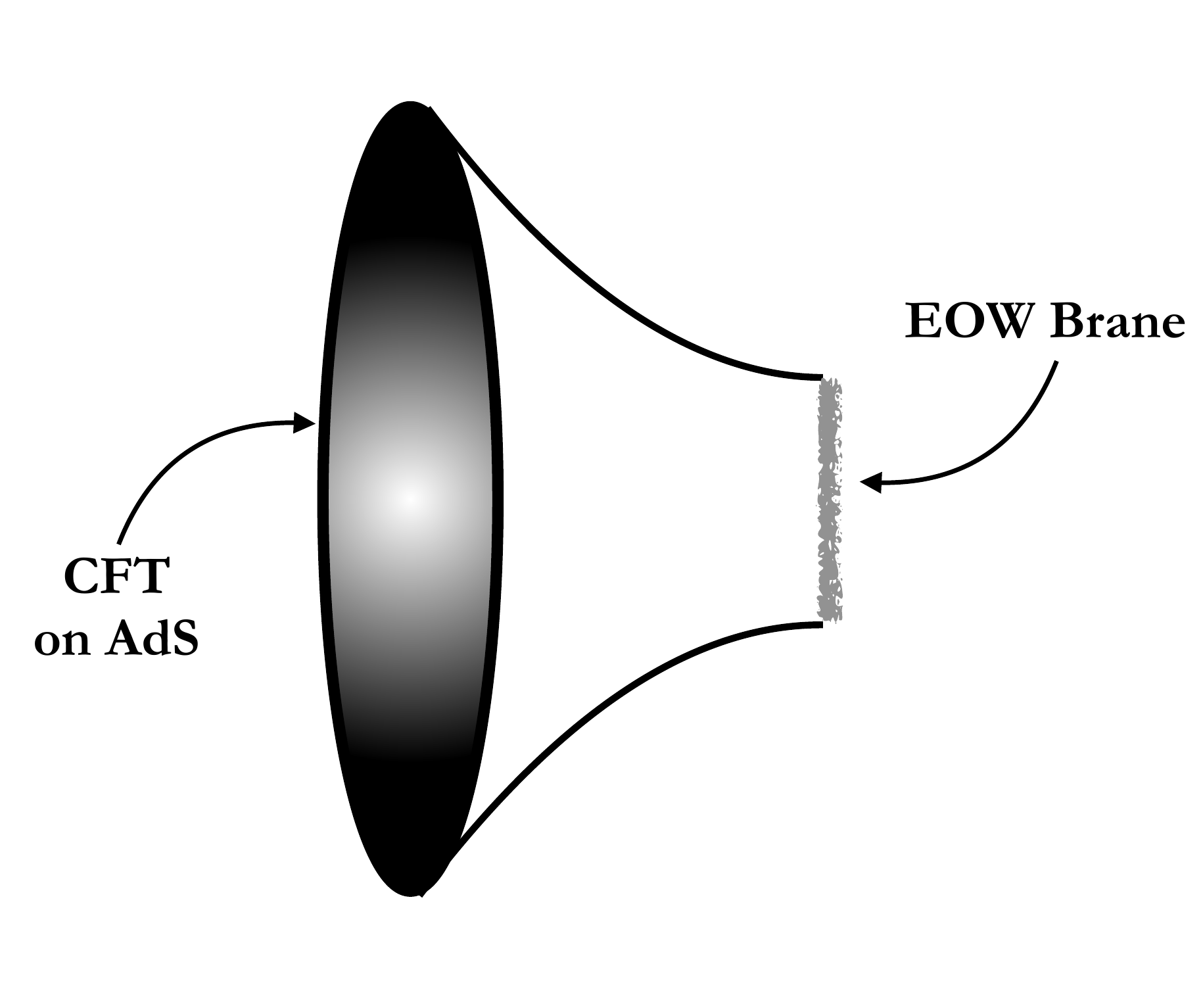}\hspace{0.5cm}
\includegraphics[width = 6.5cm]{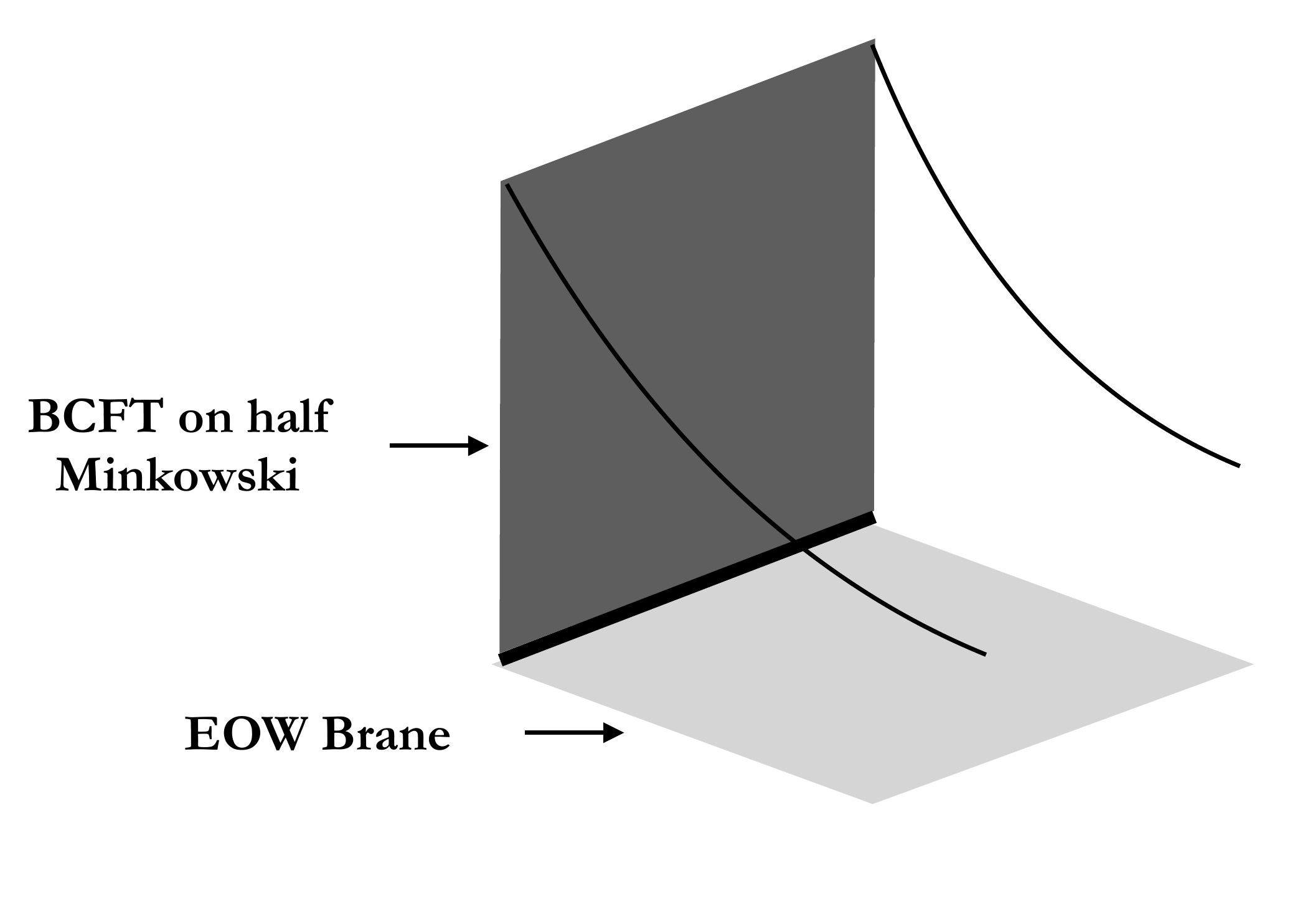}
\caption{\footnotesize{Single-AdS-boundary solution vs. flat BCFT.} }\label{fig:single-side}
\end{center}
\end{figure}

\subsection{Discussion and open problems}

The bulk geometric picture of the solutions we discuss in this work is rather clear. However, several questions remain open concerning their holographic interpretation. Furthermore, there are several extensions of the analysis concerning fluctuations and correlation functions of both local and non-local operators (e.g Wilson lines) which would be interesting to explore in the future. Below we give a  list of  open questions which we think are important to investigate in future work.
\begin{itemize}
\item One interesting question about  the AdS wormhole solutions is the nature of the ``side'' boundary or interface: this is the $d$-dimensional  surface parametrized by the holographic coordinate $u$ and the $d-1$ boundary coordinates of the fixed-$u$ AdS slices (the surface in red  in  left figure \ref{fig:CFT}). In global coordinates, this surface is mapped to a $d-1$-dimensional surface (the product of the time axis and the equator of the boundary sphere $S_{d-1}$) and its role can be interpreted  in the context of interface CFTs. However, it is not at all straightforward to give it a holographic interpretation if we stay in AdS slicing. The root of the problem is that the usual Fefferman-Graham expansion does not work as we approach this surface. As a consequence, holographic renormalization cannot be performed via the standard counterterms. One can nevertheless ask the question of whether it makes sense to insert operators on the side boundary, i.e. to consider non-zero sources (which may be non-trivial functions of $u$)  in the asymptotics of the bulk fields as we approach the slice boundary. Although in  global slicing such sources  would be singular on the equator of the boundary sphere, there seems to be nothing wrong in principle in considering them in AdS slicing. This would correspond to operators inserted on the interface.

\item Another question concerns the nature of single-boundary solutions. Although consistent, the construction we propose here is rather  {\it ad hoc,} since the fact that geometries  terminate  in the interior is not a consequence of the field equations.  One may therefore ask,  are the one-sided geometries we consider here the only possibility, or should one consider more general ways of terminating the solution in the IR? Furthermore, how should one understand  the apparent freedom to choose different boundary conditions for fluctuating probe fields, and how does this freedom extend to the full non-linear case?
    A study of the exact top-down solution of the type in \cite{DH2} would probably help clarify this issue.

\item One important problem is that of the stability of our solutions under small perturbations.  In  holography, the  existence of a wormhole solution connecting different boundary regions means that the partition function of a product of the two boundary CFTs does not  factorize. However, there may be a loophole to this argument, in that the wormhole may not be stable, \cite{MM,van,Santos}. If that is the case, it should not be included as a semiclassical saddle point in the path integral. It is therefore crucial to analyze the spectrum of small perturbations around the solutions and study their stability. It is known that,  in minimally-coupled Einstein-dilaton theories,  flat-slicing holographic RG flows with a regular IR are stable under perturbations (see e.g. \cite{exotic}). Whether this extends to RG flows with AdS-sliced bulk requires a detailed (and non-trivial) analysis which should be addressed in future work.
\item In a two-boundary solution connected by a wormhole, an interesting problem is the behavior of non-local observables, such as Wilson loops \cite{BKP}.
This same problem is of interest also in single boundary solutions obtained for the two boundary solutions.
Although the theories studied here have a conformal fixed point when considered in flat space, the scale factor in the wormhole solution has a turning  point like in {\em confining} holographic theories, in which the Wilson loop obeys an area law, \cite{BKP}. It would be very interesting therefore to analyse the behavior of Wilson loops expectation value in these solutions, because for large enough loop size these may effectively behave as in a confining theory. Similarly, one can analyse   correlators between Wilson loop operators inserted on opposite boundaries and analyse their factorisation properties. Similar questions concern other non-local observables like the entanglement entropy.

\item In this work we have limited ourselves to theories whose RG flows, in flat space-time, connect a UV and an IR conformal fixed point. One may extend the analysis to theories whose infra-red is gapped: this includes most notably confining theories. These may be modelled holographically by Einstein-Dilaton theories like those considered in this work, with the difference that the scalar runs to infinity in the IR \cite{ihqcd}. The existence of a mass gap in the flat space theory then requires the potential to be steep enough in the large-dilaton region. When considering AdS slicing, the scale  factor and/or  the dilaton may bounce before one reaches the IR endpoint and a new boundary may open up as in the examples studied here.  One therefore may expect the competition between the IR physics (governed by the confinement scale) and the effect of the boundary curvature. This is an interesting question for holographic QCD-like theories on negatively curved space-times.

 \item  The study of axion solutions in the two boundary case is of primordial interest as axions are typically dual to instanton densities in dual QFTs.
 According to \cite{CW} one should expect qualitative changes in the behavior of such solutions in the theories studied here. This in turn should translate into the properties of the instanton fluid in the dual gauge theory. For example, in holographic flows connecting AdS vacua, it can be shown in general that even turning a non-trivial $\theta$-angle, the vev of the dual operator vanishes, \cite{Ham}. In the presence  of AdS slices, this is not expected to be true anymore. On the other hand, it is well known that in the presence of axions wormhole solutions exist without negative curvature slices, \cite{gs}.

\item The study of the on-shell action in the context with two boundaries. Now the action depends on two sets of sources, and appropriate holographic renormalization must be applied to both boundaries, \cite{BKP}.
    The interest of this action is trying to construct a measure of distance between theories in the spirit of C and F functions. However, this case is radically different from standard RG flows and this is what makes this investigation more exciting. This tool may also be helpful with the study of the cobordism conjecture in  quantum gravity\footnote{We thank Irene Valenzuela for a stimulating discussion on this issue.}, \cite{co}.

    \item Last but not least, the phenomenon of flow fragmentation provides an intriguing possibility of introducing a topology and a topological algebra on the space of holographic RG flows. It is interesting to explore it further, and study the type of algebra it implies as well as its implications for gravity and holography.

\end{itemize}

\subsection{The structure of the paper}

This paper is organized as follows. In Section \ref{sec:setup} we present the Einstein-dilation setup, give a qualitative description of the AdS-sliced two-boundary solutions  and provide their classification depending on the nature of the two boundaries. In Section \ref{sec:2par} we study the space of solutions in detail and identify the relevant boundary parameters on which they depend. In a case study with a given (but  generic)  dilaton potential, we construct the space of solution numerically and we identify boundaries in parameter space. The solutions considered in this section  depend on two out of the three dimensionless parameters of the boundary theories. In Section \ref{limflo} we analyse in detail special cases which lie on the boundaries separating different regions of solution space.  In Section \ref{3p} we generalise the two-parameter solutions found earlier and  explain how to construct geometries which depend on all three dimensionless parameters of the boundary theory.
In Section  \ref{sec:single} we discuss how one may construct single-sided solutions with AdS-slicing. Some technical details are left to the Appendix.

\section{Setup} \label{sec:setup}

We consider Einstein dilaton theory in $d+1$ bulk dimensions. This is a minimal setup where to study holographic RG flows  driven by a single relevant operator.
\be\label{Abulka}
S_{Bulk}=M_P^{d-1}\int du\, d^d x \sqrt{-g}\Big(R^{(g)}-\frac12 g^{ab}\partial_a\f\partial_b\f-V(\f)\Big)\,.
\ee
In this paper we shall be interested in the ground state of the dual quantum field theories (QFTs) defined on negatively curved $d$-dimensional space-times.  We therefore choose the following domain wall metric and scalar field
\be\label{Abulksol}
\f=\f(u)\,,\qquad ds^2=du^2+e^{2A(u)}\z_{\m\n} dx^\m dx^\n \,.
\ee
Here, $A(u)$ is a scale factor that depends on the holographic coordinate $u$ only and $\z_{\m\n}$ is a metric
 on a constant curvature $d$-dimensional space-time with negative curvature.
 Although, at some point, we shall  focus  on a maximally symmetric manifold, namely an AdS space, the flows we describe are valid for any constant negative curvature manifold without necessarily maximal symmetry.

As a consequence of constant curvature we have
\be
R^{(\z)}_{\m\n}=\kappa\z_{\m\n}\,,\qquad R^{(\z)}=d\kappa\,, \qquad \kappa= - \frac{(d-1)}{\alpha^2}\,,
\label{alp}\ee
where $\a$ is the curvature length scale .

Varying the action \eqref{Abulka} with respect to the metric and the scalar field gives rise to the following equations of motion
\be
\label{AEOM1} 2(d-1) \ddot{A} + \dot{\f}^2 + \frac{2}{d} e^{-2A} R^{(\zeta)} =0 \,,
\ee
\be
\label{AEOM2} d(d-1) \dot{A}^2 - \frac{1}{2} \dot{\f}^2 + V - e^{-2A} R^{(\zeta)} =0 \,,
\ee
\be
\label{AEOM3} \ddot{\f} +d \dot{A} \dot{\f} - V' = 0 \,.
\ee
Here, derivatives
with respect to $u$ will be denoted by a dot while derivatives with respect to $\f$ will be
denoted by a prime.

Such flows were studied sporadically earlier, but a systematic study has been established in \cite{exotic, exotic2,C,F} where the first order formalism was also systematically developed and several exotic RG flows found. In particular, in \cite{C}, the RG flows of holographic QFTs both on positive and negative curvatures were studied, but the second case was not fully studied. General properties for such flows were found. Here, we shall study in more detail  flows on QFTs  defined on negative constant curvature spaces and in particular on AdS$_d$, and analyze in detail a generic example.

 The example will involve a bulk scalar potential that contains two maxima and a minimum. This is expected to locally describe a general potential with many more maxima and minima.

 The expansion of the potential around a maximum $\f_{m}$ can be written as
\begin{equation}
 V(\f)= -\frac{d(d-1)}{\ell^2}- \frac{m^2}{2}(\f-\f_m)^2+\mathcal{O}((\f-\f_m)^3)\,.
 \end{equation}
 If the scalar is constant and located at the maximum, the solution of equations of motion is a $d+1$ dimensional AdS$_{d+1}$ space-time sliced by AdS$_d$ slices.

It should be clear that
  $\alpha$ in (\ref{alp})  is the radius of the fiducial d-dimensional slice metric $\zeta_{\m\n}$, and is an auxiliary parameter.
  $\ell$ above is the radius of AdS$_{d+1}$
and is a true parameter of the gravitational theory, being related to the value of the scalar potential at the relevant extremum.

\begin{figure}[!t]
\centering
\includegraphics[width =9cm]{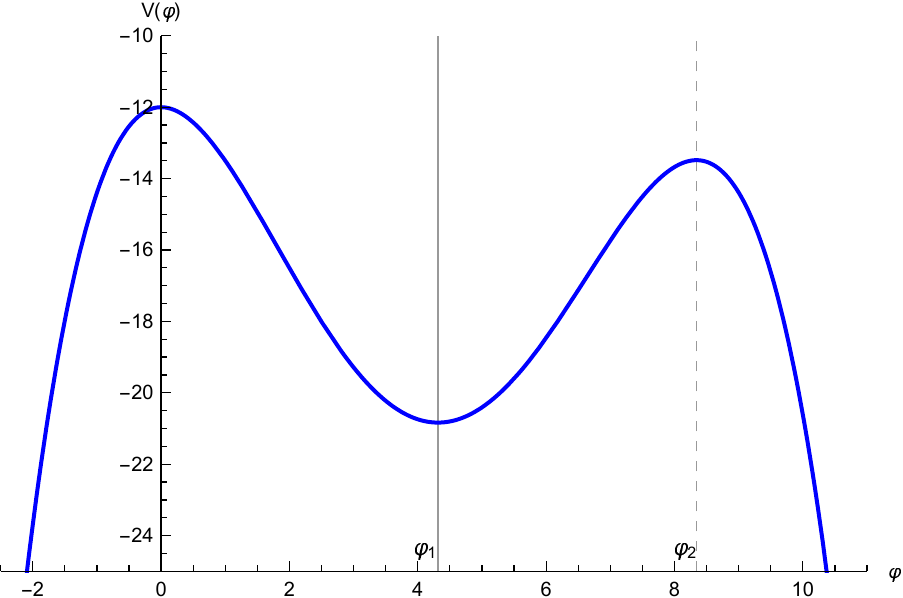}
\caption{\footnotesize{The scalar potential \protect\eqref{Apot} plotted for the specific values, $\ell_L=1, \ell_R= 0.94, \D_L=1.6$ and $\D_R=1.1$. The minimum is located at $\f_1=4.31$. There are two maxima at $\f=0$ and $\f_2=8.34$.}}\label{potential}
\end{figure}

For the numerical analysis that is going to follow, we shall pick $d=4$ and parametrize a  quartic scalar potential as
\be \label{Apot}
 V(\f)=-\frac{12}{\ell_L^2}+\frac{\D_L(\D_L-4)}{2\ell_L^2}\f^2\!-\!\frac{(\f_1+\f_2)\D_L(\D_L-4)}{3\ell_L^2\f_1\f_2} \f^3\!+\!\frac{\D_L(\D_L-4)}{4\ell_L^2\f_1\f_2} \f^4\,,
\ee
where $\f_1$ and $\f_2$ are defined as
\begin{align}\label{Aextrm}
&\f_1=\frac{12\ell_R^2 \sqrt{\ell_R^2-\ell_L^2}\D_L(\D_L-4)}{\sqrt{\ell_R^2\D_L(\D_L-4)-\ell_L^2\D_R(\D_R-4)}\big(\ell_R^2\D_L(\D_L-4)+\ell_L^2\D_R(\D_R-4)\big)}\,,\nn\\
&\f_2=\frac{12\sqrt{\ell_R^2-\ell_L^2}}{\sqrt{\ell_R^2\D_L(\D_L-4)-\ell_L^2\D_R(\D_R-4)}}\,.
\end{align}
In figure \ref{potential} we have sketched this potential for specific fixed values.

The above potential then has the following properties:
\begin{itemize}

\item There are two maxima for this potential, one is located at $\f=0$ and the other at $\f=\f_2$. We also have a minimum of the potential at $\f=\f_1$.

\item  At $\f=0$, the AdS length scale is $\ell=\ell_L$ and at $\f=\f_2$ it is $\ell=\ell_R$.

\item At $\f=0$ the mass square of the scalar field is equal to $m_L^2=\frac{4\D_L-\D_L^2}{\ell_L^2}$ and  at  $\f=\f_2$ we have  $m_R^2=\frac{4\D_R-\D_R^2}{\ell_R^2}$.

\end{itemize}

\subsection{The first order formalism}

To interpret solutions to the equations of motion \eqref{AEOM1} - \eqref{AEOM3} in terms of first order flows, it would be convenient to rewrite the second-order Einstein equations as a set of first-order equations. We define the following set of variables
\be\label{AWST}
\dot{A}(u) \equiv  -\frac{1}{2(d-1)}  W(\f(u))\,,\quad
\dot{\f}(u)\equiv  S(\f(u))\,, \quad
R^{(\zeta)}e^{-2 A(u)} \equiv  T(\f(u))\,.
\ee
In terms of these new functions the equations of motion  \eqref{AEOM1} - \eqref{AEOM3} become
\bsq
\begin{align}
\label{AEOM4} S^2 - SW' + \frac{2}{d} T &=0 \, , \\
\label{AEOM5} \frac{d}{2(d-1)} W^2 -S^2 -2 T +2V &=0 \, , \\
\label{AEOM6} SS' - \frac{d}{2(d-1)} SW - V' &= 0 \, .
\end{align}
\esq
From the definition  (\ref{AWST}) we also have the identity
\be
\frac{T'}{T}=\frac{W}{(d-1)S}\,.
\ee
We can partially solve this system by eliminating $T$ algebraically so that we are left with the following equations
\bsq
\begin{align}
\label{AEOM7} \frac{d}{2(d-1)} W^2 + (d-1) S^2 -d S W' + 2V &=0 \, , \\
\label{AEOM8} SS' - \frac{d}{2(d-1)} SW - V' &= 0 \, .
\end{align}
\esq
In appendix \ref{afirst} we analyze the equations and derive the inequality
\be
W^2\leq \frac{d W'^2}{ 2}-\frac{4(d-1)}{d}V\,.
\ee
We can also derive second-order equations for the functions $W$ and $S$:
\begin{align}
\big(d \big(d (W')^2 &-2 W^2\big)-8 (d-1)
   V\big) \big(4 (d-1) V'+W' \left((d+2) W-2 d
   W''\right)\big)^2 \nn \\
&=\big(-2 d^2 (W')^2 W''+4 (d-1)V' W'-8 (d-1) V \left(W-W''\right)\nn \\
&+2 d W^2 W''+d (d+2) W (W')^2-2 d W^3\big)^2\,,
\end{align}
and
\begin{align}
&2(d-1)S^3S'' +\frac{2(d-1)(d+2)}{d}V'SS'-\frac{2(d-1)}{d}S^2S'^2\nn \\
&
-(d-1)S^4-2(V+(d-1)V'')S^2-\frac{2(d-1)}{d}(V')^2=0\,.
\end{align}

\subsection{General features of the solutions}

We shall discuss here the near boundary expansion of solutions around an extremum of the potential as well as the algorithm for the construction of the solutions.

We take the extremum to be the UV fixed point at the maximum at $\f=0$.\footnote{The different asymptotics in all cases are presented in appendix \ref{exp1}.}  Close to this maximum, as
$\f \rightarrow 0^+$, the expansion of $W$ and $S$ are given by (this is known as the minus branch asymptotic solution)
\bsq
\begin{align}
\label{Wminus}
W&=\frac{2(d-1)}{\ell}+\frac{\D_-}{2\ell}\f^2+\frac{\mathcal{R}}{d\ell}\f^{\frac{2}{\D_-}}+\frac{C}{\ell}\f^{\frac{d}{\D_-}}+\cdots\,,\\
S&=\frac{\D_-}{\ell}\f + \frac{Cd}{\D_- \ell}\f^{\frac{d}{\D_-}-1}+\cdots\,,
\label{Sminus}
\end{align}
\esq
where $\mathcal{R}$ and $C$ are constants of integration of the first order system, (\ref{AEOM7}), (\ref{AEOM8}),  and we have defined
\be\label{deltapm}
\D_\pm=\frac{d}{ 2}\pm\sqrt{\frac{d^2}{ 4} - m^2\ell^2}\,.
\ee
The interpretation of the integration constants of the original system in $A$ and $\f$, depends on the type of solution. For example using the relations \eqref{AWST} and from expansions \eqref{Wminus} and \eqref{Sminus} close to the UV fixed point at $\f=0$ we obtain
\bsq
\begin{align}
\label{Aphimsol} \f(u) &= \f_- \ell^{\Delta_-}e^{\Delta_-u / \ell} + \frac{C d \, |\f_-|^{\Delta_+ / \Delta_-}}{\Delta_-(d-2 \Delta_-)} \, \ell^{\Delta_+} e^{\Delta_+ u /\ell} + \ldots \, , \\
\label{AAmsol} A(u) &= {A}_- -\frac{u}{\ell} - \frac{\f_-^2 \, \ell^{2 \Delta_-}}{8(d-1)} e^{2\Delta_- u / \ell}  -\frac{\mathcal{R}|\f_-|^{2/\Delta_-} \, \ell^2}{4d(d-1)} e^{2u/\ell} \\
\nonumber & \hphantom{=} \ - \frac{\Delta_+ C |\f_-|^{d/\Delta_-} \, \ell^d}{d(d-1)(d-2 \Delta_-)}e^{du/\ell} +\ldots \,,
\end{align}
\esq
where $\f_-$ identifies as the source for the scalar operator $\mathcal{O}$ in the boundary field theory associated with $\f$, and the vacuum expectation value of $\mathcal{O}$ depends on $C$ and is given by
\begin{align}\label{Avev}
\langle \mathcal{O} \rangle_- = \frac{Cd}{\Delta_-} \, |\f_-|^{\Delta_+ / \Delta_-} \, .
\end{align}
Using the definition of $T$ in (\ref{AWST}), its relation to $W$ and $S$ in (\ref{AEOM4}) and the UV expansions of $A, \f$ and $W ,S$ in (\ref{Aphimsol}), (\ref{AAmsol}),  (\ref{Wminus}) and (\ref{Sminus}) we obtain
\be\label{ERR}
R^{(\zeta)}e^{-2A_-}=\mathcal{R}|\f_-|^{2/ \Delta_-}\,.
\ee
To have more intuition on $\mathcal{R}$,  we define the induce metric $\g_{\m\n}$ on a $d$-dimensional slice at constant $u$,  which is given by
\be
\gamma_{\m\n}=e^{2A(u)}\z_{\m\n}\,.
\ee
Using  this definition, we can evaluate the scalar curvature on each slice as
\be
R^{(\g)}=e^{-2A(u)}R^{(\z)}\,.
\ee
On the other hand,  according to the expansion of \eqref{AAmsol}, the UV boundary is asymptotically located at $u\rightarrow -\infty$ (or at $u \rightarrow +\infty$, see appendix \ref{exp1}). Consequently we can associate an induced metric and scalar curvature on the UV boundary as \be
\g^{UV}_{\mu\nu}=\lim_{u\rightarrow -\infty} e^{\frac{2u}{\ell}}\gamma_{\m\n}\,,\qquad R^{UV}=\lim_{u\rightarrow -\infty} e^{-\frac{2u}{\ell}}R^{(\gamma)}\,.
\ee
Using the above definitions and the expansion relation for the scale factor in \eqref{AAmsol}, we obtain
\be \label{RUVZ}
R^{UV}=e^{-2A_-} R^{(\z)}\,.
\ee
By comparing \eqref{ERR} and \eqref{RUVZ}, the integration constant $\mathcal{R}$ is related to the curvature $R^{UV}$ of the manifold on which the UV QFT is defined by
\be \label{AUVcurve}
\mathcal{R} =  R^{UV} |\f_-|^{-2/\Delta_-}\,.
\ee

We now introduce some terms that are useful in describing the properties of the solutions.
The first term is {\em $\f$-bounce} which is  a point where the monotonicity of $\f$ changes (i.e. $\dot\f=0$).
Such points were called simply bounces in \cite{exotic}.

The next term is {\em $A$-bounce} which is  a point where the monotonicity of $A$ changes  (i.e. $\dot A=0$).
Both $A$-bounces and $\f$-bounces are defined AWAY from the extrema of the  potential.

$A$-bounces and $\f$-bounces  can also be, in principle, degenerate i.e.  the relevant second derivative can also vanish.
However, we find that this cannot happen for $\f$-bounces, as in that case the flow stops. Although it can happen in principle for $A$-bounces the only examples we have so far of such behavior appears in otherwise singular flows, and therefore their existence in regular flows is still open. We also define an {\em IR-bounce} to mean an A-bounce and $\f$-bounce happening at the same point.

One general feature of the regular solutions for the RG equations, in the case where the curvature of the slice is negative,  is the existence of the A-bounces, where the derivative $\dot A$ changes sign.
For each regular solution with negative curvature slices,  there exists at least one A-bounce where $W=0$.\footnote{As we shall see in section \ref{multi}, there exist solutions with multiple A-bounces, but we shall not study them extensively in this paper. Such solutions have appeared in other contexts, \cite{C1,C2}.}

At an A-bounce, the scale factor reaches a minimum. We denote the value of $\f$ at which there is an A-bounce ($\dot A=0$), by $\f_0$.
The expansion of the solution around this point is given in appendix \ref{BIT}.
As we deduce from equations \eqref{AturnW} - \eqref{AturnT}, the only free parameter at each generic point $\f=\f_0$ is
$$S_0=\dot\f\Big|_{\f=\f_0}\;.$$

As a result, by choosing $\f_0$ and $S_0$ as the initial values for our system of differential equations, we can solve the equations of motion numerically
 {\it at both sides of $\f_0$}.
  Generically, the solutions  continue until they arrive at a UV fixed point (extremum) of the potential.
   Another possibility is the $\f$ runs to
  infinity. Such solutions can be shown to be singular (and therefore unacceptable), \cite{C}. As we shall see, most of the flows studied in this paper will turn out to be regular. In particular, the flows depicted in the colored regions of figure \ref{Moduli} are all regular, but some flows outside are singular.

  We also expect that as an flow reaches a UV fixed-point, the solutions for $W$ and $S$ have the standard near-boundary expansions presented in appendix \ref{exp1}. In this way, the source and vev data of the solutions  $\mathcal{R}$ and $C$ can be found as a function of $\f_0$ and $S_0$.
  It was already shown in \cite{C}, that curved RG flows can never end at a minimum of the potential (so that the minimum is an IR Fixed point). However, with very special RG flows, the + branch can start at a minimum (which becomes an AdS boundary).

We should emphasize, that from equation \eqref{AturnT}, every point $\f=\f_0$ can be an A-bounce as far as we consider negative curvature solutions,  $T<0$.
Since at every point we have considered $V<0$,  then this equation does not put any constraint on the values of $S_0\neq 0$, and the expansions \eqref{AturnW} - \eqref{AturnT} are valid at every point and specifically in the interval between two maxima of the potential, $0<\f_0<\f_2$. There is an infinite number of flows that cross through a specific A-bounce but we do not know which one is regular or not until we solve the equations of motion.

We should also remind the reader that even  when $S_0=0$,  the A-bounce is not a fixed point and  the flow does not stop here.
The properties and expansion of the solution close to this point are described in appendix \ref{BIT}.

We conclude that,  up to an integration of the first order flow equations (\ref{AWST}),  the space of solutions can be constructed out of two real parameters, $\f_0$, the position of an A-bounce, and $S_0$ the derivative $\dot \phi$ at that point. Once this pair is chosen, a unique solution for the superpotentials $W,S,T$ can be constructed\footnote{This assumes that there are no further A-bounces and this will be the case for the moduli space of solutions we consider in this paper. In the general case, it is less clear what is the best organisation and labelling of the solutions.}. If it is everywhere regular, it is an acceptable holographic flow.
Such solutions terminate at an AdS boundary on one side of the flow and another AdS  boundary at the other side of the flow.
Therefore, they are characterized by two sets of sources and two sets of vevs for the operators included in this paper (stress tensor and a scalar operator).
In the interface picture, the two CFTs at the two ends of the flow correspond to two QFTs separated by an interface and are both driven by the scalar operator dual to $\f$.

\subsection{A classification and characterization of flows between extrema of the potential}

To classify the solutions of the flow equations \eqref{AEOM7} and \eqref{AEOM8}, we need to examine
 the solutions of this system in the vicinity of the extremal points of the potential.
 This has been already done in \cite{C} in the general case where the curvature
 of the slices is non-zero. We have summarized the results in appendix \ref{exp1} and we have given the structure of the expansion in the previous subsection.

Near a maximum of the potential, there are two branches of solutions known as the $-$ and the $+$ branch.

$\bullet$ The $-$ branch contains the generic solutions that contain both source and vev.

$\bullet$ The $+$ branch contains only the special solutions for which the source vanishes.

For both types of solutions above,  the metric has an AdS boundary at the maximum. We  denote these asymptotics as $Max_{\pm}$.

Near a minimum of the potential, we also have the $+$ and $-$ branches of solutions.

$\bullet$ The $-$ branch contains the generic solution but does not exist for non-zero slice curvature. It exists only for flat slices
and in that case, it describes the IR-end of a  flow.

$\bullet$ The $+$ branch contains the special solution, and in principle can exist for both flat and curved slices. The bulk metric has an AdS boundary in this case and the solution describes a UV fixed-point perturbed by the vev of an irrelevant operator.

 We  denote these asymptotics as $Min_{\pm}$. We remind the reader that three of these asymptotics, namely  $Max_{\pm}$ and $Min_{+}$ are associated with AdS boundaries and therefore to QFT UV fixed points,
 while one, $Min_-$,  to a shrinking slice geometry and therefore to an IR Fixed point.
 The $+$ branch solutions, as they contain fewer integration constants, exist only in fine-tuned cases. Moreover, as shown in \cite{C}
 the $Min_{-}$ solution does not exist, when the (dimensionless) curvature of the slice ${\cal  R}$, defined in (\ref{AUVcurve}), is non-zero.

Therefore the nature of a complete flow, between extrema of the potential, is specified by giving the type of solution  near its beginning and its end.
For, various values of the dimensionless curvature, defined in (\ref{ERR}) we have the following possibilities.

\begin{itemize}

\item ${\cal R}=0$. In this case, we have three possibilities, $(Max_-,Min_-)$ that is the generic RG flow driven in the UV by a relevant coupling,
$(Max_+,Min_-)$, is an RG flow driven in the UV by the vev of a relevant operator, and $(Min_+,Min_-)$ is driven in the UV by the vev of an irrelevant operator.
Moreover, in this case, all flows start and end at the extrema of the potential.

\item ${\cal R}>0$. In this case, although flows can start at an extremum of the potential, (both maxima as $Max_{\pm}$ and minima as $Min_{+}$), they always end at intermediate points, not at extrema, \cite{C}.
The end is always an IR end-point where the slice volume vanishes.

 \item ${\cal R}<0$. In this case, it is not possible for a flow to be regular and end at intermediate points (non-extrema of the potential), as there is no slicing of flat space with AdS slices.
 Therefore, all regular flows must start and end at an extremum of the potential. As the asymptotic solution $Min_-$ does not exist when ${\cal R}\not=0$, \cite{C}, we have in total the following options, all of them having two AdS boundaries and being therefore either wormholes or interface solutions:

$(Max_-,Max_-)$, $(Max_-,Max_+)$ and its reverse.

$(Max_+,Max_+)$, $(Max_-,Min_+)$ and its reverse.

$(Max_+,Min_+)$ and its reverse, and $(Min_+,Min_+)$.

 Note also that for the $Max_+$ and $Min_+$ asymptotics, ${\cal R}\to -\infty$.

 Taking into account the fact that the $Max_+, Min_+$ asymptotics are fine-tuned (they have half the adjustable integration constants), it is clear that the generic solutions will be of the
  $(Max_-,Max_-)$ type.
 Fine-tuning of the potential or the integration constants is needed for the
 $(Max_-,Max_+)$ and $(Max_-,Min_+)$ to exist, and double fine-tuning needed for $(Max_+,Max_+)$, $(Max_+,Min_+)$  and  $(Min_+,Min_+)$ to exist.

In this paper, beyond the generic  $(Max_-,Max_-)$ solutions, we shall find in special points of the  space of solutions, also $(Max_-,Max_+)$  (e.g. the $a_2$  boundary in figure \ref{typeAmoduli})  and $(Max_-,Min_+)$ solutions (e.g. the $a_4$  boundary in figure \ref{typeAmoduli}). Inversely, ($Max_+$, $Max_-$) appear for example on the borders of the moduli space of figure \ref{Moduli} (e.g. the $a_1$  boundary in figure \ref{typeAmoduli}).

 We  find also the double fine-tuned $(Max_+,Max_+)$ solutions, which appear  for example at the point where the $a_1$ and $a_2$ boundaries intersect  in figure \ref{typeAmoduli}.
There are also the double fine-tuned  $(Max_+,Min_+)$ solutions in our sample that appear for example  at the intersection point of the $b_1$ and $b_2$ boundaries  in figure \ref{typeBmoduli}.

We do not have examples of the  $(Min_+,Min_+)$ solutions. The reason may be  our choice of potential as it has a single minimum\footnote{In principle, such a flow could appear also with a single minimum as it could start and end at the same minimum.}.
\end{itemize}

\section{Two-parameter flows}  \label{sec:2par}

In this section, we discuss the regular solutions that are characterized by the two parameters $(\f_0, S_0)$ and have two boundaries. From the holographic RG flow point of view,
 the two QFTs living on these boundaries are the fixed points of the RG equations.

 Here we have two main groups of solutions:

1) The flows that interpolate  between one  UV fixed point of the potential (say $\f=0$ in \eqref{Apot}) and a distinct  UV fixed-point (like $\f=\f_2$ in \eqref{Apot}).

2) The flows that start at a UV fixed point and end at the same UV fixed point, but with different vevs and dimensionless curvatures.

Here, we classify the possible flows by solving the equations of motion numerically. Our results show the relationship between different parameters of the boundary QFTs on both sides of the flow. To begin with, we consider the potential in figure \eqref{potential} with the following specific values (which are typical):

 Our potential is plotted in figure \ref{potential} and has three extrema.

\begin{figure}[!t]
\begin{center}
\includegraphics[width = 12cm]{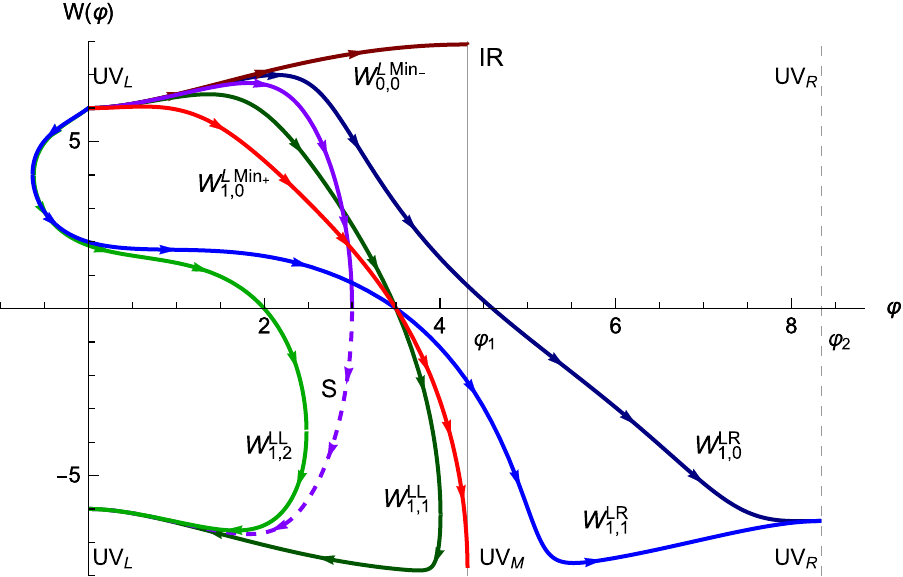}
\caption{\footnotesize{Various flows with one fixed point at the $UV_L$ boundary. We take the start of the flow ( $W>0$) to be  located on the maximum of the potential at $\f=0$. Both $W^{LR}_{1,0}$ and  $W^{LR}_{1,1}$  flows reach the same $UV_R$ fixed point at $\f_2$, the difference is the existence of a $\f$-bounce for the type $W^{LR}_{1,1}$ solution. Similarly, the  $W^{LL}_{1,1}$ and  $W^{LL}_{1,2}$ flows return to the $UV_L$ fixed point. The S type of solution is a flow that has an intermediate IR-bounce. The dashed curve is the mirror image of the original flow. $W^{L Min_+}_{1,0}$ solutions describe the flows that have a UV fixed point at the minimum $\f_1$.  $W^{L Min_-}_{0,0}$ is an exception that connects to an IR fixed point and it will be explained later in this paper.}}\label{AAllW}
\end{center}
\end{figure}

\begin{itemize}
\item We have two maxima of the potential, one at $\f=0$ that we shall call $UV_{L}$ as it is always a UV fixed point. The other maximum is at $\f_2=8.34$ which we shall denote as $UV_R$.
 There is also a minimum located at $\f_1=4.31$  that we shall denote by $Min$.

\item The AdS length scale at $\f=0$ is $\ell_L=1$ and at $\f_2$, it is equal to $\ell_R= 0.94$. Moreover, at the minimum, this length scale  is $\ell_M=0.76$.

\item At the UV fixed points,  one needs to know the value of the conformal dimension $\D_-$ in \eqref{ACfact}. For the specific potential \eqref{Apot}, $\D_L=1.6$ at $\f=0$ and $\D_R=1.1$ at $\f_2$. At the minimum of the potential, we shall need the value of $\D_+$ according to the expansions near the UV fixed point. This value is  $\D_M=4.37$ for our specific choice of potential.
\end{itemize}
The strategy to find all possible  flows is to start from a generic A-bounce at $\f_0$ and scan all possibilities by changing the value of $S_0$. Varying $\f_0$ and $S_0$ one should obtain all possible regular solutions.
In figure \eqref{AAllW} we have shown seven types of flows that start at the $UV_L$ boundary ($\f=0$ as UV fixed point) with $W>0$ and end at another extremum.

To describe them we introduce the notation $W^{LR}_{m_A,m_{\f}}$ which denotes a flow that starts at $UV_L$ ends at $UV_R$, has $m_A$ A-bounces and $m_{\f}$ $\f$-bounces.
Similarly $W^{LL}_{m_A,m_{\f}}$ denotes a flow that starts at $UV_L$ ends at $UV_L$, has $m_A$ A-bounces and $m_{\f}$ $\f$-bounces.
Finally $W^{LM}_{m_A,m_{\f}}$ is a flow that starts at $UV_L$ and ends at the minimum and has $m_A$ A-bounces and $m_{\f}$ $\f$-bounces.

We choose the convention that all our flows start at $UV_L$. Similar results can be obtained for flows starting at $UV_R$.
We have the following classes of flows.

\begin{itemize}

\item {$W^{LR}_{1,0}$} : The flow starts at $UV_L$ ends at the $UV_R$ fixed point at $\f=\f_2$ and has a single $A$-bounce but no $\f$-bounces\footnote{The term ``bounce" was introduced in \cite{exotic} to indicate the place where a change of direction for $\f$ is happening during the flow. Here we call it $\f$-bounce to distinguish it from points where the scale factor changes monotonicity (that we call A-bounces).}.

\item {$W^{LL}_{1,1}$}: The flow starts at $UV_L$ and returns to the $UV_L$ fixed point. It has a single A-bounce and a single $\f$-bounce.

\item  {$W^{LL}_{1,2}$}: The flow starts at $UV_L$ and returns to $UV_L$ with a single A-bounce and 2 $\f$-bounces.

\item  {$W^{LR}_{1,1}$}: The flow starts at $UV_L$ and ends at $UV_R$. It has a single A-bounce and a single $\f$-bounce.

\item { $W^{LMin_+}_{1,0}$}: The flow starts at $UV_L$ and ends at a  UV fixed point at the minimum of the potential ($Min_+$ asymptotics). It has a single $A$-bounce and no $\f$-bounces.

\item  { $W^{LMin_-}_{0,0}$}: This class starts at $UV_L$ and ends as an  IR fixed point at the minimum of the potential ($Min_-$ asymptotics).
This exists  as a limiting flat-sliced solution. It appears a piece of a complete flow.

\item Type S (Symmetric solutions): For this class of solutions, $\dot \f$ vanishes at the same place where $\dot A=0$.
 We shall call this type of point an {\it{IR-bounce}}.  Such flows were analyzed in detail in \cite{C}.
    For such flows, one can introduce a tensionless brane at $\f_0$ and consider only one half of the flow. We shall comment on this in the concluding section.

\end{itemize}

Some solutions have more $\f$-bounces but we restrict ourselves in the present paper to the type of solutions above.

Such flows can describe  wormhole solutions as well as nontrivial conformal interfaces.
To be considered as bonafide wormholes,  the two AdS boundaries at the two endpoints of the flow must be isolated boundaries.
This can only happen only if the constant negative curvature slices are compact.
In $d=2$, we may consider a compact higher-genus Riemann surface as a constant curvature slice metric and in that case, such solutions are genuine wormholes.
If in higher-dimensional examples one can orbifold AdS$_{d}$ to produce a compact manifold of constant negative curvature, then such cases also are genuine wormholes.
In three dimensions such compact manifolds exist, \cite{3c}.
Constant negative curvature compact manifolds also exist in four dimensions, \cite{4c}.

In all other cases, where the slices are non-compact, notably AdS$_d$, then the proper interpretation of such holographic solutions are as two distinct QFTs, on two copies of AdS$_d$ with an intersection along a common boundary. Such theories can be considered as limits of Janus-type solutions, \cite{janus}.

\subsection{The space of solutions}
\begin{figure}[!t]
\begin{center}
\includegraphics[width = 15cm]{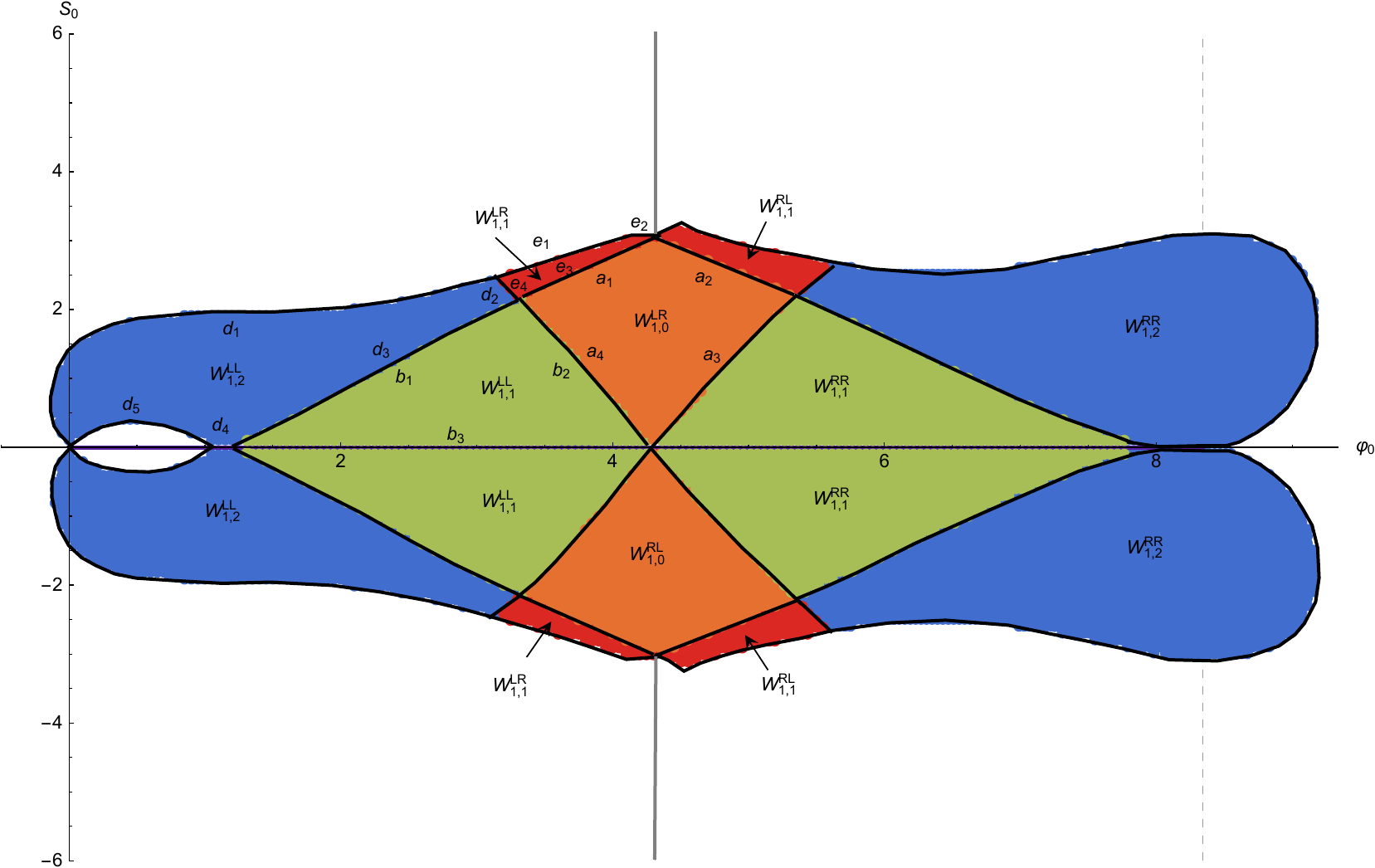}
\caption{\footnotesize{The connected space of solutions found with one A-bounce. The orange, green, blue and red regions correspond to  $W^{LR}_{1,0}$,  $W^{LL}_{1,1}$,  $W^{LL}_{1,2}$ and  $W^{LR}_{1,1}$ solutions respectively. The horizontal line of $S_0=0$ is the space of type S solutions. The diagonal lines (curves) between orange and green or red and blue regions are the space of  $W^{L Min_+}_{1,1}$ solutions. The vertical dashed lines show the location of the potential minimum $\f_1$ in the middle and the potential maximum $\f_2$ on the right. The   $W^{L Min_-}_{0,0}$ class of solutions does not explicitly appear in this moduli space but emerge in more intricate ways that will be explained in section \protect\ref{limflo}}}\label{Moduli}
\end{center}
\end{figure}

The space of solutions of each type is sketched in figure \eqref{Moduli}. The horizontal axis is $\f_0$ and the vertical one is $S_0$.
The whole $\f_0$, $S_0$ plane contains all solutions in our ansatz. The colored parts are the solutions that have been found explicitly and analyzed numerically.
They contain flows from $UV_{L,R}$ to $UV_{L,R}$ with at most two $\f$-bounces, and flows from $UV_L$ to $UV_R$ with at most one $\f$-bounce.
We believe that this is the full set of solutions with only one $A$-bounce but we do not have proof of this.

1) The top-\textbf{orange} region, labeled  $W^{LR}_{1,0}$,  is the space of the  $W^{LR}_{1,0}$ curves. At any generic A-bounce point  $\f_0$, if we choose $S_0$ from this region then there is a flow that crosses $\f_0$ and has two UV fixed points. One fixed point is a UV QFT that lives on the boundary at $\f=0$, which we indicate by $UV_L$. The other fixed point is located at the $UV_R$ at $\f=\f_2$. For simplicity from now on, we call the $UV_L$ with $W>0$, the top $UV_L$  and $UV_R$ with $W<0$, the bottom $UV_{R}$. In a symmetric way, the bottom-orange region is the moduli space of the solutions that have two similar fixed points. One at $UV_R$ with $W>0$ (top $UV_R$) and the other at $UV_L$ with $W<0$ (bottom $UV_{L}$).

2) The left-\textbf{green} region is the space of the  $W^{LL}_{1,1}$ solutions. Every flow with initial values in this region has two UV fixed points at the same place, i.e. both points are located at $\f=0$, one is living at the top $UV_L$ and the other at the bottom $UV_L$. The right-green region is the space of flows from top $UV_R$ to bottom $UV_R$.

3) The horizontal line $S_0=0$ is the space of the type S solutions that start at the left or right UV fixed points and have an ``IR-bounce" defined below equation (\ref{AUVcurve}).
Note that solutions with $\f_0=\f_1$, the minimum, are singular solutions and  do not belong to the space of S solutions.

4) The left(right)-\textbf{blue} region is the space of the  $W^{LL}_{1,2}$ solutions. It contains those flows which have two fixed points, both on the $UV_L$ ($UV_R$) but with two $\f$-bounces.

5) The top(bottom)-\textbf{red} region is the  space of the  $W^{LR}_{1,1}$ solutions. These flows are similar to  $W^{LR}_{1,0}$ with  two distinct fixed points, one at the top $UV_L$ and the other at bottom $UV_R$ (or top $UV_R$ and bottom $UV_L$), but with an extra $\f$-bounce.

6) The diagonal black lines (curves) between orange-green or red-blue regions are the one-dimensional space of the  $W^{L Min_+}_{1,0}$ solutions. In this type, a UV fixed point is at the minimum of the potential at $\f=\f_1$ (type + solution) and the other UV fixed point is located on one of the maxima of the potential.

7) The leftover regions on this diagram either correspond to the multi-$\f$-bounce solutions or the singular solutions. For example in figure \ref{Moduli}, the hollow white region on the left contains the flows with two fixed points at $UV_L$ but with three $\f$-bounces.

\subsection{How the QFTs on UV fixed points are related to each other?} \label{sec:parameters}
As we already mentioned, in this section, we are dealing with flows which have two UV fixed points at their ends. At each fixed point there is an associated  QFT on the negative curvature space, with a source which is coupled to an operator $\mathcal{O}$. We can interpret this system as two QFTs at the two AdS boundaries which are connected through a wormhole in the bulk, or   as two theories that are connected by a defect depending on whether the negative curvature slice has a boundary or not  along the philosophy of Janus solutions.  This system has the following properties:

\begin{enumerate}
\item   In a single boundary problem, the equations have a single independent  source for each bulk field. The rest of the integration constants are fixed by bulk regularity in terms of these parameters.
Here the situation is as in asymptotically AdS wormholes: there is no constraint from bulk regularity, and the number of boundary sources is doubled (one on each boundary for each dynamical field).

Each QFT at a UV boundary is defined by two parameters (coupling constants): The (constant) curvature of the negative constant curvature space  on which this QFT is living, $R^{UV}$, and $\f_-$ which is the source of the scalar operator $\mathcal{O}$. Each theory on its own, has therefore one dimensionless parameter, $\mathcal{R}$ defined in (\ref{AUVcurve}).  For two boundary QFTs (which we call QFT$_i$ and QFT$_f$) on AdS$_d$, we therefore have four independent dimensionful parameters,
\be
R^{UV}_i\,, \; R^{UV}_f\,, \; \f_-^{(i)}\,,  \; \f_-^{(f)}\,.
\ee
From these,  we can make three dimensionless parameters:

$\bullet$ Two dimensionless curvatures, $\mathcal{R}_i$ and  $\mathcal{R}_f$.
\be \label{dimRif}
\mathcal{R}_i=R^{UV}_i (\f_-^{(i)})^{-2/\D_-^i}\,,\qquad \mathcal{R}_f=R^{UV}_f (\f_-^{(f)})^{-2/\D_-^f}\,.
\ee

$\bullet$ The ratio of the two relevant coupling constants which can be found from the asymptotic sources of the $\f$ field
\be\label{Eratio}
\xi\equiv \frac{(\f_-^{(i)})^{1/\Delta_-^i}}{(\f_-^{(f)})^{1/\Delta_-^f}}\,.
\ee

In our problem, a solution for the  superpotential equations (\ref{AEOM7} - \ref{AEOM8}) is fixed by the two independent parameters $\f_0,S_0$. These are mapped to  $\mathcal{R}_i$ and
 $\mathcal{R}_f$ which enter the asymptotics of $W(\f)$ close to each fixed point at subleading order, as one can see in equation (\ref{Wminus}).

On the contrary, the scalar sources  $\f_-^{(i)},  \f_-^{(f)}$ do not appear in the superpotentials: they are determined instead by integrating the first order equations (\ref{AWST}). Notice however, that these have only one integration constant, not two: once we solve for $\f(u)$, the integration constant of the $A(u)$ flow equation (a shift in $A$)  is fixed by the third equation of the system (\ref{AWST}) in such a way that it cancels the dependence of the solution on the fiducial curvature $R^{(\zeta)}$. Therefore integrating the first order system (\ref{AWST}) does not affect the dimensionless curvatures and determines  at most one independent combination of   $\f_-^{(i,f)}$.

It is convenient to identify the third independent integration constant in the bulk solution  with a shift in the coordinate $u$ by a fixed amount $u_0$. This affects the dimensionless ratio (\ref{Eratio}) but not the dimensionless curvatures (\ref{dimRif}), as we argue momentarily.
First, notice that since the first order system (\ref{AWST}) has no explicit dependence on $u$, starting from any pair  $(A(u), \f(u))$ which is a solution,  the pair of functions  $(A(u-u_0), \f(u-u_0))$ is again a solution\footnote{A shift in $u$  is a bulk diffeomorphism  and therefore one may think it leads to an equivalent solution. This is not the case, since it is a {\em large} diffeomorphism as it acts non-trivially on the asymptotic boundary quantities defined at $u\to \pm \infty$.}  for any constant $u_0$. As one can see from the near-boundary asymptotics (\ref{AAmsol}) for $u\to -\infty$ (and from their counterpart as $u\to +\infty$ obtained by sending $u\to -u$), a shift in the coordinate $u$ has the effect:
\be\label{shift1}
u\to u-u_0 \; \Rightarrow \; \f_-^{(i)} \to \f_-^{(i)}e^{-\Delta_-^i u_0/\ell_i}, \; \f_-^{(f)} \to \f_-^{(f)}e^{\Delta_-^f u_0/\ell_f}\,,
\ee
which corresponds to a scaling of $\xi$:
\be \label{shift2}
u\to u-u_0 \; \Rightarrow\;  \xi \to \xi \, \exp\left[-2u_0\left(\frac{1}{\ell_i} + \frac{1}{ \ell_f}\right)\right]\,,
\ee
$\ell_{i,f}$ are the AdS scales of the potential extrema at the two end-points of the flow, see section \ref{sec:setup}.

Since the change (\ref{shift1})
  does not affect $\mathcal{R}_{i,f}$, it  also changes the boundary curvatures in opposite ways:
\be \label{shift3}
u\to u-u_0 \; \Rightarrow \;  R^{UV}_i \to   R^{UV}_i \,e^{-\frac{2u_0}{ \ell_i}}, \;  R^{UV}_f \to  R^{UV}_f\, e^{\frac{2u_0}{ \ell_i}}.
\ee

The integration constant $u_0$ is rather trivial as it is a diffeomorphism in the bulk, therefore its only effect is a trivial rescaling of $\xi$. It is therefore convenient to fix it once and for all (for example by in such a way that the $A$-bounce is at $u=0$, i.e.  $\f(u=0) = \f_0$). If we operate this choice, the third parameter dimensionless parameter (\ref{Eratio}) is not fixed  and therefore,  the three dimensionless field theory parameters in (\ref{dimRif}) and (\ref{Eratio}) are not independent. If needed, the value of  (\ref{Eratio}) can be changed by reinstating $u_0\neq 0$.

As we have seen,  we can  identify three of the four boundaries dimensionful  CFT parameters with integration constants of  solutions of the form (\ref{Abulksol}). This ansatz has not enough free parameters to change all four boundary sources independently. This can be also understood from the fact that the bulk equations of motion \eqref{AEOM1} and \eqref{AEOM3} for $\f$ and $A$ are  second-order differential equations, but we have also a first-order constraint \eqref{AEOM2} and therefore,  in total,  there are three constants of integration, not four. We shall see later  how to enhance our ansatz so that we have { four} independent parameters as expected from the dual QFTs  in section \ref{3p}.

\item From the leading terms in the near-boundary expansion of the solutions  in  \eqref{Wminus} and \eqref{Sminus}, we can read $\mathcal{R}$ and $C$ for each QFT on the boundary.
As we have already mentioned, $\mathcal{R}$ is the dimensionless curvature and $C$ controls the vev of the scalar operator.
 By another integration, we obtain  two more constants from \eqref{Aphimsol} and \eqref{AAmsol}, $\f_-$ and $A_-$ or equivalently $R^{UV}$ from \eqref{RUVZ}.
 Moreover, we have the relation \eqref{AUVcurve}, so in total, we have three independent parameters for each QFT.
  We should notice that $\mathcal{R}$ and $C$ are dimensionless parameters while the mass dimension of $R^{UV}$ and $\f_-$ are two and $\D_-$ respectively.

\item All  flows here contain two fixed points, $UV_{i}$ and $UV_f$. Their (dimensionless) parameters are $(\mathcal{R}_i, C_i)$ and $(\mathcal{R}_f, C_f)$. The  solutions impose two relations
\be\label{CRiRf}
C_i=C_i(\mathcal{R}_i,\mathcal{R}_f)\,,\quad C_f=C_f(\mathcal{R}_i,\mathcal{R}_f)\,.
\ee
which fix the two vevs as a function of the two sources.

\item For $(+)$ type solutions, as discussed in Appendix \eqref{exp1} and detailed in \cite{C}, the value of the source is zero. In this case, the leading behavior of the scalar field near the boundary is $\f (u)= \f_+ e^{-\frac{\Delta_+ u}{\ell}}$, where $\f_+$ is related to the vev of the dual scalar operator. In this case, the dimensionless curvature is given by
\begin{equation}
 \mathcal{R}=R^{UV}|\f_+| ^{-2/\Delta_+}.
 \end{equation}
Flows represented by $(+)$ type solutions are called the vev flows since the source is zero.
\end{enumerate}

Since the rigid  shift in $u$ (\ref{shift1}) can be performed trivially on any solution, in the rest of this section we  shall not use this freedom  and concentrate on solutions characterized by the two non-trivial free parameters which enter the superpotentials:  $(\f_0,S_0)$ in the bulk, or equivalently $(\mathcal{R}_i, \mathcal{R}_f)$ on the boundary. In these solutions, these parameters will completely fix the quantity $\xi$ in (\ref{Eratio}).

\subsection{The numerical strategy}
 We analyze all possible flows related to the space of solutions in figure \ref{Moduli} to extract the information about the quantum field theories on the UV fixed points of the theory. This information includes the values of dimensionless curvature $\mathcal{R}$ and the constant $C$ in terms of $\f_0$ and $S_0$, the coordinates of the space of solutions. To read these values, we should insert the numerical values of $W$ and $S$ for each flow into the analytic series expansions at a specific cut-off $\f-\f_m=\epsilon$, where $\f_m$ is the location of the UV fixed point. For our numerical proposes, we present the results with a cut-off near the UV boundaries at $\epsilon=10^{-6}$. We have checked that varying this does not induce important errors.

Practically we pick a value for $\f_0$ and we choose two initial values at $\f=\f_0$ to solve the RG equations \eqref{AEOM7} and \eqref{AEOM8}, these are $W(\f_0)=0$ and $S(\f_0)=S_0$. By numerical calculations for two generic QFT$_i$ and QFT$_f$ on the two UV fixed points we expect to find
\be
\mathcal{R}_i=\mathcal{R}_i(\f_0,S_0)\,,\qquad \mathcal{R}_f=\mathcal{R}_f(\f_0,S_0)\,.
\ee

To find the last dimensionless parameter we use the same strategy. To obtain the value of $\f_-$ on each side we can integrate $\dot{\f}=S$ near each cut-off boundary and then by using the expansion of $\f(u)$ we can read the value of the couplings.

\begin{figure}[!ht]
\centering
\includegraphics[width=10cm]{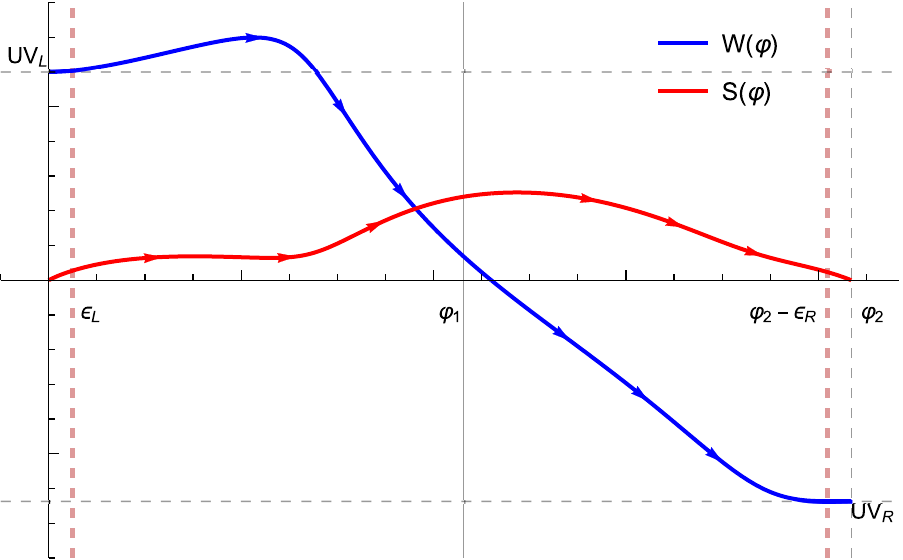}
\caption{\footnotesize{An example of the  $W^{LR}_{1,0}$ solution with the two cut-offs (red dashed lines)}}\label{CC}
\end{figure}

To do this, consider for example figure \ref{CC}. Here we consider a flow of type  $W^{LR}_{1,0}$, which in this case $UV_i=UV_L$ and $UV_f=UV_R$. We consider the left UV boundary is located at  $\f=\epsilon_L$ and the right one at $\f=\f_2-\epsilon_R$. Here $\epsilon_L$ and $\epsilon_R$  are two cut-offs and at the end, we shall send both cut-offs to zero.

Equivalently, in the holographic coordinate $u$, we can consider two cut-offs at $u=-\Lambda_L$ and $u=\Lambda_R$.
From equation $\dot{\f}=S$ we can integrate
\be
\int_{u_0}^{u(\f)} du = \int_{\f_0}^\f \frac{d\f}{S(\f)}\,,
\ee
where $\f_0$ is an arbitrary point. For example as $\f\rightarrow 0$, we expect that $u\rightarrow -\infty$ according to the asymptotic behaviors in \eqref{Aphimsol}. Now suppose that in the above equation $\f_0=\f_m\pm\epsilon$
where $\epsilon$ is a cut-off and $\f_m$ is the location of the UV fixed ($+$ for $UV_L$ fixed point and $-$ for $UV_R$ fixed point), after the integration we obtain
\be\label{uintg}
u(\f) =\int_{\f_m\pm\epsilon}^\f \frac{d\f}{S(\f)}+s\frac{\ell}{\D_-}\log \epsilon{\Big|}_{\epsilon\rightarrow 0}\,,
\ee
where  $s=+1$ for fixed points at $W>0$ and $s=-1$ for $W<0$. In obtaining the above relation we have assumed that

\be\label{cuts}
\Lambda_L=-\frac{\ell_L}{\D_-^L}\log\epsilon_L\,,\qquad \Lambda_R=-\frac{\ell_R}{\D_-^R}\log\epsilon_R\,.
\ee
The above assumptions are reliable as far as the point $\f$ in \eqref{uintg} is very close to the UV boundary.

Now using  \eqref{uintg} we can integrate near each boundary to obtain $u(\f)$ and then by using the expansion of $\f(u)$ near each UV fixed point, we can read the value of $\f_-$. Finally from the definition of \eqref{Eratio}, we find the ratio of the two relevant couplings.

\subsection{Connecting different fixed points}
\subsubsection{The  $W^{LR}_{1,0}$ solutions}\label{TAS}

\begin{figure}[!ht]
\centering
\begin{subfigure}{0.59\textwidth}
\includegraphics[width=1\textwidth]{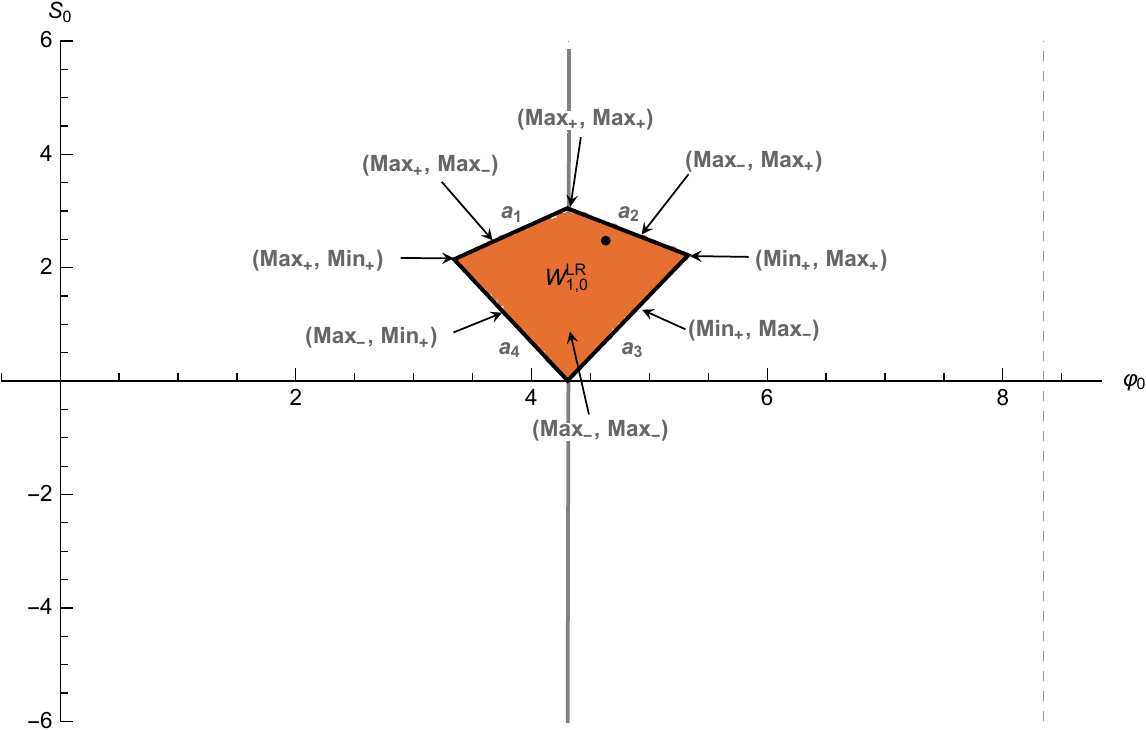}
\caption{}\label{typeAmoduli}\end{subfigure}
\centering
\begin{subfigure}{0.4\textwidth}
\includegraphics[width=1\textwidth]{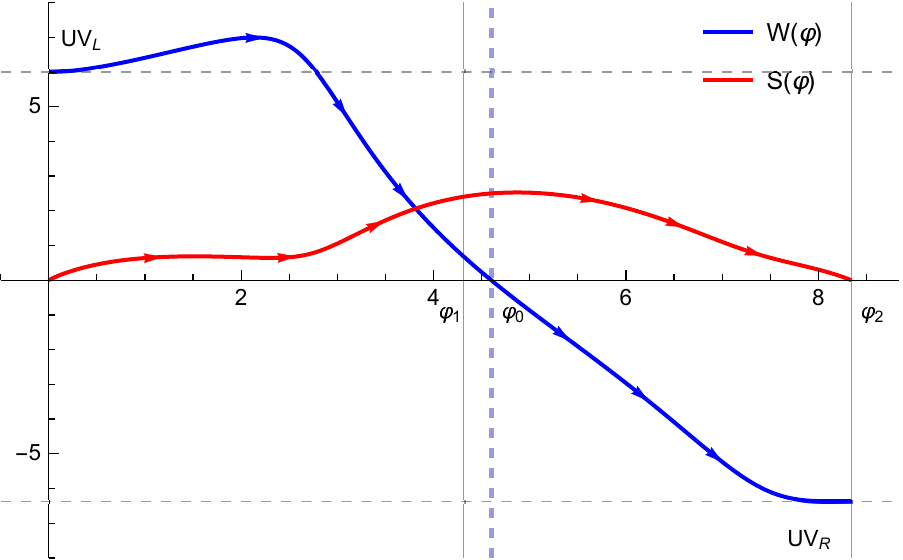}
\caption{\vspace*{0cm}}\label{AUVLRa}
\end{subfigure}

\caption{\footnotesize{(a):  The space of the  $W^{LR}_{1,0} \in (Max_-,Max_-)$ solutions. For each $a_1$ to $a_4$ boundary and corner of this region, we have different types of solutions. (b): The blue and red curves describe an example of  $W^{LR}_{1,0}$ flow corresponding to the black dot in figure (a). This flow connects the $UV_L$ fixed point at $\f=0$ to the $UV_R$ at $\f=\f_2$. The blue dashed line indicates the location A-bounce. The vertical solid line at $\f_1$ shows the minimum of the potential.}}
\end{figure}
In this section, we look at the $W^{LR}_{1,0}$ solutions. The parameter space of these solutions is given in figure \ref{typeAmoduli}. The points inside this region are the initial $(\f_0, S_0)$ values for $(Max_-, Max_-)$ flows. We also discuss the points on the boundary of the orange region in this section.
Figure \ref{AUVLRa} describes the holographic flow between two different UV QFTs, one is located at $UV_{L}$ ($W>0$) and the other at $UV_R$ ($W<0$). This specific solution in figure \ref{AUVLRa} belongs to a specific point in the space of solutions of the  $W^{LR}_{1,0}$ flows (the black dot in figure \ref{typeAmoduli}).
Both QFTs live on constant negative curvature  space-times which are slices of the five-dimensional bulk space-time and are localized at the maxima of the potential.
Each QFT is described  independently by the coupling of the operator dual to the scalar field and also by the curvature of the 4d negative curvature space  on which it lives.

Near the maximum of the potential at $\f=0$, $W$ (blue curve) and $S$ (red curve), as functions of $\f$, have  the expansions in  \eqref{WLU} and \eqref{SLU}. Similarly,
 in the vicinity of the  maximum at $\f=\f_2$, we have the  \eqref{WRD} and \eqref{SRD} expansions. The constants of integrations on both sides are related to the vev of the scalar operator $\mathcal{O}$ and the curvature of the manifold at the UV through the equations \eqref{Avev} and \eqref{AUVcurve}.
Due to the definition of $\beta$-function in \eqref{Abeta}, at the left UV fixed point $\b\rightarrow 0^-$ but at the right UV fixed point $\b\rightarrow 0^+$.

\begin{figure}[!t]
\centering
\begin{subfigure}{0.49\textwidth}
\includegraphics[width=1\textwidth]{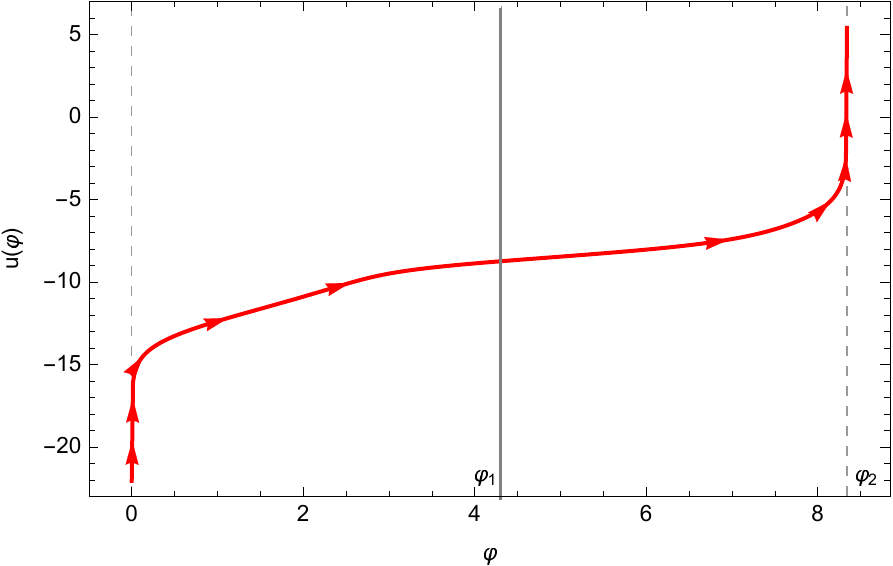}
\caption{}\label{AUVLRb}
\end{subfigure}
\centering
\begin{subfigure}{0.49\textwidth}
\includegraphics[width=1\textwidth]{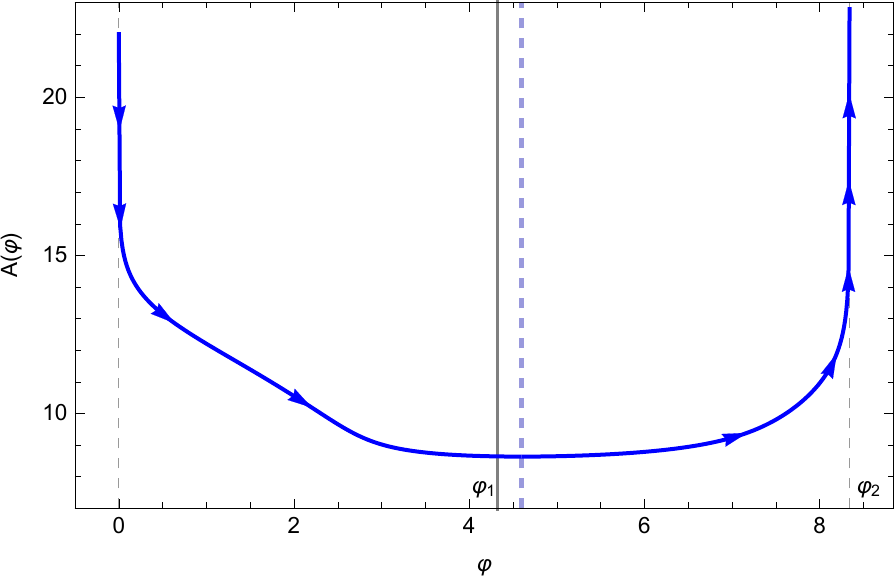}
\caption{}\label{AUVLRc}
\end{subfigure}
\caption{\footnotesize{ (a): The holographic coordinate $u$ as a function of $\f$. At $UV_L$ ($\f=0$) it tends to $ -\infty$ and at $UV_R$ ($\f=\f_2$) to $+\infty$. (b): The scale factor has an A-bounce at $\f=4.6$ (the blue dashed line) near $\f_1$, the minimum of the potential.}}
\end{figure}

In figures \ref{AUVLRb} and \ref{AUVLRc}  we plot the holographic coordinate $u$ and the scale factor $A$ as functions of the running scalar $\f$.
 As we observe in diagram \ref{AUVLRb}, the left QFT  (QFT$_{L}$) as a UV fixed point of the flow is located asymptotically at $u\rightarrow -\infty$. Similarly the QFT$_{R}$ is located at the point $u\to +\infty$. Both these behaviors are consistent with the boundary expansions of $\f(u)$ in \eqref{phiLU} and \eqref{phiRD}.

In figure \ref{AUVLRc} we observe that the scale factor of the geometry has an A-bounce (at the blue dashed line where $W=0$). $A(\f)$ tends to infinity on both sides as we approach the UV fixed points (boundaries).

\begin{figure}[!t]
\centering
\begin{subfigure}{0.49\textwidth}
\includegraphics[width=1\textwidth]{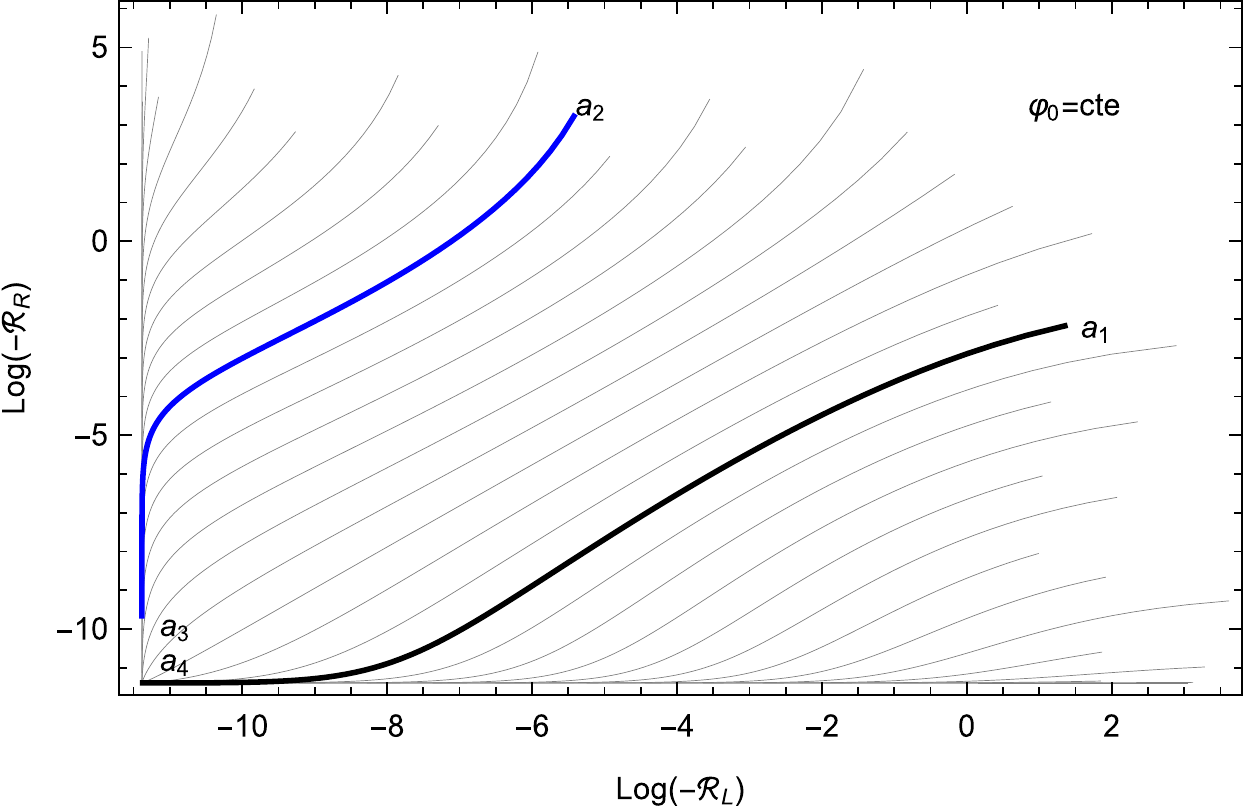}
\caption{}\label{RLRRfi0}
\end{subfigure}
\centering
\begin{subfigure}{0.49\textwidth}
\includegraphics[width=1\textwidth]{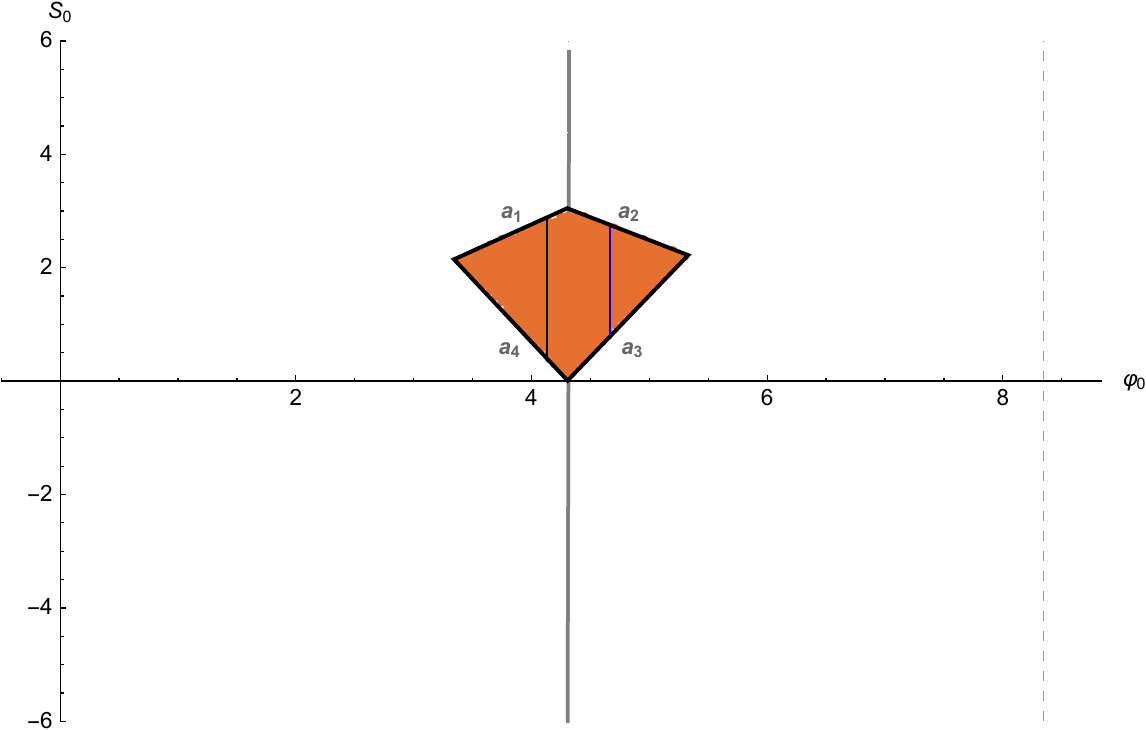}
\caption{}\label{typeAbb}
\end{subfigure}
\caption{\footnotesize{The relation between left and right dimensionless curvatures at fixed $\f_0$ (while varying $S_0$). For example, the points on the blue curve in figure (a) are in a one-to-one relation with the points on the blue vertical line in figure (b) at $\f_0=\frac{93}{20}$. Similarly the black curve is at $\f_0=\frac{83}{20}$.}}
\end{figure}

\begin{figure}[!t]
\centering
\begin{subfigure}{0.49\textwidth}
\includegraphics[width=1\textwidth]{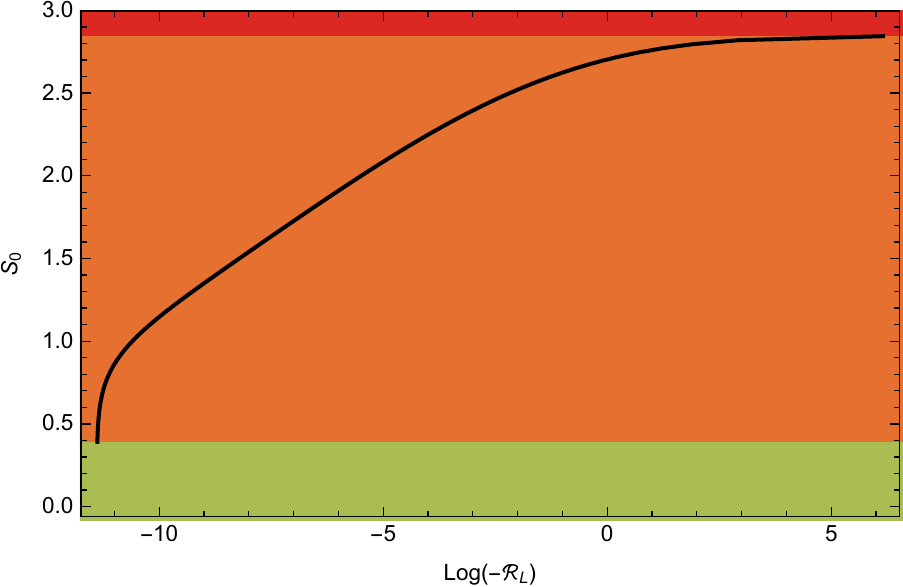}
\caption{}\label{RLRRfi01}
\end{subfigure}
\centering
\begin{subfigure}{0.49\textwidth}
\includegraphics[width=1\textwidth]{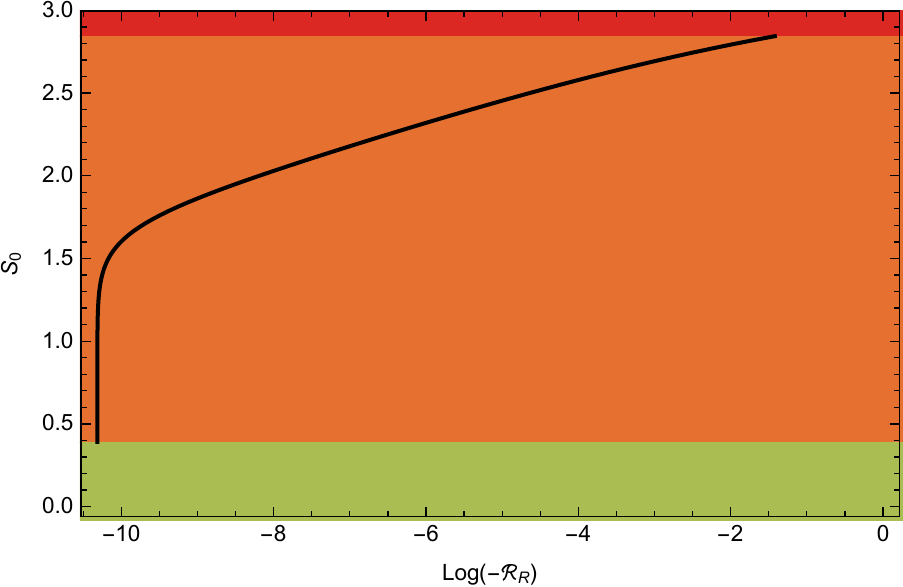}
\caption{}\label{RLRRfi02}
\end{subfigure}
\centering
\begin{subfigure}{0.49\textwidth}
\includegraphics[width=1\textwidth]{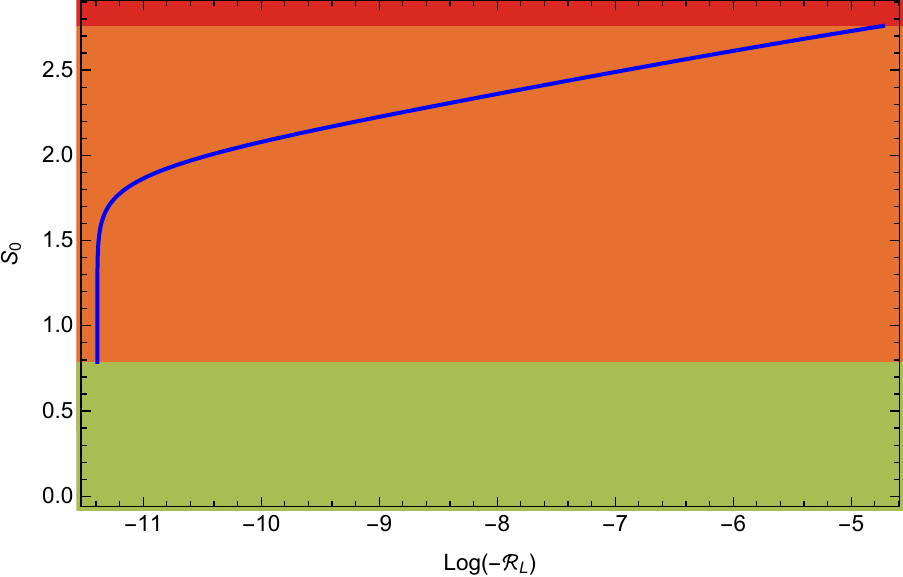}
\caption{}\label{RLRRfi03}
\end{subfigure}
\centering
\begin{subfigure}{0.49\textwidth}
\includegraphics[width=1\textwidth]{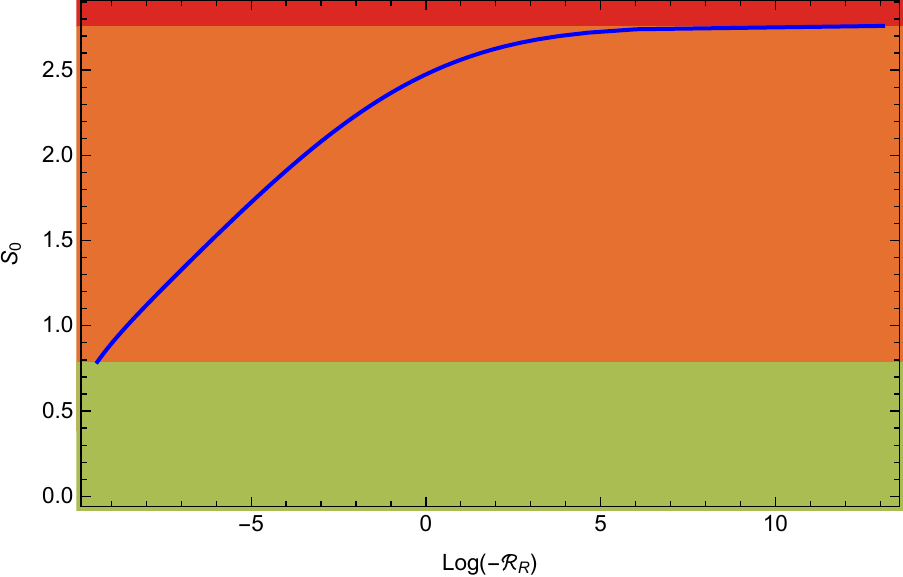}
\caption{}\label{RLRRfi04}
\end{subfigure}
\caption{\footnotesize{The dependence of the left/right dimensionless curvature at  $\f_0=\frac{83}{20}$ for black curve, and at  $\f_0=\frac{93}{20}$ and blue curve, in figure \protect\ref{RLRRfi0} as a function of $S_0$.
The color code is the same as in the moduli space figure \protect\ref{Moduli}.}}
\end{figure}

In type  $W^{LR}_{1,0}$ solutions, the relation between $\mathcal{R}_L$ and $\mathcal{R}_R$ and the parameters of the solution $(S_0,\f_0)$ are presented in figure \ref{RLRRfi0}. In this figure,  we present the behavior of $\log(-\mathcal{R}_R)$ as a function of $\log(-\mathcal{R}_L)$ as we move along a typical constant line $\f_0=constant$ in the space of solutions shown in figure \ref{typeAbb} by varying $S_0$.
These curves are bounded by either the $a_1$ and $a_4$ boundaries (e.g. the black curve) or between $a_2$ and $a_3$ boundaries (e.g. the blue curve) shown on figure \ref{typeAmoduli}.
The behavior of these two typical blue and black curves as a function of $S_0$ are shown in figures \ref{RLRRfi01} to \ref{RLRRfi04} independently.

In summary, by numerical analysis of the curves in the  $W^{LR}_{1,0}$ region, we have the following results for dimensionless curvatures as we move towards the boarder of orange region, see figure \ref{typeAmoduli}:\footnote{We  should note that the $a_1$ ($a_2$) boundary is the joint border between $W^{LR}_{1,1}$ ($W^{RL}_{1,1}$) and $W^{LR}_{1,0}$ regions of the space of solutions (see figure \ref{Moduli}) in our figures and similarly, $a_3$ ($a_4$) is the boundary  between the $W^{LR}_{1,0}$ ($W^{RL}_{1,0}$) and $W^{LL}_{1,1}$ ($W^{RR}_{1,1}$) regions.}

$\bullet$ $(Max_+,Max_-)$ solutions: $\mathcal{R}_L \rightarrow -\infty$\footnote{The limit $\mathcal{R}\to -\infty$ indicates a very strongly curved manifold, or that the relevant coupling constant $\f_-\to 0$. When the relevant coupling vanishes, then $\mathcal{R}\to \pm\infty$ and such solutions are vev driven flows of a CFT on a curved space.} as we move towards the $a_1$ boundary (in figure \ref{RLRRfi01} near the red-orange border the slope of the curve goes to zero at large values) but  $\mathcal{R}_R$ asymptotes to a finite value (in figure \ref{RLRRfi02} the slope of the curve has a nonzero finite value near the border).

$\bullet$ $(Max_-,Min_+)$ solutions: As we move towards the $a_4$ boundary, $\mathcal{R}_L$ asymptotes  to a finite value  (in figure \ref{RLRRfi01} near the orange-green border, the slope of the curve has a nonzero finite value) but  $\mathcal{R}_R \rightarrow 0$ (in figure \ref{RLRRfi02} the slope of the curve goes to infinity at small values).\footnote{ $\mathcal{R}\to 0$ when either the manifold becomes flat or the relevant coupling $\f_-\to\infty$.}

$\bullet$ $(Max_-,Max_+)$ solutions: $\mathcal{R}_L$ asymptotes  to a finite value as we move towards the $a_2$ boundary (in figure \ref{RLRRfi03} near the red-orange border the slope of the curve goes to a finite value) but  $\mathcal{R}_R \rightarrow -\infty$ (in figure \ref{RLRRfi04} the slope of the curve asymptotes to zero at large values).

$\bullet$ $(Min_+,Max_-)$ solutions: $\mathcal{R}_L\rightarrow 0$ as we move towards the $a_3$ boundary (in figure \ref{RLRRfi03} near the orange-green border the slope of the curve goes to infinity at small values) but  $\mathcal{R}_R $ asymptotes  to a finite value (in figure \ref{RLRRfi04} the slope of the curve is a nonzero finite value).

$\bullet$ There are also three specific types of solutions on the corners of the figure \ref{typeAmoduli}:

1. $(Max_+,Min_+)$ solution at the joining point of $a_1$  and $a_4$ boundaries. At this corner, $\mathcal{R}_L\rightarrow -\infty$ and $\mathcal{R}_R\rightarrow 0$.

2. $(Max_+,Max_+)$ solution at the joining point of $a_1$  and $a_2$ boundaries. At this point both $\mathcal{R}_L,\mathcal{R}_R \rightarrow -\infty$.

3. $(Min_+,Max_+)$ solution at the joining point of $a_2$  and $a_3$ boundaries. Here we observe $\mathcal{R}_L\rightarrow 0$ and $\mathcal{R}_R\rightarrow -\infty$.

\begin{figure}[!ht]
\centering
\begin{subfigure}{0.49\textwidth}
\includegraphics[width=1\textwidth]{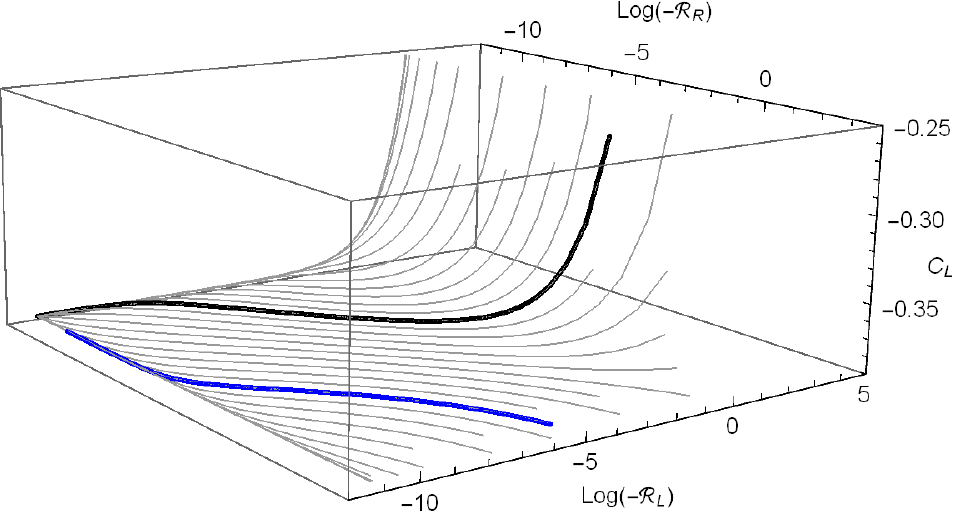}
\caption{}\label{CLRLRR3D}
\end{subfigure}
\begin{subfigure}{0.49\textwidth}
\includegraphics[width=1\textwidth]{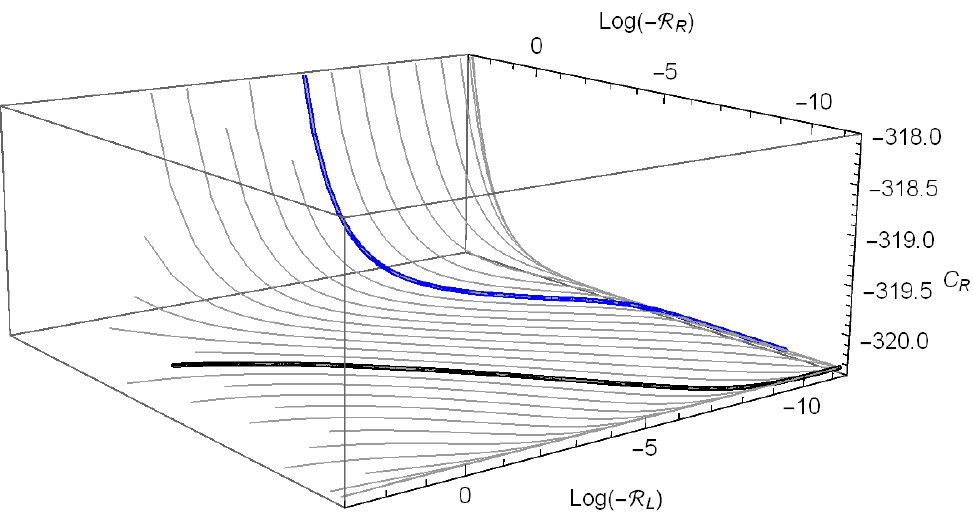}
\caption{}\label{CRRLRR3D}
\end{subfigure}
\caption{\footnotesize{As we change $S_0$ along a $\f_0=constant$ line in the orange region in figure \protect\ref{typeAbb} we can read $C_L$ and $C_R$ in terms of the dimensionless curvatures. For example in above figures, the black curve has $\f_0=\frac{83}{20}$ and the blue one has $\f_0=\frac{93}{20}$.}}
\end{figure}

As we already discussed, the solutions determine uniquely the values of the vev-related constants $C$ for the left and right QFTs. The relation between $C_L$ and $C_R$ with $\mathcal{R}_L$ and $\mathcal{R}_R$ are plotted in figures \ref{CLRLRR3D} and \ref{CRRLRR3D}.

\begin{figure}[!ht]
\centering
\begin{subfigure}{0.49\textwidth}
\includegraphics[width=1\textwidth]{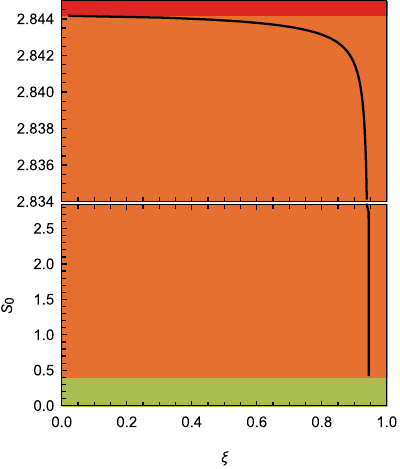}
\caption{}\label{xrl}
\end{subfigure}
\centering
\begin{subfigure}{0.49\textwidth}
\includegraphics[width=1\textwidth]{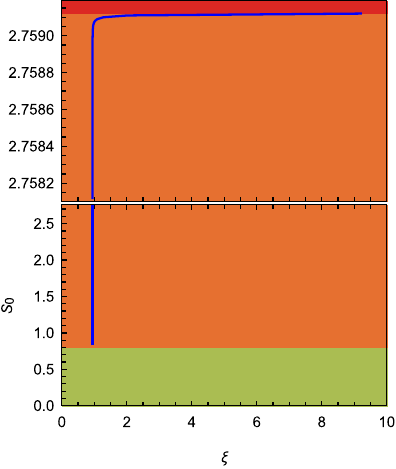}
\caption{}\label{xrr}
\end{subfigure}
\caption{\footnotesize{ The ratio $\xi$ of the relevant couplings for  $W^{LR}_{1,0}$ solutions. The black/blue curve in figure (a)/(b) correspond to the black/blue curve in figure \protect\ref{RLRRfi0}. The vertical constant part of both curves is at  $\xi\simeq 0.944$.}}
\end{figure}

Finally, we can find the ratio of relevant couplings of the left and right QFTs, $\xi$,  for various values of $\f_0$ and $S_0$ in the space of the  $W^{LR}_{1,0}$ solutions. In figures \ref{xrl} and \ref{xrr} we have found the values of the dimensionless parameter $\xi$ defined in \eqref{Eratio}, in terms of $S_0$ for two typical black and blue curves in figure \ref{typeAbb} with fixed values of $\f_0$. According to relation \eqref{ERR} and the behavior
of the dimensionless curvatures in figures \ref{RLRRfi01} to \ref{RLRRfi04}, we can explain the behavior of these two curves as follows:


1) Close to the $a_1$ boundary, near the red-orange border in figure \ref{xrl}, since $\mathcal{R}_L\rightarrow -\infty$ then $\f_-^{(L)} \rightarrow 0$ and we expect $\xi\rightarrow 0$.

2) Close to the $a_2$ boundary,  near the red-orange border in figure \ref{xrr}, since $\mathcal{R}_R\rightarrow -\infty$ then $\f_-^{(R)} \rightarrow 0$,  we expect (and obtain) $\xi\rightarrow \infty$.

3) Close to  the $a_4$  boundary  (orange-green border in figure \ref{RLRRfi02}), we showed that $\mathcal{R}_R\rightarrow 0$. We shall also show in section \ref{crossAtoB} that in this situation, the part of the flow which connects to $UV_R$ turns into the $W_-$ solution (see figure \ref{AUVIR} and \ref{Atino}) which is a flat-slice solution. This means that in the left-hand side of equation \eqref{ERR} we have $R^{(\z)}\rightarrow 0$. This implies that $\f_-^{(R)}$ has a finite value. Therefore, we expect to have a finite value for $\xi$ too. This is shown numerically in figure \ref{xrl} near the orange-green border.

4) The same behavior as the previous case is realized near the $a_3$ boundary, figure \ref{xrr} near the orange-green border. There is a difference  though: in this case, $\mathcal{R}_L\rightarrow 0$ so that $\f_-^{(L)}$ will be finite close to the $a_3$ boundary.

\subsection{Connecting one fixed point with itself}
\subsubsection{Type  $W^{LL}_{1,1}$ solutions}\label{green}
\begin{figure}[!ht]
\centering
\begin{subfigure}{0.59\textwidth}
\includegraphics[width=1\textwidth]{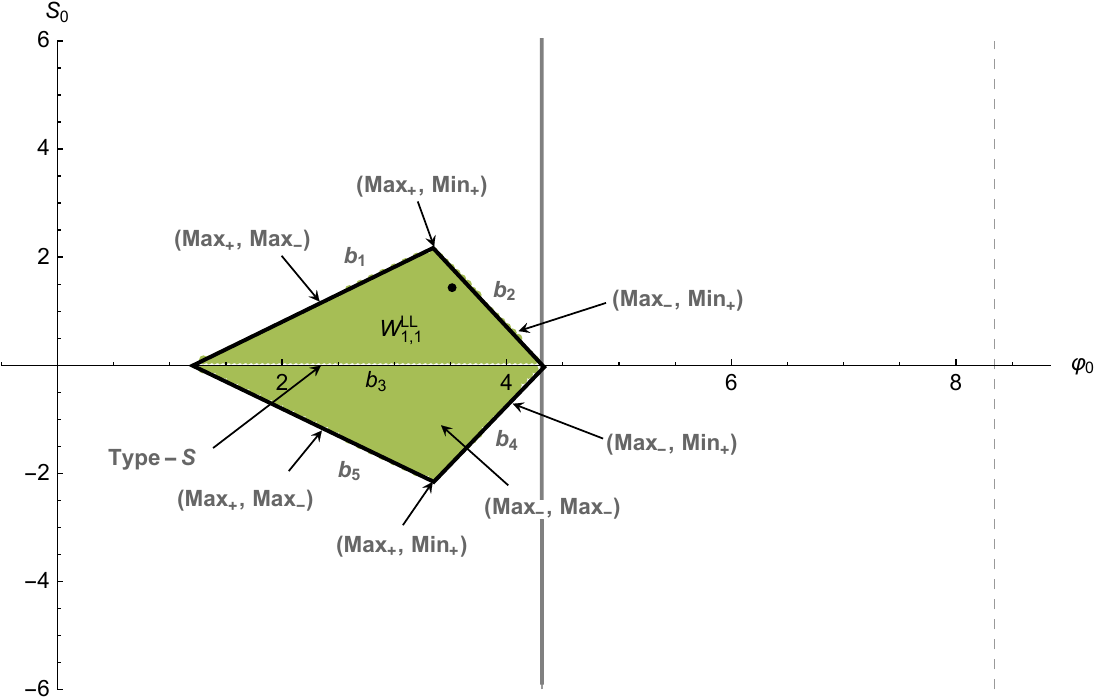}
\caption{}\label{typeBmoduli}
\end{subfigure}
\centering
\begin{subfigure}{0.4\textwidth}
\includegraphics[width=1\textwidth]{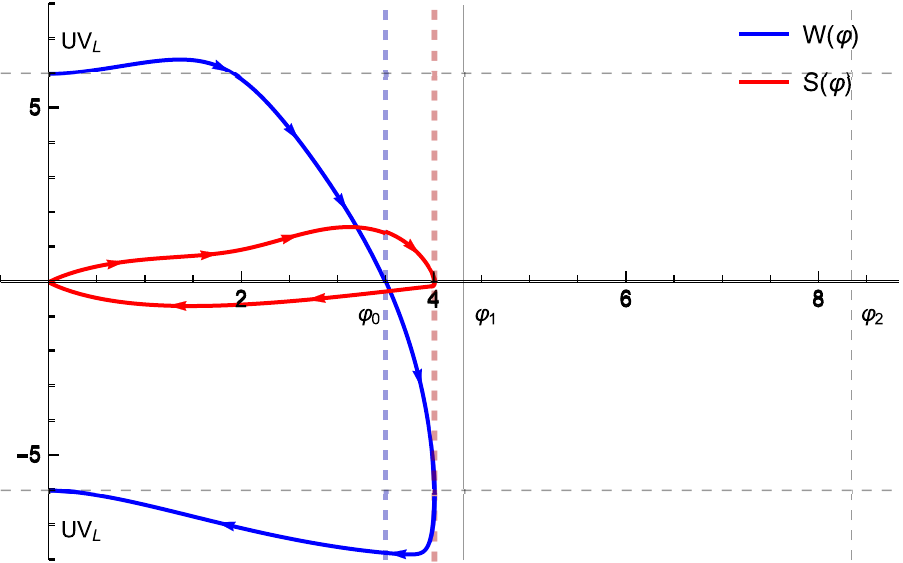}
\caption{\vspace*{0cm}}\label{AUVLLa}
\end{subfigure}
\caption{\footnotesize{(a): The space of the  $W^{LL}_{1,1} \in (Max_-,Max_-)$ solutions. For each $b_1$ to $b_5$ boundary and corner of this region, we have different types of solutions. (b): The blue and red curves describe a flow (represented by the black dot of diagram (a)) that connects the top $UV_L$ boundary to the bottom $UV_L$ at the same fixed point. Here the blue and red dashed lines show the location of the A-bounce and $\f$-bounce points respectively. }}
\end{figure}

Among the solutions of $(Max_-,Max_-)$ we have the $W^{LL}_{1,1}$ solutions. The space of these solutions is shown in figure \ref{typeBmoduli}.
 Here the $\f_0$ axis is  excluded as it belongs to the type S solutions. For every point with $S_0>0$ in this region, there is a flipped solution with the same value of $S_0$ but negative.
As an example of these solutions (corresponding to the black dot in figure \ref{typeBmoduli}), we have the figure \ref{AUVLLa}. This flow is connecting two QFTs that are both located on the same fixed point but at different AdS boundaries, after a $\f$-bounce. For both QFTs, the $\beta$-function \eqref{Abeta} vanishes, $\b\rightarrow 0^-$.
If we flip these points with respect to the $\f_0$ axis, (i.e. take $S_0\to -S_0$) the curves of the flow will flip similarly, i.e. $W\to -W$ and $S\to -S$.
\begin{figure}[!ht]
\centering
\begin{subfigure}{0.49\textwidth}
\includegraphics[width=1\textwidth]{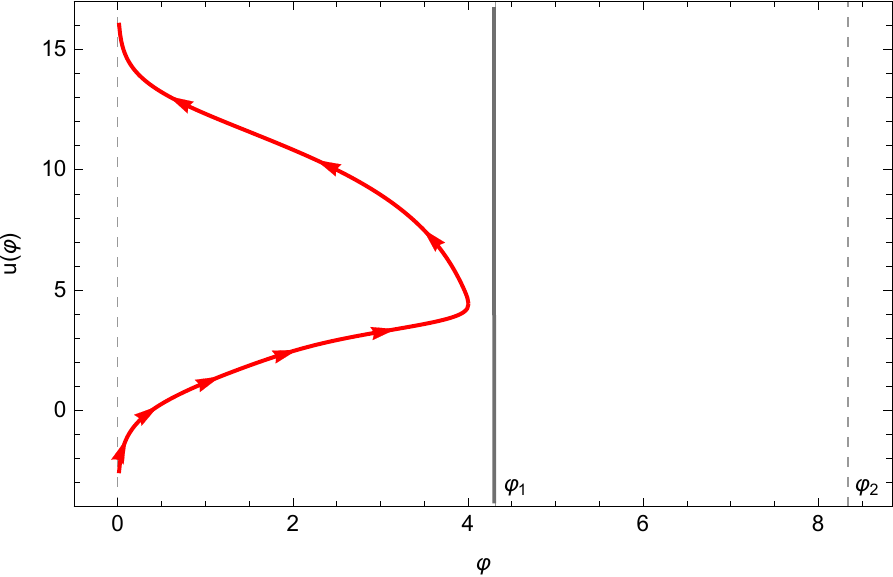}
\caption{}\label{AUVLLb}
\end{subfigure}
\centering
\begin{subfigure}{0.49\textwidth}
\includegraphics[width=1\textwidth]{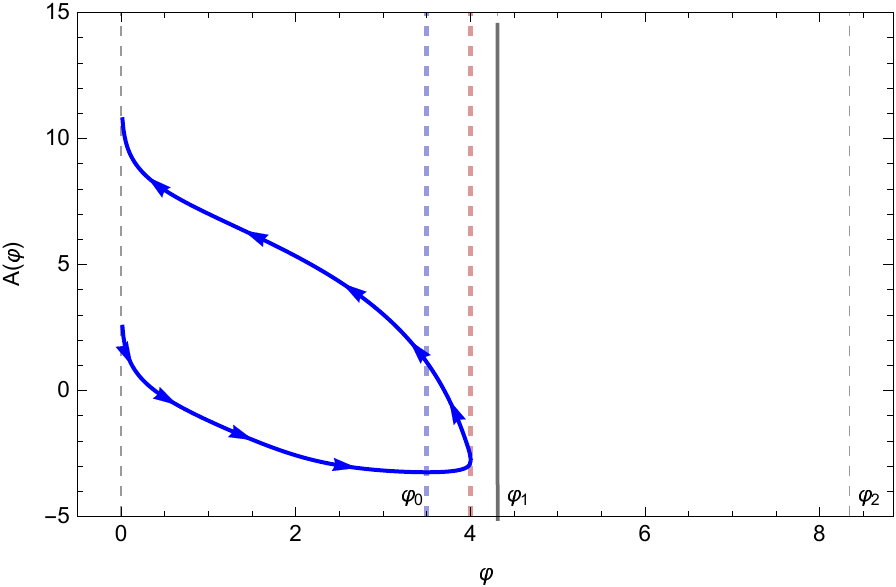}
\caption{}\label{AUVLLc}
\end{subfigure}
\caption{\footnotesize{(a): The holographic coordinate at top $UV_L$ tends to $-\infty$ and at bottom $UV_L$ to $+\infty$. (b): The scale factor has an A-bounce at $\f_0=3.5$ (blue dashed line) and a $\f$-bounce at $\f=4.0$ (red dashed line).}}
\end{figure}

The expansions of the scalar field and scale factor near the $UV_L$ at $W>0$ are given by equations \eqref{phiLU} and \eqref{ALU} and to reach this boundary we should send $u\rightarrow -\infty$. On the contrary, to reach the $UV_L$ boundary at $W<0$ we need $u\rightarrow \infty$, this can be seen in the expansions \eqref{phiLD} and \eqref{ALD}. In figure \ref{AUVLLa} the direction of the  flow is chosen to be from $-\infty$ to $+\infty$.

The dependence of the holographic coordinate and scale factor in terms of $\f$ are presented in figures \ref{AUVLLb} and \ref{AUVLLc}.
The geometry has an A-bounce (when $W(\f)=0$, indicated by a blue dashed line in figure \ref{AUVLLc} or \ref{AUVLLa}). Moreover, there is a $\f$-bounce in the flow (when $S(\f)=0$, the red dashed line).
This geometry can be described as a Janus interface between the same UV theories at different couplings and different curvatures.

\begin{figure}[!ht]
\centering
\begin{subfigure}{0.49\textwidth}
\includegraphics[width=1\textwidth]{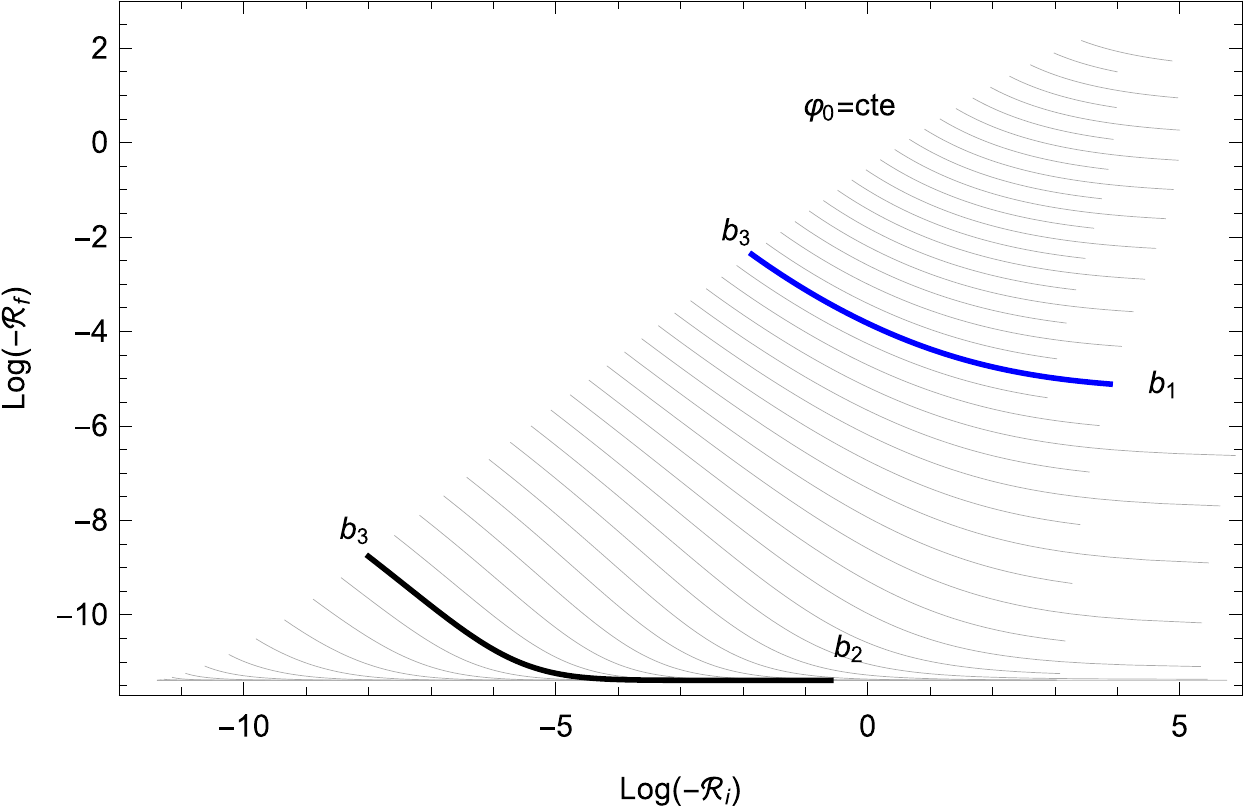}
\caption{}\label{RLRLS0}
\end{subfigure}
\centering
\begin{subfigure}{0.49\textwidth}
\includegraphics[width=1\textwidth]{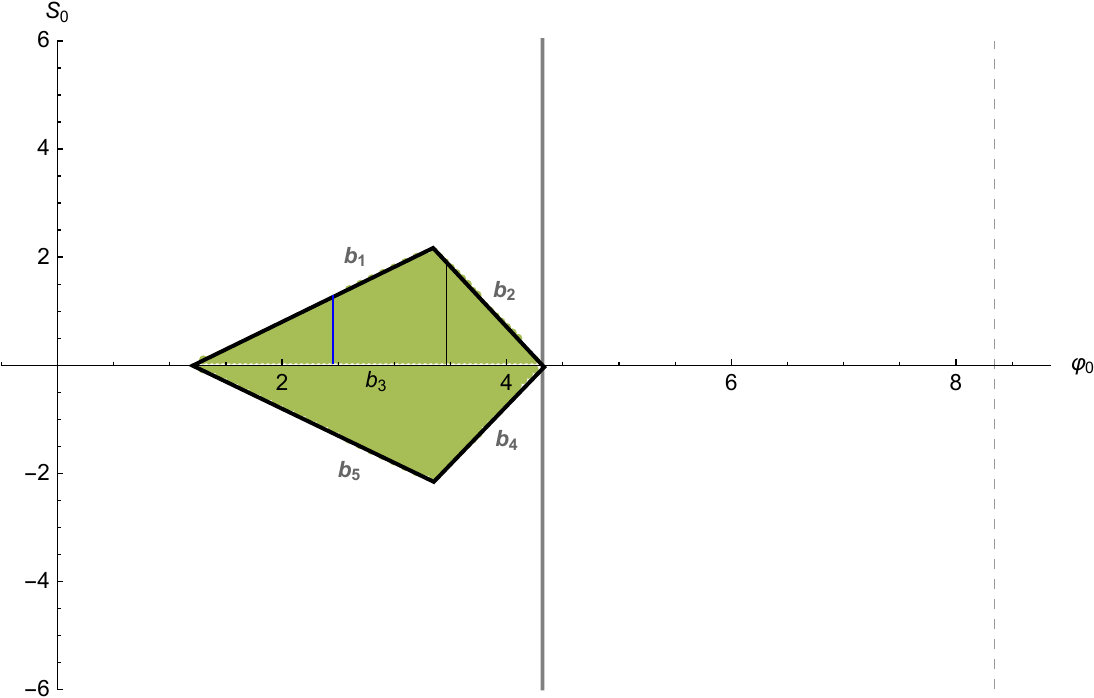}
\caption{}\label{typeBbb}
\end{subfigure}
\caption{\footnotesize{The relation between two dimensionless curvatures at fixed $\f_0$. The blue curve is a typical curve between two boundaries  $b_1$ and $b_3$, and the black curve is also a typical curve between $b_2$ and $b_3$ boundaries. These curves are in a one-to-one relation with the blue and black lines  in figure (b).}}
\end{figure}

To read the dimensionless curvatures, in figure \ref{RLRLS0} we have calculated numerically the relation between dimensionless curvature $\mathcal{R}_i$ for the top $UV_L$ boundary and $\mathcal{R}_f$ for the bottom $UV_L$.
In this figure we portray the behavior of $\log(-\mathcal{R}_f)$ as a function of $\log(-\mathcal{R}_i)$ as we move along a typical constant line $\f_0=cte$ in the moduli space in figure \ref{typeBbb}. The curves are bounded between the $b_1-b_3$ (e.g. the blue curve) or $b_2-b_3$ (e.g. the red curve) boundaries.

\begin{figure}[!ht]
\centering
\begin{subfigure}{0.49\textwidth}
\includegraphics[width=1\textwidth]{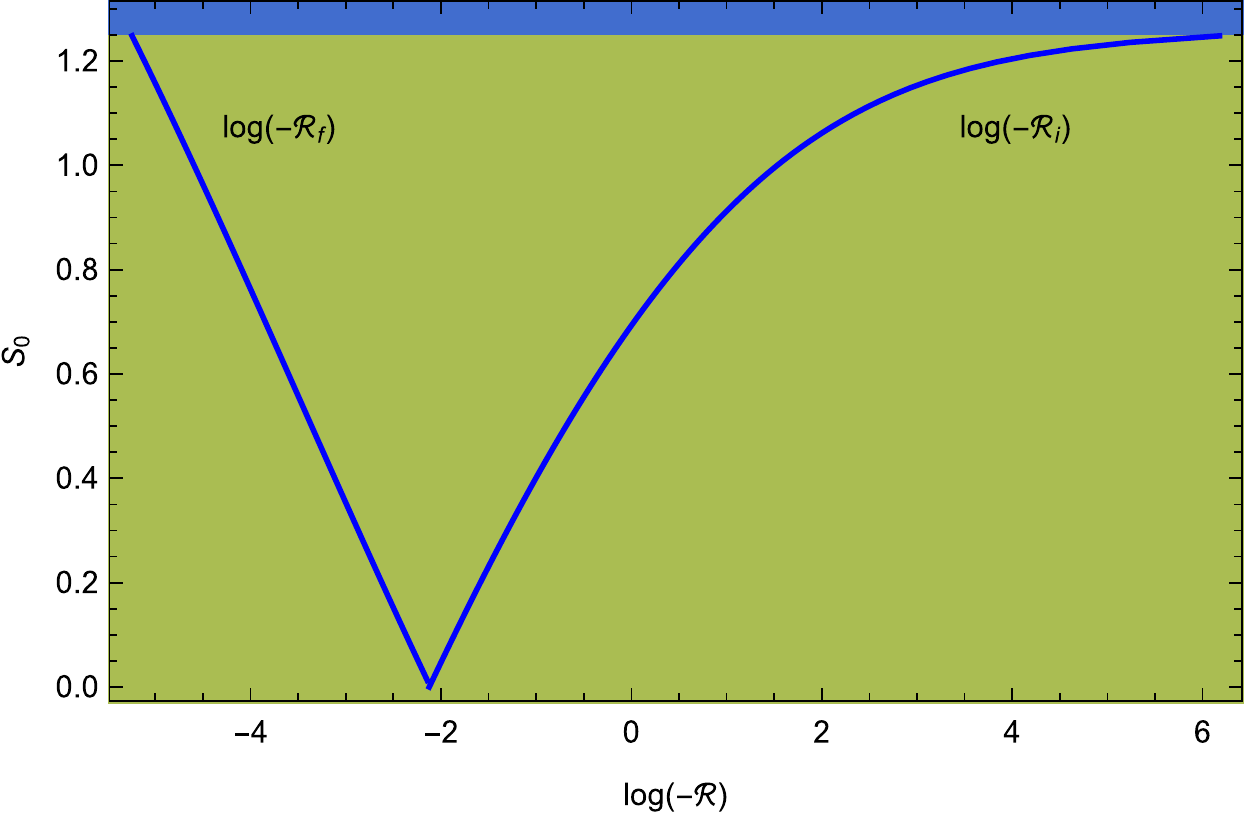}
\caption{}\label{LLfi01}
\end{subfigure}
\centering
\begin{subfigure}{0.49\textwidth}
\includegraphics[width=1\textwidth]{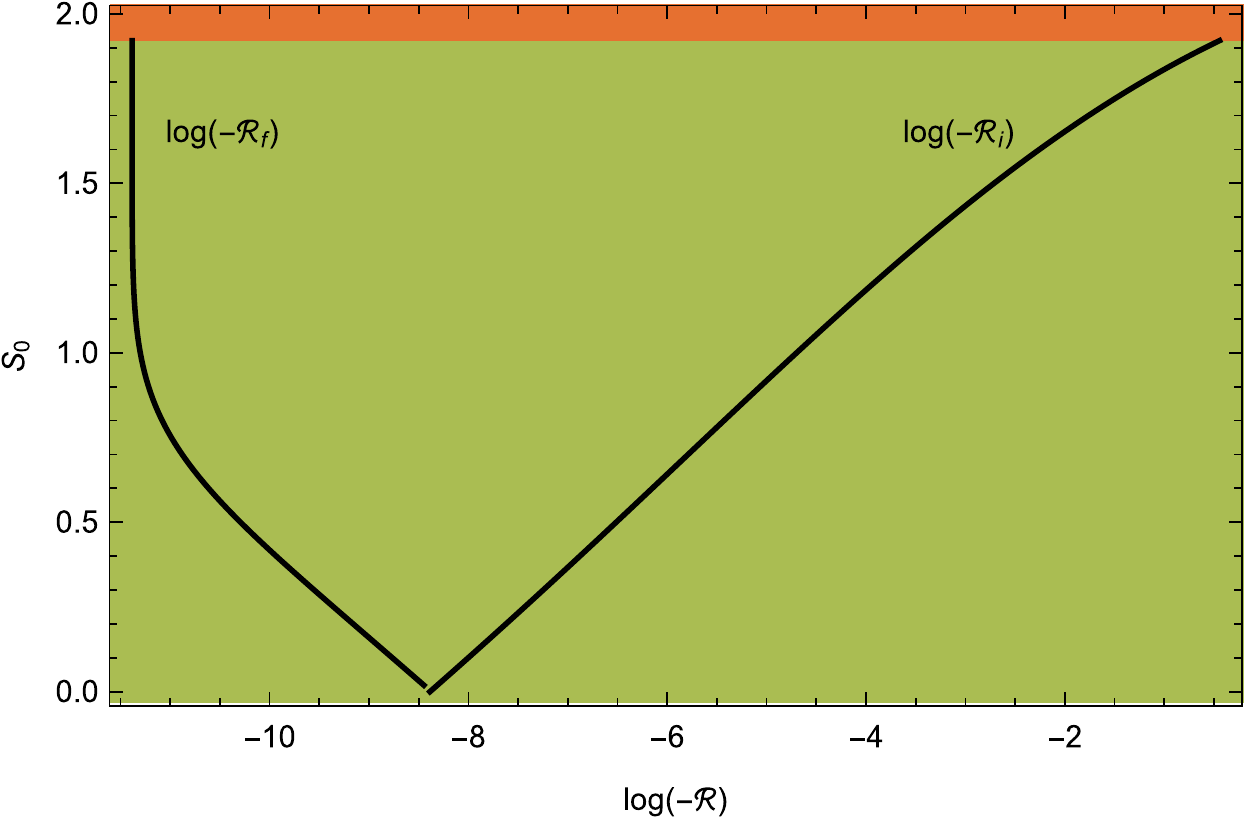}
\caption{}\label{LLfi02}
\end{subfigure}
\caption{\footnotesize{The dimensionless curvatures vs $S_0$, for blue curve with $\f_0=\frac{49}{20}$, and black curve $\f_0=\frac{69}{20}$, in figure \protect\ref{RLRLS0}. The horizontal axis is the logarithm of the dimensionless curvature, either $\mathcal R_{L}$ or $\mathcal R_R$.}}
\end{figure}
To read the behavior of dimensionless curvatures of the two UV boundaries easily, we have redrawn the blue and black curves in figure \ref{RLRLS0} in terms of $S_0$ in figures \ref{LLfi01} and \ref{LLfi02}.

By going close to the boundaries of $W^{LL}_{1,1}$ region, we observe the following solutions (see figure \ref{typeBmoduli}):

$(Max_+,Max_-)$ solutions: As we move towards the $b_1$ boundary (blue-green border in figures \ref{LLfi01}) $\mathcal{R}_i \rightarrow -\infty$ but $\mathcal{R}_f$ remains finite.

$(Max_-,Min_+)$ solutions:  As we move towards the $b_2$ boundary (orange-green border in figures \ref{LLfi02}) $\mathcal{R}_f \rightarrow 0$ but $\mathcal{R}_i$ remains finite.

Type-S solutions: Close to the $b_3$ boundary ($S_0=0$ in the above figures) both $\mathcal{R}_i$ and $\mathcal{R}_f$ curvatures are finite and they become equal to each other.
 The equality of dimensionless curvatures comes back to the fact that at the $b_3$ boundary, the RG solutions of type $W^{LL}_{1,1}$ turn into type S which are symmetric solutions around the center of the flow, see section \ref{typeCsec}.

In addition to the above solutions, at that corner where $b_1$ and $b_2$ boundaries are joined we have a solution of $(Max_+,Min_+)$ type where $\mathcal{R}_i \rightarrow -\infty$ and $\mathcal{R}_f \rightarrow 0$.

\begin{figure}[!ht]
\centering
\begin{subfigure}{0.49\textwidth}
\includegraphics[width=1\textwidth]{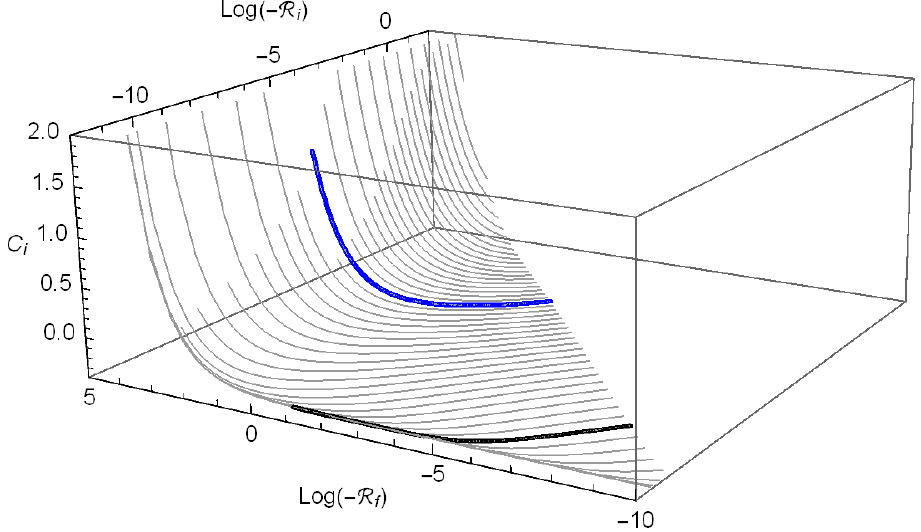}
\caption{}\label{CLif3D}
\end{subfigure}
\begin{subfigure}{0.49\textwidth}
\includegraphics[width=1\textwidth]{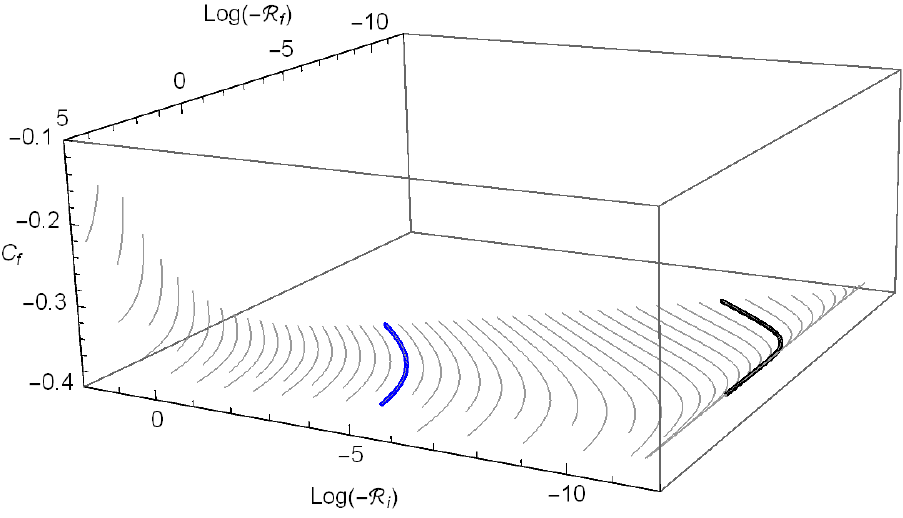}
\caption{}\label{CRif3D}
\end{subfigure}
\caption{\footnotesize{As we change $S_0$ along a $\f_0=constant$ line in the green region in figure \protect\ref{typeBbb} we can read $C_L$ and $C_R$ in terms of the dimensionless curvatures. For example in above figures, the blue curve is at $\f_0=\frac{49}{20}$ and the black one at $\f_0=\frac{69}{20}$.}}
\end{figure}

As we already discussed, $C_i$ and $C_f$ on each UV boundary are not independent parameters but concrete functions of the sources, and this can be seen in the plots of figure \ref{CLif3D} and \ref{CRif3D}.

\begin{figure}[!ht]
\centering
\begin{subfigure}{0.49\textwidth}
\includegraphics[width=1\textwidth]{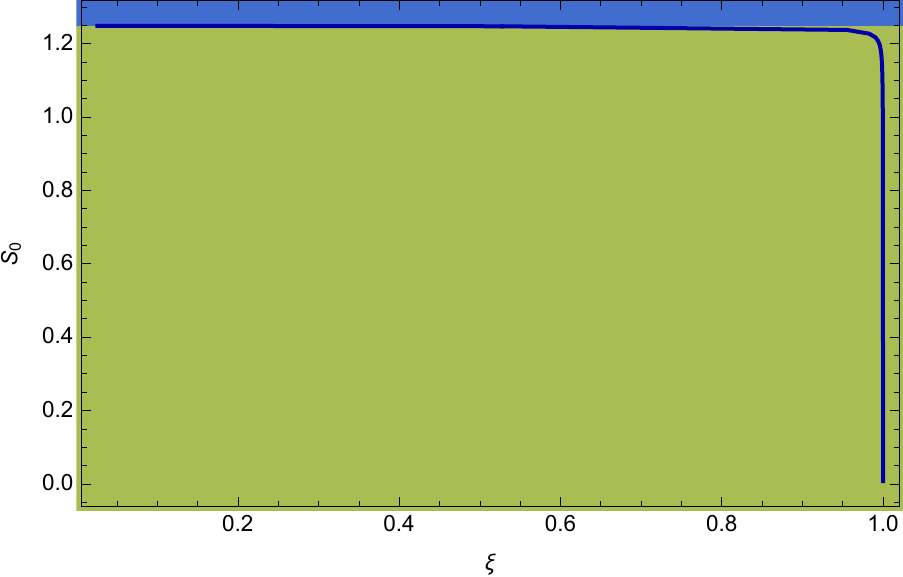}
\caption{}\label{xiru}
\end{subfigure}
\centering
\begin{subfigure}{0.49\textwidth}
\includegraphics[width=1\textwidth]{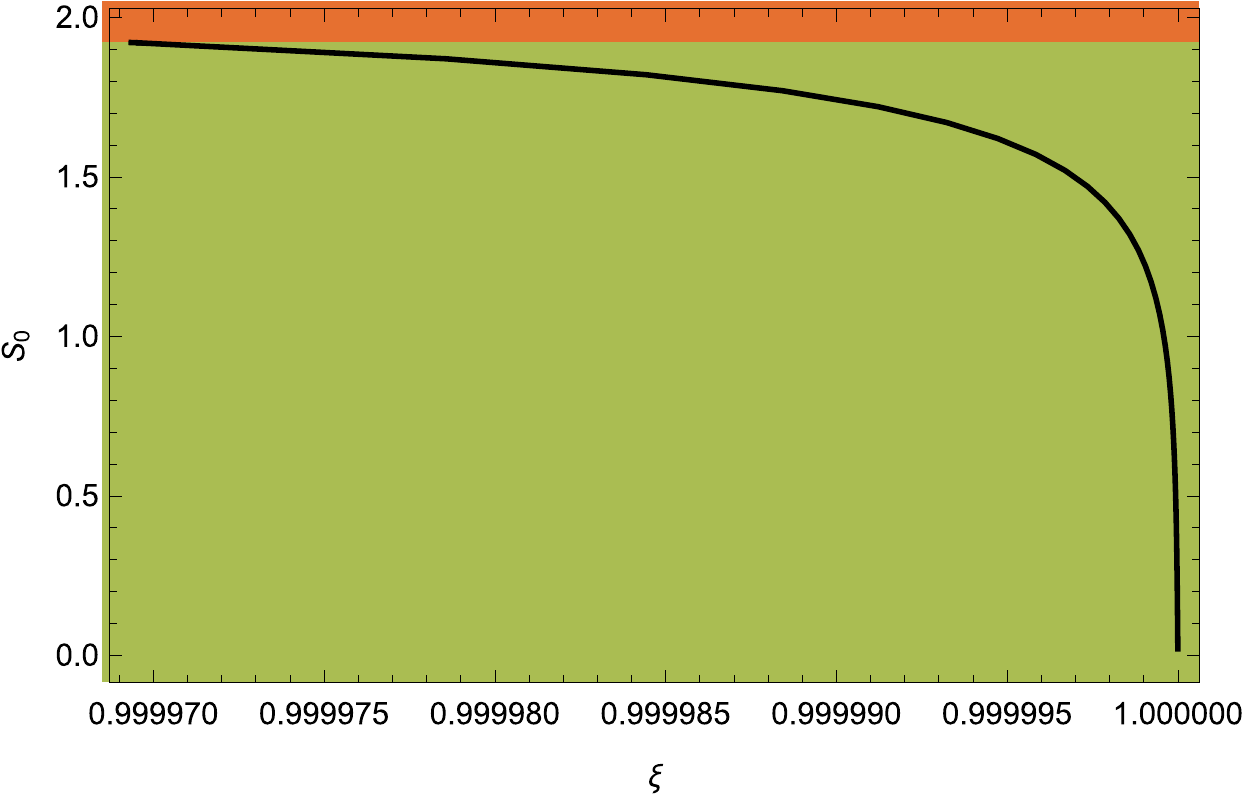}
\caption{}\label{xird}
\end{subfigure}
\caption{\footnotesize{The ratio of the couplings vs $S_0$ for the two typical blue and black curves  shown in figure \protect\ref{RLRLS0}.}}
\end{figure}

Since we have two independent coupling constants associated with the source of $\f$, we expect to see a nontrivial behavior for the ratio of these two couplings.
In figures \ref{xiru} and \ref{xird} we have sketched the behavior of the  $\xi$ in relation \eqref{Eratio} as we move in the space of the $W^{LL}_{1,1}$ solutions  along a constant $\f_0$ line. In summary we have the following properties for this ratio:

1) Very close to the $b_1$ boundary (blue-green border in figure \ref{xiru}) this ratio vanishes because $\mathcal{R}_i \rightarrow -\infty$ and therefore $\f_-^{(i)} \rightarrow 0$.

2) Close to the $b_2$ boundary (orange-green border in figure \ref{xird}) the ratio has a finite value. The reason is the same as what we explained in the $W^{LR}_{1,0}$ case near the $a_4$ boundary ($a_4$ and $b_2$ boundaries are the same).

3) Near the $b_3$ boundary at $S_0=0$, since both $\mathcal{R}_i$ and $\mathcal{R}_f$ becomes equal we expect and find that $\xi \rightarrow 1$.

\subsubsection{Type-S solutions} \label{typeCsec}
\begin{figure}[!ht]
\centering
\begin{subfigure}{0.51\textwidth}
\includegraphics[width=1\textwidth]{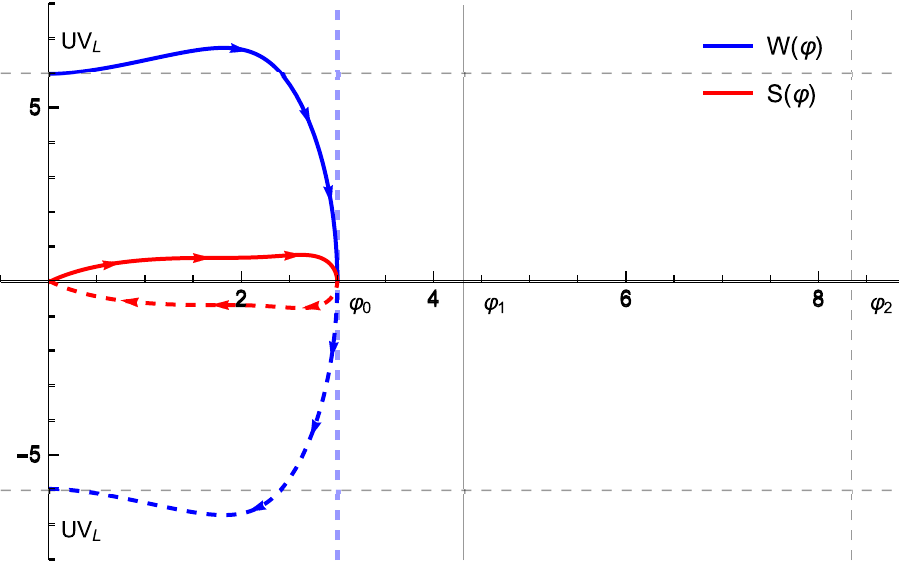}
\caption{}\label{AUVLLIRa}
\end{subfigure}
\centering
\begin{subfigure}{0.49\textwidth}
\includegraphics[width=1\textwidth]{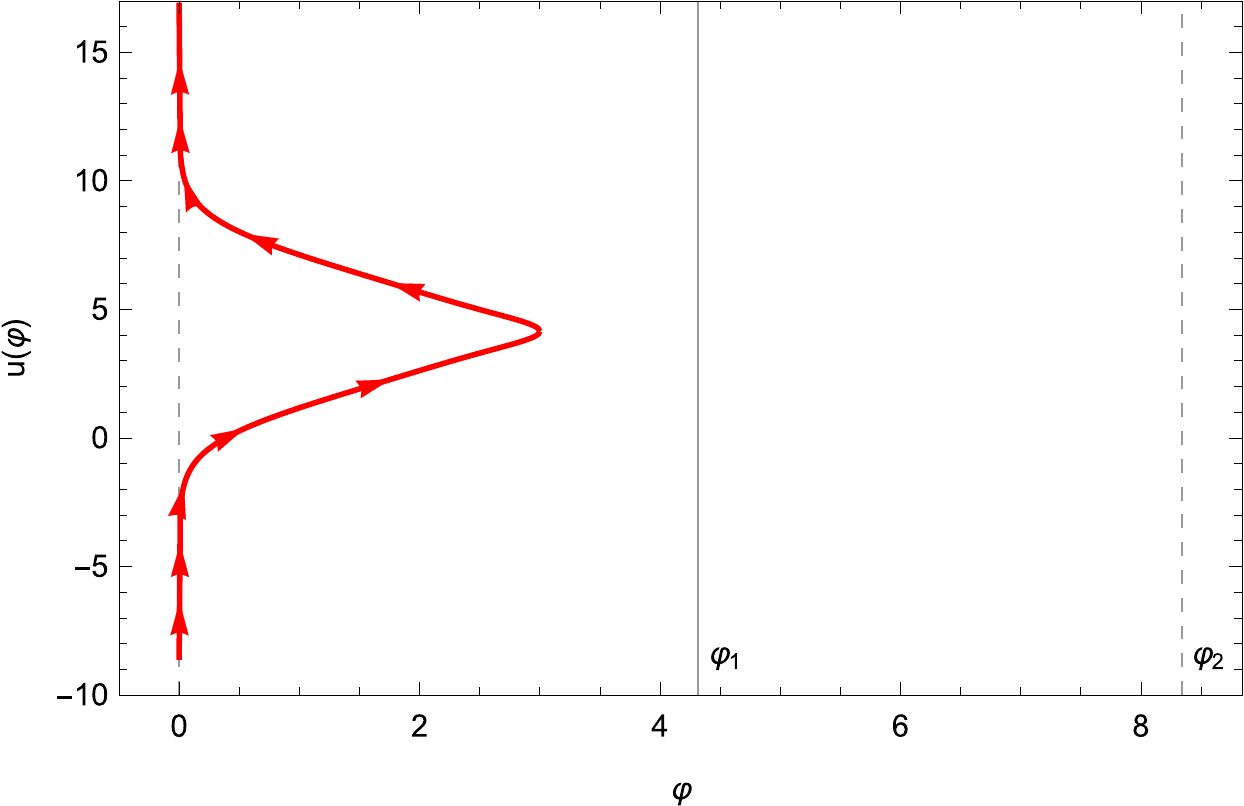}
\caption{}\label{AUVLLIRb}
\end{subfigure}
\centering
\begin{subfigure}{0.49\textwidth}
\includegraphics[width=1\textwidth]{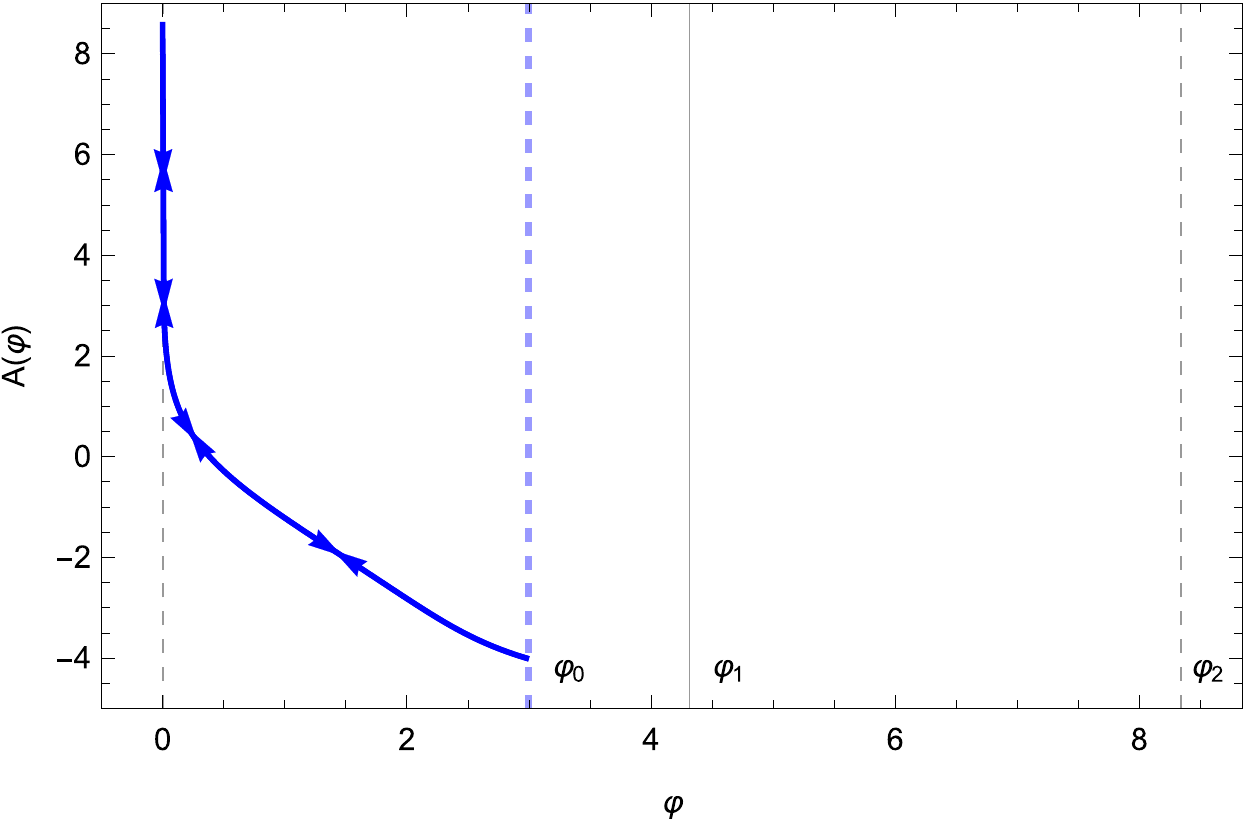}
\caption{}\label{AUVLLIRc}
\end{subfigure}
\caption{\footnotesize{Type S solution. (a): The blue and red curves describe $W,S$ of the  flow that starts from the $UV_L$ boundary and reaches a special  ``IR-bounce" (at $\f=\f_0$, the blue dashed line): this is an A-bounce  of the scale factor as well as a $\f$-bounce for $\f$ at the same place. The flow returns to the same UV fixed point  through the mirror image of the original flow (b): The holographic coordinate at the  $UV_L$ fixed point, tends to $-\infty$ and at
the IR-bounce  it reaches  a finite value. After that,  it  tends to $+\infty$ and returns to the $UV_L$ fixed point again. (c): The scale factor has an A-bounce at $\f_0=3.0$ (at the blue dashed line) and is symmetric on the two sides of the flow.}}
\end{figure}

Type-S solutions are holographic flows that have a $UV_L$ fixed point on one side, $W>0$, and reach an IR turning point, in which both $W$ and $S$ are zero, see figure \ref{AUVLLIRa}. The flow does not stop here and it returns to the same $UV_{L}$ with $W<0$ through a mirror image of the original flow (dashed curves in figure \ref{AUVLLIRa}).
The special point about such solutions is that the scale factor has a minimum, and the scalar field turns around at the same bulk point.
 Therefore these solutions are left-right symmetric as far as the flow is concerned.

The space of solutions is the $S_0=0$ axis. In fact, for any generic fixed point $\f_0$ in this space, the expansions of $W$ and $S$ are given by \eqref{AIREW} and \eqref{AIRES} which are characterized by a single parameter $S_1$. At an extremum of the potential, where $S_1=\pm\sqrt{2V_1}=0$, the solutions are singular,  therefore we should exclude the minima of the potential from the space of regular solutions. For $\f_0<\f_1$ the flow connects $UV_L$ fixed point to itself, but for $\f_1<\f_0$ it connects $UV_R$ to itself.

As we see from diagram \ref{AUVLLIRc}, the scale factor reaches a minimum and the geometry of the space-time at this point described by \eqref{IRendmetr} is a regular piece of the geometry.
Figure \ref{AUVLLIRb} shows the behavior of the holographic coordinate.   In all the $W^{LL}_{1,1}$ solutions, the geometry, looks like a wormhole that connects two UV boundaries at $u\rightarrow \pm \infty$ but these boundaries belong to the same UV fixed point. But the type-S solutions are symmetric and therefore the couplings on the two boundary theories are equal, i.e.
$\xi=1$.

Due to the symmetry of the solutions, we always have $\mathcal{R}_i=\mathcal{R}_j$, (see figures \ref{LLfi01} or \ref{LLfi02} where close to $S_0=0$ both curves arrive at a single point). Consequently, this symmetry implies that the ratio of two couplings is $\xi=1$. This behavior can be seen for example in figures \ref{xiru} and \ref{xird} where both black or blue curves reach $\xi=1$ as we approach $S_0=0$.

\subsection{Solutions with extra $\f$-bounces}
\subsubsection{ The $W^{LL}_{1,2}$ solutions}\label{blue}
\begin{figure}[!ht]
\centering
\begin{subfigure}{0.59\textwidth}
\includegraphics[width=1\textwidth]{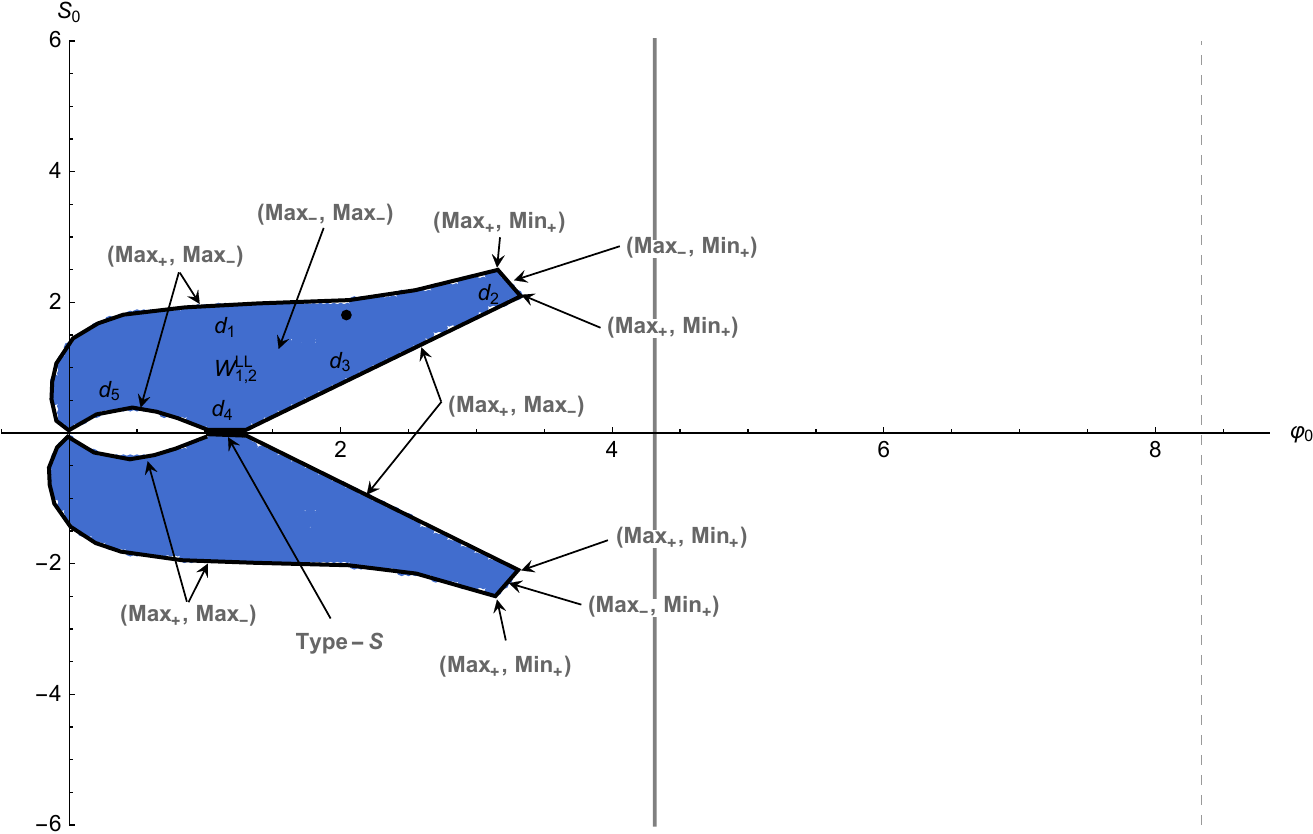}
\caption{}\label{typeDmoduli}
\end{subfigure}
\centering
\begin{subfigure}{0.4\textwidth}
\includegraphics[width=1\textwidth]{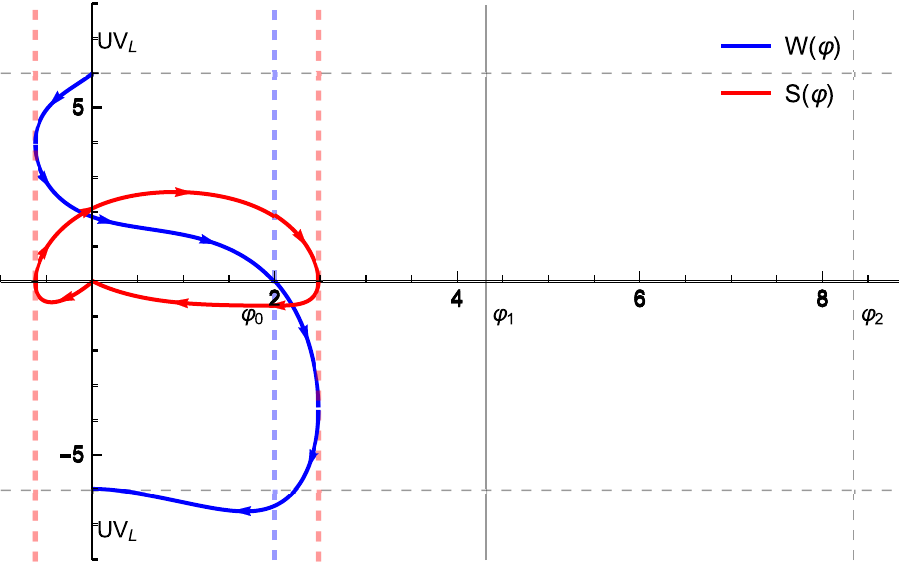}
\caption{\vspace*{0cm}}\label{A2BLLa}
\end{subfigure}
\caption{\footnotesize{(a): The space of the $W^{LL}_{1,2}$ solutions is the upper blue region. The black dot represents the specific solutions of diagram (b). The lower blue region corresponds to the solutions with an extra $\f$-bounce near the bottom $UV_L$. (b): The blue and red curves for $W,S$, describe a flow that connects the $UV_L$ fixed point to itself but after two $\f$-bounces. The locations of the $\f$-bounces are indicated by red dashed lines.}}
\end{figure}

The space of the $W^{LL}_{1,2} \in (Max_-,Max_-)$ solutions is sketched in figure \ref{typeDmoduli}, which also includes a region with $\f<0$. The RG solution in figure \ref{A2BLLa} corresponds to the black dot in this space as is shown in figure \ref{typeDmoduli}. For every solution in the upper part of this space, there is a flipped solution in the lower part. The flipped solutions have an extra $\f$-bounce point near the bottom $UV_L$ boundary.

In figure \ref{A2BLLa}, a QFT on the top $UV_L$ boundary with $\mathcal{R}_i$ dimensionless curvature is connected via a flow after two $\f$-bounces to another QFT on the bottom $UV_L$ boundary but with $\mathcal{R}_f$ curvature. The flow first moves to the left (note that if we enlarge the region very close to the $\f=0$ then $W'\rightarrow 0$) and after a $\f$-bounce at the negative values of $\f$, it returns to the region $\f>0$. Another $\f$-bounce changes the direction of the flow towards the left UV fixed-point again.
This means that the two UV theories that are at the end-points of this flow have opposite sign relevant couplings.

\begin{figure}[!t]
\centering
\begin{subfigure}{0.49\textwidth}
\includegraphics[width=1\textwidth]{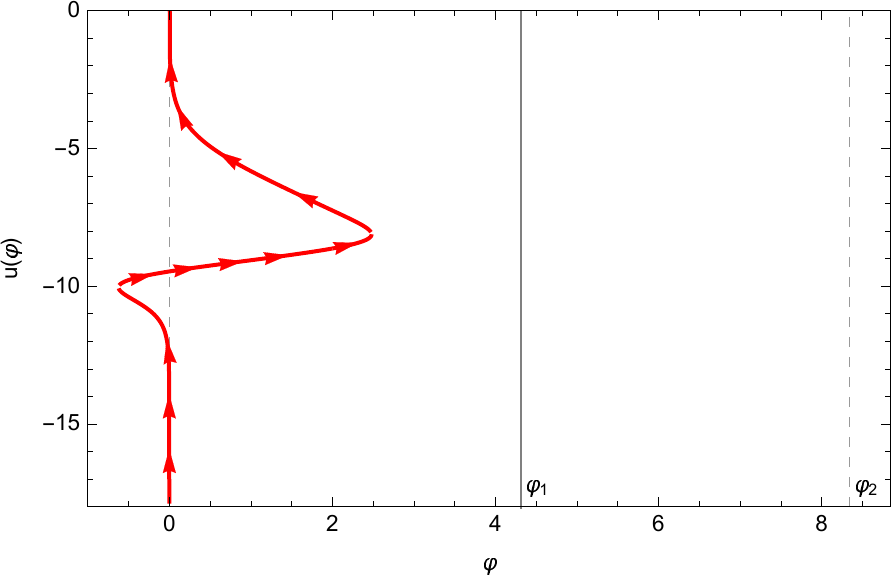}
\caption{}\label{A2BLLb}
\end{subfigure}
\centering
\begin{subfigure}{0.49\textwidth}
\includegraphics[width=1\textwidth]{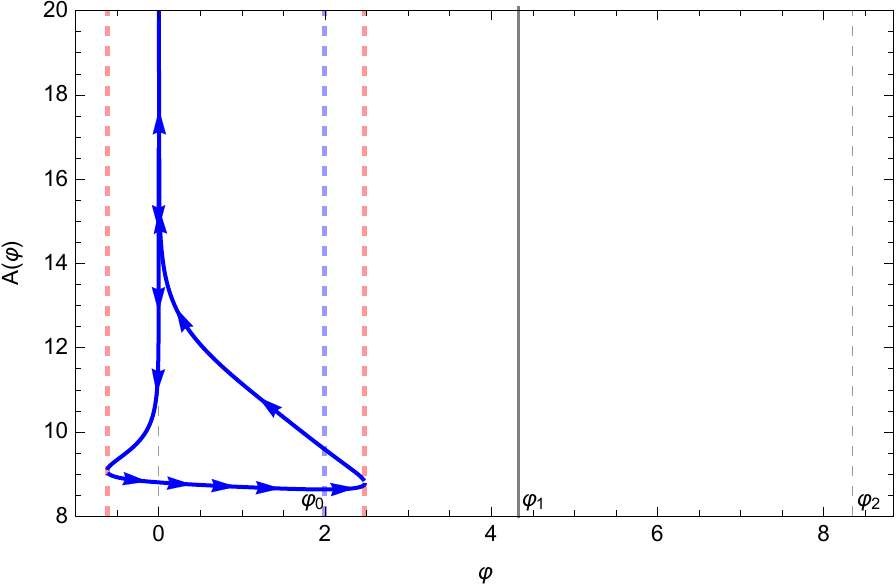}
\caption{}\label{A2BLLc}
\end{subfigure}
\caption{\footnotesize{(a): The holographic coordinate at top $UV_L$ boundary tends to $-\infty$ and for bottom $UV_L$ to $ +\infty$. (b): The scale factor has an A-bounce at $\f=2.0$, the blue dashed line. The first $\f$-bounce on the left occurs at $\f=-0.62$ and the second one at $\f=2.48$, the red dashed lines.}}
\end{figure}

Figures \eqref{A2BLLb} and \eqref{A2BLLc} show the behavior of the holographic coordinate and scale factor in terms of $\f$ for the specific flow of figure \ref{A2BLLa}. Again the Euclidean geometry of space-time describes a wormhole with a minimum length at the point where the scale factor has an A-bounce, $\f=2.0$.

\begin{figure}[!t]
\centering
\begin{subfigure}{0.49\textwidth}
\includegraphics[width=1\textwidth]{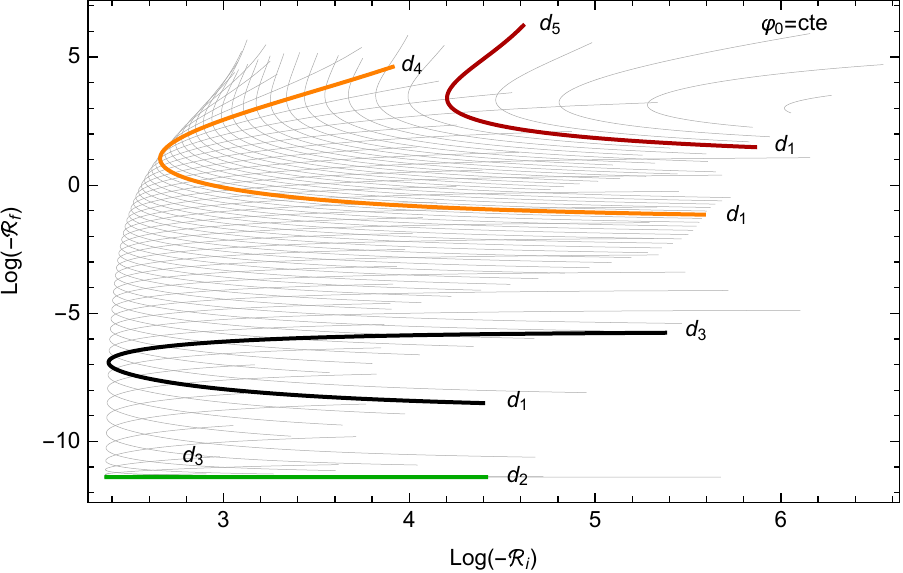}
\caption{}\label{RLRLBlue}
\end{subfigure}
\centering
\begin{subfigure}{0.49\textwidth}
\includegraphics[width=1\textwidth]{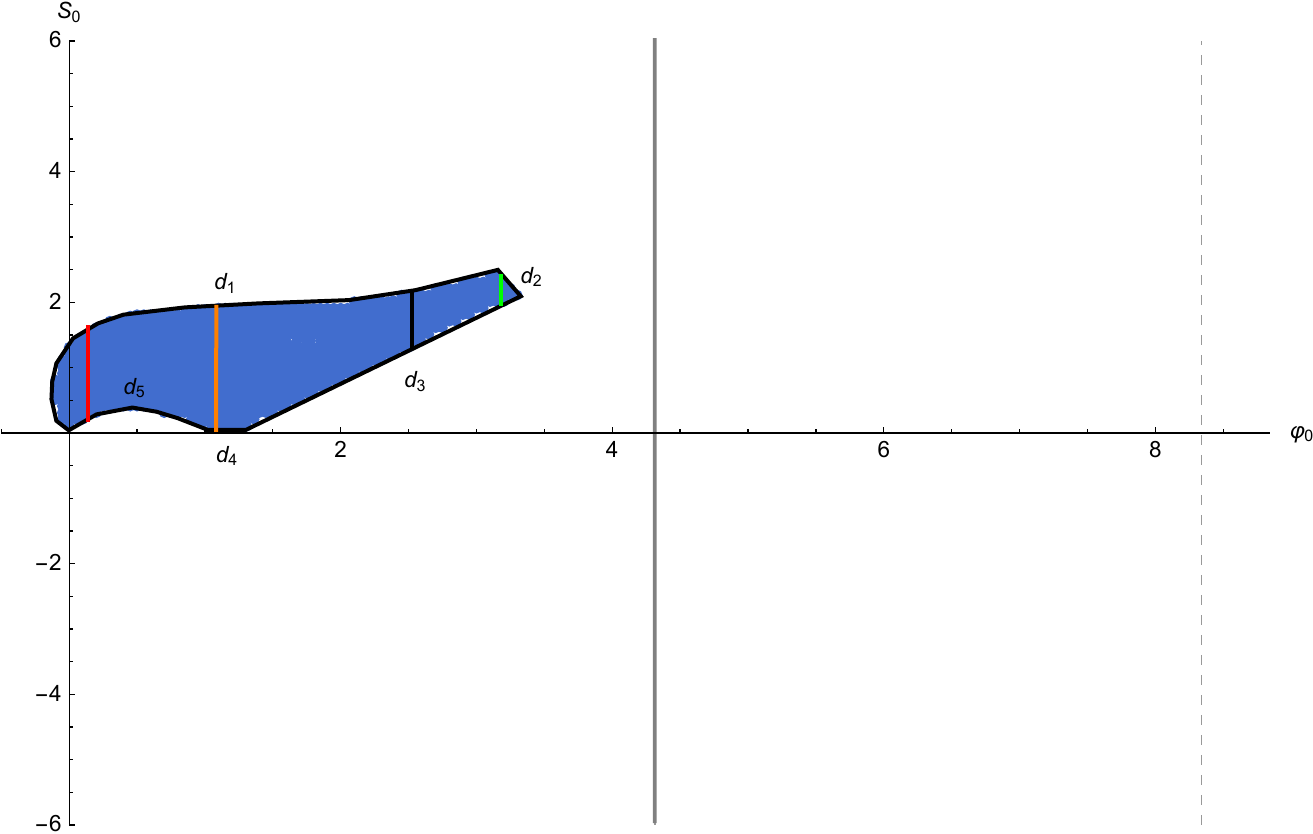}
\caption{}\label{typeDbb}
\end{subfigure}
\caption{\footnotesize{In $W^{LL}_{1,2}$ region we can read the behavior of $\mathcal{R}_f$ in terms of $\mathcal{R}_i$ for different constant values of $\f_0$. The colored curves represent the generic behavior of the dimensionless curvatures as we move between different boundaries of the space of solutions in figure (b). For the red curve $\f_0=\frac{1}{10}$, the orange curve $\f_0=\frac{11}{10}$, the black curve $\f_0=\frac{25}{10}$ and the green curve $\f_0=\frac{31}{10}$.}}
\end{figure}

\begin{figure}[!hb]
\centering
\begin{subfigure}{0.32\textwidth}
\includegraphics[width=1\textwidth]{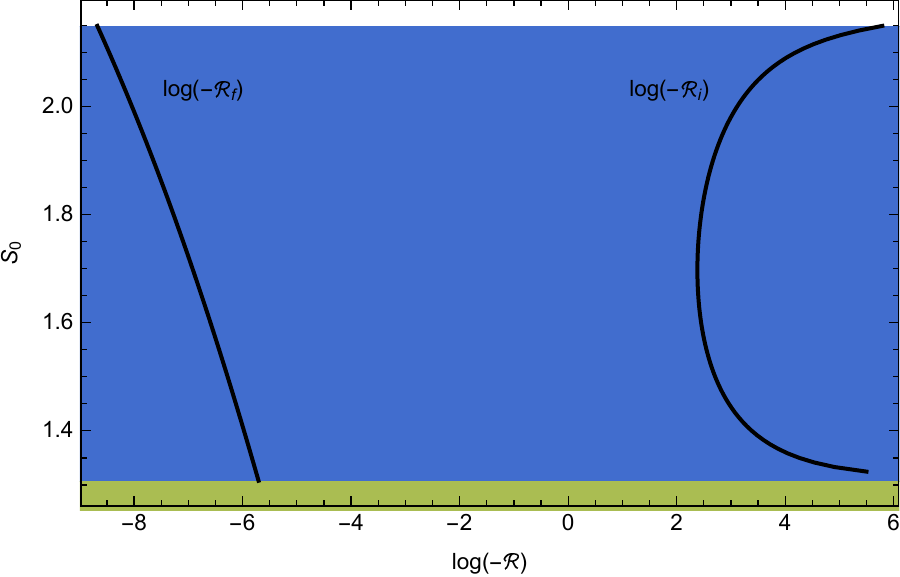}
\caption{}\label{d1d3}
\end{subfigure}
\centering
\begin{subfigure}{0.32\textwidth}
\includegraphics[width=1\textwidth]{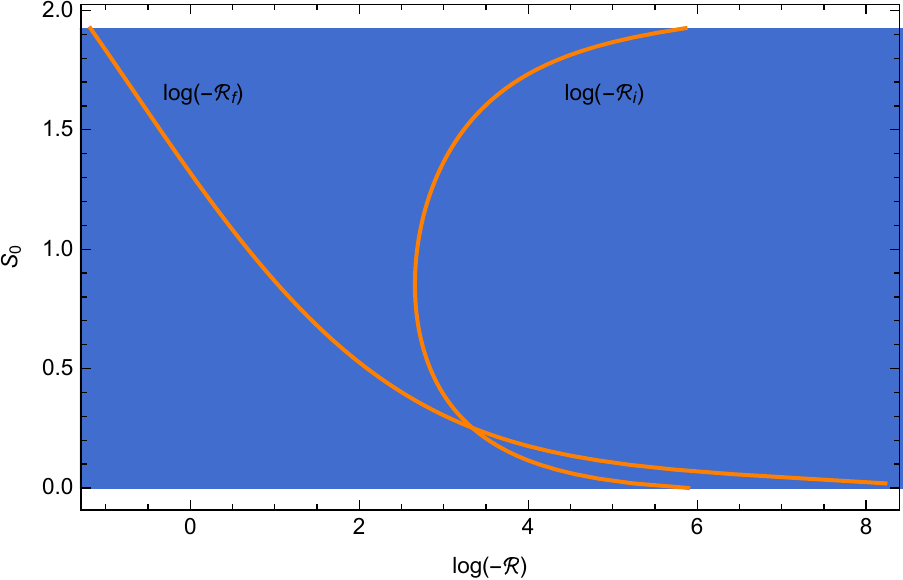}
\caption{}\label{d1d4}
\end{subfigure}
\centering
\begin{subfigure}{0.32\textwidth}
\includegraphics[width=1\textwidth]{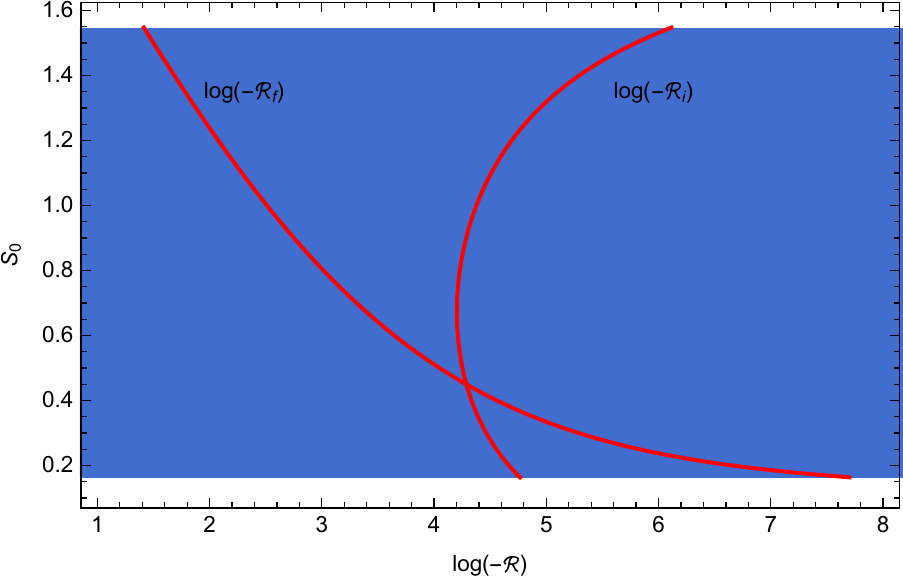}
\caption{}\label{d1d5}
\end{subfigure}
\centering
\begin{subfigure}{0.32\textwidth}
\includegraphics[width=1\textwidth]{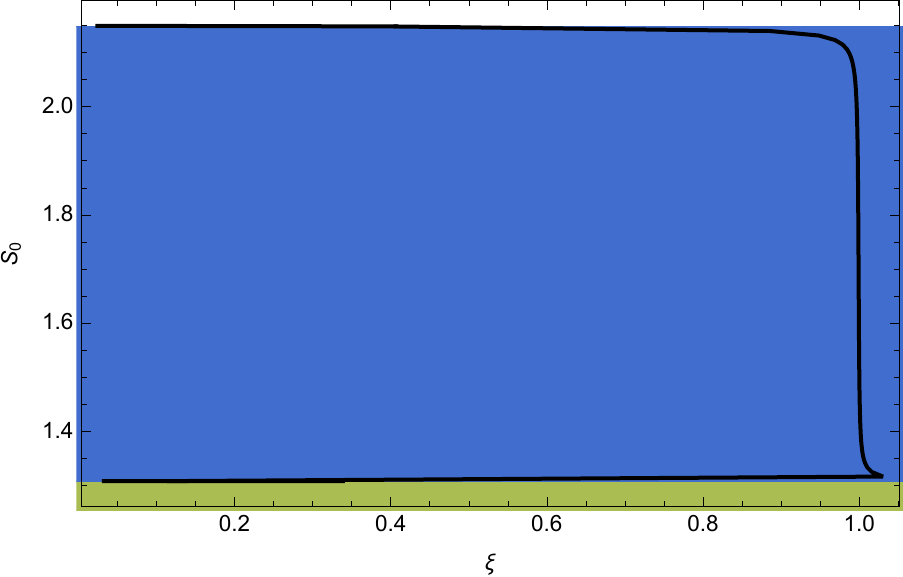}
\caption{}\label{d1d3x}
\end{subfigure}
\centering
\begin{subfigure}{0.32\textwidth}
\includegraphics[width=1\textwidth]{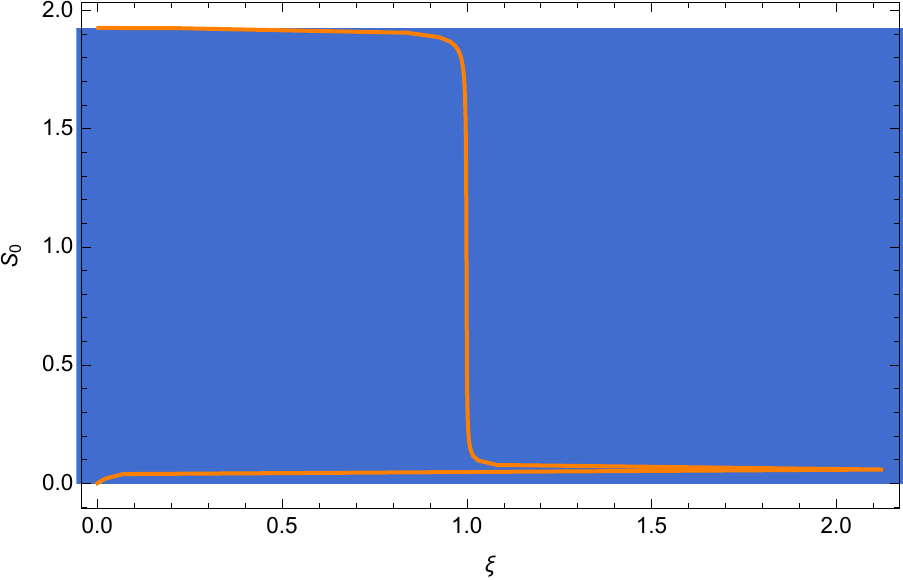}
\caption{}\label{d1d4x}
\end{subfigure}
\centering
\begin{subfigure}{0.32\textwidth}
\includegraphics[width=1\textwidth]{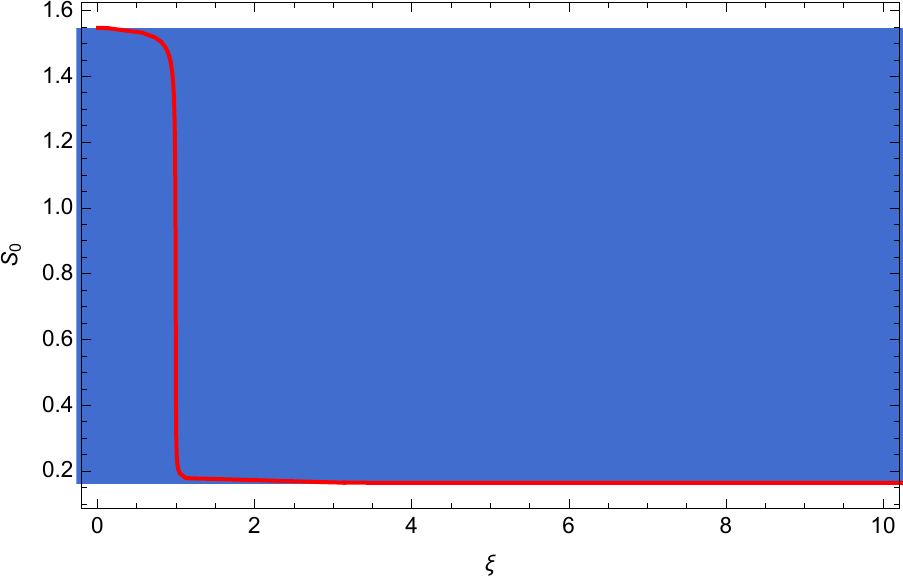}
\caption{}\label{d1d5x}
\end{subfigure}
\caption{\footnotesize{Dimensionless parameters of the $W^{LL}_{1,2}$ solutions with $\f_0=\frac{25}{10}$, $\f_0=\frac{11}{10}$ and $\f_0=\frac{1}{10}$ (black, orange and red curves in figure \protect\ref{RLRLBlue}).
Notice that in figures (d) and (e) at the bottom of the graphs, both curves first go to right and then return to $\xi=0$.}}
\end{figure}

We can determine the corresponding parameters of each QFT i.e. $(\mathcal{R}_i, C_i)$
and $(\mathcal{R}_f, C_f)$ by comparing the numerical solutions of equations of motion with the expansions \eqref{WLU}, \eqref{SLU}, \eqref{WLD} and \eqref{SLD}. The relation between dimensionless curvatures is sketched for constant values of $\f_0$ in figure \ref{RLRLBlue}. We observe that their functional dependence is different from type $W^{LL}_{1,1}$ (figure \ref{RLRLS0}) due to the existence of the extra $\f$-bounce at $\f<0$.  To show the behavior of these curves in different regions of the space of $W^{LL}_{1,2}$ solutions better, we have specified four different groups of solutions in figure \ref{RLRLBlue} which are placed between different boundaries of this space. According to each group, we have the following properties close to the boundaries of the space of $W^{LL}_{1,2}$ solutions (figure \ref{typeDmoduli}):

1) Close to the $d_1$ boundary, $\mathcal{R}_i \rightarrow -\infty$ but $\mathcal{R}_f$ remains finite, this is the properties of $(Max_+,Max_-)$ solutions. This can be seen for different regions in the space of solutions, for example see the upper part of figures \ref{d1d3}, \ref{d1d4} and \ref{d1d5}. The reason is that beyond the $d_1$ boundary, there exist other solutions with even more $\f$-bounces. In other words, whenever a new $\f$-bounce is happening around the left fixed point at $W>0$ a new region like the $W^{LL}_{1,2}$ region is created and at its new boundary we observe that $\mathcal{R}_i \rightarrow -\infty$. Because of the divergence of $\mathcal{R}_i$, the ratio of two relevant couplings vanishes, $\xi\rightarrow 0$, as seen in the upper part of figures \ref{d1d3x}, \ref{d1d4x} and \ref{d1d5x}.

2) Close to the $d_3$ boundary,  $\mathcal{R}_i \rightarrow -\infty$ while $\mathcal{R}_f$ remains finite (see the blue-green border of figure \ref{d1d3}). The reason for the divergence of $\mathcal{R}_i$ is similar to the $d_1$ boundary. As we move from type $W^{LL}_{1,1}$ to type $W^{LL}_{1,2}$, an extra $\f$-bounce appears at the negative values of $\f$. This means that at the boundary of these two regions ($d_3$ or $b_1$) the sign of the coupling of the QFT on the top-left UV boundary is changing. So at this boundary $\f_-^{(i)}$ vanishes and according to the equation \eqref{Eratio} this means that $\xi\rightarrow 0$. This behavior can be seen close to the blue-green border of figure \ref{d1d3x}.

\begin{figure}[!ht]
\centering
\begin{subfigure}{0.49\textwidth}
\includegraphics[width=1\textwidth]{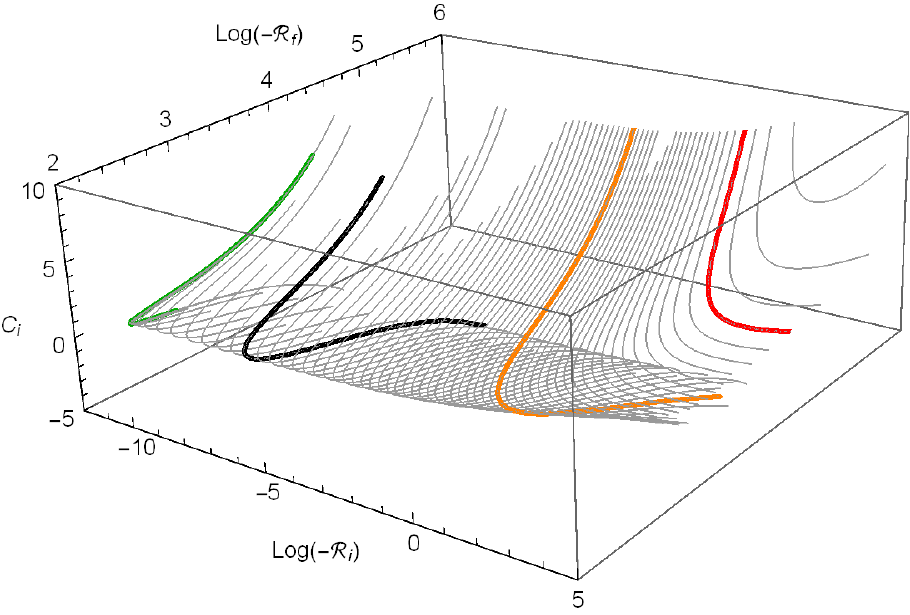}
\caption{}\label{CLif3DB}
\end{subfigure}
\begin{subfigure}{0.49\textwidth}
\includegraphics[width=1\textwidth]{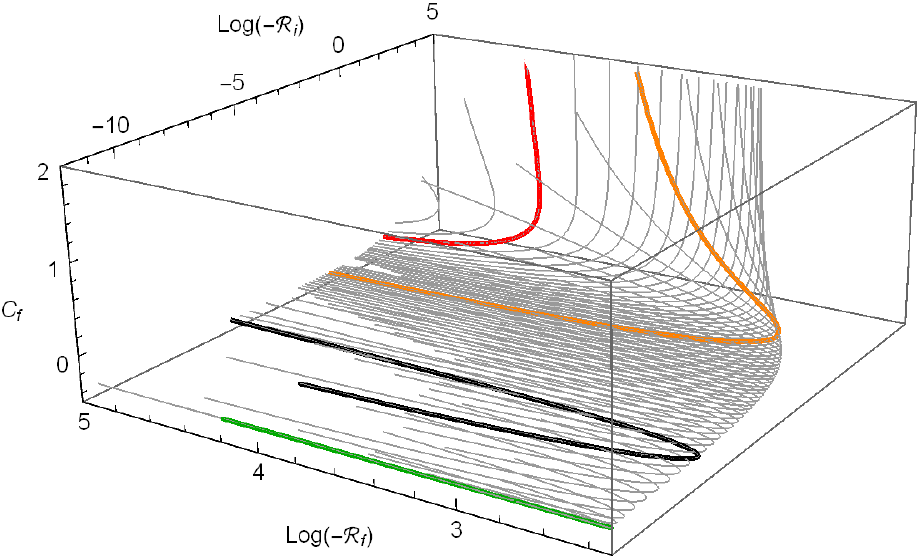}
\caption{}\label{CRif3DB}
\end{subfigure}
\caption{\footnotesize{As we change $S_0$ along a $\f_0=constant$ line in the blue region in figure \protect\ref{typeDbb} we can read $C_i$ and $C_f$ in terms of the dimensionless curvatures. For example in above figures, the red curve is at $\f_0=\frac{1}{10}$, the orange curve at $\f_0=\frac{11}{10}$, the black curve  at $\f_0=\frac{25}{10}$ and the green one at $\f_0=\frac{31}{10}$.}}
\end{figure}

3) Close to the $d_4$ boundary, we expect the same behavior as near the $d_3$ boundary. Since near the $\f_0$ axis the solutions become symmetric, we expect that the value of $\mathcal{R}_f$ is comparable to $\mathcal{R}_i$ (see figure \ref{d1d4} near $S_0=0$). The ratio of the  two relevant couplings is plotted in figure \ref{d1d4x}.

4) According to figure \ref{d1d5} and close to the $d_5$ boundary (the lower part of figure), $\mathcal{R}_f \rightarrow -\infty$ and $\mathcal{R}_i$ remains finite. This boundary is a joint border between the $W^{LL}_{1,2}$ solutions and solutions that have multiple $\f$-bounces, but with a $\f$-bounce that appears on the left of the UV boundary at $W<0$. This can be confirmed in figure \ref{d1d5x} which near the $d_5$ boundary the value of $\xi\rightarrow +\infty$.

5) Near the $d_2$ boundary $\mathcal{R}_i \rightarrow 0$ and $\mathcal{R}_f$ remains finite, which are the properties of the $(Max_-,Min_+)$ solutions. This is  expected again because the border belongs to type $W^{L Min_+}_{1,0}$ solutions that we shall discuss in section \ref{crossAtoB}.

Finally, figures \ref{CLif3DB} and \ref{CRif3DB} show the behavior of $C_i$ and $C_f$ in terms of dimensionless curvatures. Each curve in these figures belongs to solutions with the same value of $\f_0$ in the space of the solutions. For example the colored curves correspond to colored lines in figure \ref{typeDbb}.

\subsubsection{The $W^{LR}_{1,1}$ solutions}\label{red}
\begin{figure}[!ht]
\centering
\begin{subfigure}{0.59\textwidth}
\includegraphics[width=1\textwidth]{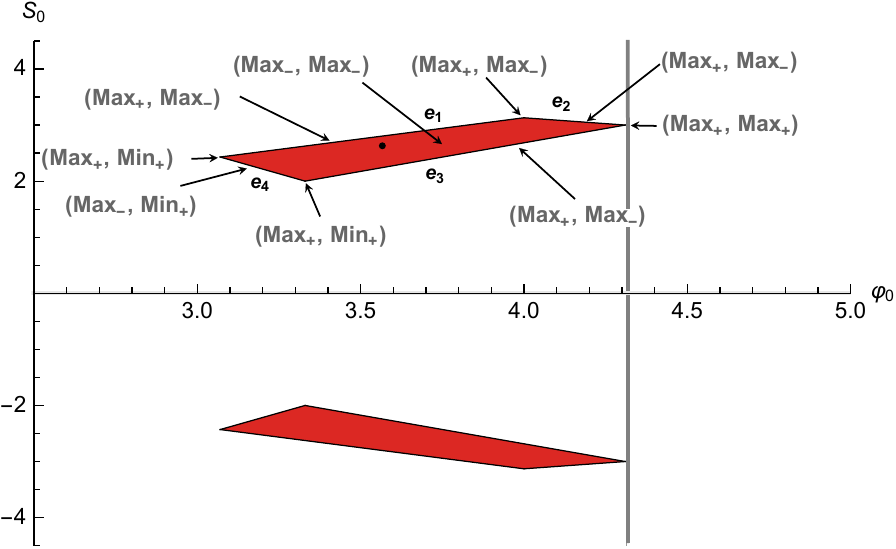}
\caption{}\label{typeEmoduli}
\end{subfigure}
\centering
\begin{subfigure}{0.4\textwidth}
\includegraphics[width=1\textwidth]{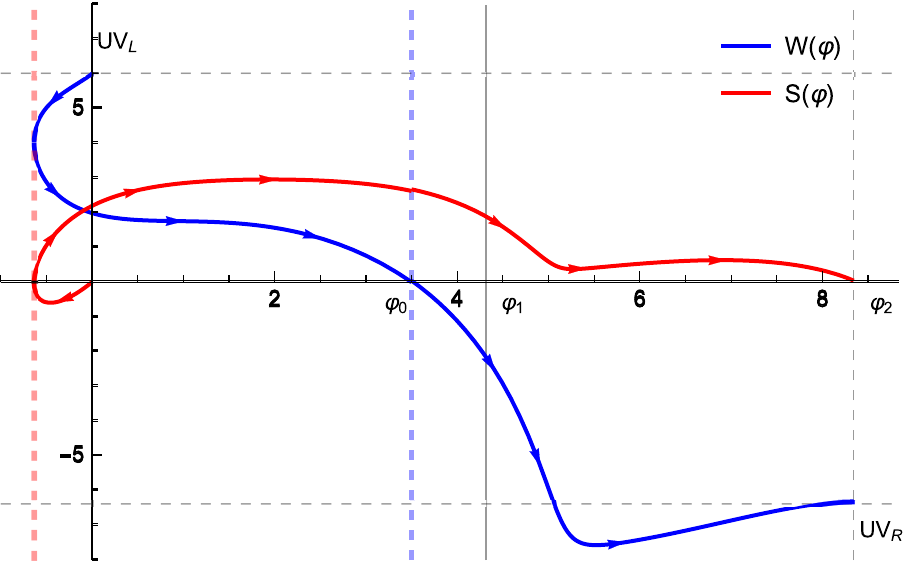}
\caption{
}\label{ABLRa}
\end{subfigure}
\caption{\footnotesize{(a): A zoomed picture of the  space of the $W^{LR}_{1,1}$ solutions. The black dot represents the flow in the diagram (b). (b): The  flows of type $W^{LR}_{1,1}$ are between the $UV_L$ boundary and  $UV_R$. There is a $\f$-bounce at $\f<0$, the red dashed line.  Notice that the red region at $S_0<0$ in figure (a) is the space of solutions with an extra $\f$-bounce near $UV_L$ but at $W<0$.}}
\end{figure}

\begin{figure}[!ht]
\centering
\begin{subfigure}{0.49\textwidth}
\includegraphics[width=1\textwidth]{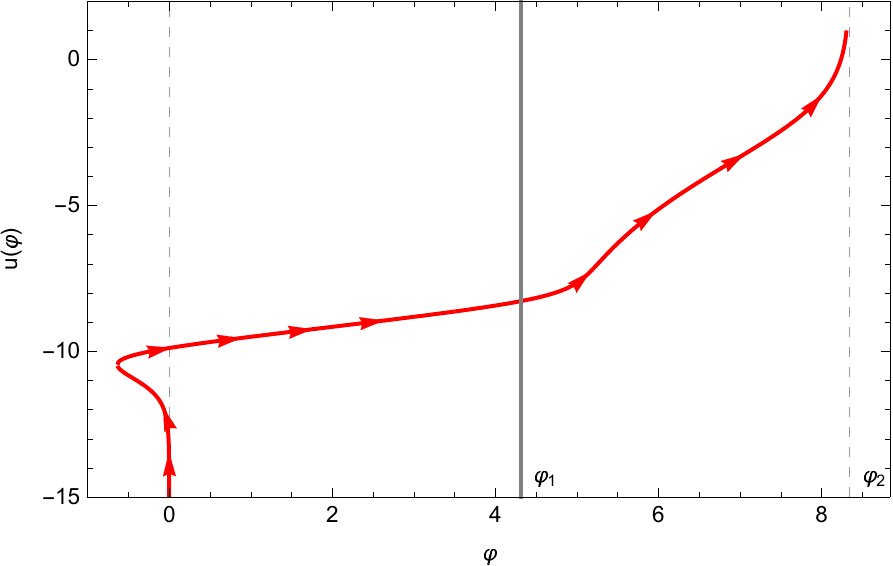}
\caption{}\label{ABLRb}
\end{subfigure}
\centering
\begin{subfigure}{0.49\textwidth}
\includegraphics[width=1\textwidth]{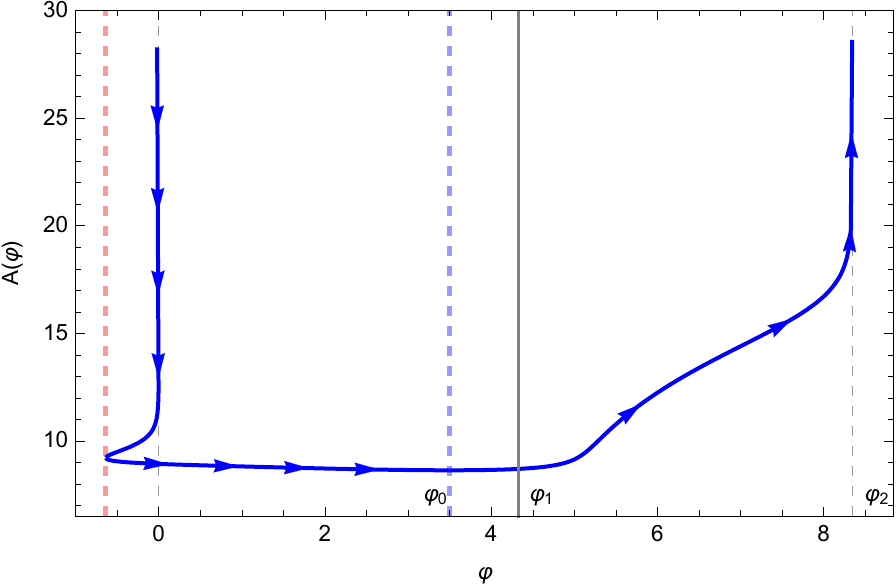}
\caption{}\label{ABLRc}
\end{subfigure}
\caption{\footnotesize{(a): The holographic coordinate at $UV_L$ boundary tends to $-\infty$ and at $UV_R$ to $ +\infty$. (b): The scale factor has an A-bounce at $\f_0=3.5$, the blue dashed line. A $\f$-bounce occurs at $\f=-0.64$, the red dashed line.}}
\end{figure}

The space of solutions  $W^{LR}_{1,1}\in (Max_-,Max_-)$ is sketched in figure \ref{typeEmoduli}.
Figure \ref{ABLRa}, shows a flow (related to the black dot in figure (a)) with two fixed points one on the $UV_L$ boundary ($W>0$) and the other one on the $UV_R$ ($W<0$). The distinction between this type and type $W^{LR}_{1,0}$ solution is the existence of an extra $\f$-bounce at $\f<0$.
Figures \eqref{ABLRb} and \eqref{ABLRc} show the behavior of holographic coordinate and scale factor in terms of $\f$ related to the solutions in figure \ref{ABLRa}.

\begin{figure}[!t]
\centering
\begin{subfigure}{0.49\textwidth}
\includegraphics[width=1\textwidth]{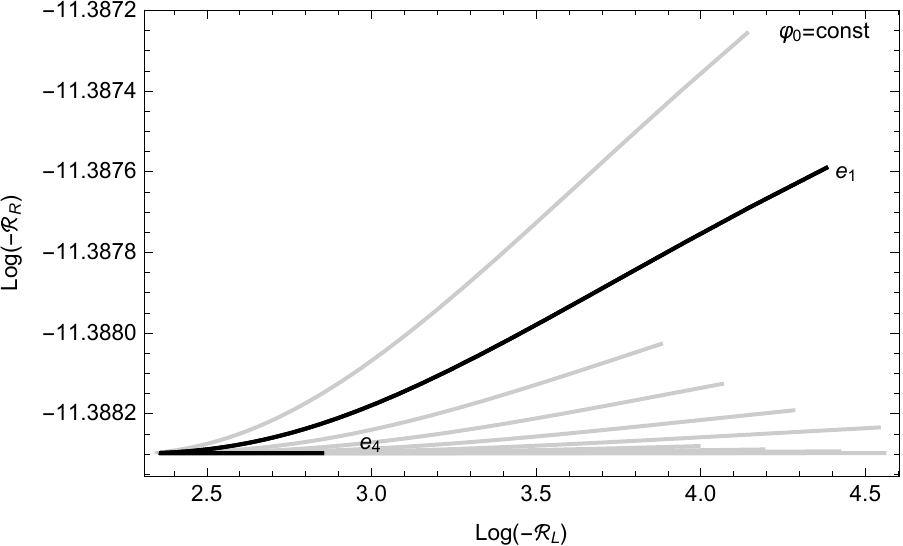}
\caption{}\label{RLRRB2}
\end{subfigure}
\centering
\begin{subfigure}{0.49\textwidth}
\includegraphics[width=1\textwidth]{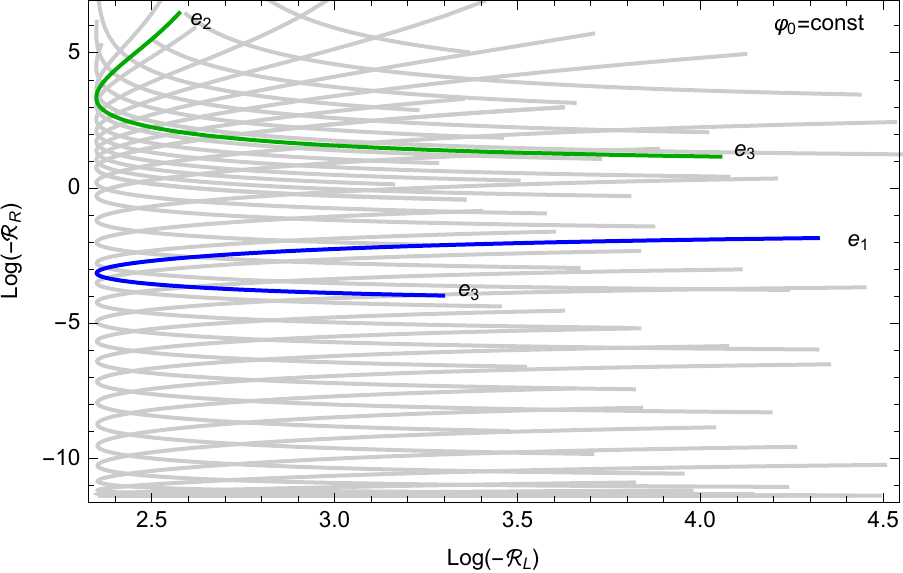}
\caption{}\label{RLRRB1}
\end{subfigure}
\caption{\footnotesize{$\mathcal{R}_R$ vs $\mathcal{R}_L$ for the $W^{LR}_{1,1}$ solutions at fixed $\f_0$. In figure (a) all curves belong to the points between $e_1$ and $e_4$ boundaries in figure \protect\ref{typeEmoduli}. In figure (b) curves are sketched either between $e_1$ and $e_4$ boundaries (e.g. the blue curve at $\f_0=4$) or between $e_2$ and $e_3$ boundaries (e.g. the green curve at $\f_0=4.245$).}}
\end{figure}

The relation between  $\mathcal{R}_L$ and $\mathcal{R}_R$ for different constant values of $\f_0$ are presented in figure  \ref{RLRRB2} for solutions between $e_1$ and $e_4$ boundaries and in figure \ref{RLRRB1} for flows in the space of solutions between $e_1$ and $e_3$ boundaries (e.g. the blue curve) or  $e_2$ and $e_3$ boundaries (e.g. the green curve).
\begin{figure}[!t]
\centering
\begin{subfigure}{0.32\textwidth}
\includegraphics[width=1\textwidth]{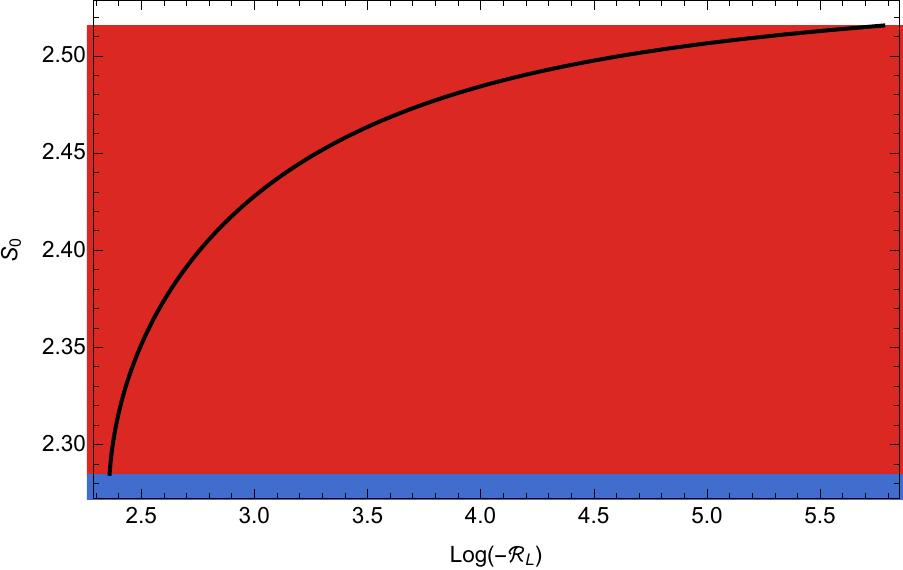}
\caption{}\label{RB1}
\end{subfigure}
\centering
\begin{subfigure}{0.32\textwidth}
\includegraphics[width=1\textwidth]{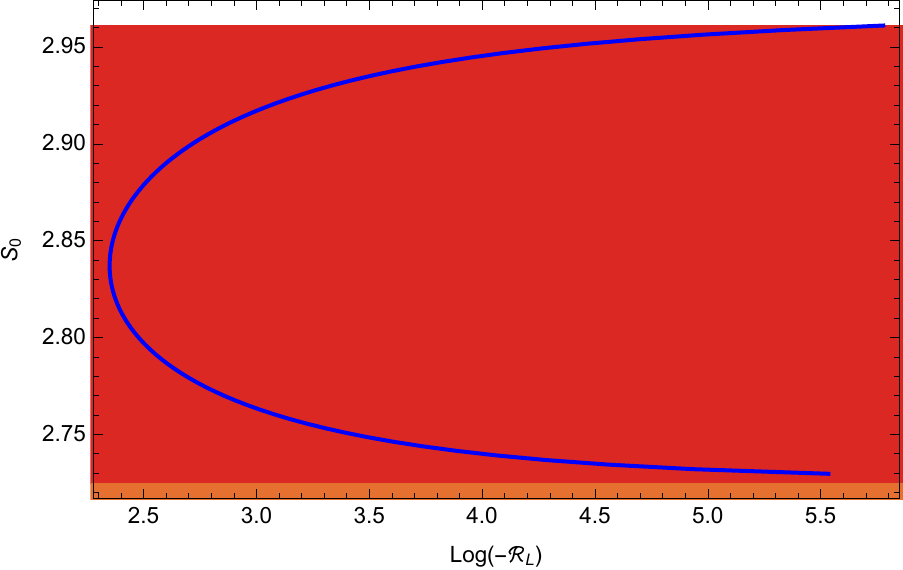}
\caption{}\label{RO1}
\end{subfigure}
\centering
\begin{subfigure}{0.32\textwidth}
\includegraphics[width=1\textwidth]{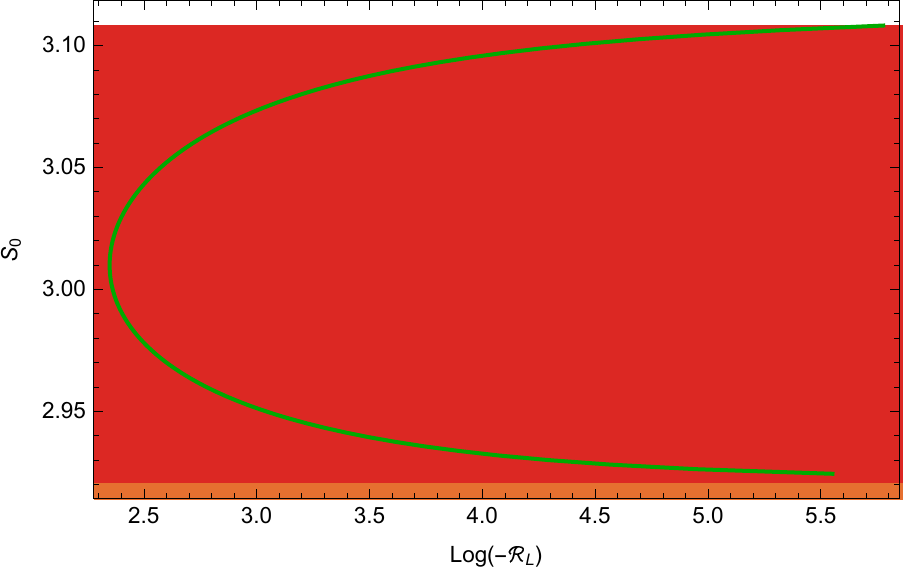}
\caption{}\label{ERO1}
\end{subfigure}
\centering
\begin{subfigure}{0.32\textwidth}
\includegraphics[width=1\textwidth]{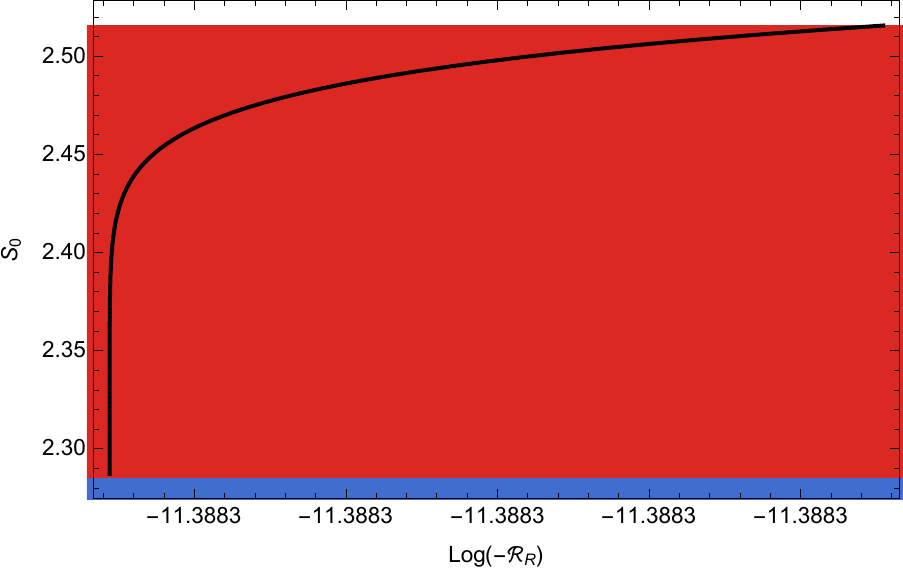}
\caption{}\label{RB2}
\end{subfigure}
\centering
\begin{subfigure}{0.32\textwidth}
\includegraphics[width=1\textwidth]{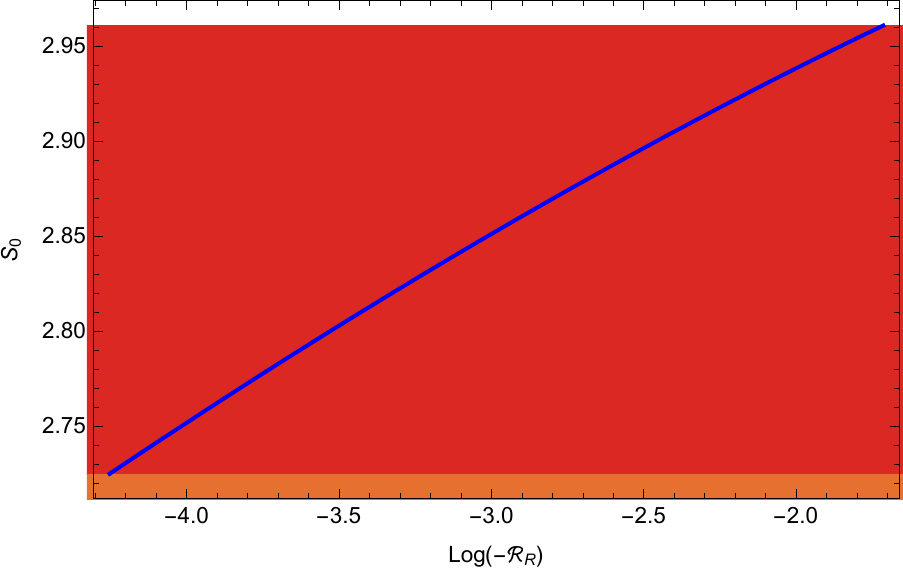}
\caption{}\label{RO2}
\end{subfigure}
\centering
\begin{subfigure}{0.32\textwidth}
\includegraphics[width=1\textwidth]{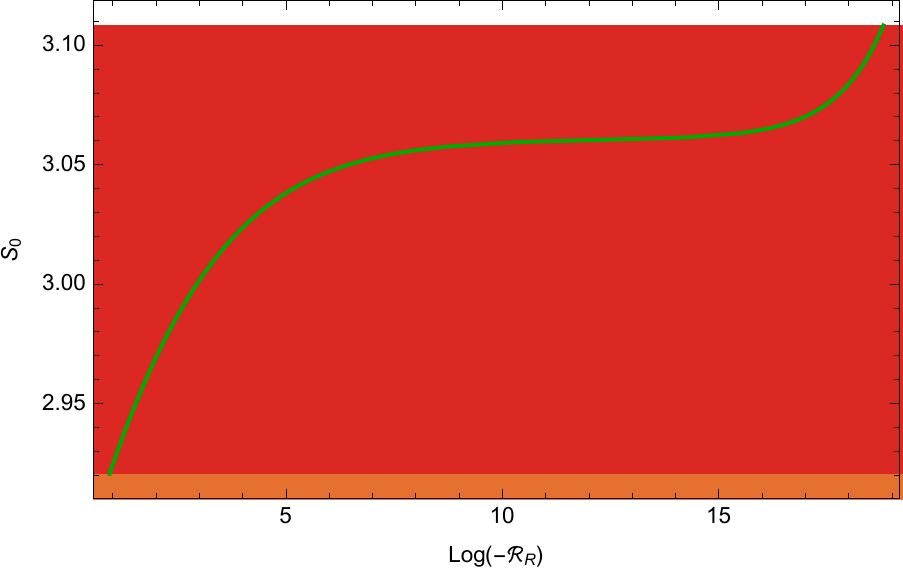}
\caption{}\label{ERO2}
\end{subfigure}
\caption{\footnotesize{The dimensionless curvatures  vs $S_0$ for three generic cases, $\f_0=3.25$ (a) and (d) diagrams, $\f_0=4$ (b) and (e) diagrams and $\f_0=4.245$ (c) and (f) diagrams. By comparing (c) and (b) or (f) and (e) we observe that the qualitative features of the $e_1$ and $e_2$ boundaries are similar and therefore we should be interpret them as a single boundary.}}
\end{figure}

\begin{figure}[!ht]
\centering
\begin{subfigure}{0.49\textwidth}
\includegraphics[width=1\textwidth]{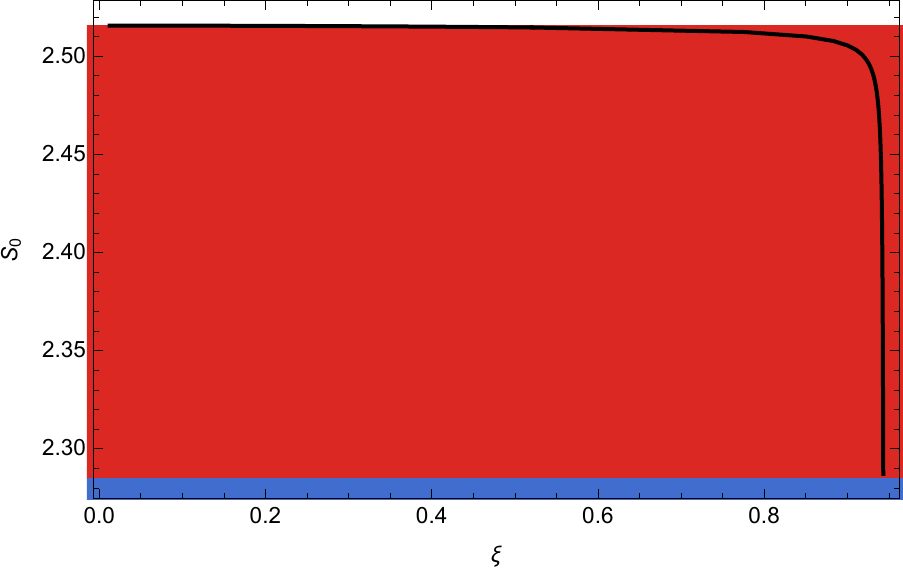}
\caption{}\label{RB3}
\end{subfigure}
\centering
\begin{subfigure}{0.49\textwidth}
\includegraphics[width=1\textwidth]{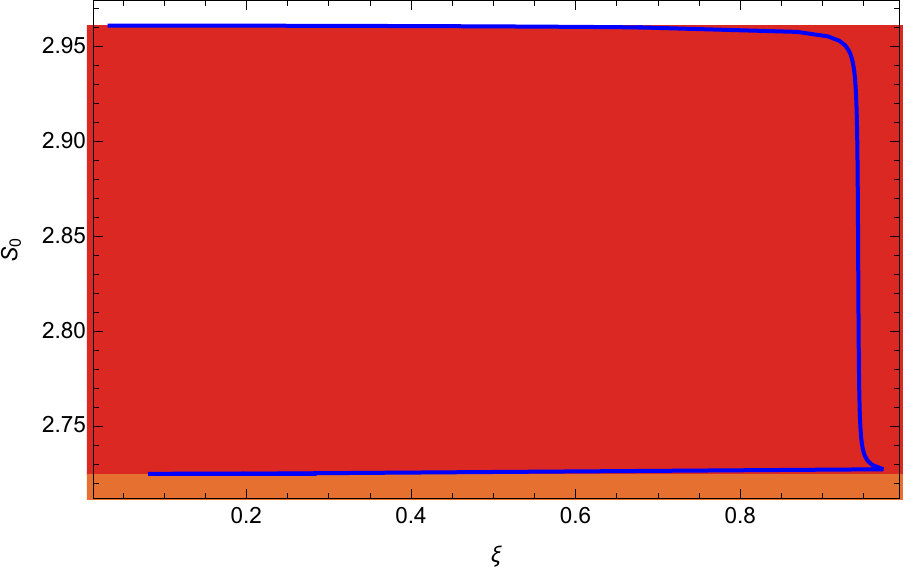}
\caption{}\label{RO3}
\end{subfigure}
\caption{\footnotesize{The ratio of the relevant couplings for two fixed values $\f_0=3.25, 4$ as we change the value of $S_0$.}}
\end{figure}

To see the general behavior of dimensionless parameters more explicitly, we have considered specific values of $\f_0$ in the space of $W^{LR}_{1,1}$ solutions. Figures \ref{RB1}, \ref{RB2} and \ref{RB3} belongs to  $\f_0=3.25$ and describe solutions between $e_1$ and $e_4$ boundaries.  Figures \ref{RO1}, \ref{RO2} and \ref{RO3} are sketched at  $\f_0=4$ and are related to solutions between $e_1$ and $e_3$ boundaries. Moreover, by comparing figure \ref{ERO1} (\ref{ERO2}) for solutions between $e_2$ and $e_3$ boundaries and figure \ref{RO1} (\ref{RO2}), we observe that $e_2$ boundary is of the same type of $e_1$ i.e. $(Max_+,Max_-)$.

We can summarize the results as follows (since $e_2$ has the same properties as $e_1$ boundary we just discuss $e_1$ here):

1) Close to the $e_1$  boundary $\mathcal{R}_L\rightarrow -\infty$, see the top part of figures \ref{RB1} or \ref{RO1}. These solutions belong to $(Max_+,Max_-)$ solutions. Beyond this boundary there are other types of flows with more $\f$-bounces. On the contrary, the value of $\mathcal{R}_R$ remains constant near this boundary, figures \ref{RB2} or \ref{RO2}, and it shows that as the flow reaches the $UV_R$ boundary no extra $\f$-bounce would appear.

2) As we move towards the $e_4$ boundary, the red-blue border in figure \ref{RB1} or \ref{RB2}, $\mathcal{R}_L$ has a finite value but $\mathcal{R}_L\rightarrow 0$ (the slop of curve becomes vertical). These solutions belong to $(Max_-,Min_+)$ solutions.  The reason for this behavior is the existence of the $W^{L Min_+}_{1,0}$ solutions in the joint border type $W^{LR}_{1,1}$ and $W^{LR}_{1,0}$ regions.

3) As we move closer to the $e_3$ boundary (the red-orange border in figure \ref{RO1} or \ref{RO2}), again $\mathcal{R}_L\rightarrow -\infty$. This is obvious because the $W^{LR}_{1,1}$ solutions have an extra $\f$-bounce with respect to the $W^{LR}_{1,0}$ solutions near their left UV fixed point. This is not happening on the right UV fixed point so  $\mathcal{R}_R$ remains finite.

From the properties above, we may  understand the behavior of $\xi$ in figures \ref{RB3} and \ref{RO3}. Close to $e_1$ boundary since $\mathcal{R}_L \rightarrow -\infty$ we expect that $\f_-^{(L)}\rightarrow 0$ and therefore $\xi\rightarrow 0$ as well. Close to the $e_4$, we approach to  $W^{LMin_+}_{1,0}$ solutions, $\f_-^{(R)} \neq 0$ and  $\xi$ obtains  a finite value. On the other hand, close to the $e_3$ boundary we have the same situation as $e_1$ therefore again we see that $\xi\rightarrow 0$.

\begin{figure}[!ht]
\centering
\begin{subfigure}{0.49\textwidth}
\includegraphics[width=1\textwidth]{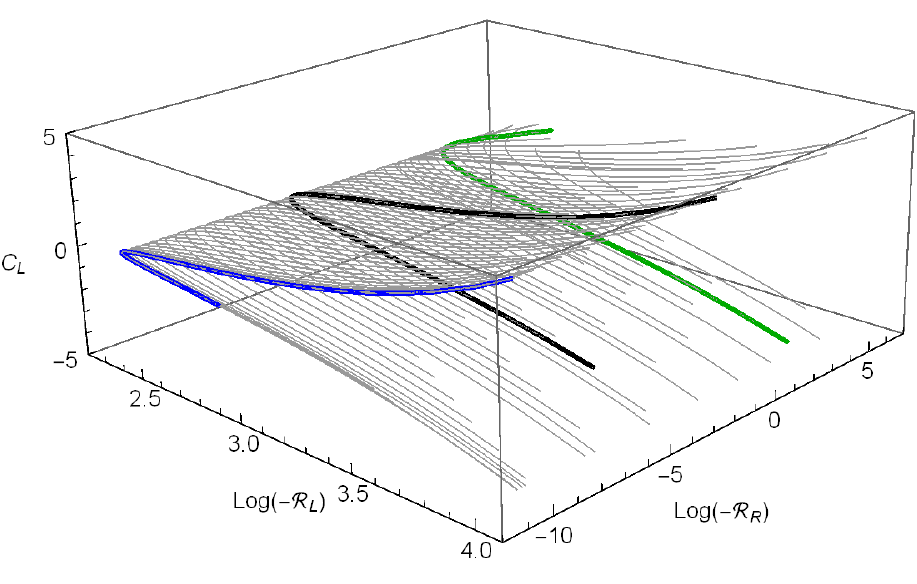}
\caption{}\label{CLif3DE}
\end{subfigure}
\begin{subfigure}{0.49\textwidth}
\includegraphics[width=1\textwidth]{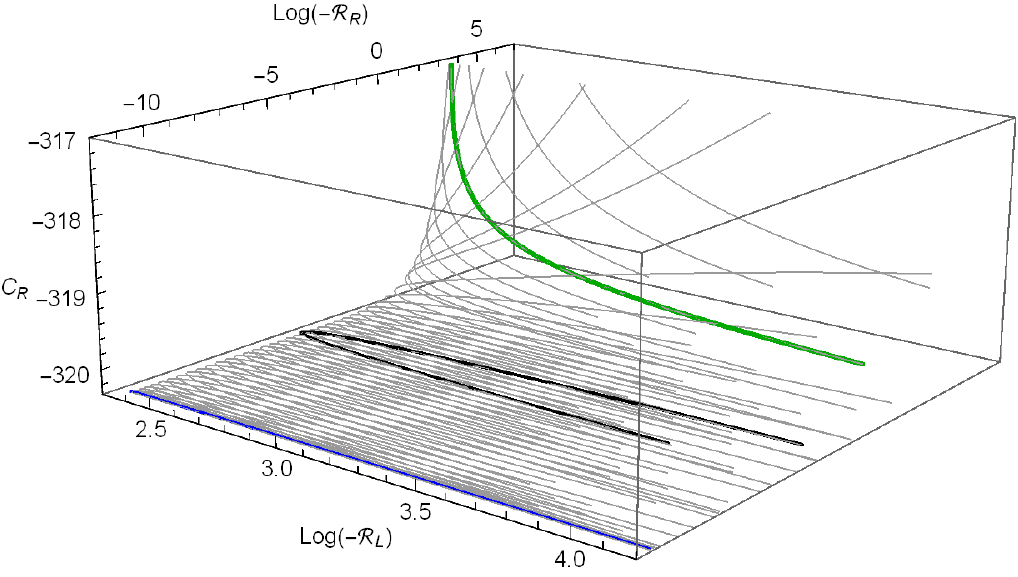}
\caption{}\label{CRif3DE}
\end{subfigure}
\caption{\footnotesize{As we change $S_0$ along a $\f_0=constant$ line in the red region in figure \protect\ref{typeEmoduli} we can read $C_L$ and $C_R$ in terms of the dimensionless curvatures. For example in the above figures, the black curve is at $\f_0=3.25$ and the black one at $\f_0=4$.}}
\end{figure}

As the final property of type $W^{LR}_{1,1}$, we can plot the dependence  of $C_L$ and $C_R$ in terms of the left and right dimensionless curvatures, see the diagrams in figures \ref{CLif3DE} and \ref{CRif3DE}.

\subsubsection{Multi-$\f$-bounce and multi-A-bounce solutions\label{multi}}

\begin{figure}[!ht]
\centering
\begin{subfigure}{0.49\textwidth}
\includegraphics[width=1\textwidth]{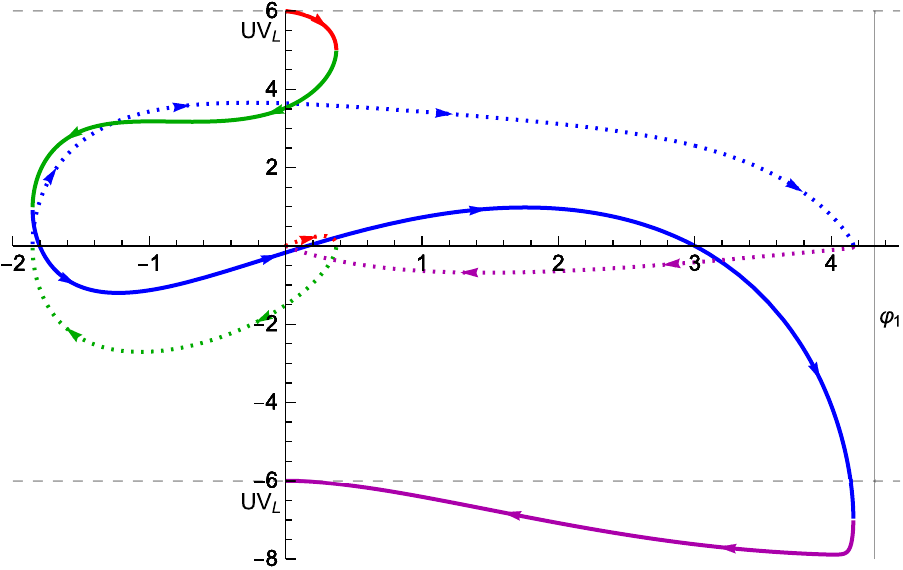}
\caption{}\label{B3}
\end{subfigure}
\centering
\begin{subfigure}{0.49\textwidth}
\includegraphics[width=1\textwidth]{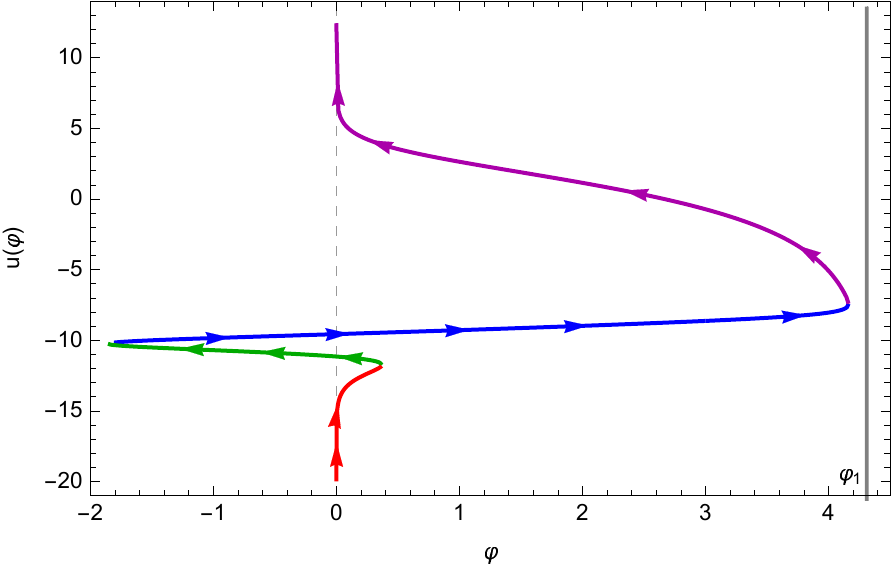}
\caption{}\label{M1}
\end{subfigure}
\centering
\begin{subfigure}{0.49\textwidth}
\includegraphics[width=1\textwidth]{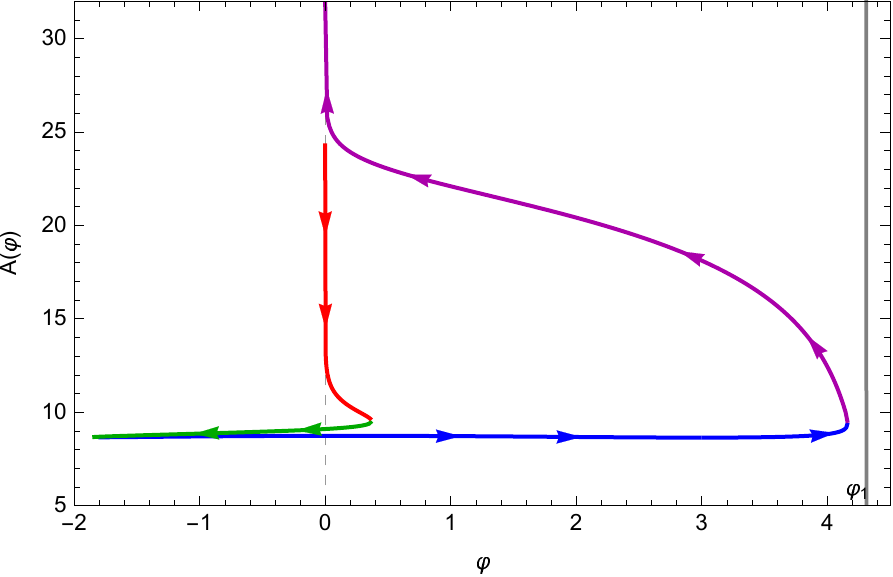}
\caption{}\label{M2}
\end{subfigure}
\centering
\begin{subfigure}{0.49\textwidth}
\includegraphics[width=1\textwidth]{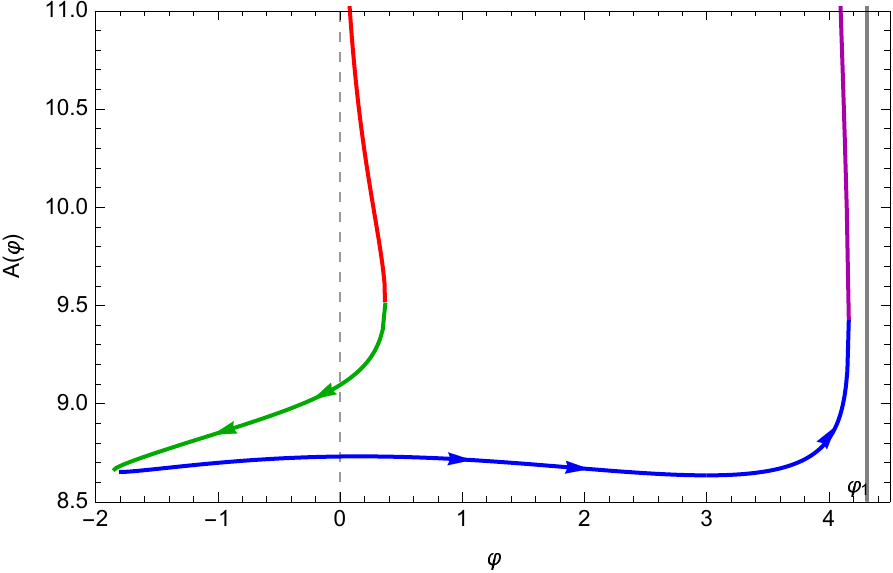}
\caption{}\label{M3}
\end{subfigure}
\centering
\begin{subfigure}{0.65\textwidth}
\includegraphics[width=1\textwidth]{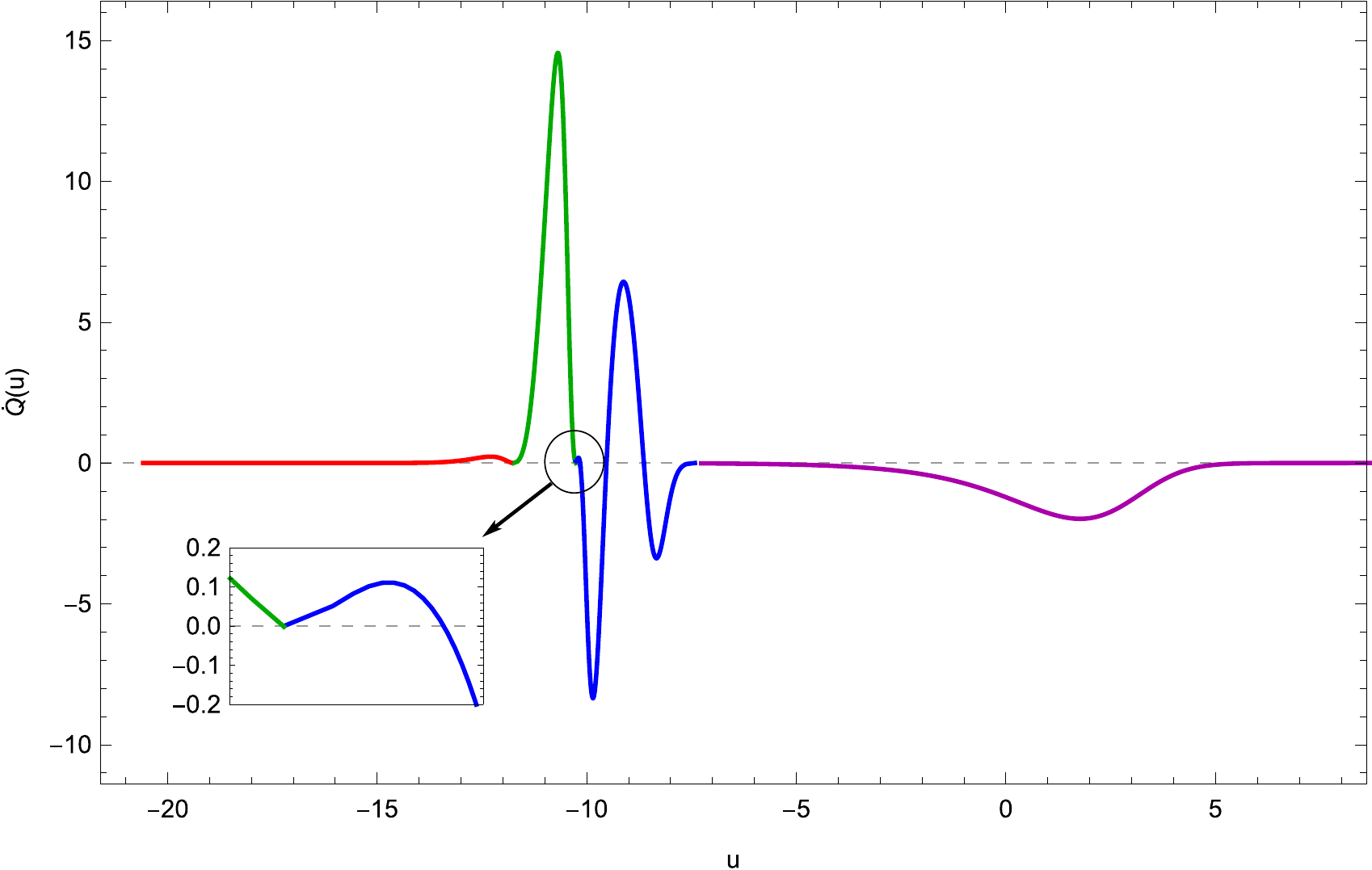}
\caption{}\label{M4}
\end{subfigure}
\caption{\footnotesize{(a): An example of a multi-$\f$-bounce solution, $W^{LL}_{3,3}$. The solid line is $W(\f)$ and dotted line is $S(\f)$.  In this case, a  flow connects two UV boundaries on the left UV fixed point after three $\f$-bounces. Unlike the previous cases, the geometry here has three A-bounces. (b)and (c) show the behavior of holographic coordinate and scale factor in terms of $\f$. Figure (d) is the magnification of the bottom of figure (c). It shows that there are three A-bounces for this flow. (e): The roots of $\dot{Q}$ in equation \protect\eqref{qdot} shows the location of $\f$-bounces where the color of the graph is changed and the location of A-bounces where the blue part of the curve crosses the $u$ axis.}}
\end{figure}

In addition, in the above solutions, one may find geometries that connect the UV QFTs
with more than two $\f$-bounces. An example of such solutions with three $\f$-bounces
 is sketched in figure \ref{B3}. For example,  for solutions similar to the specific solution
  in figure \ref{B3},  this space is located on the left hollow white region (surrounded by the blue region)
   in figure \ref{Moduli}.  The space of these multi-$\f$-bounce solutions is the white regions in figure \ref{Moduli}.

The holographic coordinate and scale factor as a function of $\f$ are given in figures \ref{M1} and \ref{M2}. By zooming at the bottom of figure \ref{M2} we can see that this  flow has three A-bounces, see figure \ref{M3}.

To  describe the behavior of the flow better, it would be proper to introduce the following variables,
\be\label{qdot}
Q(u)=\frac12 \dot{\f}^2-V\geq 0\,,\qquad \dot Q=\frac{d}{2(d-1)}W S^2\,.
\ee
The variable $Q$ and its first $u$ derivative across the flow, track precisely $A$ bounces and $\f$-bounces, as $\dot Q$ vanishes at both, as seen from (\ref{qdot}).
Moreover, $\dot Q$ changes sign at $A$-bounces but not $\f$-bounces.

We have plotted $\dot{Q}$ in terms of $u$ in figure \ref{M4} for the specific flow in figure \ref{B3}. The zero points of $\dot Q$ of the graph show the points that either $S=0$ ($\f$-bounces) or $W=0$ (A-bounces).  Near the
left and right of each $\f$-bounce, $\dot{Q}$ have the same sign (at these points the color of the graph changes), but for the A-bounces $\dot{Q}$ changes its sign (this happens just for the blue part of the graph).

In general, and for a generic  potential, we expect that most of the  full $(S_0,\f_0)$ plane of parameters will be filled with regular solutions unlike the cases with single boundaries.
Typically, the set of regular solutions corresponds to finite values for $\f_0$ and $S_0$, but as these values become larger and larger, eventually solutions tend to run to $\f\to\pm\infty$ and the solutions become singular.
It would be interesting to test this conjecture in top-down scalar potentials, like in \cite{C1,C2}.

\section{Special flows on the inter-region boundaries}\label{limflo}

We have seen that the moduli space of solutions involves several distinct boundaries that separate classes of solutions that have different geometric characteristics. These involve the type of theories that are at the end-points of the flow, as well as the number of $\f$-bounces (i.e. the number of times the scalar changes direction along the flow).

\subsection{Crossing from the $W^{LR}_{1,0}$ to the $W^{LL}_{1,1}$  region}\label{crossAtoB}

\begin{figure}[!ht]
\centering
\begin{subfigure}{0.65\textwidth}
\includegraphics[width=1\textwidth]{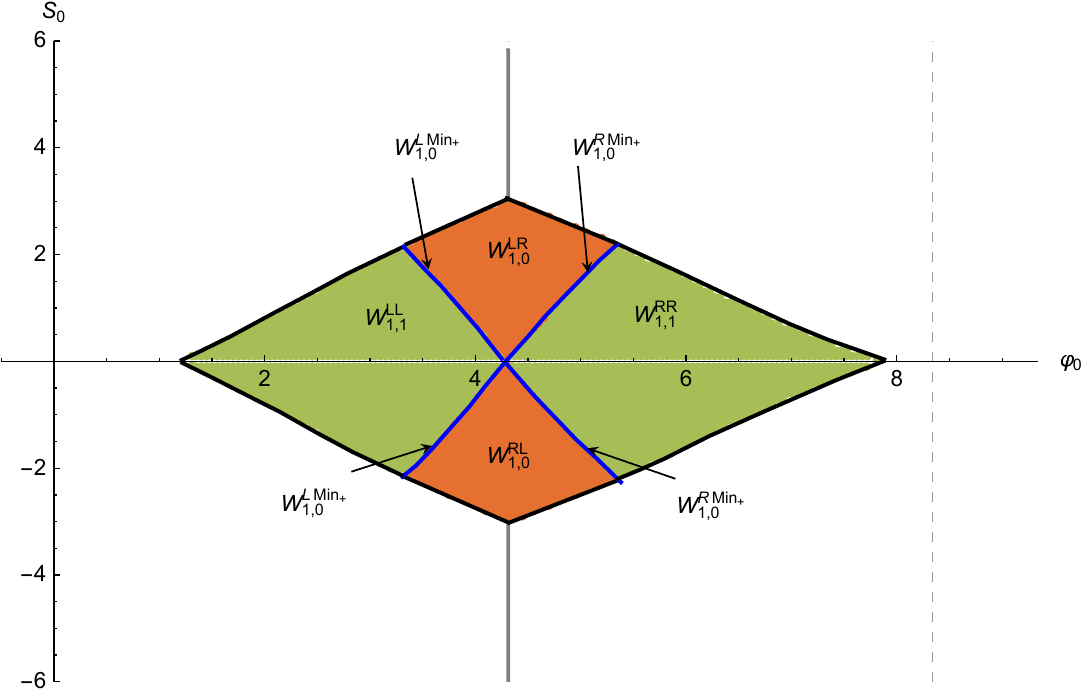}
\end{subfigure}
\caption{\footnotesize{The space of solutions of $W^{LR}_{1,0}$ (the orange region), $W^{LL}_{1,1}$ (the green region) and $W^{L Min_+}_{1,0}$ (the blue diagonal lines).}}
\label{Ftype}
\end{figure}

\begin{figure}[!ht]
\centering
\begin{subfigure}{0.49\textwidth}
\includegraphics[width=1\textwidth]{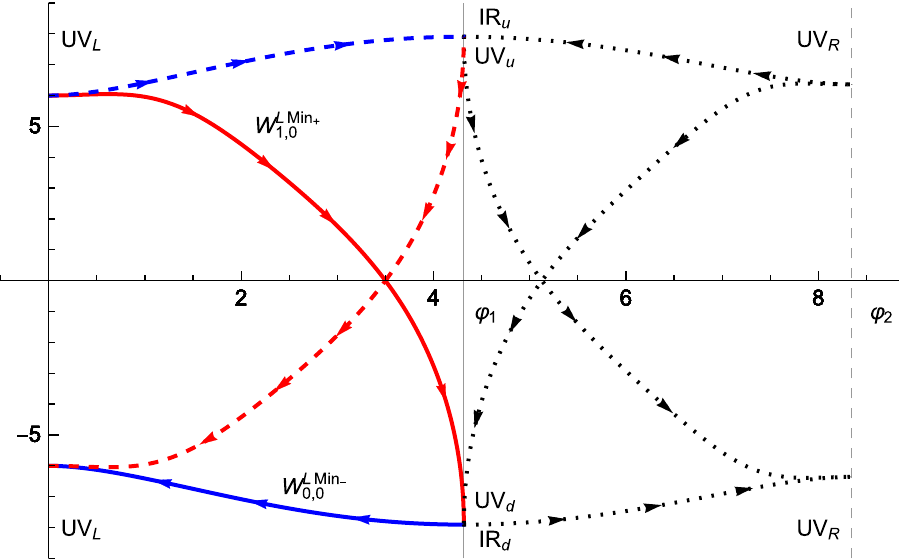}
\caption{}\label{AUVIR}
\end{subfigure}
\centering
\begin{subfigure}{0.49\textwidth}
\includegraphics[width=1\textwidth]{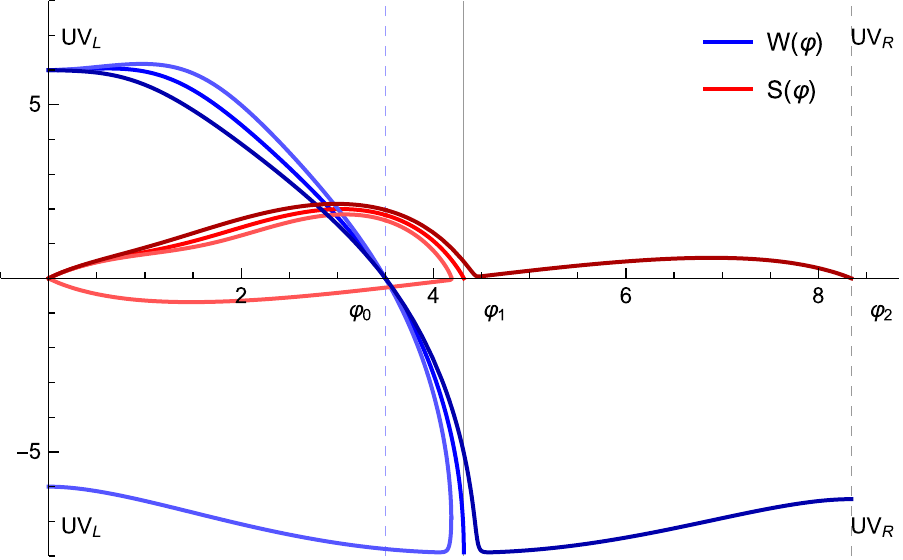}
\caption{}\label{Atino}
\end{subfigure}
\caption{\footnotesize{(a): An example of a  flow between a maximum and a minimum. For the solid curves, $(Max_-,Min_{+})$  is a flow between a UV fixed point at maximum $\f=0$ and another UV  fixed point at the minimum $\f=\f_1$. For the $(Max_{-},Min_{-})$ part of the solution, the minimum is an IR fixed point.
The dashed curves show the flipped image of the solid curves.
The black dotted curves are other possible flows with the same UV fixed points.
(b): At a fixed $\f_0$ when the value of $S_0$ is exactly on the border of type $W^{LR}_{1,0}$ and type $W^{LL}_{1,1}$, we have the $W^{L Min_+}_{1,0}$ branch solution (the middle flow). If we increase or decrease the value of $S_0$ we have the $W^{LR}_{1,0}$ or $W^{LL}_{1,1}$ solutions respectively.}}
\end{figure}

On the border between type $W^{LR}_{1,0}$ and type $W^{LL}_{1,1}$ regions shown in figure \ref{Moduli}, one may find a degenerate type of  flow, shown in figure \ref{Ftype}.
The standard $W^{LL}_{1,1}$ solutions start as usual at the $UV_L$ fixed point and end on the same UV$_{L}$
on the other side. But exactly at the boundary, this flow splits into two clearly distinct pieces: one is a flow from the $UV_L$ to the intermediate minimum of the potential at $\f_1$. This is
a $+$ flow near the minimum and therefore the minimum becomes a new AdS boundary, see figure \ref{AUVIR}. Overall this is a $(Max_-,Min_+)$ type of flow. This is the $W^{L Min_+}_{1,0}$ flow we defined in section \ref{sec:2par}.

The remaining flow is a $(Max_-,Min_-)$ flow starting at the $UV_L$ fixed point (UV boundary) and ending at the minimum of the potential with $-$ branch flow. This is possible because for this limiting solution $\mathcal R_R=0$.
This flow is what we defined as a $W^{L Min_-}_{0,0}$ type flow in section \ref{sec:2par}.
This degeneration into the $W^{L Min_+}_{1,0}$ $+$ $W^{L Min_-}_{0,0}$ flow is shown graphically in figure \ref{Atino}.

As we approach the minimum of the potential from the left, we can find the expansion of $W$ and $S$. In this case, there is just one branch (the $+$ branch) for negative curvature slices. As $\f\rightarrow \f_1^-$ for $W<0$, we have
\bsq
\begin{align}\label{AminsolW}
W_+&=-\frac{2(d-1)}{\ell}-\frac{\D_+}{2\ell}(\f-\f_1)^2-\frac{\mathcal{R}}{d\ell}|\f-\f_1|^{\frac{2}{\D_+}}+\cdots\,,\\
S_+&=\frac{\D_+}{\ell}|\f-\f_1|+\cdots\,,\label{AminsolS} \\
T_+ &=\frac{\mathcal{R}}{\ell^2}|\f-\f_1|^\frac{2}{\D_+}+\cdots\,. \label{AminsolT}
\end{align}
\esq
where $\D_+=d/2+\sqrt{d^2/4 + m^2\ell^2}>0$. The expansions of the scalar field and scale factor  are given by
\bsq
\begin{align}
\label{Aphimin} \f(u) &=\f_1- \f_+ \ell^{\Delta_+}e^{-\Delta_+ u / \ell}  + \cdots \, , \\
\label{AAmin} A(u) &= {A}_+ +\frac{u}{\ell} - \frac{\f_+^2 \, \ell^{2 \Delta_+}}{8(d-1)} e^{-2\Delta_+ u / \ell}  -\frac{\mathcal{R}|\f_+|^{2/\Delta_+} \, \ell^2}{4d(d-1)} e^{-2u/\ell} +\cdots \,,
\end{align}
\esq
On the other hand, as shown in \cite{C},  the minus branch solution  near a minimum of the potential exist only if  $R^{UV}=0$, and since $T=0$ in that case,   this  implies that $S=W'$ and we have
\bsq
\begin{align}\label{AminsolWm}
W_-&=-\frac{2(d-1)}{\ell}-\frac{\D_-}{2\ell}(\f-\f_1)^2+\cdots\,,\\
S_-&=-\frac{\D_-}{\ell}|\f-\f_1|+\cdots\,,\label{AminsolSm}
\end{align}
\esq
where $\D_-=d/2-\sqrt{d^2/4 + m^2\ell^2}<0$.
We also obtain
\bsq
\begin{align}
\label{Aphiminm} \f(u) &=\f_1- \f_- \ell^{\Delta_-}e^{-\Delta_- u / \ell}  + \cdots \, , \\
\label{AAminm} A(u) &= {A}_- +\frac{u}{\ell} - \frac{\f_-^2 \, \ell^{2 \Delta_-}}{8(d-1)} e^{-2\Delta_- u / \ell} +\cdots \,.
\end{align}
\esq
For $W^{L Min_+}_{1,0}$ and $W^{L Min_0}_{0,0}$ solutions we have the following properties:

1) The space of solutions is a one-dimensional curve $S_0=S_0(\f_0)$ which is the joint border of type $W^{LR}_{1,0}$ and $W^{LL}_{1,1}$ regions.
 This is shown as blue diagonal lines in figure \ref{Ftype}. A similar type of solution also exists in the joint border of $W^{LR}_{1,1}$ and $W^{LL}_{1,2}$ regions, see section \ref{crossEtoAD}.

2) The $(Max_{-},Min_+)$ branch of the solution: There is a $UV_L$ fixed point at the maximum $\f=0$ and the QFT at this point is characterized by the parameters $\mathcal{R}_L$ (source) and $C_L$ (vev). At the minimum of the potential, at $\f_1$,  there is another UV fixed point (associated to the $W_+$ branch solution there) and it can be reached as $u\rightarrow +\infty$ so that $e^{A(u)}\rightarrow\infty$\footnote{We should note that for a fixed point at $W>0$, the minimum is a UV fixed point too but at $u\rightarrow -\infty$.}. The QFT at this point has only one dimensionless  parameter $\mathcal{R}$, because the $+$ branch is a flow purely driven by a vev. This parameter can be thought of  as the ratio of the curvature to the vev of the irrelevant operator (that is a modulus in such a case).
Since $\f_-=0$ for $W_+$ branch, equation \eqref{Aphimin}, then according to the relation \eqref{Eratio} the value of $\xi\rightarrow \infty$ for $(Max_{-},Min_+)$ solutions.

3) The $(Max_-,Min_-)$ branch of the solution: The flow connects a UV fixed point to an IR one.  There is a $UV_L$ fixed point at the maximum $\f=0$ but with $\mathcal{R}=0$ (as we discussed already, close to the $a_4$ in type $W^{LR}_{1,0}$ or equivalently $b_2$ in type $W^{LL}_{1,1}$, $\log(-\mathcal{R}_f) \rightarrow -\infty$). The minimum $\f_1$ is an IR fixed point at $u\rightarrow -\infty$ and $e^{A(u)}\rightarrow 0$.
Since there is no boundary at the IR fixed point, no QFT lives here so the ratio of couplings is meaningless.

To show the behavior of the above branches,  we have sketched $W$ in figure \ref{AUVIR}.
To draw this figure, we have used the potential in figure \ref{potential} which at the minimum $\f_1=4.31$ the AdS length scale is $\ell=0.76$ and moreover at that minimum, $\D_+=4.25$ and $\D_-=-0.25$.
The dashed curves show the flows between left UV fixed points and the fixed point at a minimum of the potential with $W>0$ near the fixed point. Other independent flows with a right UV fixed point are sketched with black dotted curves in this figure.

In figure \ref{Atino} we have shown two flows very close to the $(Max_-,Min_+)$ flow (which the one in the the middle of the three flows). At fixed $\f_0$, as we decrease $S_0$ away from its value on the border, a $\f$-bounce appears and the flow returns to the left UV. But when we increase $S_0$ a little, the flow continues until it reaches the right UV fixed point.

\begin{figure}[!ht]
\centering
\begin{subfigure}{0.49\textwidth}
\includegraphics[width=1\textwidth]{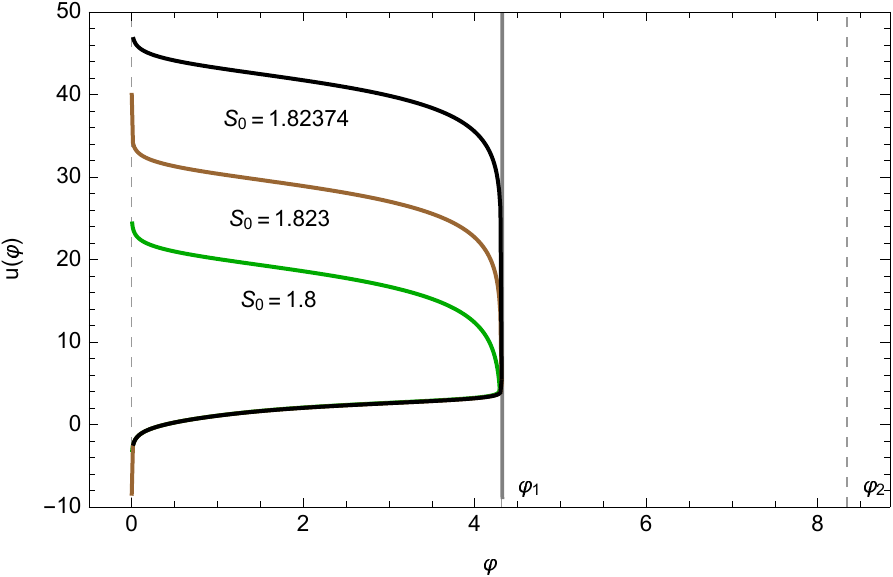}
\caption{}\label{NF1}
\end{subfigure}
\centering
\begin{subfigure}{0.49\textwidth}
\includegraphics[width=1\textwidth]{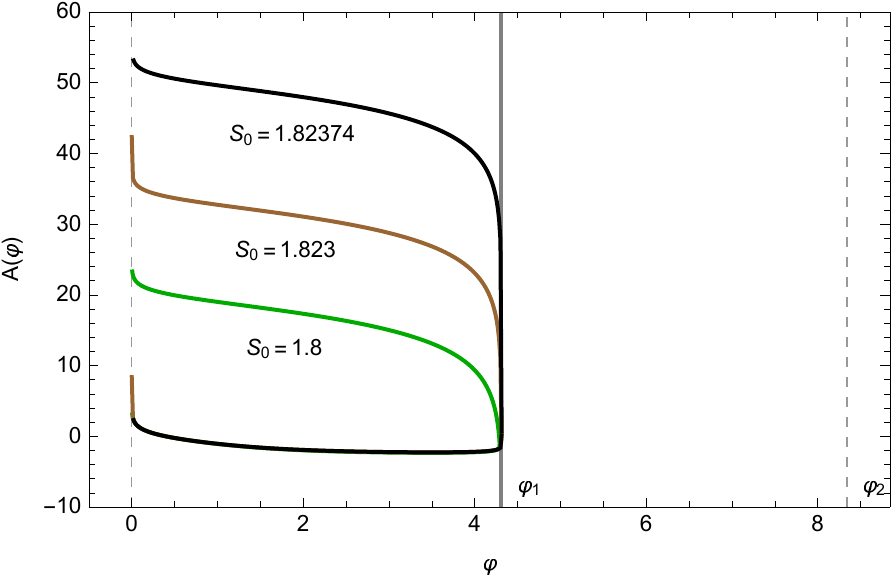}
\caption{}\label{NF2}
\end{subfigure}
\caption{\label{33}\footnotesize{The behavior of (a): the holographic coordinate and (b): the scale factor at $\f_0=3.5$ as the flow solutions in type $W^{LL}_{1,1}$ move towards the $b_2$ boundary (as $S_0$ increases) where the $W^{L Min_+}_{1,0}$ solution exists.}}
\end{figure}

To follow what is happening to the geometry as we approach the boundary, we have drawn the holographic coordinate and scale factor of three different RG solutions in figures \ref{NF1} and \ref{NF2}. In these figures, all curves are sketched for a fixed value $\f_0=\frac72$ in the space of solutions. As we increase the value of $S_0$ towards the $b_2$ boundary where the $W^{L Min_+}_{1,0}$ solutions exist, we move from the green curve to the black curve.
\begin{figure}[!ht]
\centering
\begin{subfigure}{0.49\textwidth}
\includegraphics[width=1\textwidth]{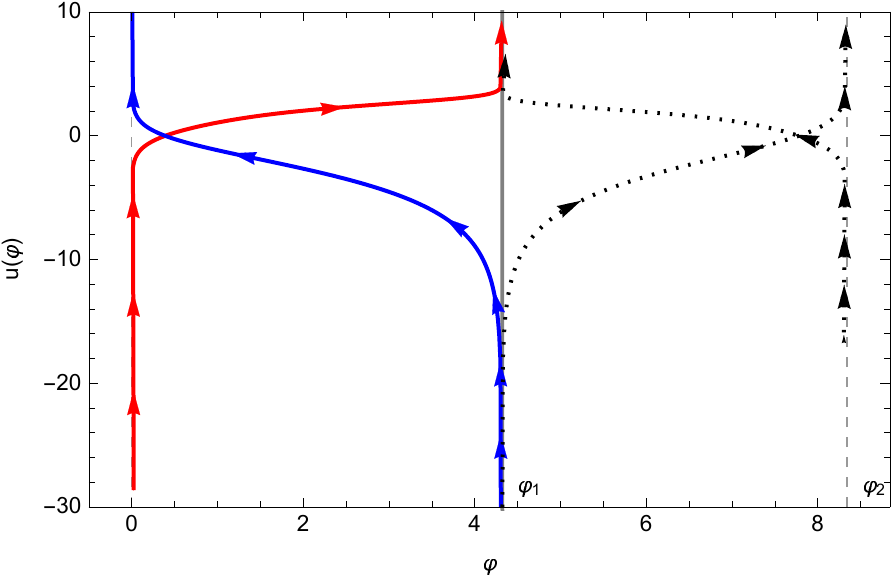}
\caption{}\label{AF2}
\end{subfigure}
\centering
\begin{subfigure}{0.49\textwidth}
\includegraphics[width=1\textwidth]{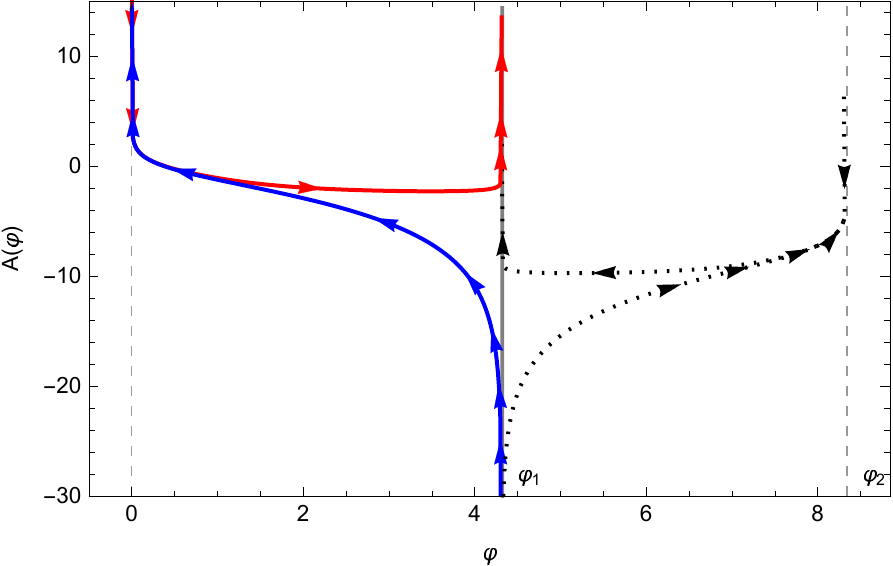}
\caption{}\label{AF3}
\end{subfigure}
\caption{\footnotesize{The behavior of the holographic coordinate and scale factor  in terms of $\f$ for the $W^{L Min_+}_{1,0}$ and $W^{L Min_-}_{0,0}$ flows. The red curve belongs to the $W^{L Min_+}_{1,0}$ branch and the blue one to the $W^{L Min_-}_{0,0}$ branch in figure \protect\ref{AUVIR}.}}
\end{figure}

Figures \ref{AF2} and \ref{AF3} show the behavior of the holographic coordinate $u(\f)$ and scale factor $A(\f)$ when a $W^{L Min_+}_{1,0}$ solution appears.
 The red curves describe the $(Max_-,Min_+)$ branch, a flow between two UV fixed points one on the maximum
 and the other on the minimum of the potential. The blue curve shows that the $(max_{-},Min_{-})$ branch is
 a flow between a UV fixed point at maximum $\f=0$ and an IR fixed point at the minimum $\f=\f_1$.
\begin{figure}[!t]
\centering
\begin{subfigure}{0.51\textwidth}
\includegraphics[width=1\textwidth]{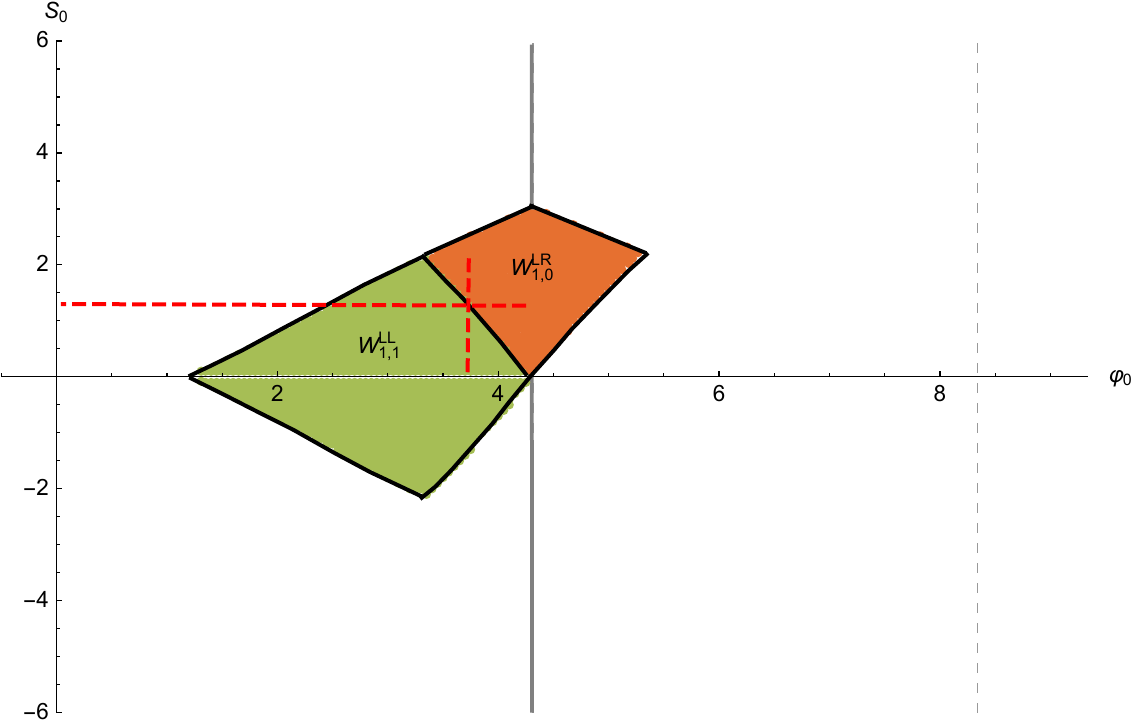}
\caption{}\label{crossAB}
\end{subfigure}
\centering
\begin{subfigure}{0.49\textwidth}
\includegraphics[width=1\textwidth]{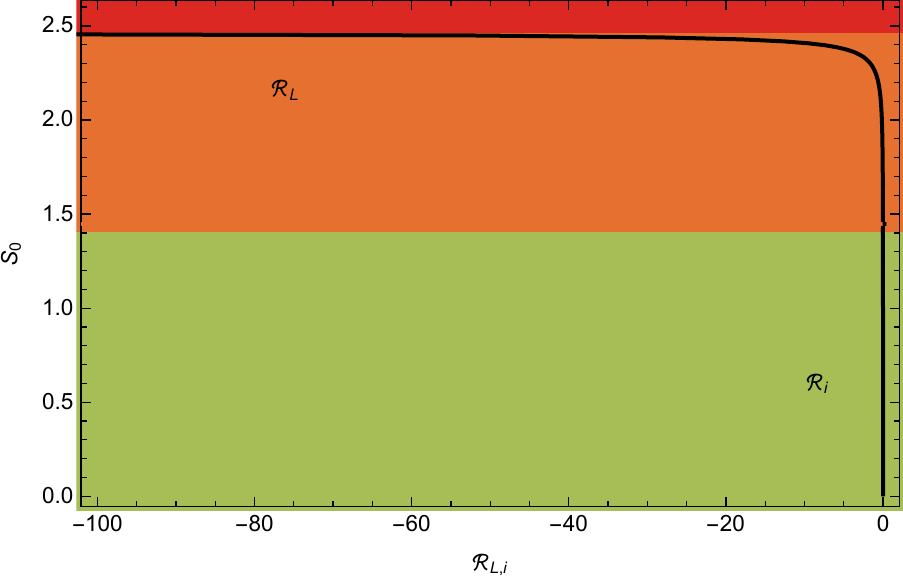}
\caption{}\label{crossR}
\end{subfigure}
\centering
\begin{subfigure}{0.49\textwidth}
\includegraphics[width=1\textwidth]{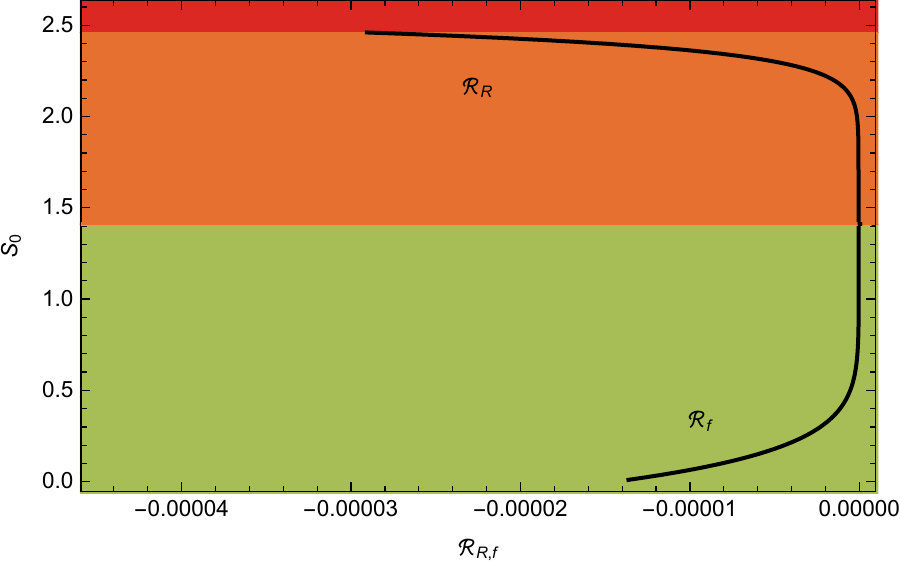}
\caption{}\label{crossR1}
\end{subfigure}
\caption{\footnotesize{(a): Shows the space of solutions in type $W^{LR}_{1,0}$ and $W^{LL}_{1,1}$ region and their joint boundary. (b): The behavior of dimensionless curvature for left UV boundary when we move from region $W^{LR}_{1,0}$ to  $W^{LL}_{1,1}$. (c) The dimensionless curvatures on the $UV_f$ or $UV_R$ fixed points of the flow tend to zero as one approaches the boundary.}}
\end{figure}

We now consider a fixed value for $\f_0$ and try to change the value of $S_0$ near the joint border of type $W^{LR}_{1,0}$ and type $W^{LL}_{1,1}$, see figure \ref{crossAB} as we move along the vertical red dashed line at $\f_0=3.7$. In summary, we have the following properties close to the joint border:

1) Close to the boundary (around $S_0=1.4$) that part of the  flow which connects to the $UV_L$ with $W>0$ gradually turns into the $(Max_-,Min_+)$ branch or the $W^{L Min_+}_{1,0}$ solutions.  At the boundary, the value of the dimensionless curvature for the $UV_L$ fixed point of $(Max_-,Min_-)$ branch is given by the value of the curvatures ($\mathcal{R}_L=\mathcal{R}_i$) at the joint boundary (see the orange-green border in figure \ref{crossR}).

2) As we move towards the joint border, the value of the dimensionless curvatures $\mathcal{R}_R$ and $\mathcal{R}_f$ both tend to zero, see figure \ref{crossR1}
This is expected because as we move closer and closer to the border, this part of the flow turns into the $Max_{-}, Min_-)$ branch or $W^{L Min_-}_{0,0}$ solutions and it corresponds to the solutions with flat slices.

\begin{figure}[!ht]
\centering
\begin{subfigure}{0.49\textwidth}
\includegraphics[width=1\textwidth]{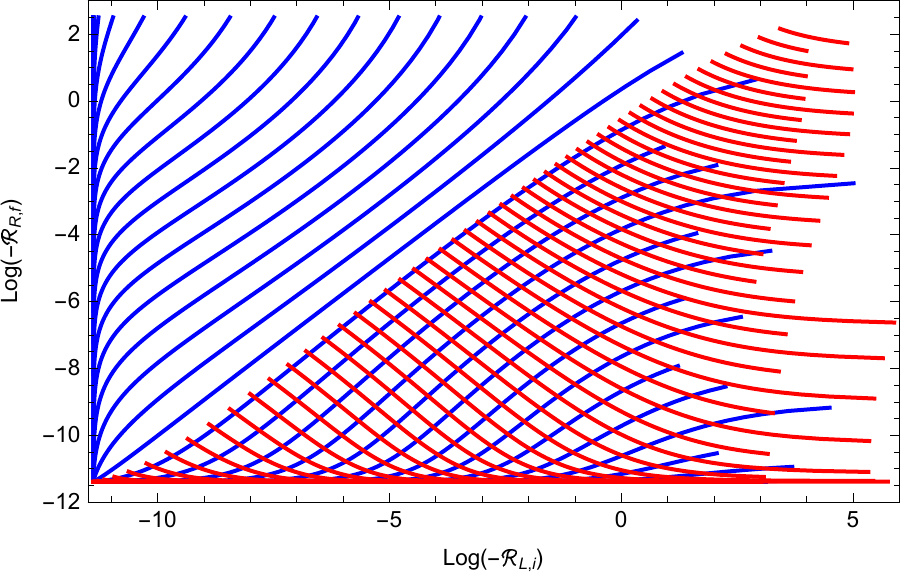}
\end{subfigure}
\caption{\footnotesize{Type  $W^{LR}_{1,0}$ (blue curves) and type  $W^{LL}_{1,1}$ (red curves) overlap of dimensionless curvatures. The curves are demonstrated for different constant values of $\f_0$.}}\label{typeABRR}
\end{figure}

In figure \ref{typeABRR}, we have shown the overlap of values for dimensionless curvatures in type $W^{LR}_{1,0}$ and $W^{LL}_{1,1}$. The curves are drawn for different fixed values of $\f_0$.

It is clear from figure \ref{AUVIR}, that in the $(Max_{-},Min_{+})$ part of the flow, there are two parts: a standard one, starting at $UV_L$ and another one near the minimum of the potential.
In this second part that becomes asymptotically long, $\f$ is very near the minimum, and stays there for a long period, but the scale factor changes very fast.
If we look at the scalar equation (\ref{AEOM3}) in this regime, the last term is essentially negligible and the approximate equation to solve is
\be
\ddot\f+d\dot A\dot\f=0\ar \dot\f\sim e^{-dA}\,.
\ee
Substituting in the other equations (\ref{AEOM1}), (\ref{AEOM2}) we find that $\dot\f^2$ is subleading to the other terms and can be dropped.
From (\ref{AEOM2}) the main contribution comes from the value of the potential at the minimum and therefore
\be
\dot A\simeq \frac{1}{\ell}\ar A=\frac{u}{\ell}\,,
\ee
and space is AdS near that point with a boundary close by. As the curvature terms in the equations are subleading, we arrive at the solution which we would find in a flat-sliced case, arriving at the minimum of the potential with a plus branch solution. However, this does not imply that the solution near this AdS boundary has zero boundary curvature, but as usual, near AdS boundaries, the curvature although a source, is subleading
compared to the source of the metric.

However, on the other side of the $\f$-bounce, the $(Max_{-},Min_-)$ part,  the story is different. According to the plots above, the scale factor decreases slowly to a minimum and this minimum approaches zero as we approach the transition solution. Again here, the solution is seen to be a $-$ branch solution arriving at the minimum and the curvature of the slice is subleading.
The two solutions meet at the $\f$-bounce.

The key property in this class of solutions is that, because of parameters, there is an intermediate regime in which the slice curvature is subleading to the other terms in the equations.

\subsection{Walking}

The solutions near the boundary between the  $W^{LR}_{1,0}$ and  $W^{LL}_{1,1}$ regions, studied above, are examples of ``walking" solutions.
As can be seen in figure \ref{33}, in the middle of the flow, near the intermediate minimum of the potential,
 the scalar flows very slowly while the scale factor changes at the same time by many orders of magnitude.
This is a classic example of ``walking" associated with an intermediate fixed point. For all flows that are
not exactly on the boundary of $W^{LR}_{1,0}$ and  $W^{LL}_{1,1}$, this minimum is never reached, and eventually the flow departs
and returns to $UV_{L}$\footnote{ An example with such a type of flow, with an intermediate near fixed point
 has been observed in a non-Lorentz-invariant example in \cite{Chris}.}.

Walking here is triggered by being close to the boundary between the two classes of solutions, and which
 describes the transition between ending in $UV_{L}$ and ending at $UV_{R}$.
 This is distinct from the walking described in the class of models of V-QCD, \cite{vqcd}, where the
 walking obtained there, below the conformal windows is controlled by parameters\footnote{In that case
 the parameter is the ratio of flavors to colors.} that appear in the bulk action. Here,
 the parameter is a boundary condition of the equations.

\subsection{Flow fragmentation and the generation of a new boundary}

The generic solution discussed in this paper can be interpreted as having two boundaries.  In our ansatz, they are located at $u\to \pm \infty$.
If the slices are compact constant curvature hyperbolic surfaces, then the solutions are standard wormholes with two boundaries.
If the slices are copies of AdS$_d$, then the solutions are describing interfaces between the two copies of the boundary at
$\pm \infty$, which form parts of the total boundary of the bulk solutions.

\begin{figure}[!t]
\centering
\begin{subfigure}{1\textwidth}
\includegraphics[width=1\textwidth]{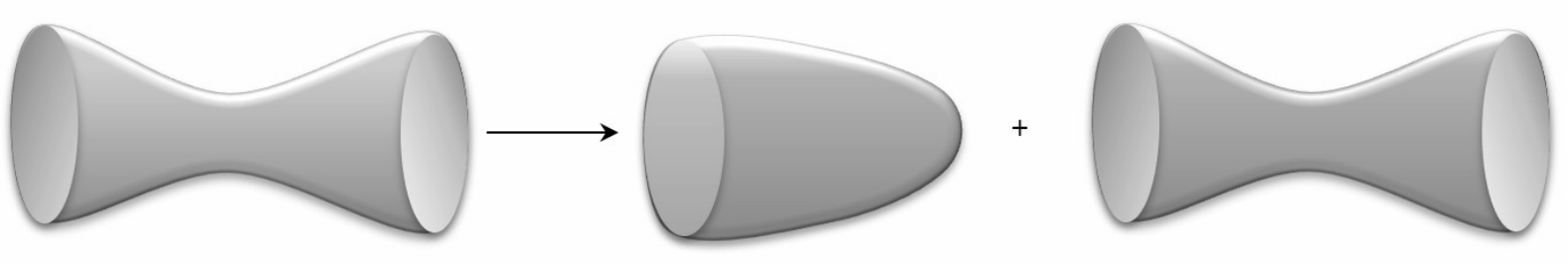}
\end{subfigure}
\caption{\footnotesize{Schematic representation of flow-fragmentation and the generation of an extra boundary. }}\label{frag}
\end{figure}

However, as we described above, the degeneration of the single wormhole solution as we approach the boundary between
 regions $W^{LR}_{1,0}$ and  $W^{LL}_{1,1}$ leads to a composite solution that is composed of a direct sum of two solutions.
The first, the $-+$ part is a solution with two boundaries. The first boundary is always at $UV_{L}$ and exists always.
 The second boundary appeared in the limit, and is at the minimum of the potential.
The $--$ part of the solution is a solution with a single boundary at $UV_{L}$ that existed always. It has
 developed now an end-point at the minimum of the potential, where the size of the slice has
shrunk to zero and the solution ends.
Therefore, at the degeneration limit, at the minimum of the potential both a boundary and an  end of
 space are generated for the two pieces of the solution.
 We call this limiting phenomenon, {\em flow-fragmentation}. It has been also observed in other curved flows in \cite{C}.
 We denote it schematically in figure \ref{frag}.

The existence of flow-fragmentation suggests that in the gravity landscape, the number of boundaries of the wormholes is not a property
 that is topologically distinct, but the moduli space of solutions involves a fractionalization/degeneration
  of solutions with a different number of boundaries.
The phenomenon suggests that there might exist a (topological) algebra of flows,   but whether this is true, or what are the rules is not clear yet.

Flow-fragmentation is an  interesting phenomenon, but it is not yet clear to us what the implications of this phenomenon
 for holography are, and what is the interpretation for  Euclidean wormholes.

\begin{figure}[!b]
\centering
\begin{subfigure}{0.49\textwidth}
\includegraphics[width=1\textwidth]{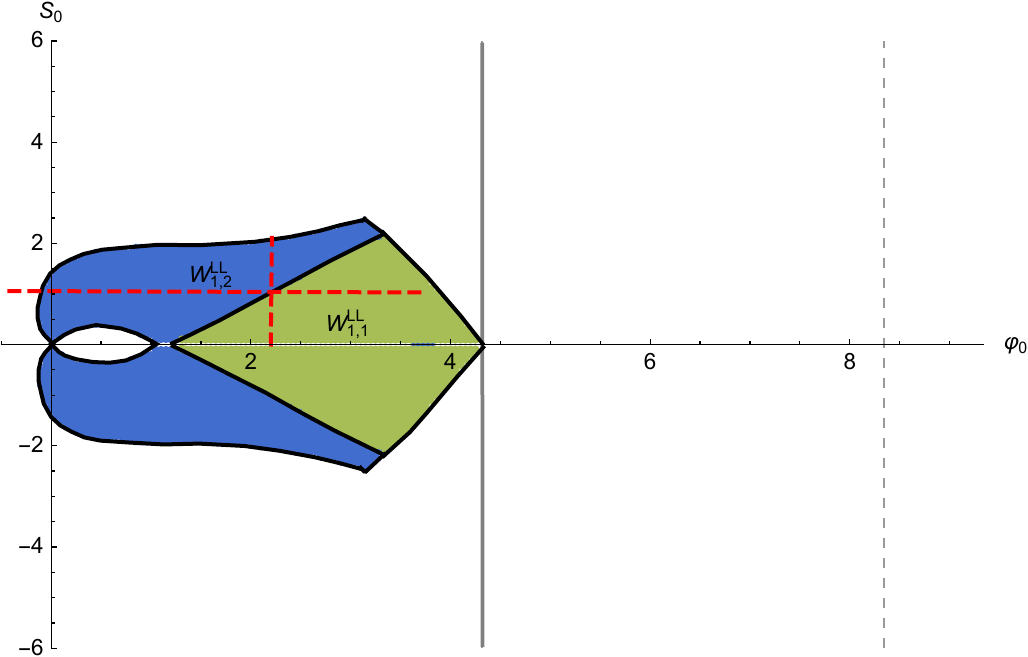}
\caption{}\label{crossBD}
\end{subfigure}
\centering
\begin{subfigure}{0.49\textwidth}
\includegraphics[width=1\textwidth]{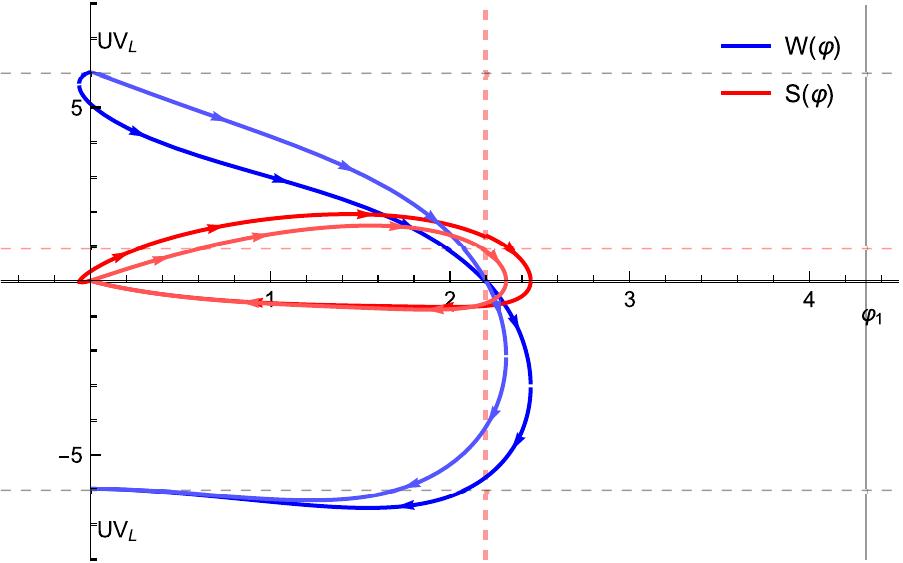}
\caption{}\label{TDB}
\end{subfigure}
\centering
\caption{\footnotesize{(a): Type $W^{LL}_{1,1}$ and $W^{LL}_{1,2}$ space of solutions. (b): Two  flows of type $W^{LL}_{1,2}$ and type $W^{LL}_{1,1}$ at $\f_0=2.2$ (red vertical dashed line) near the border (a point where the red horizontal and vertical dashed lines are joined). The blue curve with two bounces belongs to the $W^{LL}_{1,2}$ region and the flow with one bounce is a $W^{LL}_{1,1}$ solution.}}
\end{figure}

\subsection{Crossing from type $W^{LL}_{1,1}$ to type $W^{LL}_{1,2}$ region}\label{blgr}
To find the behavior of the dimensionless parameters more explicitly, we consider moving along a vertical line from the upper border of $W^{LL}_{1,2}$ type space of solutions down to the $\f_0$ axis in the type $W^{LL}_{1,1}$ region, see figure \ref{crossBD}. By crossing this border, the relevant coupling of the QFT on the $UV_{L}$ boundary with $W>0$ changes from negative to positive values. This can be seen in figure \ref{TDB} where the blue curve with an extra bounce belongs to the $W^{LL}_{1,2}$ region and the  flow without the extra bounce is a $W^{LL}_{1,1}$ type solution.
\begin{figure}[!t]
\centering
\begin{subfigure}{0.49\textwidth}
\includegraphics[width=1\textwidth]{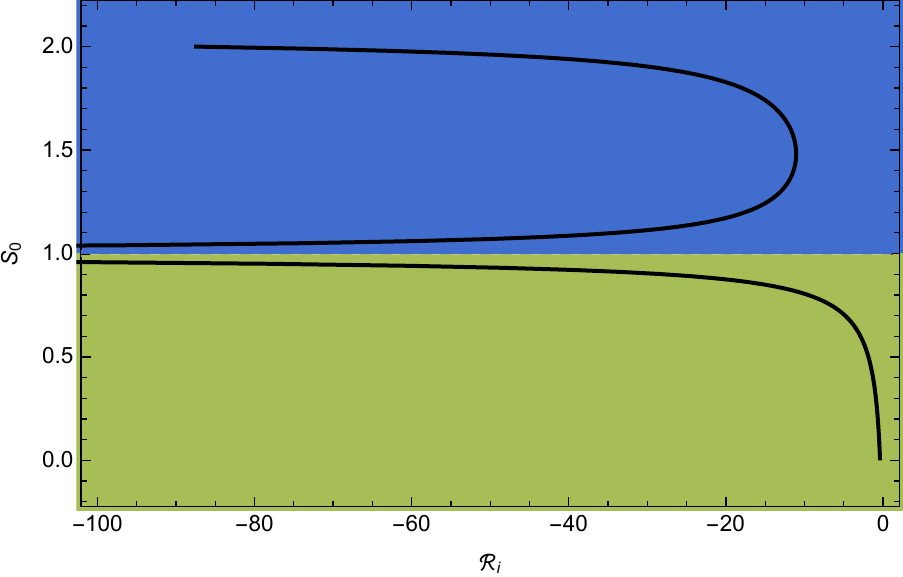}
\caption{}\label{crossii}
\end{subfigure}
\centering
\begin{subfigure}{0.49\textwidth}
\includegraphics[width=1\textwidth]{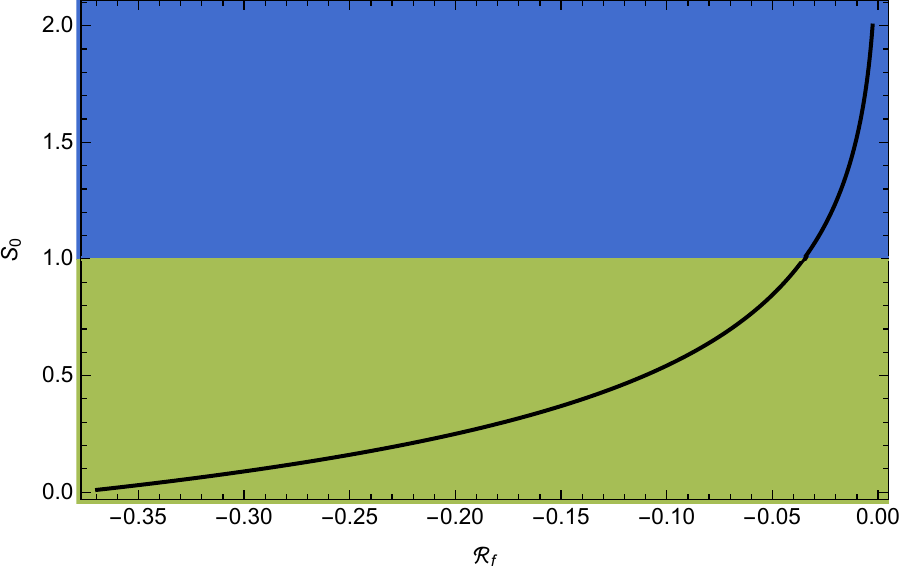}
\caption{}\label{crossff}
\end{subfigure}
\caption{\footnotesize{(a): Space of type $W^{LL}_{1,1}$ and $W^{LL}_{1,2}$ solutions and their joint boundary. We move from $W^{LL}_{1,1}$ region to $W^{LL}_{1,2}$ region on a constant $\f_0$. (b,c): The behavior of dimensionless curvature $\mathcal{R}_i$ when we cross the boundary from region $W^{LL}_{1,1}$ or $W^{LL}_{1,2}$ on a constant $\f_0=2.2$. The horizontal red dashed line denotes the location of the boundary nearly at $S_0\approx 0.95$. }}
\end{figure}

Results are summarized as follows:

1) As we already explained in type $W^{LL}_{1,2}$ solutions, close to the border of blue and green regions in figure \ref{crossii}, on both sides $\mathcal{R}_i \rightarrow -\infty$ while $\mathcal{R}_f$ remains finite and continuous (see figure \ref{crossff}). The reason is the existence of an extra $\f$-bounce point at negative values of $\f$ for the $W^{LL}_{1,2}$ solution. At this border, the coupling of the left QFT vanishes and according to equation \eqref{ERR}, the ratio of two dimensionless couplings also vanishes.
Therefore, this is a solution that is pure vev on the initial $UV_{L}$ side.

2) On the upper bound of the blue region again a divergence for $\mathcal{R}_i$ curvature is happening because the region beyond the blue region belongs to other types of  flows with multi-$\f$-bounce (more than three) solutions.

3) On the lower bound of the green region $\mathcal{R}_i\approx\mathcal{R}_f$ because it is close to the S-type region where the solutions are symmetric.

\subsection{Crossing from $W^{LR}_{1,1}$ to $W^{LR}_{1,0}$ and $W^{LL}_{1,2}$ regions}\label{crossEtoAD}
\begin{figure}[!ht]
\centering
\includegraphics[width=0.5\textwidth]{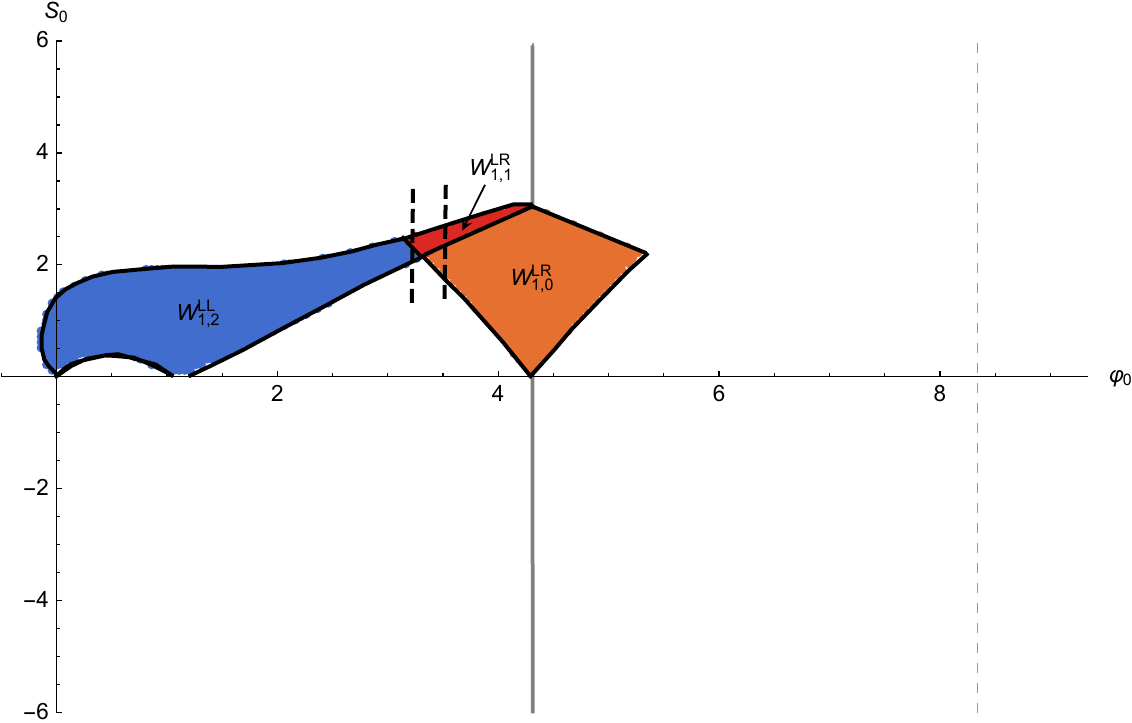}
\caption{Type $W^{LR}_{1,0}$, $W^{LL}_{1,2}$ and $W^{LR}_{1,1}$ space of solutions.}\label{crossEAD}
\end{figure}

To complete our analysis we also consider moving on a constant line $\f=\f_0$ in the type $W^{LR}_{1,1}$ space of solutions, into the type $W^{LR}_{1,0}$ or type $W^{LL}_{1,2}$ regions, see figure \ref{crossEAD}.
\begin{figure}[!t]
\centering
\begin{subfigure}{0.49\textwidth}
\includegraphics[width=1\textwidth]{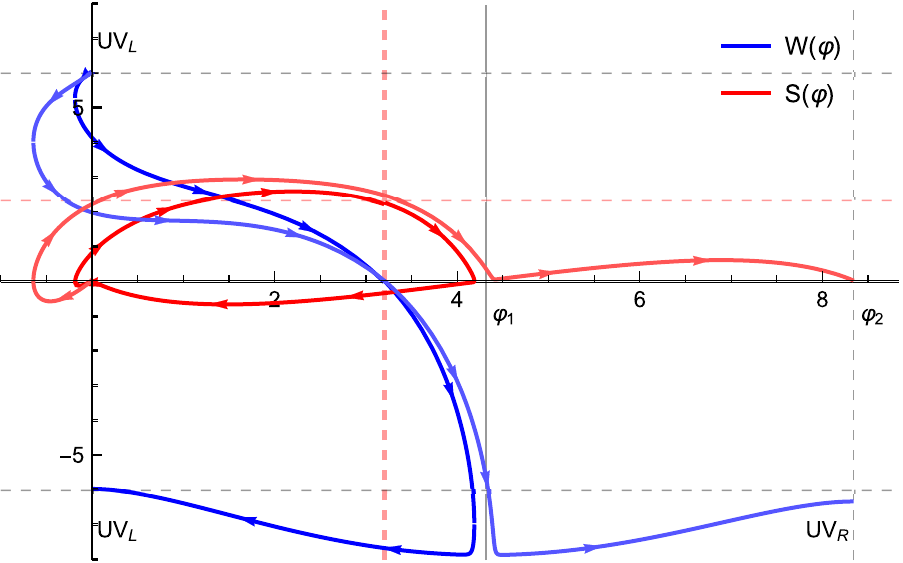}
\caption{}\label{TED}
\end{subfigure}
\centering
\begin{subfigure}{0.49\textwidth}
\includegraphics[width=1\textwidth]{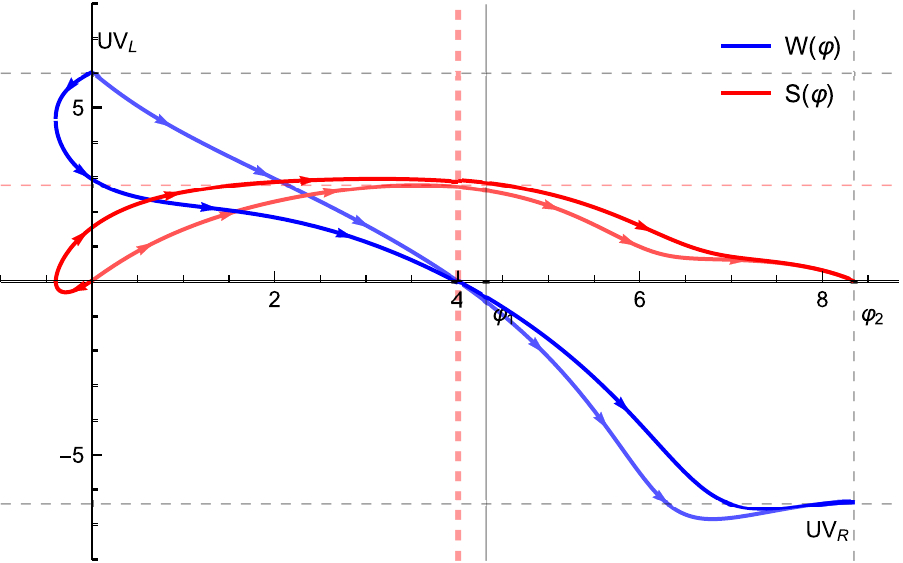}
\caption{}\label{TEA}
\end{subfigure}
\caption{\footnotesize{ (a): An example of flows of type $W^{LR}_{1,1}$ (ends on UV$_R$) and type $W^{LL}_{1,2}$ (ends on UV$_L$) with $\f_0=3.2$ and $S_0$ near the border of these regions. (b): An example of flows of type $W^{LR}_{1,0}$ and $W^{LR}_{1,1}$  with $\f_0=4$.}}
\end{figure}
\begin{figure}[!t]
\centering
\begin{subfigure}{0.48\textwidth}
\includegraphics[width=1\textwidth]{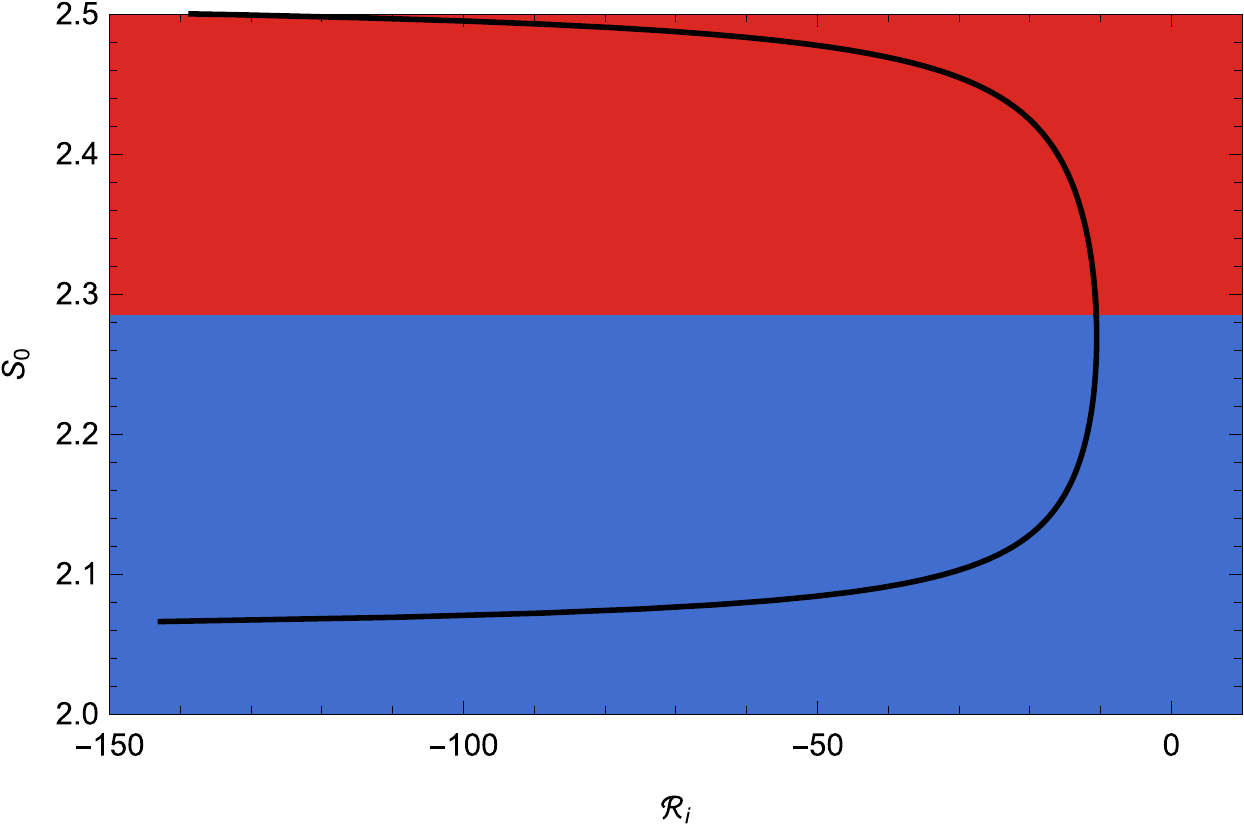}
\caption{}\label{crossED1}
\end{subfigure}
\centering
\begin{subfigure}{0.51\textwidth}
\includegraphics[width=1\textwidth]{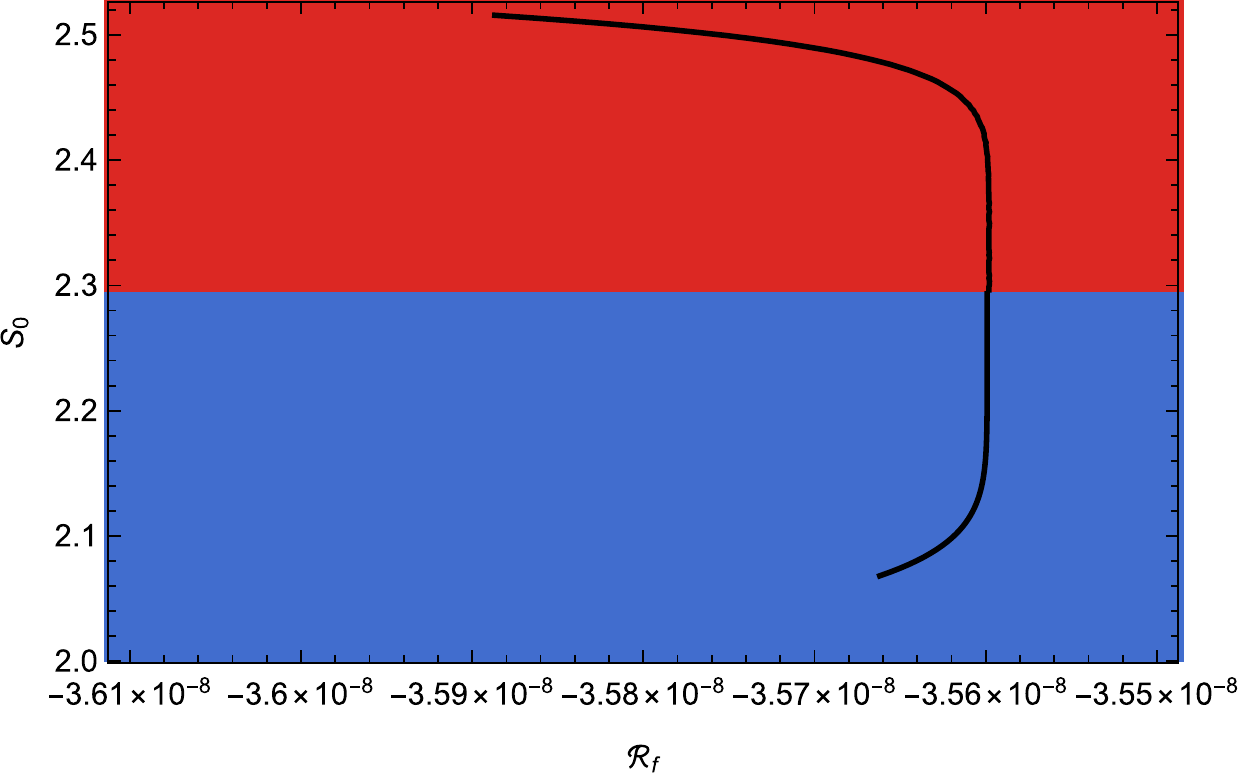}
\caption{}\label{crossED2}
\end{subfigure}
\centering
\begin{subfigure}{0.49\textwidth}
\includegraphics[width=1\textwidth]{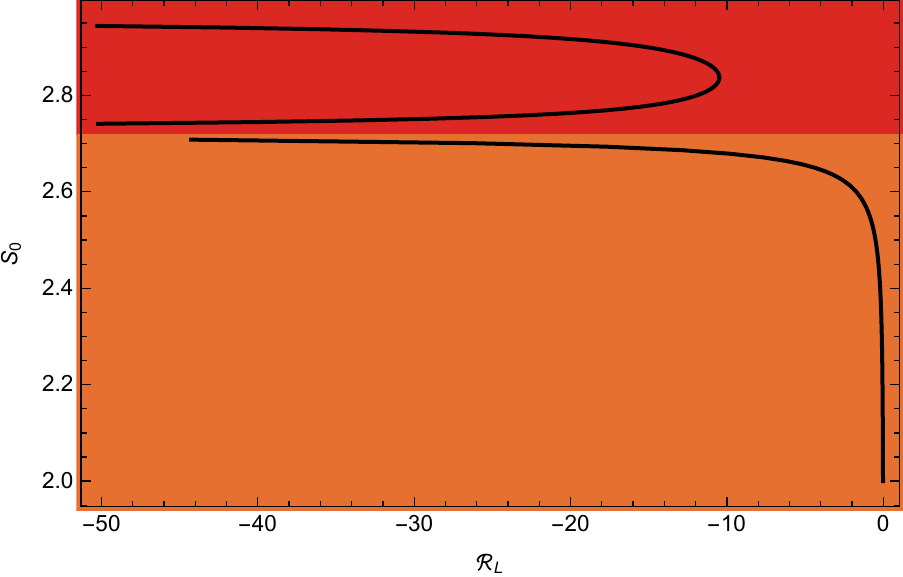}
\caption{}\label{crossEA1}
\end{subfigure}
\centering
\begin{subfigure}{0.49\textwidth}
\includegraphics[width=1\textwidth]{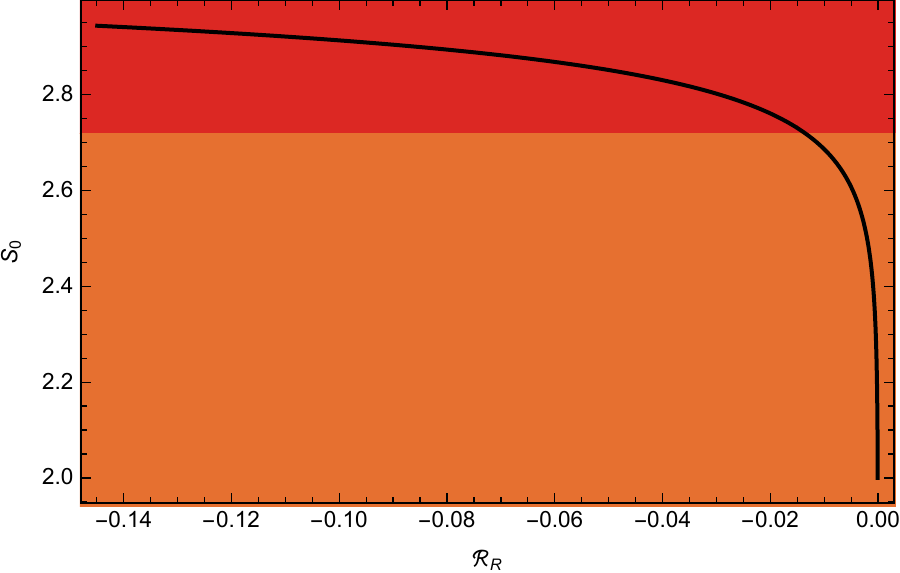}
\caption{}\label{crossEA2}
\end{subfigure}
\caption{\footnotesize{Dimensionless curvatures vs. $S_0$ as we cross from type $W^{LR}_{1,1}$ to type $W^{LR}_{1,0}$ region in figures (a) and (b) and as we cross from type $W^{LR}_{1,1}$ to type $W^{LL}_{1,2}$ region, figures (c) and (d).}}
\end{figure}
As an example in figure \ref{TED} we have sketched two  flows with the same $\f_0=3.2$ but different values of $S_0$. In this case, both  flows have a bounce at negative values of $\f$ but the final UV fixed point is different. Similarly figure \ref{TEA} shows two  flows with $\f_0=4$, one in the type $W^{LR}_{1,0}$ and the other in type $W^{LL}_{1,2}$ region. According to these  flows, we observe the following properties:

1) As we cross the border into the type $W^{LL}_{1,2}$ region,  that part of the  flow which connects to the $UV_L$ with $W>0$ gradually tends to the $W^{L Min_+}_{1,0}$ branch or the $W^{L Min_+}_{1,0}$ type solutions. At the boundary, the value of the dimensionless curvature for the $UV_L$ fixed point of $W^{L Min_+}_{1,0}$ branch is given by the value of the curvatures at the joint boundary, see figure \ref{crossED1}.
On the other hand, the value of the dimensionless curvature $\mathcal{R}_f$ tends to zero from both sides of the border, see figure \ref{crossED2} (note that the distance to zero is due to the finite numerical cut-off). As we move closer to the border the flow tends to the $W^{L Min_+}_{1,0}$ solutions.

2) If we move towards the $W^{LR}_{1,0}$ region the situation would be different. Since in type $W^{LR}_{1,1}$ we have an extra $\f$-bounce at $\f<0$ near the left UV fixed point, we expect at the UV boundary $\f_-^{(L)}$ change its sign and it becomes zero when we choose $S_0$ on the border between the red and orange regions. Therefore we see $\mathcal{R}_L$ tend to $-\infty$ in figure \ref{crossEA1} on both sides of the border. Since there is no similar $\f$-bounce near the right UV fixed point the value of $\mathcal{R}_R$ reaches a finite value at the border, see figure \ref{crossEA2}.

\subsection{Corners}\label{a3a4cor}
\begin{figure}[!t]
\centering
\begin{subfigure}{0.51\textwidth}
\includegraphics[width=1\textwidth]{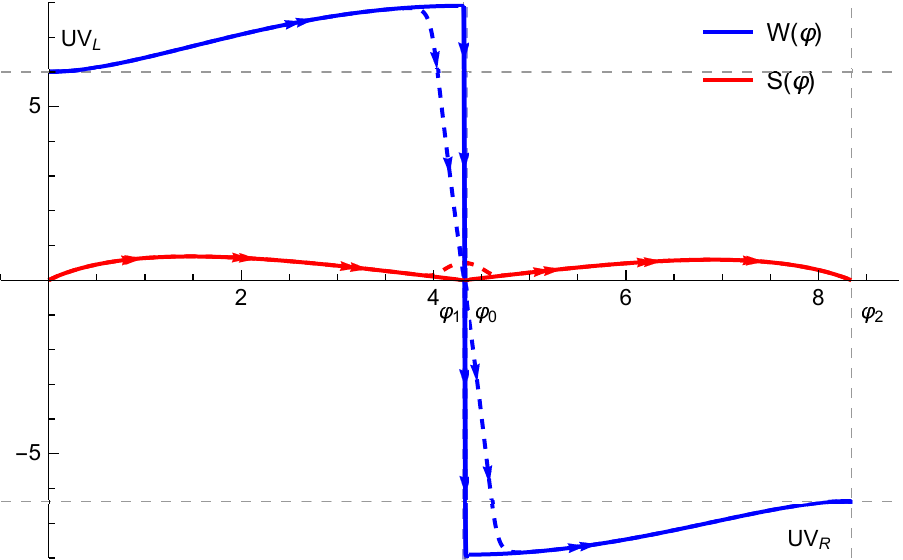}
\caption{}\label{NM1}
\end{subfigure}
\centering
\begin{subfigure}{0.49\textwidth}
\includegraphics[width=1\textwidth]{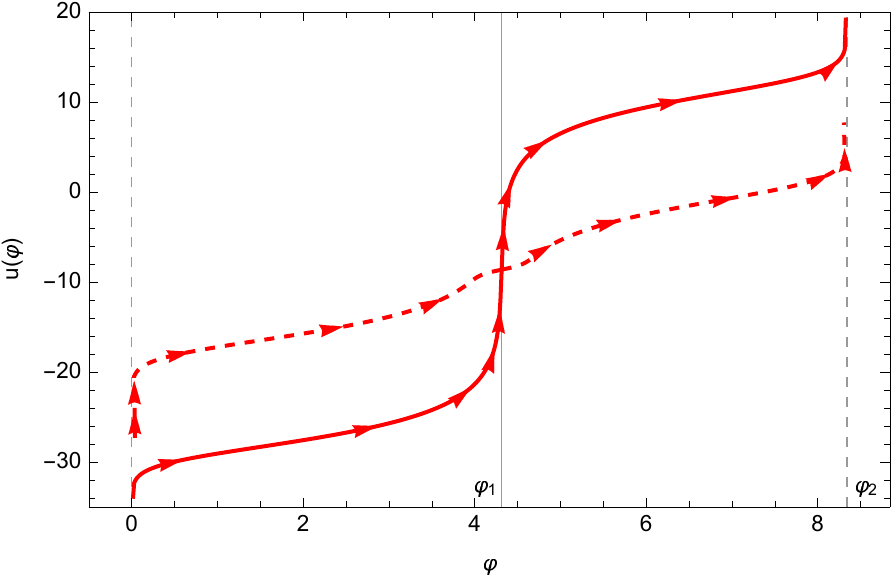}
\caption{}\label{NM2}
\end{subfigure}
\centering
\begin{subfigure}{0.49\textwidth}
\includegraphics[width=1\textwidth]{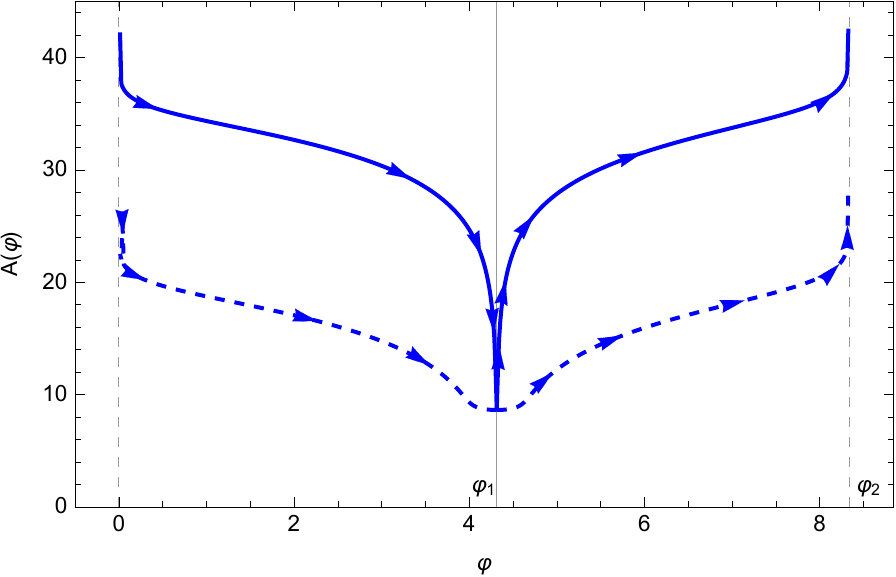}
\caption{}\label{NM3}
\end{subfigure}
\caption{\footnotesize{Along the fixed line $\f_0=\f_1$ i.e. the minimum of the potential, see figure \protect\ref{typeAmoduli}, if we decrease the value of $S_0$ down to zero,  gradually the dashed curves in all figures above move toward the solid curves. In the above curves the dashed curves have $S_0=0.5$ and the solid ones $S_0=0.01$.}}
\end{figure}

Consider we are moving inside the orange region in figure \ref{typeAmoduli}. All solutions inside this region correspond to the $W^{LR}_{1,0}$ or $(Max_-, Max_-)$ solutions. We are now looking to the deformation of the flows as we move towards the corners of this region i.e. the points where boundaries are intersected:

$\bullet$ $a_3\cap a_4$ corner:
As a specific point in the space of solutions, let's move towards the $S_0=0$ along the constant $\f_0=\f_1$ i.e. the minimum of the potential. This is depicted in figure \ref{NM1}, where the dashed curve gradually changes to a solid curve as we decrease $S_0$.
The behavior of holographic coordinate and scale factor are sketched in figures \ref{NM2} and \ref{NM3}. These figures show that as the corner the $(Max_-, Max_-)$ solution splits into three different solutions. The first one is the flat solution $(Max_-,Min_-)$ stretched between the $UV_L$ boundary and the minimum of the potential. The second one is an $AdS$ solution located at the minimum of the potential $\f=\f_1$ and the third one is a flat solution $(Min_-,Max_-)$ which is stretched between the minimum of the potential and $UV_R$ boundary.
\begin{figure}[!ht]
\centering
\includegraphics[width=0.6\textwidth]{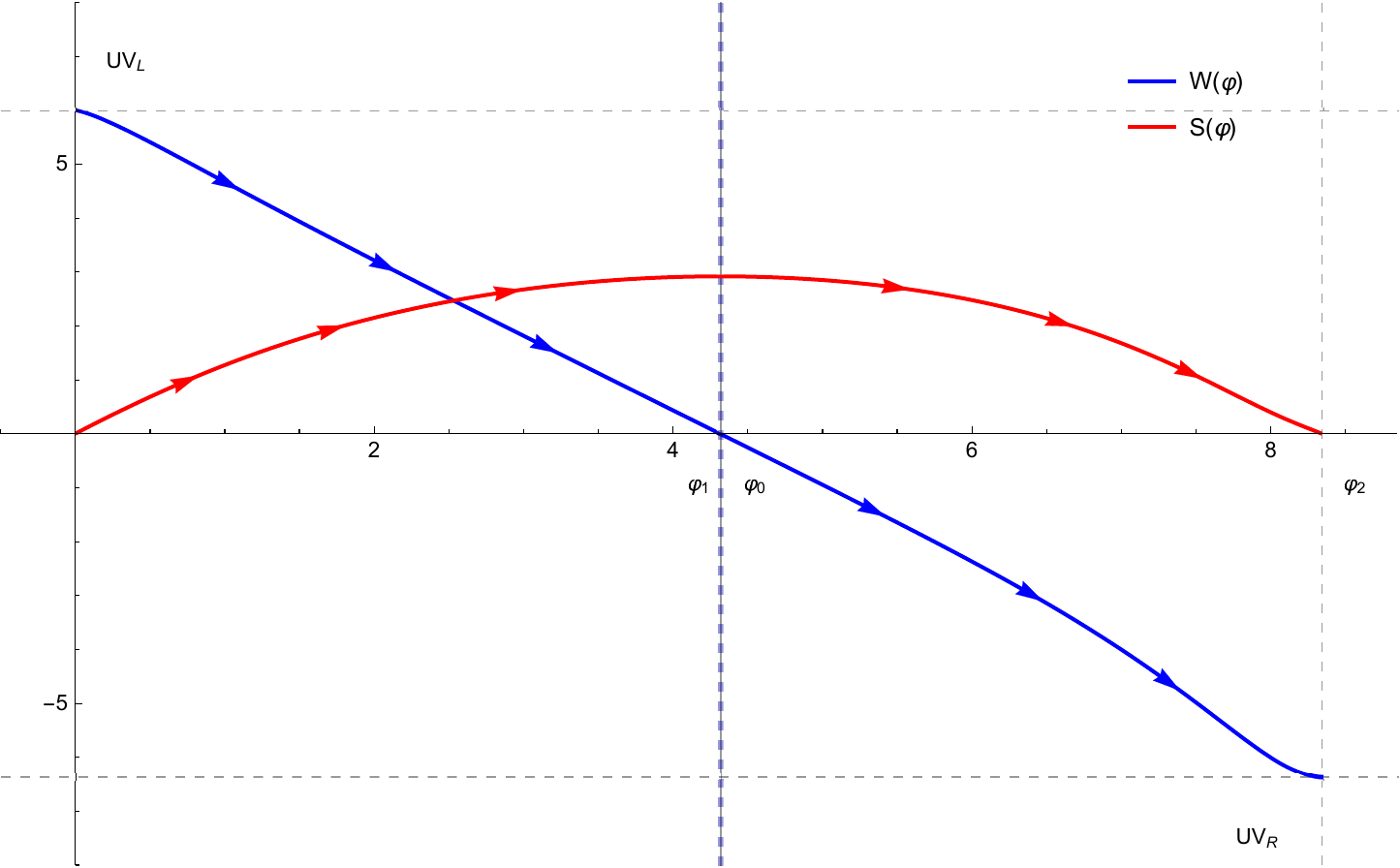}
\caption{\footnotesize{When we move towards the top corner of the orange region in figure \protect\ref{typeAmoduli}, the $(Max_-,Max_-)$ solution gradually tends to the $(Max_+,Max_+)$ solutions.}}\label{cornu}
\end{figure}

$\bullet$ $a_1\cap a_2$ corner: If we move towards the $S_0\approx 2.98$ along the constant $\f_0=\f_1$ i.e. the minimum of the potential, then the $(Max_-, Max_-)$ solution gradually tends to the $(Max_+, Max_+)$ solution which both $\mathcal{R}_L$ and $\mathcal{R}_R$ go to $-\infty$. This is sketched in figure \ref{cornu}.

\begin{figure}[!t]
\centering
\begin{subfigure}{0.49\textwidth}
\includegraphics[width=1\textwidth]{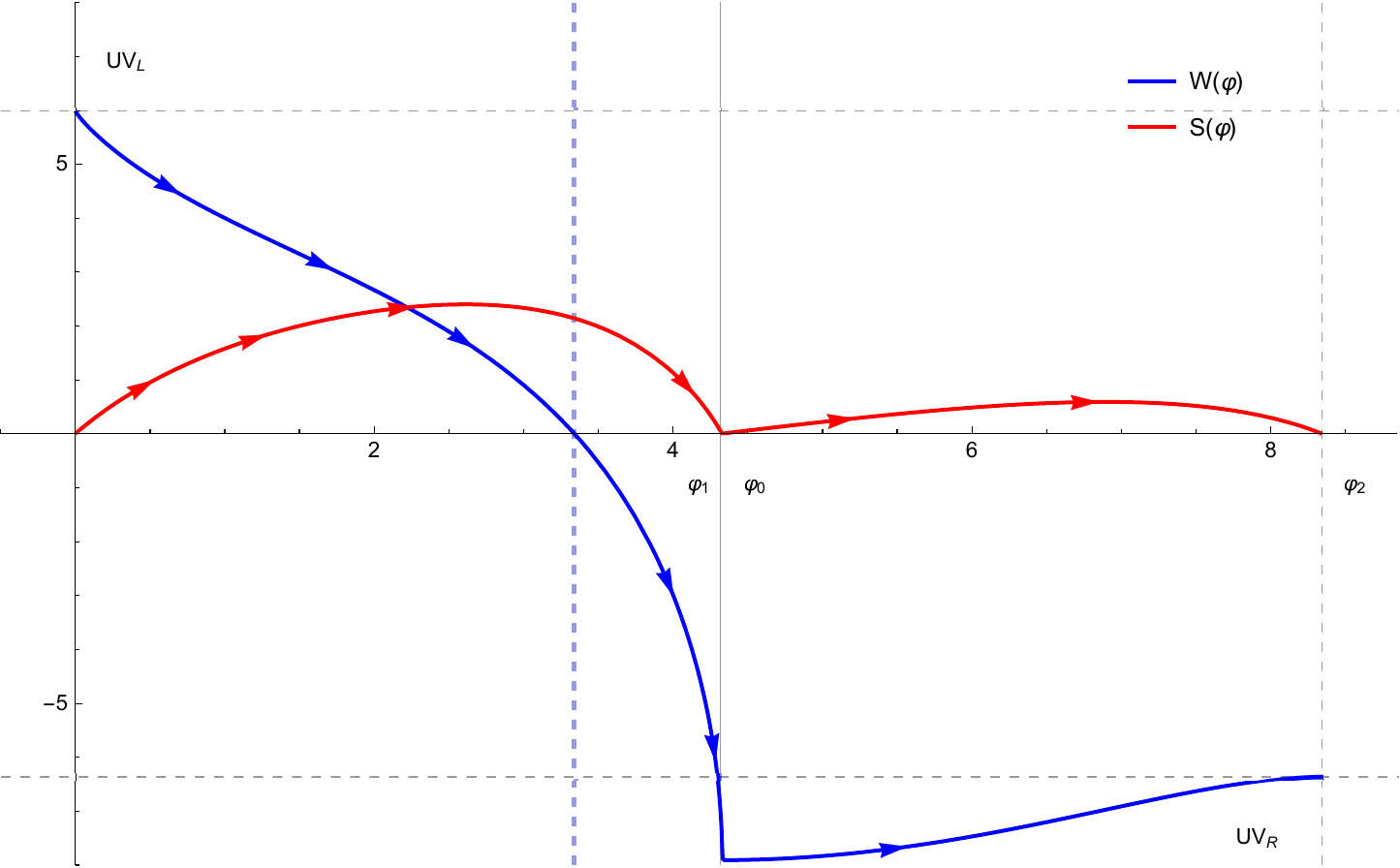}
\caption{}\label{cornl}
\end{subfigure}
\centering
\begin{subfigure}{0.49\textwidth}
\includegraphics[width=1\textwidth]{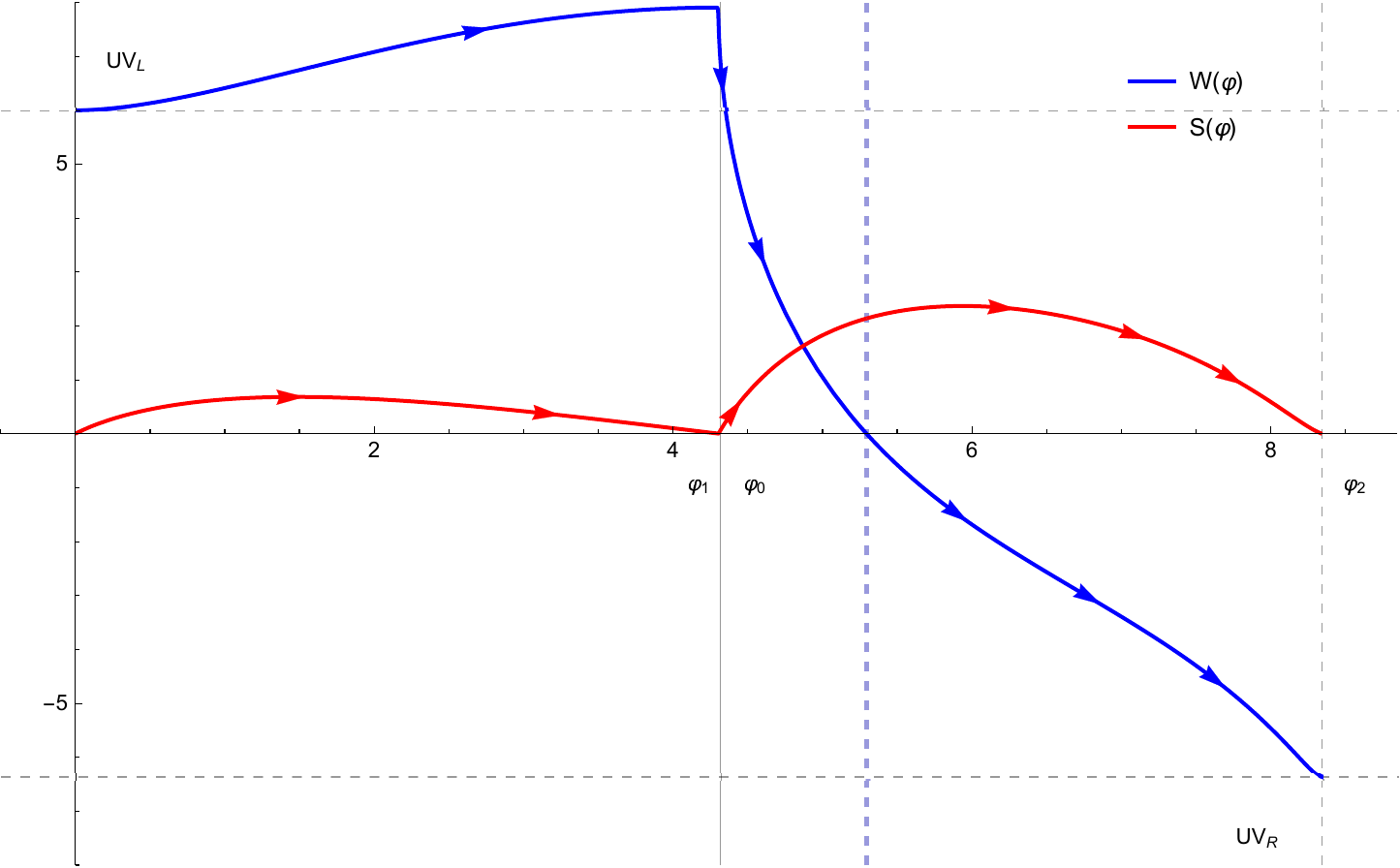}
\caption{}\label{cornr}
\end{subfigure}
\caption{\footnotesize{As we move towards the left or right corners of the orange region in figure \protect\ref{typeAmoduli} the $(Max_-,Max_-)$ solution gradually tends to above solutions. (a): Splitting of solution on the left or $a_1\cap a_4$ corner. (b): Splitting on the right or $a_2\cap a_3$ corner.}}
\end{figure}

$\bullet$ $a_1\cap a_4$ corner: If we move towards the $S_0\approx 2.19$ and $\f_0\approx 3.32$ we arrive the corner where two $a_3$ and $a_4$ boundaries are intersected. In this case, the $(Max_-, Max_-)$ solution gradually splits into the $(Max_+, Min_+)$ and the flat solution $(Min_-,Max_-)$. This is sketched in figure \ref{cornl}.

$\bullet$ $a_2\cap a_3$ corner: If we move towards the $S_0\approx 2.19$ and $\f_0\approx 5.33$ we arrive the corner where two $a_2$ and $a_3$ boundaries are intersected. In this case, the $(Max_-, Max_-)$ solution gradually splits into the flat solution $(Max_-,Min_-)$ and $(Min_+, Max_+)$. This is sketched in figure \ref{cornr}.

\subsection{Summary of the results}
\begin{figure}[!t]
\centering
\begin{subfigure}{0.49\textwidth}
\includegraphics[width=1\textwidth]{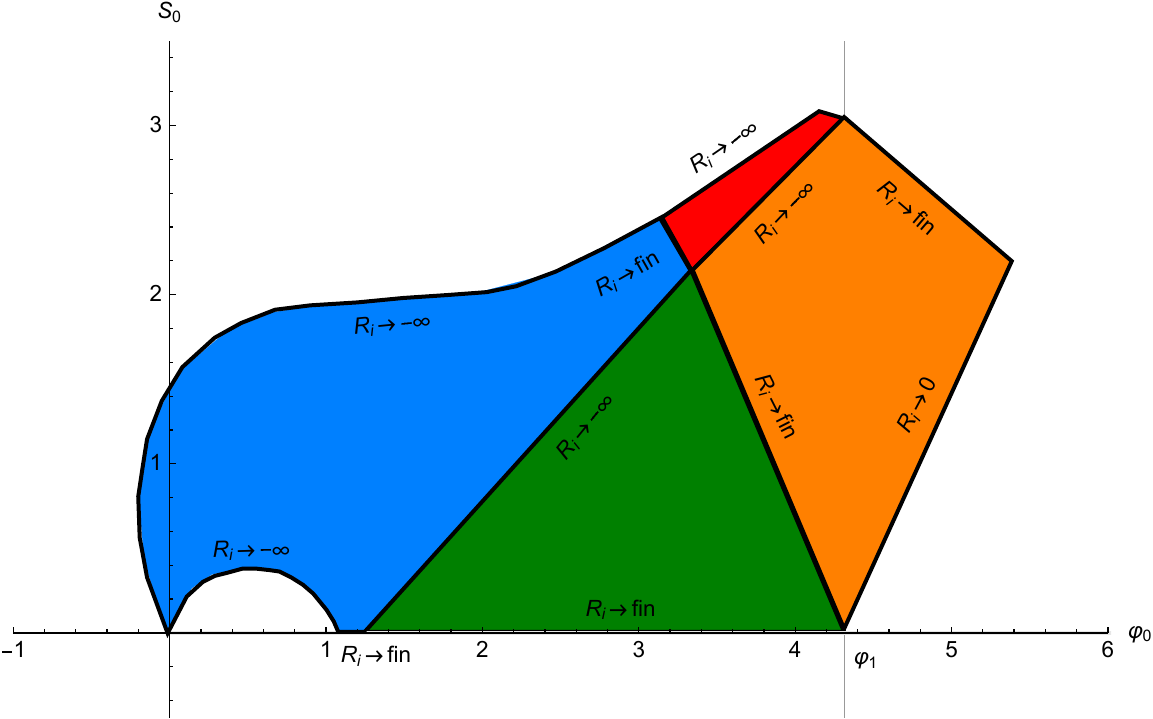}
\caption{}\label{sum1}
\end{subfigure}
\centering
\begin{subfigure}{0.49\textwidth}
\includegraphics[width=1\textwidth]{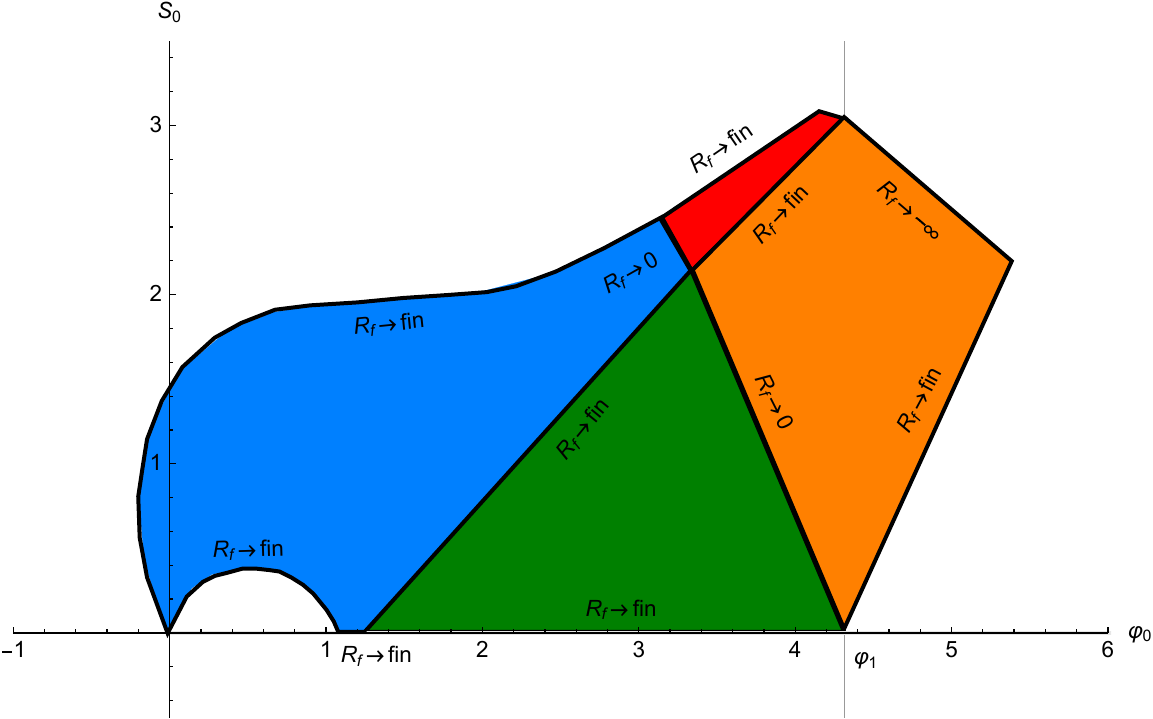}
\caption{}\label{sum2}
\end{subfigure}
\caption{\footnotesize{(a) and (b): The behavior of $\mathcal{R}_i$ and $\mathcal{R}_f$ at boundaries.}}
\end{figure}

In this section, we have summarised all the previous results in the table \ref{tab:tab1}.
\begin{table}[!ht]
\begin{center}
    \begin{tabular}{| c | c | c | c | c | c |}
    \hline
Type & $\mathcal{R}_i$ & $\mathcal{R}_f$ & $\xi$ & Region & Section\\ \hline
\multirow{2}{*}{$(Max_-, Max_-)$} & \multirow{2}{*}{fin} & \multirow{2}{*}{fin} & \multirow{2}{*}{fin}& \multirow{2}{*}{Colored regions} &  \ref{TAS}, \ref{green},\\
&  &  &  &  & \ref{blue}, \ref{red} \\ \hline

\multirow{2}{*}{$(Max_+, Max_-)$ }& \multirow{2}{*}{$-\infty$ }& \multirow{2}{*}{fin} & \multirow{2}{*}{0} & $a_1, b_1,$ & \ref{crossEtoAD}, \ref{blgr}, \\
&  &  &   & $(d_1, d_5), (e_1, e_2)$ & \ref{blue}, \ref{red}
\\ \hline

$(Max_-, Max_+)$ & fin & $-\infty$  & $\infty$ & $a_2$ & \ref{TAS}\\ \hline

$(Max_-, Min_+)$ & fin & fin  & $\infty$ & \multirow{3}{*}{$a_4,  d_2$} &\multirow{3}{*}{\ref{crossAtoB}, \ref{crossEtoAD}} \\
$\oplus$ &  &  &  &  &  \\
$(Min_-, Max_-)$  & 0 & 0  & - & & \\ \hline

$(Max_-, Min_+)$ & 0 & fin & $\infty$ & \multirow{3}{*}{$a_3$} &  \multirow{3}{*}{\ref{TAS}} \\
$\oplus$ &  &  &  &  &  \\
$(Min_-, Max_-)$  & 0 & 0 & - & & \\ \hline

S & fin & fin  & 1 & $b_3, d_4$ & \ref{typeCsec} \\ \hline
$(Max_+, Max_+)$ & $-\infty$ & $-\infty$  & - & $a_1\cap a_2$ & \ref{a3a4cor}\\ \hline

$(Min_+, Max_+)$ & fin & $-\infty$  & - & \multirow{3}{*}{$a_2\cap a_3$} & \multirow{3}{*}{\ref{a3a4cor}}\\
$\oplus$ &  &  &  &  &  \\
$(Max_-, Min_-)$  & 0 & 0  & - & &\\ \hline

$(Max_+,Min_+)$ & $-\infty$ & fin  & - & \multirow{3}{*}{$a_1\cap a_4$} & \multirow{3}{*}{\ref{a3a4cor}}\\
$\oplus$ &  &  &  &  &  \\
$(Min_-, Max_-)$  & 0 & 0  & - & & \\ \hline

$(Max_-,Min_-)$  & 0 & 0  & - & \multirow{5}{*}{$a_3\cap a_4$ } & \multirow{5}{*}{\ref{a3a4cor}}\\
$\oplus$ &  &  &  &  &  \\
$AdS_{Min}$  & fin & fin  & fin & &\\
$\oplus$ &  &  &  &  &  \\
$(Min_-, Max_-)$  & 0 & 0  & - & & \\ \hline
\end{tabular}
\caption{\label{tab:tab1}\footnotesize{In this table we address the properties of different types of solutions in the space of the solutions. For example $a_1\cap a_2$ means the intersection point of $a_1$ and $a_2$ boundaries.
The boundaries in the fifth column are defined in figure \protect\ref{sum0}, or \protect\ref{Moduli}. The information of the second and third columns are also portrayed in figures \protect\ref{sum1}, \protect\ref{sum2}.
}}
\end{center}
\end{table}
\begin{figure}[!b]
\centering
\begin{subfigure}{0.49\textwidth}
\includegraphics[width=1\textwidth]{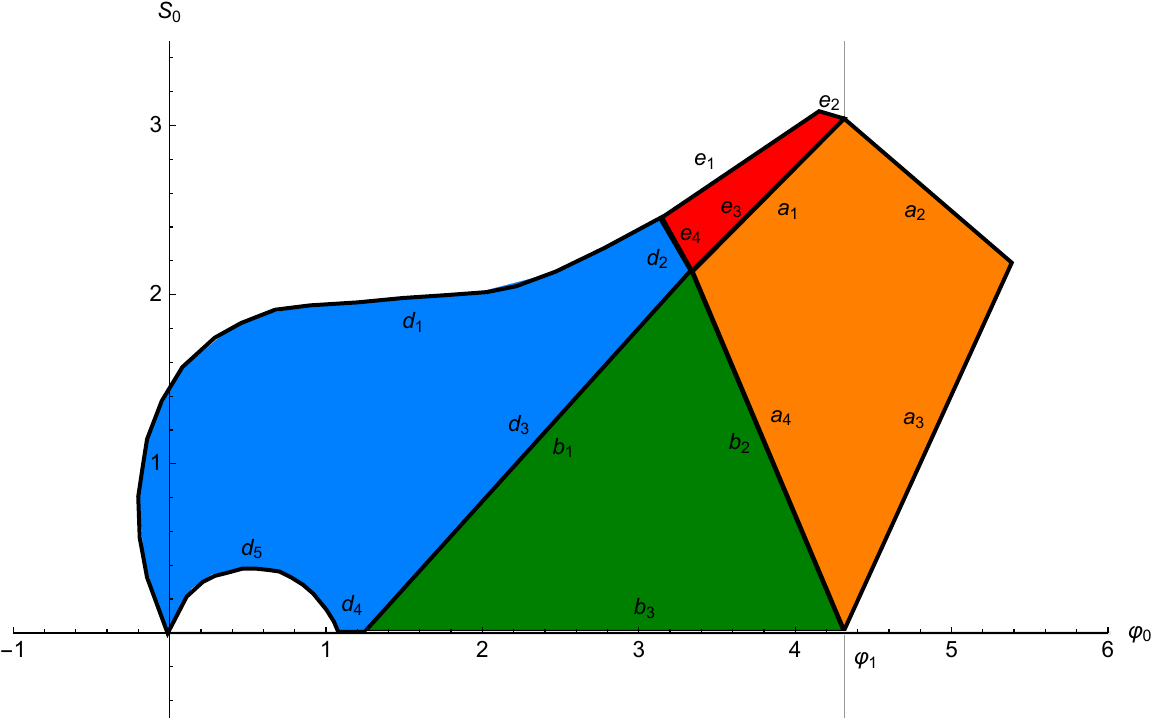}
\caption{}\label{sum0}
\end{subfigure}
\centering
\begin{subfigure}{0.49\textwidth}
\includegraphics[width=1\textwidth]{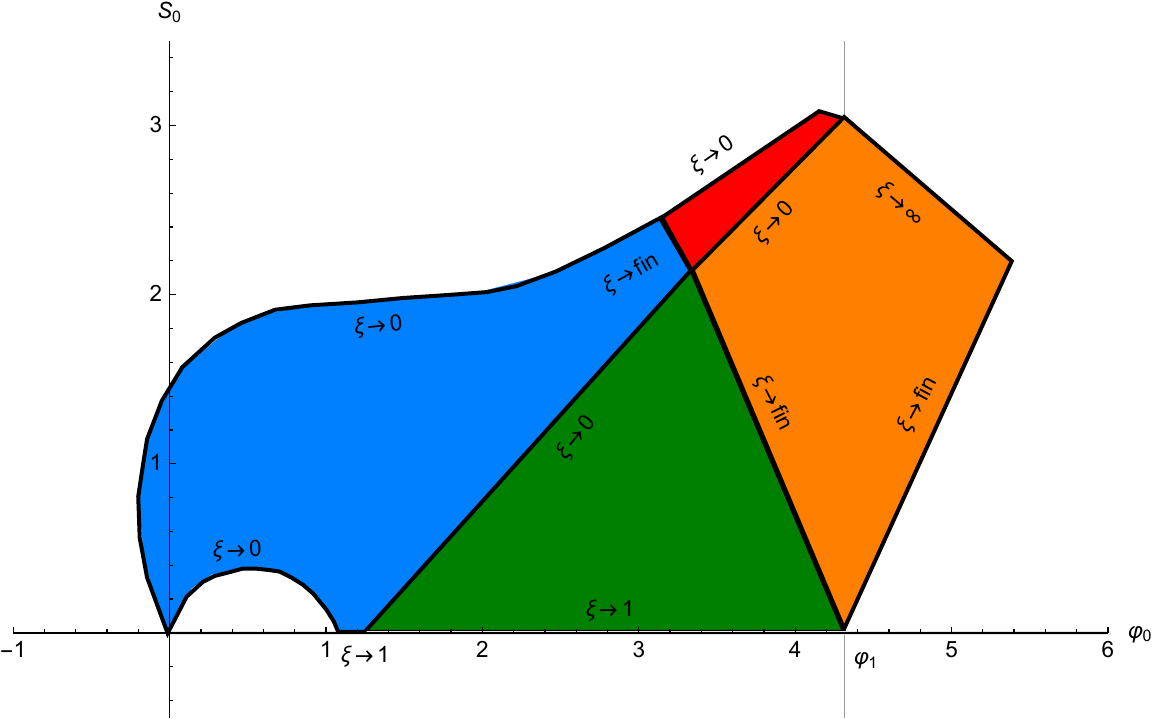}
\caption{}\label{sum3}
\end{subfigure}
\caption{\footnotesize{(a) Space of solution with its boundaries. (b): The ratio of two relevant couplings, $\xi$, near the boundaries.}}
\end{figure}
\begin{itemize}

\item
In the first column different possible solutions related to the space of solution in figure \ref{Moduli} are given.
\item
The second and third columns give the information on dimensionless curvatures for the initial and final boundaries of the RG solutions. To have a better view, we have shown this information on the boundaries in figures \ref{sum1} and \ref{sum2}.
\item
The fourth column gives the ratio of two relevant couplings of the boundary QFTs. This ratio for example does not exist for $(Min_-, Max_-)$ or $(Max_-,Min_-)$ because $Min_-$ is an IR endpoint. The behavior of this ratio near the boundaries is sketched in figure \ref{sum3}.
\item
The last two columns show the region of validity  and the section which we have discussed that type of solution.
\end{itemize}

\section{Four-parameter solutions\label{3p}}

In a boundary theory which is the product of two QFTs, there are in principle four dimensionful parameters, namely the two UV curvatures $R^{UV}_i, R^{UV}_f$ and the two UV relevant couplings  $\f_-^{(i)},  \f_-^{(f)}$.

 In the previous sections, we have extensively explored a three-parameter family of solutions. As discussed  in subsection \ref{sec:parameters}, in the bulk, the three independent parameters are $\f_0$ (the position in field space of the A-bounce),  $S_0$ (the derivative of the scalar field at the A-bounce), and $u_0$, the position of the $A$-bounce. As we have seen, these three ``bulk'' parameters can be traded for three  dimensionless ``boundary'' parameters, for example, the dimensionless curvatures ${\cal R}_{i.f}$ at the initial and final boundary and the ratio of the initial and finite couplings $\xi$ defined in equation (\ref{Eratio}).

In this section we show how to extend the solution spaces by adding the missing fourth   parameter, without having to solve the Einstein equation from the start: the new  solutions are constructed piecewise using  as building blocks the  solutions we have already encountered.

This completes the solution space, as now a generic point in the boundary field theory parameter space can be, in principle, associated with a bulk solution\footnote{More precisely, the space of bulk solutions has the same dimensionality as the space expected in the boundary theory. The correspondence may not be one-to-one, i.e. there may exist choices of the boundary parameters for which no bulk regular solution exists.}.

We begin to illustrate the construction in the case of pure AdS gravity, then we generalize it to the scalar flows.

\subsection{Pure gravity}

We consider pure Euclidean gravity in $d+1$-dimensions with a negative cosmological constant,
\be
\Lambda  ={ -\frac{d(d-1)}{ \ell^2}}\,.
\ee
The Einstein equation is solved by an   AdS$_d$-sliced AdS$_{d+1}$, which  is discussed in detail in Appendix \ref{app:ads-slicing}. We
 first write a  generalization of the metric (\ref{GVH5}):
\be \label{AdSbc1}
ds^2 = du^2 + \cosh^2\left({\frac{u-u_0}{\ell}}\right) ds^2_{d,\ell}\,.
\ee

The position of the A-bounce in the metric (\ref{AdSbc1}) is at
$u_0$. This is now left  arbitrary, as it is an integration constant of Einstein's equations.
In fact, $u_0$  can be set to
zero by a large diffeomorphism $u \to u+u_0$. However, this changes the
boundary condition at both $u = \pm \infty$ and  therefore affects the source of the metric.
Therefore, $u_0$ is a physical parameter of the metric as we shall see below.

The space-time (\ref{AdSbc1}) is again global Euclidean AdS$_{d+1}$, as one can see by writing the embedding space coordinates generalizing   (\ref{GVH1}) as:
\begin{align}\label{AdSbc1-iii}
 X_{-1} &= \ell \cosh(u-u_0)\, \cosh \tau \, \cosh r \,,  \\
 X_0 &= \ell \sinh(u-u_0)\,, \nn \\
 X_d &= \ell \cosh(u-u_0) \sinh \tau\, \cosh r \,, \nonumber \\
 X_i &= \ell \cosh(u-u_0)\, \sinh \tau\, \Omega_i\,, \qquad i  = 1\ldots d-1\,, \quad \sum_{i=1}^{d-1} \Omega_i^2 = 1\,, \nn
\end{align}
where $\tau$, $r$ and  $\Omega_i$ are global coordinates on Euclidean $AdS_d$ and  $\Omega_i$  are angles parametrizing $S^{d-2}$. As before,  the coordinates above describe the connected component of the manifold (\ref{GVH1}) with positive $X_{-1}$.

Now take two such metrics (\ref{AdSbc1}) with different $u_0$ and glue
them at the A-bounce, i.e. where the argument of the $\cosh$ vanishes. This results in the following metric:
\be \label{AdSbc2}
ds^2  = \left\{ \begin{array}{l}  du^2 + \cosh^2\left({\frac{u-u_L}{\ell}}\right) ds^2_{d,\ell}\,, \qquad -\infty < u < u_L\,, \\  \\
d\tilde{u}^2 + \cosh^2\left({\frac{\tilde{u}-u_R}{\ell}}\right) ds^2_{d,\ell}\,, \qquad  u_R < \tilde{u} <
    +\infty\,. \end{array} \right.
\ee
Each side is diffeomorphic to AdS$_{d+1}$ so it solves the bulk
equations.

In general, when joining two geometries at a codimension-one interface with no localized energy density, one must ensure that  Israel's junction conditions are satisfied: both the induced metric on the interface and the extrinsic curvature (roughly the derivative of the metric normal to the interface) must be continuous. Both conditions hold for the ansatz (\ref{AdSbc2}), for any value of $u_{L,R}$:
\begin{enumerate}
\item The induced $d$-dimensional metric is manifestly continuous;
\item The extrinsic curvature $K_{\mu\nu}$ satisfies:
\be \label{AdSbc3}
K_{\mu\nu} \propto
\frac{1}{ \ell}\tanh \left({\frac{u-u_L}{\ell}}\right)\,,
\ee
on the left, and a similar expression on the right. Both vanish at the
junction.
\end{enumerate}
Therefore (\ref{AdSbc2}) is a well-defined solution of the bulk Einstein's equation.
In particular, all higher derivatives are continuous.

It is not hard to  show that the space-time  (\ref{AdSbc2}) is again the full Euclidean $AdS_{d+1}$ in a different coordinate system. To see this, we can go to the embedding space $R^{1,d+1}$ with  the definition  (\ref{GVH1}), and this time we define the following local coordinates:
\bsq
\bea\label{AdSbc4-ii}
&& X_0 < 0 : \left\{\begin{array}{l} X_{-1} = \ell \cosh(u-u_L)\, \cosh \tau \, \cosh r  \\
 X_d = \ell \cosh(u-u_L) \sinh \tau\, \cosh r  \\
X_i = \ell \cosh(u-u_L)\, \sinh \tau\, \Omega_i \\
 X_0 = \ell \sinh(u-u_L) \end{array} \right. \quad u<u_L\,,  \\
&&
X_0 \geq 0 : \left\{\begin{array}{l} X_{-1} = \ell \cosh(u-u_R)\, \cosh \tau \, \cosh r  \\
 X_d = \ell \cosh(u-u_R) \sinh \tau\, \cosh r  \\
X_i = \ell \cosh(u-u_R)\, \sinh \tau\, \Omega_i \\
 X_0 = \ell \sinh(u-u_R) \end{array} \right. \quad u \geq u_R\,.
\eea
\esq
These coordinates lead to the metric (\ref{AdSbc2}) and they clearly cover the whole hyperboloid, just like the coordinates (\ref{AdSbc1-iii}).

The geometry  (\ref{AdSbc2})  has two UV boundaries with AdS$_d$ geometry at $u\to -\infty$ and $\tilde{u}\to -\infty$. The CFT  parameters are the two boundary AdS lengths $\ell_L, \ell_R$ (or equivalently the AdS curvatures $R^{UV}_L, R^{UV}_R$). Their ratio is the only dimensionless parameter.  These are the only parameters as there is no scalar field, therefore no relevant coupling deforming the two CFTs.

Near each boundary $u\to\pm \infty$, the metric (\ref{AdSbc2}) takes the form:
\be \label{AdSbc4}
ds^2 \simeq \left\{ \begin{array}{l}  du^2 + e^{-\frac{2u}{\ell}} \left[
     {\frac14} e^{\frac{2 u_L}{\ell}} ds^2_{d,\ell} + O\left(e^{\frac{2u}{\ell}}\right) \right] \,\,\,\quad u \to -\infty\,, \\
\\
d\tilde{u}^2 +  e^{\frac{2\tilde{u}}{\ell}} \left[
       {\frac14}e^{-\frac{2 u_R}{\ell}} ds^2_{d,\ell} + O\left(e^{-\frac{2\tilde{u}}{\ell}}\right) \right] \quad \tilde{u} \to +\infty\,. \end{array} \right.
\ee
From this expression we can read off the asymptotic  geometries at $u\to \pm \infty$: they are two Euclidean
AdS$_d$ space-times  with lengths
\be \label{AdSbc5}
\ell_L =   \frac{\ell}{ 2} e^{\frac{u_L}{\ell}}\,, \qquad  \ell_R  =  \frac{\ell}{ 2} e^{-\frac{u_R}{\ell}}\,.
\ee
Notice that by  freely choosing $u_L$ and $u_R$ we can change the boundary parameters independently. This is not possible if we limit ourselves to the ``global'' solution  (\ref{AdSbc1})  with $u=\tilde{u}$ and $u_L = u_R = u_0$: in this case we can change the ratio of the curvatures but not each one independently. Thus moving from the ansatz  (\ref{AdSbc1}) to the piecewise geometry (\ref{AdSbc2}) allowed us to introduce one extra parameter and match the number of source parameters of the boundary CFTs.

More generally, one could have done the gluing at a generic point (not necessarily the extremum of the Cosh function) by defining:

\be \label{AdSbc2-gen}
ds^2  = \left\{ \begin{array}{l}  du^2 + \cosh^2\left({\frac{u-u_L}{\ell}}\right) ds^2_{d,\ell}\,, \qquad -\infty < u < u_* \,, \\  \\
d\tilde{u}^2 + \cosh^2\left({\frac{\tilde{u}-u_R}{\ell}}\right) ds^2_{d,\ell}\,, \qquad  \tilde{u}_* < \tilde{u} <
    +\infty\,, \end{array} \right.
\ee
where $u_*$  is {\em arbitrary}. Then Israel's  junction conditions are satisfied if $\tilde{u}_*$ is chosen as:
\be\label{tus}
\tilde{u}_* = u_* - (u_L- u_R)\,.
\ee
Notice that (\ref{AdSbc2-gen}) can also be written as:
\be \label{AdSbc2-gen-ii}
ds^2  = \left\{ \begin{array}{ll}  du^2 + f(u) \, ds^2_{d,\ell}\,, &\qquad -\infty < u < u_* \,,\\  \\
d\tilde{u}^2 +  f(\tilde{u} - \delta) \, ds^2_{d,\ell}\,, &\qquad  u_* +\delta < \tilde{u} <
    +\infty \,,\end{array} \right.
\ee
where
\be
f(u) \equiv \cosh^2\left({\frac{u-u_L}{\ell}}\right)\,, \qquad  \delta \equiv u_R - u_L\,.
\ee

\begin{figure}[ht!]
\begin{center}
\includegraphics[width = 10cm]{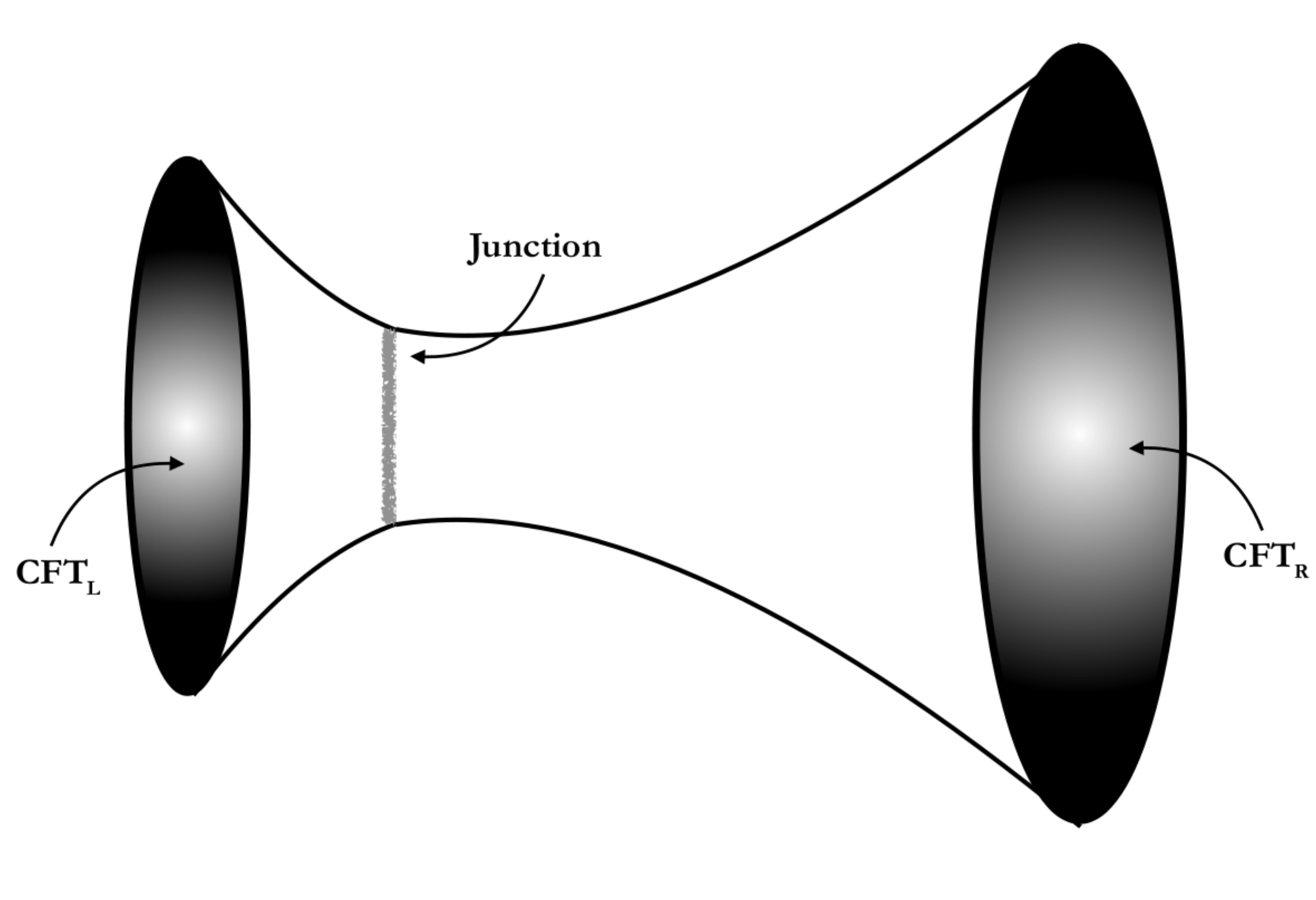}
\caption{\footnotesize{The two-parameter pure gravity solution is obtained by gluing two halves of two different solutions across a junction at fixed $u$, in such a way that the metric and the extrinsic curvature are smooth. This way one can freely choose both length scales on each UV boundary.}}\label{fig:junction}
\end{center}
\end{figure}

It is perhaps worth explaining in more detail what the  expression (\ref{AdSbc2-gen-ii}) means. The  manifold we have constructed here is schematically represented in figure \ref{fig:junction}. It is composed of two pieces: each one is a  manifold equipped with a system of coordinates and a  metric, corresponding to the upper and lower lines of equation  (\ref{AdSbc2-gen-ii}). Each of these half-manifolds has a boundary at finite radial direction, given respectively by the hypersurfaces $u=u_*$ and $\tilde{u} = u_* +\delta$. In addition, each of them has a single UV boundary, at $u\to -\infty$ and $\tilde{u} \to +\infty$.  In the interior, each component manifold is a solution of the same Einstein equation. The two boundaries are then identified, and the two pieces are glued  together to result in  the full manifold.  Since at the interface the metric and extrinsic curvature (and in fact, all higher $u$-derivatives of the metric) are continuous, the gluing can be done without adding any localized energy density.

Notice that the two half-manifolds are trivially diffeomorphic to each other, for example by the change of coordinates $\tilde{u} = u + \delta$ performed on the right half-manifold.   However, importantly, this diffeomorphism acts nontrivially
on the UV boundary at $u\to +\infty$ and  it scales  the boundary conditions (the leading order Fefferman-Graham metric)  by a factor $\exp{\delta/\ell}$. Therefore, the manifold obtained by gluing the two halves in equation (\ref{AdSbc2-gen-ii}) is physically (in the sense of the UV boundary data) different from the e.g. the one described by equation  (\ref{AdSbc1}), although they are locally diffeomorphic.

Finally, unlike the parameter $\delta$, we stress that  changing  the gluing point $u_*$ does not introduce an extra degree of freedom to the space of solutions, because it can be  reabsorbed in a diffeomorphism together with a redefinition of the parameters $u_L$ and $u_R$. Thus changing $u_*$ one moves from one solution to another in the same class.

\subsection{Einstein-Dilaton gravity} \label{sec:piecewise}
Following the previous section, a  piecewise solution which  automatically satisfies Israel's junction conditions (continuous $A(u)$ and $\f(u)$ with continuous derivatives) can be constructed using the following recipe:
\begin{enumerate}
\item Take {\em any} of the solutions from section \ref{sec:2par} connecting two AdS, with a given  scale factor and a given dilaton profile $(\bar{A}(u), \bar{\f}(u))$;
\item Define the following piecewise solution by gluing the two half-solutions:
\be \label{jumping-1}
A =  \left\{ \begin{array}{ll} \bar{A}(u) \,, & \qquad u < u_*  \,,\\  \\
\bar{A}(\tilde{u} - \delta) \,,  & \qquad   u_* +\delta < \tilde{u} <
    +\infty  \,, \end{array} \right.
\ee
\be\label{jumping-2}
\f  = \left\{ \begin{array}{ll} \bar{\f}(u) \,, & \qquad u < u_*  \,,\\  \\
\bar{\f}(\tilde{u} - \delta) \,,   & \qquad   u_* +\delta < \tilde{u} <
    +\infty   \,.\end{array} \right.
\ee
\end{enumerate}
Like the case of pure gravity discussed in the previous section, this geometry is composed of two manifolds smoothly joined at an interface, with a scalar field defined in each half and smooth at the junction.

We now  have  four  dimensionless parameters: the three we had in section \ref{sec:parameters}   for the continuous solution(namely $S_0, \f_0$ and $u_0$) plus the extra independent parameter $\delta/\ell$. If $\bar{R}_i^{UV}, \bar{R}_f^{UV}, \bar{\f}_-^i, \bar{\f}_-^f$ are  four  UV (dimensionful) parameters of the continuous solution $(\bar{A},\bar{\f})$, the   parameters of the solution  (\ref{jumping-1} - \ref{jumping-2}) are:
\be\label{transfi}
{R}^{UV}_i = \bar{{R}}^{UV}_i\,, \quad  \f^i_- = \bar{\f}^i_-\,, \quad {R}^{UV}_f = e^{2\delta/\ell} \bar{{R}}^{UV}_f\,, \quad \f^f_- =  e^{\delta \Delta_-^f/\ell} \bar{\f}^f_-\,.
\ee
The two dimensionless parameters ${\cal R}_{i,f}$ are unaffected, but the third one $\xi$ is rescaled. Notice however, that the effect  of $\delta$ on the individual couplings is different than the effect of the parameter $u_0$ in equation (\ref{shift1}, \ref{shift3}). One can then use a combination of $u_0$ and $\delta$ (or equivalently $u_L$ and $u_R$ as in the previous subsection) to separately change $\f^i_-$ and $\f^f_-$.

Therefore, this extra parameter completes the full space of expected  independent couplings of the solutions.
 As in the case of pure gravity, the choice of the gluing point $u_*$ is not an extra parameter in the space of solutions (for simplicity, it can be taken to be the turning point of the scale factor).

As a final remark, notice that  dimensionless vev parameters, $C_{i,f}$ are insensitive to $\delta$, but according to \eqref{Avev} the vevs of boundary operators change as
\be
\langle\mathcal{O}_i\rangle=\langle\bar{\mathcal{O}}_i\rangle\,,\qquad \langle\mathcal{O}_f\rangle=\langle\bar{\mathcal{O}}_f\rangle e^{\delta \Delta_+^f/\ell}  \,.
\ee

\section{Single-boundary solutions} \label{sec:single}

In the previous sections, we have seen that hyperbolic-slicing holographic  flows connect two asymptotic AdS regions, and can be interpreted as the holographic dual to two coupled CFTs defined on AdS, each with a relevant deformation turned on.
This begs the question: what is the holographic description of a {\em single} CFT on AdS? In this section, we put forward a possible answer.

In the family of solutions we have studied, there is a special class of geometries which can help answer this question: these are the type S solutions,  corresponding to the points on the horizontal axis in the parameters space shown in figure  (\ref{Moduli}). In these solutions, the scale factor and the dilaton turn around at the same point. Notice that they asymptote to the same UV extremum on the left and the right.

Type S solutions are $Z_2$-invariant under $(u-u_0) \to -(u-u_0)$. Therefore,
to obtain a single-sided solution, one can simply take a quotient by $Z_2$  of a type S solution. The resulting geometry is regular at the fixed point $u = u_0$ since at this point {\em both} the scale factor and scalar field radial derivative vanish: this means there is no need to introduce a localized dilaton potential term at the IR end-of-space $u=u_0$, to absorb boundary terms in the variation of the action. Furthermore, these solutions have the correct number of dimensionless parameters to describe a single CFT: since $S_0 = 0$, they are parametrized by the single dimensionless parameter $\f_0$, which can be traded for ${\cal R}$ on the boundary. This is the only dimensionless combination one can form from the boundary curvature and deformation mass parameter $\f_-$ in a single CFT.

Thus, one possibility is that the holographic dual of a single QFT on AdS is the $Z_2$-symmetric quotient of a left-right symmetric two-boundary solution.

In general, this prescription  breaks part of the conformal symmetry on AdS. To see this, it is convenient to analyze the situation in the simplest case in which the dilaton is treated as a probe, and the bulk theory is a pure gravity with a negative cosmological constant. In  this case, the appropriate two-sided solution is AdS$_{d+1}$ in Ads-slicing coordinates, given in  equation (\ref{AdSbc1}), in which we can set $u_0=0$ for simplicity:
\be \label{sing1}
ds^2  = du^2 + (\cosh^2 \frac{u}{ \ell}) ds^2_{d,\ell}\,, \qquad u \in (-\infty,+\infty)\,.
\ee
For concreteness, we  will write the slice metric in Poincar\'e coordinates,
\be \label{sing2}
ds^2_{d,\ell} = \frac{\ell^2 }{ z^2} \left(dz^2 + d\vec{x}^2_{d-1}\right)\,,
\ee
and we take the slice-AdS length to be equal to the bulk-AdS length  $\ell$ for simplicity.  The single-sided solution is obtained by restricting the range of $u$ to $(-\infty, 0]$.

To study correlation functions of dimension-$\Delta$  operators $O$ in the $d$-dimensional boundary theory on $AdS_d$, we add a probe scalar of a given mass $m^2 = \Delta(\Delta-d)$, whose action we take as:
\be \label{sing3}
S = -\frac{1}{ 2\ell^{d-1}} \int du \int d^d x \sqrt{-g}\left[ g^{ab}\de_a \f \de_b \f + m^2 \f^2\right]\,.
\ee
where $g$ is the metric tensor in equation (\ref{sing1}).

Boundary correlators are  obtained by the usual holographic prescription: solve the bulk field equation with given source $\f_-$ in the UV, evaluate the action on-shell and differentiate with respect to the source.  Holographic correlators in the geometry (\ref{sing1}) were studied in \cite{hinter}. Here, we borrow their results and refer the reader to that work for details.

Let us first describe the full two-boundary geometry. In this case there is a  copies of the CFT on  each boundary $u\to \pm \infty$, and  we can turn on two independent source $\f_-^{(-)}$ and $\f_-^{(+)}$  on the two $AdS_d$ space-times at $u\to \pm \infty$. Therefore  there are four types of boundary correlators,
\be \label{sing4}
G_{\pm \pm}(z,z'; \vec{x}-\vec{x}')  = \langle O_{\pm} (z,\vec{x})  O_{\pm} (z',\vec{x}') \rangle\,,
\ee
and the on-shell action will be a boundary term  of the form, schematically:
\be\label{sing5}
S_{on-shell} = \int  \left(\f_-^{(-)}G_{--}\f_-^{(-)} + \f_-^{(+)}G_{++}\f_-^{(+)} + \f_-^{(-)}G_{-+}\f_-^{(+)} + \f_-^{(+)}G_{+-}\f_-^{(-)} \right)\,.
\ee
Out of the four correlators in (\ref{sing4}) only two are independent, since by the $Z_2$-symmetry of the background $G_{--} = G_{++}$ and $G_{-+} = G_{+-}$.

In a single-sided solution, there is a single operator $O(z,\vec{x})$ and correspondingly a single source: one must give a prescription to relate the two sources  $\f_-^{(-)}$ and $\f_-^{(+)}$. The correlator one obtains will then be a linear combination of $G_{--}$ and $G_{-+}$. In particular, the single-sided solution one obtains by a $Z_2$-orbifold around $u=0$ correspond to picking a $Z_2$-even solution for the bulk scalar field, $\f(-u,z,\vec{x}) = \f(u,z,\vec{x})$. This is equivalent to introducing an IR brane at $u=0$ with a Neumann boundary for the scalar field. In this case, the boundary correlator is \cite{hinter}:
\be \label{sing6}
 \langle O (z,\vec{x})  O (z',\vec{x}') \rangle_{N} = \frac{1}{ 2^\Delta} \left[\frac{1}{ \left(\cosh L -1 \right)^\Delta } + \frac{1}{ \left(\cosh L +1 \right)^\Delta }  \right]\,,
\ee
where we have defined:
\be\label{sing7}
\cosh L = 1 + \frac{(z-z')^2 + |\vec{x} - \vec{x}'|^2  }{ { 2}z z'}\,.
\ee
We have:
\begin{align} \label{sing7-ii}
&\cosh L - 1 = \frac{(z-z')^2 + |\vec{x} - \vec{x}'|^2  }{ { 2}z z'} \equiv {\cal C}(z,\vec{x},z',\vec{x}')\,, \\
&\cosh L +1 = \frac{(z+z')^2 + |\vec{x} - \vec{x}'|^2 }{ { 2}z z'}\nn\,.
\end{align}
The operator $O$ in the theory with Neumann boundary conditions is related to the operators of the $\pm$ CFTs by:
\be
O(z,\vec{x})\equiv \frac{1}{ \sqrt{2}}(O_{+}(z,\vec{x})+O_{-}(z,\vec{x}))\,.
\ee

The quantity   ${\cal C}(z,\vec{x},z',\vec{x}')$ in (\ref{sing7-ii}) is the {\em conformal distance} between the points  $(z,\vec{x},z',\vec{x}')$ on the boundary AdS$_d$. It has the interesting property that it is covariant under coordinate transformations generated by conformal killing vectors  of $AdS_d$ (which exist in the same number as flat space conformal killing vectors) \cite{Alvarez:2020pxc} .  In other words, the conformal distance behaves just  like the Euclidean distance in flat space under conformal transformations. On the contrary,  the geodesic distance (which is given by  $L$ in (\ref{sing7})) does not satisfy this property.

In a conformally invariant quantum field theory on $AdS_d$, one then expects the  correlator to be  conformally covariant in a similar way as its flat space counterpart and given by:
\be \label{sing10}
\langle O (z,\vec{x})  O (z',\vec{x}') \rangle_{conf} =   \frac{1}{ 2^\Delta} \frac{1}{ \left[{\cal C}(z,\vec{x},z',\vec{x}') \right]^\Delta} =  \frac{1}{ 2^\Delta} \frac{1}{ \left(\cosh L -1 \right)^\Delta }\,,
\ee
i.e. the first term in (\ref{sing6}).  The presence of the second term  in  (\ref{sing6})  breaks conformal invariance. This is related to the fact that only certain boundary conditions at the $z\to 0$ boundary of $AdS_d$ preserve full conformal invariance \cite{porrati}, and the holographic single-sided solution defined above using Neumann boundary conditions is not compatible with these conformal boundary conditions.

One can construct different, more general single-sided solutions if one abandons the idea of imposing $Z_2$-symmetry on the two-sided geometry:  simply cut off the space by a general IR brane and assume {\em some} boundary at $u=0$. For example, if one were to  impose Dirichlet boundary conditions for the scalar $\f$  at $u=0$, one would find \cite{hinter}:
\be \label{sing11}
 \langle O (z,\vec{x})  O (z',\vec{x}') \rangle_{D} = \frac{1}{ 2^\Delta} \left[\frac{1}{ \left(\cosh L -1 \right)^\Delta } - \frac{1}{ \left(\cosh L +1 \right)^\Delta } \right]\,.
\ee
This also  differs from the conformal result (\ref{sing10}).

One  may then ask the question of whether there exists a definition of IR boundary conditions such that, at least in the probe scalar case, the holographic correlator reproduces the equation (\ref{sing10}). The answer is affirmative but, as we  show below, these ``conformal'' IR boundary conditions cannot be encoded in a local brane  action at $u=0$.

The most general, local,   linear boundary condition can be obtained by adding to the action (\ref{sing3})  a quadratic term  localized  on  the brane,
\be\label{sing12}
S_{IR} = -\frac{\mu }{ 2} \int_{u=0} \sqrt{-\gamma} \f^2\,,
\ee
where $\gamma_{\mu\nu}$ is the induced metric at $u=0$ and $\mu$ is a constant.  Upon varying $S + S_{IR}$ one obtains the bulk equation and the IR condition:
\be \label{sing13}
\frac{\de \f }{ \de u}(z,\vec{x},u=0) = \mu \, \f(z,\vec{x}, u=0)\,.
\ee
From equations (\ref{sing6}) and (\ref{sing11}) it is clear that the conformal boundary correlator (\ref{sing10})  is given by the linear combination  of the Neumann and Dirichlet results:
\be \label{sing14}
\langle O (z,\vec{x})  O (z',\vec{x}') \rangle_{conf} = \frac12\left[ \langle O (z,\vec{x})  O (z',\vec{x}') \rangle_{D} + \langle O (z,\vec{x})  O (z',\vec{x}') \rangle_{N} \right]\,.
\ee
Recall that the holographic boundary correlator is, schematically, given by the derivative of the bulk-to-boundary Green's function, i.e. a solution of the bulk field equation with unit source, obeying the corresponding boundary condition:
\be \label{sing15}
\langle O (z,\vec{x})  O(z',\vec{x}') \rangle_{D, N} \sim (\de_u K_{D,N})(z,z'; \vec{x}- \vec{x}', u=0)\,,
\ee
where on the left-hand side  $K_{D,N}$ are the  bulk-to-boundary Green's functions obeying respectively Dirichlet and Neumann boundary conditions at the IR brane. Therefore, the linear combination (\ref{sing14}) is obtained by choosing $\mu$ in (\ref{sing12}) so that the  boundary conditions at $u=0$ pick the following bulk-to-boundary Green's function:
\be \label{sing16}
K_{conf} = \frac12\left(K_D + K_N\right)\,.
\ee
This combination is characterized by the  property:
\be
(\de_u K_{conf}) (u=0)  = \frac{(\de_u K_D) }{ K_N }\Big|_{u=0} \, K_{conf}(u=0)\,.
\ee
Recalling that $K_{conf}$ is a solution  of the bulk field equation, and comparing with (\ref{sing13}) suggests that we should  identify the parameter $\mu$ with the quantity:
\be \label{sing16-i}
\mu =  \frac{(\de_u K_D) }{ K_N }\Big|_{u=0}\,.
\ee
This cannot be correct, however, because the right-hand side is not a constant but a non-trivial function of the boundary points. This implies that it is impossible to obtain a conformal correlator by any choice of  local action at the IR boundary of the type (\ref{sing12}).

If we are willing to use a non-local action,  the way to proceed is to go to  ``momentum space'' along the slice, i.e. by decomposing the $d+1$ Klein-Gordon equation
\be \label{sing17}
\Box_{d+1} \f = m^2 \f\,,
\ee
 into eigenmodes  of the slice Laplacian,  namely writing $\f(u,z,\vec{x}) = F_\nu(u) \phi_{\nu}(z,\vec{x})$ and, using the metric in the form (\ref{sing1}), writing equation (\ref{sing17}) as:
\bsq
\begin{align} \label{sing18-i}
&\left[ \de_u^2 + d\, \tanh u \de_u + \cosh^{-2}u \left(\nu^2 - \frac{(d-1)^2 }{ 4} \right) \right] F_\nu(u) =  m^2  F_\nu(u)\,, \\
&\Box_d \phi_\nu(z,\vec{x}) = \left(\nu^2 - \frac{(d-1)^2 }{ 4} \right) \phi_\nu(z,\vec{x})\,, \label{sing18-ii}
\end{align}
\esq
In the form above, the (complex) number $\nu$ determines the eigenvalue of the slice-AdS laplacian. The boundary condition at $u=0$ is then imposed on the function $F$,  and its  general form
\be \label{sing19}
F'_\nu(u=0) = \mu(\nu) F_{\nu}(u=0)\,,
\ee
 where $\mu$ is, in general, a function of the slice Laplacian eigenvalue   $\nu$. One  then  writes the brane-localized action as a decomposition over  a complete set of eigenmodes (parametrized by $\nu$ and spatial momentum conjugate to $\vec{x}$) , i.e schematically
\be\label{sing20}
S_{IR} =  \int d\nu \int \frac{d^{d-1} k }{ (2\pi)^{d-1} } \, \mu(\nu) \,| \tilde{\f}(\nu, \vec{k}) |^2\,,
\ee
where  $\tilde{\f}(\nu, \vec{k})$ are ``Fourier modes'' of the field $\f(u=0)$, adapted to the eigenvalue equation in AdS$_d$.
One can then go back to position space using the AdS analogue of the Parseval formula for the Fourier transform, which results in a non-local action on $AdS_d$
\be\label{sing21}
S_{brane} = \int dz dz' d^{d-1}x\, d^{d-1}x' \, \mu(z,z', \vec{x}-\vec{x}') \f(z,\vec{x}) \f(z',\vec{x}')\,,
\ee
where $\mu(z,z', \vec{x}-\vec{x}') $ is obtained from $\mu(\nu)$ by an appropriate integral transform. The interested reader can find the technical details in Appendix \ref{app:transforms}. There, it is also shown that the expression of $\mu$  in momentum space which results in the conformal correlator is:
\def\G{\Gamma}
\be\label{sing22}
\mu(\nu) = \frac{2\tan(\frac{\pi}{ 2}(\nu - \frac12))}{ \pi^2}  \left|\G\left(\frac34 + \frac{\nu}{2} + \frac{\g}{2}\right)\right|^2\left|\G\left(\frac34 + \frac{\nu}{ 2} - \frac{\g}{ 2}\right)\right|^2(\cos(\pi\g) +\sin(\pi \nu))\,.
\ee

\section*{Acknowledgements}\label{ACKNOWL}
\addcontentsline{toc}{section}{Acknowledgements}

We thank D. Anninos, P. Betzios, V. Niarchos, K. Papadodimas, O. Papadoulaki, M. Van Raamsdonk, C. Rosen ,
 I. Valenzuela and T. van Riet for useful discussions and suggestions.
The work of A. G. is supported by Ferdowsi University of Mashhad under grant 2/57225 (1401/03/21).

\appendix
\renewcommand{\theequation}{\thesection.\arabic{equation}}
\addcontentsline{toc}{section}{Appendix\label{app}}
\section*{Appendix}

\section{AdS-slicing vs. global coordinates} \label{app:ads-slicing}
Here we want to relate the  $R \times S^{d-1}$-slicing (usually referred to as global coordinates)  and $AdS_d$-slicing  of  $AdS_{d+1}$, and the structure of the respective conformal boundaries.

 We work in Euclidean signature. The embedding space definition of $EAdS_{d+1}$ is:
\be \label{GVH1}
-X_{-1}^2 + X_0^2 + \sum_{i=1}^d X_i^2  = -\ell^2\,.
\ee
\begin{itemize}
\item {\bf $R \times S^{d-1}$-slicing}\\
It is useful to explicitly split  $S^{d-1} = [0,\pi]_\theta\times S^{d-2}$, since in hyperbolic slicing only an $S^{d-2}$ appears:
\begin{align}\label{GVH2}
 X_{-1} &= \ell  \cosh \psi \, \cosh \rho\,,  \\
 X_0 &= \ell \sinh\rho \cos\theta\,, \nonumber \\
 X_d &= \ell \sinh \psi\, \cosh \rho\,, \nonumber \\
 X_i &= \ell \sinh \rho\, \sin\theta \,\Omega_i\,, \qquad i  = 1\ldots d-1\,, \qquad \sum_{i=1}^{d-1} \Omega_i^2 = 1 \,,\nn
\end{align}
with ranges $\rho \in [0, +\infty)$, $\theta \in [0, \pi]$, $\psi \in (-\infty + \infty)$.

The metric reads:
\be \label{GVH3}
ds^2 = \ell^2 \left[d\rho^2 + \cosh^2\rho\, d\psi^2 + \sinh^2\rho \left(d\theta^2 + \sin^2\theta\, d\Omega_{d-2}^2 \right) \right]\,.
\ee
\item {\bf $EAdS_{d}$-slicing}\\
Define:
\begin{align}\label{GVH4}
 X_{-1} &= \ell \cosh u\, \cosh \tau \, \cosh r   \\
X_0 &= \ell \sinh u \nonumber \\
X_d &= \ell \cosh u \sinh \tau\, \cosh r \nonumber \\
X_i &= \ell \cosh u\, \sinh \tau\, \Omega_i \qquad i  = 1\ldots d-1\,, \qquad \sum_{i=1}^{d-1} \Omega_i^2 = 1\,,\nn
\end{align}
with ranges $u \in (-\infty, +\infty)$, $r\in [0, +\infty)$, $\tau \in (-\infty + \infty)$.

The metric reads:
\be \label{GVH5}
ds^2 = \ell^2 \left[du ^2 + \cosh^2 u\left(dr^2 + \cosh^2 r d\tau^2 + \sinh^2 r \,d\Omega_{d-2}^2 \right) \right]\,.
\ee
\end{itemize}
Both coordinate systems cover the full $EAdS_{d+1}$ manifold:  comparing (\ref{GVH2}) and (\ref{GVH4}) we see that the two sets of local coordinates cover the same range of $X_{-1} \ldots X_d$, namely the connected component with $X_{-1} >0$ of the  manifold defined by equation  (\ref{GVH1}). Therefore they are both global charts on Euclidean $EAdS_d$.

The two charts may be directly related by the following transformation:
\be \label{GVH6}
\cosh \rho = \cosh u \, \cosh r \,, \quad \cos \theta = {\frac{\sinh u}  {\left(\cosh^2 u \cosh^2 r -1 \right)^{1/2}}}\,, \quad \psi = \tau\,,
\ee
and its inverse:
\be\label{GVH7}
\sinh u  = \cos\theta \sinh \rho\,, \quad \cosh r = {\frac{\cosh\rho} {\left(\cos^2\theta \sinh^2 \rho + 1\right)^{1/2}}}\,, \quad \tau = \psi\,.
\ee
The AdS boundary is at $\rho \to +\infty$. From (\ref{GVH6}),  we can see that this corresponds in the hyperbolic slicing to taking either $u \to \pm \infty$ with $r$ finite (reaching the two disconnected $EAdS_d$ boundary slices) or $r\to +\infty$ with $u$ finite (going to the boundary along a slice).  Notice that in the latter case, $\theta \to \pi/2$.

From the last consideration, it may seem that all the slice boundaries at any $u$ collapse to the equator of the $S^{d-2}$ at $\rho \to\infty$, which is a codimension-2 hypersurface in the full manifold and a codimension-1 hypersurface of the boundary.  One may worry that this makes any non-trivial function $f(r,u)$,  which does not go to a $u$-independent constant as $r\to +\infty$, ill-defined. This question is not well-defined as it is,   since neither the $S^{d-1}$ at $\rho \to\infty$ nor the corresponding limits in hyperbolic slicing are part of the $EAdS_{d+1}$ manifold. For example, it is clearly not true that in the limit $r\to \infty$ all slices converge: take two points on two different hyperbolic slices, with coordinates $A=(u_A, r_A, \Omega_A^i)$ and     $B=(u_B, r_B, \Omega_B^i)$,  such that $r_A = r_B$, $\Omega_A = \Omega_B$, $u_A\neq u_B$. The distance between these two points is
\be \label{GVH8}
d(A,B) = \ell |u_A - u_B| \,,
\ee
as one can immediately see from  (\ref{GVH5}), and it stays finite as we move the points to the slice-boundaries  $r_{A} = r_B \to +\infty$. Therefore there is no sense in which all slices converge, and a nontrivial function of $u$ as $r\to \infty$ is certainly allowed.

The question becomes more interesting if we pose it in the conformal compactification of the space. In this case, the answer depends on which conformal compactification we use, which is related to the choice of the space where we define the dual CFT.

In global coordinates, we can go to the conformal compactification by defining:
\be \label{GVH9}
\tan \xi = \sinh \rho\,, \quad \xi \in [0, \pi/2]\,.
\ee
The metric (\ref{GVH3})  becomes:
\be \label{GVH10}
ds^2  = {\frac{1}{\cos^2\xi}}\left[d\xi^2 + d\psi^2 + \sin^2\xi \left(d\theta^2 + \sin^2\theta\, d\Omega_{d-2}^2 \right) \right]\,.
\ee
The boundary is at $\xi = \pi/2$. The conformal compactification is obtained by removing the prefactor $(\cos\xi)^{-2}$. The boundary (which now {\em does} belong to the manifold) $\xi = \pi/2$ has geometry $R_{\psi} \times S^{d-1}$. Therefore, this is the appropriate conformal compactification if we want the CFT to live on $R_{\psi} \times S^{d-1}$.
In this case, the locus $\theta = \pi/2$ with fixed  $\psi$ and fixed $\Omega_i$ is a single point on the boundary, and any regular function must have a single value at this point.

Let us now consider hyperbolic coordinates (\ref{GVH5}).  Defining the new coordinate $z$ by
\be \label{GVH11}
\tan {\frac{z}{2}} = \tanh {\frac{u}{2}}, \qquad z \in \left[-{\frac{\pi}{2}}\,, {\frac{\pi}{2}}\right]\,,
\ee
brings (\ref{GVH5}) in the form:
\be \label{GVH12}
ds^2 = {\frac{\ell^2}{\cos^2 z}} \left[dz^2 + \left(dr^2 + \cosh^2 r d\tau^2 + \sinh^2 r \,d\Omega_{d-2}^2 \right) \right]\,,
\ee

If we remove the conformal factor, we obtain the factorized space-time $[-\frac{\pi}{2},\frac{\pi}{2}] \times EAdS_d$, which has two copies of $EAdS_d$ as its boundary. This is the right conformal compactification if we want to study (two copies of) a CFT on  $EAdS_d$. In the  metric (\ref{GVH12}), , unlike the conformal compactification (\ref{GVH10}), going to the slice boundary $r\to \infty$ at fixed $\tau$ and $\Omega_i$ does not shrink the space to a point, but it leaves the segment $[-\pi/2, \pi/2]_z$. Therefore, in this conformal compactification,  points on different slices stay at a  finite distance when we move towards the slice boundary, as we noted previously.

One may argue that the conformal compactification (\ref{GVH12}) has not made the full boundary  compact, as the slice-boundaries are still of infinite size and an infinite distance away. This can be remedied by further (conformally) compactifying the manifold (\ref{GVH12}) to bring  also  the slice-boundary at a finite distance. This is easily done as in (\ref{GVH9}), but now applied to the coordinate $r$: defining
\be \label{GVH13}
\tan \beta = \sinh r\,,
\ee
brings  (\ref{GVH12}) to:
\be\label{GVH14}
ds^2 = \frac{\ell^2}{\cos^2 z\, \cos^2\beta} \left[d\tau^2 + d\beta^2 + \cos^2 \beta\, dz^2 + \sin^2\beta \,d\Omega_{d-2}^2  \right]\,,
\ee
where $\beta \in [0, \pi/2]$ and $z\in [-\pi/2, \pi/2]$. Now, it is clear that going to the slice boundary $\beta \to \pi/2$ shrinks any distance along $z$ to a point, however, the boundary doesn't have the geometry of $EAdS_d$ anymore.
Thus, this conformal compactification does not describe  CFTs on $EAdS_d$.

In fact, it is not hard to see that the conformal boundary of the manifold (\ref{GVH14}) is {\em again} $R_\tau \times S^{d-1}$. To see this, first notice that the metric  in the square brackets in (\ref{GVH14}) is locally the same as $R_\tau \times S^{d}$, where $S^d$ is written in a combination  of polar-spherical coordinates: to see this, define $S^d$ as
\be\label{GVH15}
Y_0^2 + Y_1^2 + \ldots Y_d^2 = 1\,,
\ee
and choose  polar coordinates in the $(Y_0,Y_1)$ plane and $(d-1)$-dimensional spherical coordinates in the remaining directions:
\bea\label{GVH16}
&& Y_{0} = \cos \beta\, \cos z\,,  \\
&& Y_1 =   \cos \beta  \, \sin z\,, \nn \\
&& Y_{i+1} = \sin \beta \, \, \Omega_i\,, \qquad i  = 1\ldots d-1\,, \qquad \sum_{i=1}^{d-1} \Omega_i^2 = 1\,. \nonumber
\eea
One can easily check that the metric is the same as that of the  $(\beta,z,\Omega_i)$ factor in (\ref{GVH14}).

To cover the whole $S^d$ once,  we need $\beta \in [0, \pi/2]$ and $z\in [-\pi, \pi]$. However, the manifold we arrived at in (\ref{GVH14}) has $z\in [-\pi/2, \pi/2]$, therefore it is only {\em  half} of an $S^d$: the half corresponding to $Y_0 \geq 0$. Its boundary is the intersection of $S^d$ with the hyperplane $Y_0=0$, which is  $S^{d-1}$.

\section{First order equations\label{afirst}}

In this appendix, we write the first-order equations in various forms, to elucidate their structure.

In terms of $W,S,T$ defined in  (\ref{AWST}),  the equations of motion  \eqref{AEOM1} - \eqref{AEOM3} are given by
\bsq
\begin{align}
\label{AEOM4a} S^2 - SW' + \frac{2}{d} T &=0 \, , \\
\label{AEOM5a} \frac{d}{2(d-1)} W^2 -S^2 -2 T +2V &=0 \, , \\
\label{AEOM6a} SS' - \frac{d}{2(d-1)} SW - V' &= 0 \, .
\end{align}
\esq
We also have the identity
\be
\frac{T'}{ T}=\frac{W}{ (d-1)S}\,.
\ee
We can simplify this system by eliminating $T$ algebraically so that we are left with the following equations
\bsq
\begin{align}
\label{AEOM7a} \frac{d}{2(d-1)} W^2 + (d-1) S^2 -d S W' + 2V &=0 \, , \\
\label{AEOM8a} SS' - \frac{d}{2(d-1)} SW - V' &= 0 \, .
\end{align}
\esq

The system can be converted into a second-order equation for a single variable.
Solving (\ref{AEOM7}) algebraically for $S$ gives
\be
S_{\pm}=\frac{dW'}{ 2(d-1)}\pm\frac{\sqrt{ d^2W'^2-8(d-1)V-2dW^2}}{ 2(d-1)}\,,
\ee
which indicates that for all real solutions we have the bound
\be
W^2\leq \frac{d W'^2}{ 2}-\frac{4(d-1)}{ d}V\,.
\ee
The remaining equation can be converted into a non-linear second-order differential equation for $W$
\begin{align}
\big(d \big(d (W')^2 &-2 W^2\big)-8 (d-1)
   V\big) \big(4 (d-1) V'+W' \left((d+2) W-2 d
   W''\right)\big)^2 \nn \\
&=\big(-2 d^2 (W')^2 W''+4 (d-1)V' W'-8 (d-1) V \left(W-W''\right)\nn \\
&+2 d W^2 W''+d (d+2) W (W')^2-2 d W^3\big)^2\,.
\end{align}
Another approach is to solve (\ref{AEOM8}) for $W$
\be
W=-\frac{2(d-1)}{ d}\left(\frac{V'}{ S}-S'\right)=\frac{2(d-1)}{ dS}Q'\,,\qquad Q\equiv \frac{S^2}{ 2}-V\,,
\ee
and then substitute in (\ref{AEOM7}) to obtain a second order equation for $S$
\begin{align}
&2(d-1)S^3S'' +\frac{2(d-1)(d+2)}{d}V'SS'-\frac{2(d-1)}{d}S^2S'^2\nn \\
&
-(d-1)S^4-2(V+(d-1)V'')S^2-\frac{2(d-1)}{d}(V')^2=0\,.
\end{align}

Assuming the scale factor plays the role of energy scale\footnote{This is not however clear here, as the scale factor is no longer monotonic.
A proper interpretation is obtained if we think of these solutions as two different QFTs with an interface.}, the holographic beta function can be expressed as
\be \label{Abeta}
\beta(\f)=\frac{d\f}{dA}=-2(d-1)\frac{S(\f)}{W(\f)}\,.
\ee
We can express then
\be
 S'=\frac{V'}{ S}+\frac{dW}{ 2(d-1)}=\frac{V'}{S}-\frac{dS}{ \beta}\,,\qquad
W'=\frac{2V+(d-1)S^2+2d(d-1)\frac{S^2}{\b^2}}{ dS}\,.
\ee
Differentiating we obtain
\be
\beta'=\frac{1}{ 2d}\left[\frac{2\b(\b V+d(d-1)V')}{ (d-1)S^2}+\b^2-2d(d-1)\right]\,,
\ee
from where we obtain
\be
S^2=\frac{2\b\left(\b V+{d(d-1)}{V'} \right)
}{ (d-1)\left(2d\b'-\b^2+2d(d-1)\right)}\,.
\ee
Differentiating once more, we obtain a second-order differential equation for $\beta$ in terms of the potential $V$
\be
\left[\frac{2\b\left(\b V+{d(d-1)}{V'} \right)
}{ (d-1)\left(2d\b'-\b^2+2d(d-1)\right)} \right]'+\frac{2d}{\b}\left[\frac{2\b\left(\b V+{d(d-1)}{V'} \right)
}{ (d-1)\left(2d\b'-\b^2+2d(d-1)\right)}\right]=2V'\,.
\ee


\section{Solutions near critical points}
Since the equations of motion are not exactly solvable we solve these equations numerically and it should be necessary to know the expansion of solutions in the vicinity of some specific points.
For negative curvatures (AdS slices) we have the following expansions \cite{C}:
\subsection{Asymptotic solutions near extrema}\label{exp1}

At each extremum of the potential, we have a UV fixed point.
Close to the UV fixed point at $\f=\f_m$, we consider the scalar potential has an expansion of the form ($m^2>0$ for maxima and $m^2<0$ for minima)
\begin{equation}
 V(\f)= -\frac{d(d-1)}{\ell^2}- \frac{m^2}{2}(\f-\f_m)^2+\mathcal{O}((\f-\f_m)^3)\,.
 \end{equation}
$\bullet$ Maximum of the potential:

Near the maximum of the potential, the solution of equations of motion \eqref{AEOM7} and \eqref{AEOM8} for $W$ and $S$ has two branches \cite{C}. For the minus branch as $\f\rightarrow \f_m^+$ the solutions have the following expansions
\bsq
\begin{align}\label{WLU}
W_-&=\frac{2(d-1)}{\ell}+\frac{\D_-}{2\ell}(\f-\f_m)^2+\frac{\mathcal{R}}{d\ell}|\f-\f_m|^{\frac{2}{\D_-}}+\frac{C}{\ell}|\f-\f_m|^{\frac{d}{\D_-}}+\cdots\,,\\
S_-&=\frac{\D_-}{\ell}|\f-\f_m|+\frac{Cd}{\D_-\ell}|\f-\f_m|^{\frac{d}{\D_-}-1}+\cdots\,,\label{SLU}
\end{align}
\esq
where dots stand for higher power expansion terms and $\mathcal{R}$ and $C$ are constants of integration and we have defined
\be\label{ACfact}
\D_\pm=d/2\pm\sqrt{d^2/4 - m^2\ell^2}\,.
\ee
Since $0<m^2<d^2/4\ell^2$ therefore $0< \D_- < d/2$ and $d/2< \D_+ < d$.
Moreover, the plus branch is described by the following expansions
\bsq
\begin{align}\label{Wplus}
W_+&=\frac{2(d-1)}{\ell}+\frac{\D_+}{2\ell}(\f-\f_m)^2+\frac{\mathcal{R}}{d\ell}|\f-\f_m|^{\frac{2}{\D_+}}+\cdots\,,\\
S_+&=\frac{\D_+}{\ell}|\f-\f_m|+\cdots\,.\label{Splus}
\end{align}
\esq

The plus branch as is discussed in \cite{C}, can be arrived at by a specific rescaling of the minus branch. In other words, the plus branch is the upper envelope of the family of minus branch solutions parameterized by $C$.
Given the above expansions, we can solve $\f(u)$ and $A(u)$ from \eqref{AWST} to obtain the scalar field and scale factor for the minus branch (from now on, we just discuss the minus branch so we ignore some subscripts)
\bsq
\begin{align}
\label{phiLU} \f(u) &= \f_m+\f_- \ell^{\Delta_-}e^{\Delta_-u / \ell} + \frac{C d \, |\f_-|^{\Delta_+ / \Delta_-}}{\Delta_-(d-2 \Delta_-)} \, \ell^{\Delta_+} e^{\Delta_+ u /\ell} + \ldots \, , \\
\label{ALU} A(u) &= {A}_- -\frac{u}{\ell} - \frac{\f_-^2 \, \ell^{2 \Delta_-}}{8(d-1)} e^{2\Delta_- u / \ell}  -\frac{\mathcal{R}|\f_-|^{2/\Delta_-} \, \ell^2}{4d(d-1)} e^{2u/\ell} \\
\nonumber & \hphantom{=} \ - \frac{\Delta_+ C |\f_-|^{d/\Delta_-} \, \ell^d}{d(d-1)(d-2 \Delta_-)}e^{du/\ell} +\ldots \,,
\end{align}
\esq
where $\f_-$ and $A_-$ are two constants of integration.

When $\f\rightarrow \f_m^-$ we can use the symmetries of equations of motion \eqref{AEOM7} and \eqref{AEOM8} to write the expansions again. As we see, $W\rightarrow W$, $S\rightarrow -S$ and $d\f\rightarrow -d\f$ is a symmetry of these equations so we can write the solutions as
\bsq
\begin{align}\label{WRU}
W&=\frac{2(d-1)}{\ell}+\frac{\D_-}{2\ell}(\f-\f_m)^2+\frac{\mathcal{R}}{d\ell}|\f-\f_m|^{\frac{2}{\D_-}}+\frac{C}{\ell}|\f-\f_m|^{\frac{d}{\D_-}}+\cdots\,,\\
S&=-\big(\frac{\D_-}{\ell}|\f-\f_m|+\frac{Cd}{\D_-\ell}|\f-\f_m|^{\frac{d}{\D_-}-1}+\cdots\big)\,.\label{SRU}
\end{align}
\esq
Again, we can read the expansions of $\f(u)$ and $A(u)$
\bsq
\begin{align}
\label{phiRU} \f(u) &= \f_m-\big(\f_- \ell^{\Delta_-}e^{\Delta_-u / \ell} + \frac{C d \, |\f_-|^{\Delta_+ / \Delta_-}}{\Delta_-(d-2 \Delta_-)} \, \ell^{\Delta_+} e^{\Delta_+ u /\ell} + \ldots\big) \, , \\
\label{ARU} A(u) &= {A}_- -\frac{u}{\ell} - \frac{\f_-^2 \, \ell^{2 \Delta_-}}{8(d-1)} e^{2\Delta_- u / \ell}  -\frac{\mathcal{R}|\f_-|^{2/\Delta_-} \, \ell^2}{4d(d-1)} e^{2u/\ell} \\
\nonumber & \hphantom{=} \ - \frac{\Delta_+ C |\f_-|^{d/\Delta_-} \, \ell^d}{d(d-1)(d-2 \Delta_-)}e^{du/\ell} +\ldots \,.
\end{align}
\esq
In both cases above, the UV fixed point is located at $u\rightarrow -\infty$ and the geometry is asymptotically $AdS$ space with length scale $\ell$.

For equations of motion \eqref{AEOM7} and \eqref{AEOM8} there is another symmetry as $W\rightarrow -W$, $S\rightarrow -S$ and $d\f\rightarrow d\f$. In this case
for the minus branch as $\f\rightarrow \f_m^+$ the $W$ and $S$ expansions are given by
\bsq
\begin{align}\label{WLD}
W&=-\frac{2(d-1)}{\ell}-\frac{\D_-}{2\ell}(\f-\f_m)^2-\frac{\mathcal{R}}{d\ell}|\f-\f_m|^{\frac{2}{\D_-}}-\frac{C}{\ell}|\f-\f_m|^{\frac{d}{\D_-}}+\cdots\,,\\
S&=-\frac{\D_-}{\ell}|\f-\f_m|-\frac{Cd}{\D_-\ell}|\f-\f_m|^{\frac{d}{\D_-}-1}+\cdots\,.\label{SLD}
\end{align}
\esq
To obtain the scalar field and scale factor it would be enough to replace $u\rightarrow -u$.  Therefore
 we can read the expansions of $\f(u)$ and $A(u)$ as follow
\bsq
\begin{align}
\label{phiLD} \f(u) &= \f_m+\f_- \ell^{\Delta_-}e^{-\Delta_-u / \ell} + \frac{C d \, |\f_-|^{\Delta_+ / \Delta_-}}{\Delta_-(d-2 \Delta_-)} \, \ell^{\Delta_+} e^{-\Delta_+ u /\ell} + \ldots \, , \\
\label{ALD} A(u) &= {A}_- +\frac{u}{\ell} - \frac{\f_-^2 \, \ell^{2 \Delta_-}}{8(d-1)} e^{-2\Delta_- u / \ell}  -\frac{\mathcal{R}|\f_-|^{2/\Delta_-} \, \ell^2}{4d(d-1)} e^{-2u/\ell} \\
\nonumber & \hphantom{=} \ - \frac{\Delta_+ C |\f_-|^{d/\Delta_-} \, \ell^d}{d(d-1)(d-2 \Delta_-)}e^{-du/\ell} +\ldots \,.
\end{align}
\esq
In the same way as $\f\rightarrow \f_m^-$ we have the following series expansions
\bsq
\begin{align}\label{WRD}
W&=-\frac{2(d-1)}{\ell}-\frac{\D_-}{2\ell}(\f-\f_m)^2-\frac{\mathcal{R}}{d\ell}|\f-\f_m|^{\frac{2}{\D_-}}-\frac{C}{\ell}|\f-\f_m|^{\frac{d}{\D_-}}+\cdots\,,\\
S&=\frac{\D_-}{\ell}|\f-\f_m|+\frac{Cd}{\D_-\ell}|\f-\f_m|^{\frac{d}{\D_-}-1}+\cdots\,.\label{SRD}
\end{align}
\esq
and also
\bsq
\begin{align}
\label{phiRD} \f(u) &= \f_m-\big(\f_- \ell^{\Delta_-}e^{-\Delta_-u / \ell} + \frac{C d \, |\f_-|^{\Delta_+ / \Delta_-}}{\Delta_-(d-2 \Delta_-)} \, \ell^{\Delta_+} e^{-\Delta_+ u /\ell} + \ldots\big) \, , \\
\label{ARD} A(u) &= {A}_- +\frac{u}{\ell} - \frac{\f_-^2 \, \ell^{2 \Delta_-}}{8(d-1)} e^{-2\Delta_- u / \ell}  -\frac{\mathcal{R}|\f_-|^{2/\Delta_-} \, \ell^2}{4d(d-1)} e^{-2u/\ell} \\
\nonumber & \hphantom{=} \ - \frac{\Delta_+ C |\f_-|^{d/\Delta_-} \, \ell^d}{d(d-1)(d-2 \Delta_-)}e^{-du/\ell} +\ldots \,.
\end{align}
\esq
In these two last cases, the UV fixed point is located at $u\rightarrow +\infty$ and again the geometry is asymptotically $AdS$ space-time with length scale $\ell$.

$\bullet$ Minimum of the potential:

As we approach the minimum of the potential from the left, we can find the expansion of $W$ and $S$. In this case, there is just one branch (the $+$ branch) for negative curvature slices. As $\f\rightarrow \f_1^-$ for $W<0$, we have
\bsq
\begin{align}\label{CminsolW}
W_+&=-\frac{2(d-1)}{\ell}-\frac{\D_+}{2\ell}(\f-\f_1)^2-\frac{\mathcal{R}}{d\ell}|\f-\f_1|^{\frac{2}{\D_+}}+\cdots\,,\\
S_+&=\frac{\D_+}{\ell}|\f-\f_1|+\cdots\,,\label{CminsolS} \\
T_+ &=\frac{\mathcal{R}}{\ell^2}|\f-\f_1|^\frac{2}{\D_+}+\cdots\,. \label{CminsolT}
\end{align}
\esq
where $\D_+=d/2+\sqrt{d^2/4 + m^2\ell^2}>0$. The expansions of the scalar field and scale factor  are given by
\bsq
\begin{align}
\label{Cphimin} \f(u) &=\f_1- \f_+ \ell^{\Delta_+}e^{-\Delta_+ u / \ell}  + \cdots \, , \\
\label{CAmin} A(u) &= {A}_+ +\frac{u}{\ell} - \frac{\f_+^2 \, \ell^{2 \Delta_+}}{8(d-1)} e^{-2\Delta_+ u / \ell}  -\frac{\mathcal{R}|\f_+|^{2/\Delta_+} \, \ell^2}{4d(d-1)} e^{-2u/\ell} +\cdots \,,
\end{align}
\esq
On the other hand, as shown in \cite{C},  the minus branch solution  near a minimum of the potential exist only if  $R^{UV}=0$, and since $T=0$ in that case,   this  implies that $S=W'$ and we have
\bsq
\begin{align}\label{CminsolWm}
W_-&=-\frac{2(d-1)}{\ell}-\frac{\D_-}{2\ell}(\f-\f_1)^2+\cdots\,,\\
S_-&=-\frac{\D_-}{\ell}|\f-\f_1|+\cdots\,,\label{CminsolSm}
\end{align}
\esq
where $\D_-=d/2-\sqrt{d^2/4 + m^2\ell^2}<0$.
We also obtain
\bsq
\begin{align}
\label{Cphiminm} \f(u) &=\f_1- \f_- \ell^{\Delta_-}e^{-\Delta_- u / \ell}  + \cdots \, , \\
\label{CAminm} A(u) &= {A}_- +\frac{u}{\ell} - \frac{\f_-^2 \, \ell^{2 \Delta_-}}{8(d-1)} e^{-2\Delta_- u / \ell} +\cdots \,.
\end{align}
\esq
\subsection{$\f$-Bounces, IR bounces and A-bounces}\label{BIT}
For a generic point $x=\f_0-\f$ consider an expansion for scalar potential as follow
\be
V(x)=V_0+V_1 x+V_2 x^2+\mathcal{O}(x^3)\,.
\ee

$\bullet$ $\f$-Bounce point: At this point, the expansion of the $W$, $S$ and $T$  when $x\rightarrow 0^+$ are given by
\bsq
\begin{align}\label{AboW}
W &=W_1-\frac{4(d-1)V_0+dW_1^2}{d(d-1)S_0}x^\frac12 -\frac{(d-6)W_1 (4(d-1)V_0+dW_1^2)}{6d(d-1)^2 S_0^2}x+\mathcal{O}(x^\frac32)\,,\\
S&=S_0\sqrt{x}-\frac{dW_1}{3(d-1)} x-\frac{36 (d-1) V_0+36 (d-1)^2 V_2+d (d+9) W_1^2}{36 (d-1)^2 S_0} x^{\frac32}+\mathcal{O}(x^2)\,, \label{AboS}\\
T&=\frac{4(d-1)V_0+dW_1^2}{4(d-1)}\Big(1-\frac{2W_1}{(d-1)S_0}x^\frac12+\frac{12 (d-1) V_0-(d-9) d W_1^2}{3 d (d-1)^2 S_0^2} x+\mathcal{O}(x^\frac32)\Big)\,.
\label{AboT}
\end{align}
\esq
At the $\f$-bounce point, the flow does not stop, $\ddot{\f}=S_0^2\neq0$.
The expansions at each point are parametrized by $W_1$ and the value of $S_0$ is fixed by potential, $S_0=\pm\sqrt{2V_1}$. As we see, there are two branches at this point, depending on the sign of $S_0$. In this case, the flow reaches the $\f$-bounce point $\f_0$ from one branch and then it bounces and returns from the other branch. The value of $W_1$ is not arbitrary and since $V_0<0$ it should be chosen carefully because we need to have a negative value for $T$. In fact \eqref{AboT} suggests that $W_1^2<\frac{-4(d-1)V_0}{d}$.

$\bullet$ IR-bounce: At this point, $W$ and $S$  simultaneously tend to zero, but the flow does not stop here, $\ddot{\f}\neq 0$. As $x\rightarrow 0^+$ the flow goes to the fixed point from the left. The expansions of $W$, $S$  and $T$ around this point are given by
\bsq
\begin{align}\label{AIREW}
W &=\frac{4V_0}{dS_1}\sqrt{x}+\mathcal{O}(x)\,,\\
S&=S_1\sqrt{x}+\mathcal{O}(x)\,,\label{AIRES}\\
T&=V_0+\mathcal{O}(\sqrt{x})\,.
\label{AIRET}
\end{align}
\esq
Two branches are depending on the choice of $S_1=\pm\sqrt{2V_1}$. The solution with a minus sign is the mirror image of the solution with a plus sign.
The solutions of the scalar field and metric near the IR fixed point $u=u_0$ are given by
\bsq
\begin{align}
\f(u)\approx\f_0-\frac{S_1^2}{4} (u-u_0)^2\,,\\
ds^2\approx du^2+e^{2A_0}\Big(1+\frac{(u-u_0)^2}{\ell_{IR}^2}ds^2_{AdS_{d,\a}}\Big)\,,\label{IRendmetr}
\end{align}
\esq
where $A_0$ is the integration constant and $\ell_{IR}^2=d(d-1)/|V_0|$ and $ds^2_{AdS_{d,\a}}$ is a AdS$_d$ metric with length scale $\a$.
Since $\dot{A}(u_0)=\dot{\f}(u_0) = 0$  we can continue the geometry for $u > u_0$ by gluing its mirror image around $u_0$. This geometry is regular since the metric and scalar field, as well as their derivatives, are all continuous across $u_0$. There is an exception when $V_1=0$ or when $\f_0$ is located on the minimum of the potential.

$\bullet$ A-bounce: At this point, $S(x)\neq0$ so the flow does not stop but since $W(x)=0$ the scale factor reaches a minimum (or a maximum). As $x\rightarrow 0^+$ the scale factor decreases and then after the A-bounce it increases.
The expansions of $W$ and $S$ are given by
\bsq
\begin{align}\label{AturnW}
W &=\Big(\frac{(d-1) S_0}{d} + \frac{2V_0 }{dS_0}\Big)x+\Big(\frac{d+1}{2d S_0} - \frac{V_0 }{dS_0^3}\Big)V_1 x^2+\mathcal{O}(x^3)\,,\\
S&=S_0+\frac{V_1}{S_0}x+\mathcal{O}(x^2)\,,
\label{AturnS}\\
T&=V_0-\frac{S_0^2}{2}-\frac{(S_0^2 - 2 V_0) ((d-1) S_0^2 + 2 V_0)}{4 (d-1) d S_0^2}x^2+\mathcal{O}(x^3)\,.
\label{AturnT}
\end{align}
\esq
All the above extremal points may occur when the flow approaches from the right. In this situation, we can use the above expansions for $x=\f-\f_0$.

\subsection{No $\f$-bounce at an extremum point of a potential}
Previously we have assumed that the bouncing point does not coincide with an extremum of the potential. In this section, we study the possibility of a bouncing flow when the bouncing point is an extremum of the potential. For convenience we write here again the two first-order equations that we solve:
\bsq
\begin{align}
\label{AEOM7n} \frac{d}{2(d-1)} W^2 + (d-1) S^2 -d S W' + 2V &=0 \, , \\
\label{AEOM8n} SS' - \frac{d}{2(d-1)} SW - V' &= 0 \, .
\end{align}
\esq
Near an extremum $\f=\f_1$, the potential can be expanded as:
\begin{equation}
V(\f)=V_0+\frac{V_2}{2}\left(\f-\f_1 \right)^2+\cdots\,.
\end{equation}
First, we analyze if there can be a $\f$-bounce on an extremum. For a $\f$-bounce, we demand that
\begin{equation}
\dot{\f}=0\,, \quad \ddot{\f}\neq 0\,.
\end{equation}
This can be translated to that near $\f=\f_1$
\begin{equation}
S= 0\,, \quad \frac{1}{2}\frac{d S^2}{d\f}\neq 0\,.
\end{equation}
Let us assume that near $\f=\f_1$, to the leading order:
\begin{align}
\frac{1}{2}\frac{d S^2}{d\f}=\frac{S_0^2}{2}\,,
\end{align}
where $S_0$ is a constant. Integrating this equation with the condition that $S(\f_1)=0$ we find
\begin{equation}
S=\pm S_0 \sqrt{\f-\f_1}\,.
\end{equation}
Since $SS'$ is finite, and $V'\rightarrow 0$, Eq. \eqref{AEOM8n} demands that, near $\f=\f_1$
\begin{equation}
W\sim \frac{1}{\sqrt{\f-\f_1}}\,.
\end{equation}
Therefore, near $\f=\f_1$, we can use the following expansions:
\begin{align}
& S(\f)=\sqrt{\f-\f_1}\left[S_0+S_1\sqrt{\f-\f_1}+S_2 \left(\f-\f_1 \right)+\cdots  \right]\,, \\
& W(\f)=\frac{1}{\sqrt{\f-\f_1}}\left[W_0+W_1\sqrt{\f-\f_1}+W_2 \left(\f-\f_1 \right)+\cdots  \right]\,.
\end{align}
Substituting this into Eqs. \eqref{AEOM7n}-\eqref{AEOM8n} we find:
\begin{align}
&S_0=0\,,\qquad  W_0=0\,,\qquad  W_1=\pm 2 \sqrt{\frac{(1-d)V_0}{d}}\,, \nn \\
&S_1=-\frac{1}{2} \left[-\sqrt{\frac{d V_0}{1-d}}\pm \sqrt{\frac{-dV_0+4(d-1)V_2}{d-1}} \right]\,.
\end{align}
Since $S_0=0$, we can conclude that there can not be any $\f$-bounce on an extremum of the potential. Writing
\begin{equation}
V_0=-\frac{d(d-1)}{\ell_1^2}\,,
\end{equation}
we can write
\begin{equation}
W_1=\pm \frac{2(d-1)}{\ell_1}\,, \qquad S_1=\frac{1}{2\ell_1}\left( d\pm \sqrt{d^2+4 V_2 \ell_1^2} \right)=\frac{\Delta_\pm}{\ell_1}\,,
\end{equation}
where
\begin{equation}
\Delta_\pm=\frac{1}{2}\left( d\pm \sqrt{d^2+4 V_2 \ell_1^2} \right).
\end{equation}
Using the definition
\begin{equation}
W=-2(d-1)\dot{A}\,,
\end{equation}
we find that to the leading order:
\begin{equation}
A=C\mp \frac{u}{\ell_1}\,,
\end{equation}
where $C$ is an integration constant that we can set to zero. Using
\begin{equation}
\frac{d}{2(d-1)} W^2 -S^2 -2 T +2V =0\,,
\end{equation}
we can observe that near $\f=\f_1$
\begin{equation}
T\approx 0\,.
\end{equation}
Since
\begin{equation}
T=R e^{-2 A(u)}\approx R e^{\pm \frac{u}{\ell_1}}\,.
\end{equation}
For $T\approx 0$ and $W_1=-\frac{2(d-1)}{\ell_1}$, it requires
\begin{equation}
\text{either}\ R=0\,, \qquad \text{or} \ u\rightarrow +\infty\,.
\end{equation}
For the latter case, since
\begin{equation}
e^{A(u)}\rightarrow \infty,
\end{equation}
this is a UV.
We are now going to analyze the case for which $R=0$. In this case, we have the following equation:
\begin{equation}
\frac{d}{2(d-1)} W^2 -W'^2  +2V =0\,. \label{Wflat}
\end{equation}
First, we analyze if there can be any $\f$-bounce on an extremum of the potential. We can pass from the curved case to the flat case by setting $S=W'$.  Therefore we take the following expansion:
\begin{equation}
W'(\f)=\sqrt{\f-\f_1}\left[\frac{3}{2}W_1+2W_2 \sqrt{\f-\f_1}+\frac{5}{2} W_3 \left(\f-\f_1 \right)+\cdots  \right]\,.
\end{equation}
Integrating this once we find:
\begin{equation}
W(\f)=W_0+W_1 \left(\f-\f_1 \right)^{3/2}+W_2  \left(\f-\f_1 \right)^2+\cdots\,,
\end{equation}
where $W_0$ is an integration constant. Inserting this into Eq. \eqref{Wflat}, and solving order by order we find:
\begin{equation}
W_0=\pm  2\sqrt{\frac{V_0(1-d)}{d}}\,, \quad W_1=0\,, \quad W_2=\frac{1}{8} \left[\frac{d W_0}{d-1}\pm \sqrt{16 V_2+\frac{d^2 W_0^2}{(d-1)^2}} \right]\,.
\end{equation}
Since $W_1=0$, we conclude that there can not be any bouncing solution on an extremum of the potential.
We take
\begin{equation}
V_0=-\frac{d(d-1)}{\ell_1^2}\,, \quad W_0=-\frac{2(d-1)}{\ell_1}\,.
\end{equation}
Then $W_2$ becomes:
\begin{equation}
W_2=-\frac{1}{4\ell_1} \left(d\pm \sqrt{d^2+4\ell_1^2 V_2} \right)=-\frac{\Delta_\pm}{2\ell_1}\,.
\end{equation}
Integrating
\begin{equation}
W=-2(d-1)\dot{A}\,, \qquad \dot{\f}=W'\,,
\end{equation}
we find that to the leading order:
\begin{align}
& A(u)=\frac{u}{\ell_1}+\cdots\,, \label{Aflat} \\
& \f (u)=\f_1+\f_{\pm}e^{-\frac{\Delta_\pm u}{\ell_1}}\,. \label{phiflat}
\end{align}
For a maximum, $\Delta_\pm >0$, Eq. \eqref{phiflat} implies that $u\rightarrow +\infty$. In that case, since $e^{A}\rightarrow \infty$, this corresponds to a UV. On the other hand, for a minimum $\Delta_+>0$ and $\Delta_-<0$. Therefore for $(+)$ solution we require $u\rightarrow +\infty$, which again corresponds to a UV. For $(-)$ solution, $\Delta_-<0$, which requires $u\rightarrow -\infty$ which corresponds to an IR. We can summarize all these by the following. \\
 \textit{There can not be any $\f$-bounce exactly on an extremum of a potential. A maximum always corresponds to a UV. A minimum is an IR only for a $(-)$ type flat flow. It can act as the UV for both curved and flat flow for the $(+)$ type of solution.  }

\section{Bulk equations and boundary transforms in AdS-slicing} \label{app:transforms}
In this appendix, we provide some technical details on the calculation of correlators in the presence of AdS slices, discussed in section
\ref{sec:single}.

We use a probe scalar field to study the 2-point functions on an $AdS_d$ boundary metric. Since the coordinate system (\ref{sing1}) studied in details in appendix \ref{app:ads-slicing} is not singular except at $u \rightarrow \pm \infty$, boundary conditions at both infinities are required to solve the equation of motion (\ref{sing17}).
The equation of motion in question is reproduced here
\be
\Box_{d+1} \f = m^2\f\,.
\label{corr1}
\ee
The laplacian of $AdS_{d+1}$ is written, using the coordinate system \ref{sing1}, as
\be
\ell^2\Box_{d+1} = \partial_u^2 + d \text{tanh}u\partial_u + \cosh{u}^{-2}\Box_d\,,
\label{corr2}
\ee
where $\Box_d$ is the laplacian of $AdS_d$.
The scalar $\f$ is decomposed into a basis of functions $Y^{(\vec{k}\a)}$ defined as eigenvectors of $\Box_d$. In the following, we use Poincar\'e coordinates (\ref{sing2}) for the slice $AdS_d$ to build these eigenfunctions. Their eigenvalue is parametrized as
\be
\Box_d Y^{(\vec{k},\n)}(z,\vec{x}) = \left[\n^2 - \left(\frac{d-1}{ 2}\right)^2\right]Y^{(\vec{k},\n)}(z,\vec{x})\,.
\label{corr3}
\ee
The Fourier decomposition in the AdS$_d$ basis reads,
\be
\f(u,z,\vec{x}) = \int_{\mathbb{R}^{d-1}} \frac{d\vec{k}}{ (2\pi)^{d-1}} \int_\mathbb{R} d\alpha Y^{(\vec{k},i\a)}(Z,\vec{y})\tilde{\f}(u, i\alpha, \vec{k})\,,
\label{corr4}
\ee
where $\n = i\a$ and $\tilde{\f}$ are the $AdS_d$-Fourier mode of the field $\f$.
The equation of motion (\ref{corr1}) for $\tilde{\f}$ is separable because the laplacian is replaced by the eigenvalue (\ref{corr3}). Solutions of (\ref{corr1}) are then obtained using the separation of variables
\be
\tilde{\f}(u,\n,\vec{k}) = F_\n(u)\phi(\n,\vec{k})\,.
\label{corr5}
\ee
The radial part of the equation of motion (\ref{corr1}) is then given by
\be
F_\n''(u) + d\tanh u F_\n'(u) - (\cosh{u})^{-2}\left[(\ell m)^2 - \n^2 + \left(\frac{d-1}{ 2}\right)^2\right]F_\n(u) = 0\,.
\label{corr6}
\ee
This is solved using associated Legendre functions. The most general solution is given by a linear combination of Legendre functions of the first and second kind denoted respectively by $P^m_n$ and $Q^m_n$.
\be
F_\n(u) = \lambda(\cosh{u})^{-d/2}\left[Q_{\n-1/2}^\gamma(\tanh{u}) + b P_{\n-1/2}^\gamma(\tanh{u}) \right]\,,
\label{corr7}
\ee
where $b$ and $\lambda$ are integration constants and $\g$ is defined by
\be
\g \equiv (\ell m)^2 + \frac{d^2 }{ 4}\,.
\label{corr8}
\ee
To implement Dirichlet and Neumann boundary conditions at $u=0$, the general solution (\ref{corr7}) needs to be written explicitly in terms of the integration constants at $u=0$. This is obtained using formulae \href{https://personal.math.ubc.ca/~cbm/aands/page_334.htm}{8.6} of \cite{abramowitz}. In particular, the Dirichlet condition $F_\n(0) = 0$ fixes the constant $b$ to the value
\be
b^\text{Dirichlet} = \frac{\pi}{ 2}\tan\left(\frac{\pi}{ 2}\left(\n-{\frac12} + \g\right)\right)\,,
\label{corr9}
\ee
whereas Neumann condition $F'_\a(0) = 0$ fixes $b$ to the value
\be
b^\text{Neumann} = - \frac{\pi}{ 2}\cot\left(\frac{\pi}{ 2}\left(\n-{\frac12} + \g\right)\right)\,.
\label{corr10}
\ee
The constant $\l$ is fixed by using the asymptotics of Legendre functions at the boundary $u \rightarrow - \infty$,
\be
F_\n(u) \underset{u\rightarrow -\infty}{\rightarrow}  \l e^{u \Delta_-}  2^{\frac{\Delta_+}{ 2}}\frac{\Gamma(\g)}{ \pi}\left(b\cos(\pi\n)-\frac{\pi}{ 2}\sin(\pi\n)\right)\,,
\label{corr11}
\ee
along with the unit-source boundary condition
\be
F_\n(u) \underset{u\rightarrow -\infty}{\rightarrow} = e^{ u \Delta_- }\,.
\label{corr12}
\ee
Using the values for $b^\text{Dirichlet}$ and $b^\text{Neumann}$ given in (\ref{corr10}) and (\ref{corr11}) and the values of $\l$ for each solution using (\ref{corr12}), one can obtain the particular solution (\ref{sing16}) which is a sum of Dirichlet and Neumann solutions. The result is given by
\be
b^\text{conformal} = - \frac{\pi}{ 2}\cot{\pi\g}\,.
\label{corr13}
\ee
This solution leads to the conformal correlator (\ref{sing10}).
An important remark is that a solution satisfying (\ref{corr13}) has a source on the $u\rightarrow -\infty$ boundary, but has zero source on the opposite boundary $u\rightarrow +\infty$. Indeed, the asymptotic behaviour of the general solution (\ref{corr7}) at $u\rightarrow +\infty$ is given by
\be
F_\n(u) \underset{u\rightarrow +\infty}{\rightarrow} =  \l e^{- u \Delta_-}2^{\frac{\Delta_+}{ 2}}\frac{\Gamma(\g)}{ \pi}\left(b\sin(\pi\g)+ \frac{\pi}{ 2}\cos(\pi\g)\right).
\label{corr14}
\ee
This leading term cancels in the conformal case (\ref{corr13}). Therefore, the conformal solution corresponds to a solution with zero-source on the opposite boundary.

Next, we rewrite the conformal condition (\ref{corr13}) as a condition on a brane located at $u=0$ (\ref{sing19}) at which point the scale factor has a minimum.
Taking the action for a brane at $u=0$ in (\ref{sing12}), writing it in momentum space using (\ref{corr4}), the boundary condition  for $\tilde{\f}(u=0,\n,\vec{k})$ (the position of the "IR" brane) is given by
\be
F'_\nu(u=0) = \mu(\nu) F_{\nu}(u=0)\,.
\label{corr15}
\ee
The expression of $\m(\n)$ is  obtained by taking the solution which satisfies (\ref{corr13}), and evaluating it (as well as its first derivative with respect to $u$) at $u=0$ using the formulae \href{https://personal.math.ubc.ca/~cbm/aands/page_334.htm}{8.6} of \cite{abramowitz}.
The result is given by
\be
\m(\n) = 2 \tan\left(\frac{\pi}{ 2}(\n - \frac12)\right)
\G
\left[
  \begin{tabular}{c c}
 ${\frac34} + \frac{\n}{ 2} + \frac{\g}{ 2}$, &  ${\frac34} + \frac{\n}{ 2} - \frac{\g}{2}$ \\
  ${\frac14} + \frac{\n}{ 2} + \frac{\g}{ 2}$, &  ${\frac14} + \frac{\n}{2} - \frac{\g}{ 2}$
 \end{tabular}
 \right]\,.
 \label{corr16}
\ee
This can be is simplified to the form of eq. (\ref{sing22}) using Euler's reflection formula.

\addcontentsline{toc}{section}{References}
\bibliographystyle{JHEP}

\end{document}